\documentclass[superscriptaddress,prx-quantum,nofootinbib,notitlepage]{revtex4-2}
\usepackage{graphicx}
\usepackage{amssymb,amsmath,amsthm,amsfonts}
\usepackage{thmtools}
\usepackage{mathtools}
\mathtoolsset{centercolon}
\usepackage{thm-restate} 
\usepackage{amsmath}
\usepackage{amssymb}
\usepackage{comment}
\usepackage{placeins}
\usepackage{multirow}
\usepackage{qcircuit}
\usepackage{dcolumn}
\usepackage[caption=false]{subfig}
\usepackage[colorlinks]{hyperref}
\hypersetup{
pdfstartview={FitH},
pdfnewwindow=true,
colorlinks=true,
linkcolor=blue,
citecolor=blue,
filecolor=blue,
urlcolor=blue}
\usepackage{tikz}
\usetikzlibrary{calc,backgrounds}

\usepackage{siunitx}
\usepackage{booktabs}
\usepackage{pgffor}
\usepackage{url}
\usepackage{hypcap}
\usepackage{dsfont}
\usepackage{algorithm}
\usepackage{algorithmic}
\usepackage{soul}
\usepackage{diagbox}
\usepackage{bbold}
\usepackage[capitalise]{cleveref}
\usepackage[shortlabels]{enumitem}

\usepackage[version=4]{mhchem} % chemical equations
\usepackage{siunitx} % Einheiten
\DeclareSIUnit{\atom}{atom}
\DeclareSIUnit{\lno}{f.u.}

\newcommand{\modmin}{{\ominus}}
\newcommand{\Kp}{\mathbf k'}
\newcommand{\K}{\mathbf k}
\newcommand{\Q}{\mathbf Q}

\newcommand{\R}{\mathbf r}
\newcommand{\G}{\mathbf G}

\newcommand{\vv}[1]{\ensuremath{\mathbf{#1}}}
\newcommand{\bz}{\mathcal B \! Z}

\newcommand{\ceil}[1]{\left\lceil{#1}\right\rceil}
\newcommand{\nrm}{\mathcal{N}}
\renewcommand{\L}{P}
\newcommand{\hpq}{h_{pq}}

\newcommand{\Rea}{{\rm Re}}
\newcommand{\Ima}{{\rm Im}}

\newcommand{\eq}[1]{Eq.~\hyperref[eq:#1]{(\ref*{eq:#1})}}
\renewcommand{\sec}[1]{\hyperref[sec:#1]{Section~\ref*{sec:#1}}}
\newcommand{\app}[1]{\hyperref[app:#1]{Appendix~\ref*{app:#1}}}
\newcommand{\tab}[1]{\hyperref[tab:#1]{Table~\ref*{tab:#1}}}
\newcommand{\fig}[1]{\hyperref[fig:#1]{Figure~\ref*{fig:#1}}}
\newcommand{\figa}[2]{\hyperref[fig:#1]{Figure~\ref*{fig:#1}#2}}
\newcommand{\figx}[2]{\hyperref[fig:#1]{Figure~\ref*{fig:#1}(#2)}}
\newcommand{\thm}[1]{\hyperref[thm:#1]{Theorem~\ref*{thm:#1}}}
\newcommand{\lem}[1]{\hyperref[lem:#1]{Lemma~\ref*{lem:#1}}}
\newcommand{\cor}[1]{\hyperref[cor:#1]{Corollary~\ref*{cor:#1}}}
\newcommand{\defn}[1]{\hyperref[def:#1]{Definition~\ref*{def:#1}}}
\newcommand{\alg}[1]{\hyperref[alg:#1]{Algorithm~\ref*{alg:#1}}}

\newcommand{\zetabits}{\aleph}
\newcommand{\rotbits}{\beth}
\newcommand{\rankone}{L}
\newcommand{\ranktwo}{\Xi}

\newcommand{\rotprec}{b_r}
\newcommand{\factor}{\eta} % The number of bits that can be left out when preparing equal superposition states.
\newcommand{\sel}{\textsc{select}}
\newcommand{\prep}{\textsc{prepare}}

\newcommand{\tr}{\operatorname{Tr}}

\def\ket#1{\mathinner{|{#1}\rangle}}

\newcommand{\wid}{m}
\newcommand{\dm}{d}
\newcommand{\chunk}{k}

\newcommand{\be}{\begin{equation}}
\newcommand{\ee}{\end{equation}}
\newcommand{\ba}{\begin{eqnarray}}
\newcommand{\ea}{\end{eqnarray}}

\newcommand{\nn}{\nonumber \\}

\usepackage{color}
\usepackage{xcolor}

\usepackage{listings}
\newcolumntype{L}{>{\centering\arraybackslash}m{3cm}}

\allowdisplaybreaks

\makeatletter
\renewcommand*\env@matrix[1][\arraystretch]{%
\edef\arraystretch{#1}%
\hskip -\arraycolsep
\let\@ifnextchar\new@ifnextchar
\array{*\c@MaxMatrixCols c}}
\makeatother

\newcommand{\MQ}{\affiliation{%
School of Mathematical and Physical Sciences,
Macquarie University, Sydney, NSW 2109, Australia}}

\newcommand{\BASFSE}{\affiliation{%
BASF SE, Carl-Bosch-Strasse 38, 67063 Ludwigshafen, Germany
}}

\newcommand{\QSimulate}{\affiliation{%
Quantum Simulation Technologies, Inc., Boston, 02135, United States}}

\newcommand{\Google}{\affiliation{
Google Research, Venice, CA 90291, United States}}

\newcommand{\Columbia}{\affiliation{Department of Chemistry, Columbia University, New York, NY, USA}}

\newcommand{\FSU}{\affiliation{Department of Chemistry and Biochemistry, Florida State University, Tallahassee, FL, USA}}

\begin{document}

\title{Fault-tolerant quantum simulation of materials using Bloch orbitals}

\date{\today}

\author{Nicholas C.~Rubin}
\email[Corresponding author: ]{nickrubin@google.com}
\Google

\author{Dominic W.~Berry}
\email[Corresponding author: ]{dominic.berry@mq.edu.au}
\MQ

\author{Fionn D.~Malone}
\Google

\author{Alec F.~White}
\QSimulate

\author{Tanuj Khattar}
\Google

\author{A. Eugene DePrince, III}
\Google
\FSU

\author{Sabrina Sicolo}
\BASFSE

\author{Michael K{\"u}hn}
\BASFSE

\author{Michael Kaicher}
\BASFSE

\author{Joonho Lee}
\Google
\Columbia

\author{Ryan Babbush}
\email[Corresponding author: ]{babbush@google.com}
\Google

\begin{abstract}
The simulation of chemistry is among the most promising applications of quantum computing. However, most prior work exploring algorithms for block-encoding, time-evolving, and sampling in the eigenbasis of electronic structure Hamiltonians has either focused on modeling finite-sized systems, or has required a large number of plane wave basis functions. In this work, we extend methods for quantum simulation with Bloch orbitals constructed from symmetry-adapted atom-centered orbitals so that one can model periodic \textit{ab initio} Hamiltonians using only a modest number of basis functions. We focus on adapting existing algorithms based on combining qubitization with tensor factorizations of the Coulomb operator. Significant modifications of those algorithms are required to obtain an asymptotic speedup leveraging translational (or, more broadly, Abelian) symmetries. We implement block encodings using known tensor factorizations and a new Bloch orbital form of tensor hypercontraction. Finally, we estimate the resources required to deploy our algorithms to classically challenging model materials relevant to the chemistry of Lithium Nickel Oxide battery cathodes within the surface code.
\end{abstract}

\maketitle

\tableofcontents

\section{Introduction}
Recently, first quantization quantum algorithms and constant factor resource estimation analysis for molecular systems~\cite{kassal2008polynomial, BabbushSpectra} have been adapted to materials~\cite{PRXQuantum.2.040332}. 
While the first quantization approach using a plane wave representation is attractive due to the smooth convergence to the continuum limit~\cite{PhysRevX.8.011044,  gruneis2013explicitly} a local basis representation such as atom-centered basis sets has other advantages. Similar to the molecular simulation setting, local basis functions can be advantageous when describing spatially localized phenomena such as heterogeneous catalysis or efficiently describing cusps~\cite{kato1957eigenfunctions}. 

The desire for systematically improvable electronic structure methods to treat the many examples of strongly correlated phenomena~\cite{RevModPhys.70.1039, RevModPhys.66.763, RevModPhys.75.913} in the condensed phase has recently driven the application of \textit{ab initio} wavefunction theories to the periodic setting~\cite{Pisani2003,Pisani2005,Gruneis2011,DelBen2012,DelBen2013,Booth2013,Booth2016,Mcclain2017,neufeld2022ccsdmetals, Cui2020, Zhu2020, cui2022systematic,  PhysRevB.100.045127}.
Standard treatments of symmetry in wavefunction theories~\cite{cotton1991chemical, crawford2019tensor} can be used to exploit the translational symmetry of periodic systems, thus enabling the application of post-Hartree-Fock methods to material systems. Despite these advantages, classical \textit{ab initio} treatment of such problems is limited due to the large simulation cells needed to converge to the thermodynamic limit.  This drawback has further driven the use of embedding theories~\cite{RevModPhys.68.13, PhysRevB.44.8454, inglesfield1981method, PhysRevLett.109.186404, cui2019efficient, pham2019periodic} and downfolding~\cite{zheng2018real}.  Naturally, one may ask if fault-tolerant quantum computers can alleviate the computational burden associated with {\it ab initio} simulation of solids within the local basis framework.

In this paper, we describe how to extend molecular quantum simulation algorithms of second quantization Hamiltonians represented in local basis sets to periodic systems using the qubitization framework~\cite{Lee2020, low2019hamiltonian}. Though the general structure of the algorithms is largely unchanged, introducing symmetry--\textit{i.e.}\ symmetry-adapting the block encodings--requires non-trivial modifications to realize an improvement in the asymptotic complexity. The first steps in this direction were taken in Ref.~\cite{ivanov2022quantum} using the ``sparse'' Hamiltonian representation. We provide an alternative derivation for block encodings using this representation and introduce symmetry-adapted block encodings for three other more performant tensor factorizations of the Hamiltonian: single factorization (SF), double factorization (DF) and tensor hypercontraction (THC). The result is orders of magnitude improvement in the quantum resources required to simulate materials.

For each of the four Hamiltonian representations we describe the origin of the asymptotic speedup (or lack thereof in one case), provide compiled algorithms for constant factor resource estimates, and compare the performance to non-symmetry-adapted block encodings.  We note that the derived symmetry-adapted block encodings apply to any Abelian point group symmetry with minor modifications. For SF, sparse, and DF the symmetry-adapted block encodings provide an asymptotic speedup for walk operator construction proportional to the square root of the number of $k$-points used to sample the Brillouin zone. For THC, there is no asymptotic improvement due to the linear cost of unary iteration in the block encoding. Going beyond asymptotic analysis and compiling to total Toffolis, we find that for DF and THC using symmetry-adapted block encodings provides no asymptotic speedup over their non-symmetry-adapted counterparts due to the increased number of applications of the walk operator for fixed precision phase estimation.  DF and THC are sensitive to the numerical compression of the Hamiltonian, and thus we expect the number of walk operator applications can be decreased. Furthermore, there are classical advantages to using the symmetry-adapted block encodings coming from the reduced classical complexity of representing the Hamiltonian as the system size is increased towards the thermodynamic limit.

In parallel with recent studies estimating quantum resources required to simulate high-value molecular targets~\cite{Reiher2017,Lee2020,vonBurg2020, Goints_pnas.2203533119}, we estimate the quantum resources required to simulate an open materials science problem related to the cathode structure of Lithium Nickel Oxide (LNO) batteries. The LNO systems are universally observed in the high symmetry $\mathrm{R\bar3m}$ structure which is at odds with the predicted Jahn-Teller activity of low-spin trivalent Ni \cite{Bianchini}; more background can be found in Section~\ref{sec_sub:context_LNO}.  This discrepancy combined with the difficulty of synthesizing pure LNO, the size of the unit cells~\cite{acs.chemmater.0c03442}, and potential strong correlation at the high symmetry structure~\cite{PhysRevB.84.085108} makes the LNO problem an interesting application target for quantum simulation advantage. This realistic problem frames the algorithmic improvements articulated in this paper and the prospects of the quantum advantage given modern electronic structure methods.  We find that the required resource estimates for simulating a set of benchmark systems and the LNO problem before reaching the thermodynamic limit are already substantial. In fact, the large simulation cells required to converge these calculations to the thermodynamic limit is ultimately a significant hurdle for \textit{ab initio} simulations.    

The layout of the rest of the paper is as follows: Section~\ref{sec:electronic_structure} describes the atom-centered basis sets and the Hamiltonian that we use, Section~\ref{sec:qubitization_section} describes the qubitization algorithm and the origin of the asymptotic speedup in constructing walk operators using each of the four Hamiltonian representations. Each subsection is dedicated to a particular Hamiltonian factorization and describes the qubitization algorithm and how to calculate associated parameters. Section~\ref{sec:scaling_comparison} compares all methods and extrapolates quantum resources required to simulate a diamond crystal converged towards the thermodynamic limit, and Section~\ref{sec:LNOcompare} reports the accuracy and correlation analysis of various electronic structure methods for LNO while providing estimates of quantum computing resources and runtimes.  We close with prospects for this class of methods.

\section{Electronic structure Hamiltonian of materials in Bloch orbitals}\label{sec:electronic_structure}
Though plane-wave basis sets are used in most periodic Density Functional Theory (DFT) calculations, there is a long history of local-basis methods as well. The use of a localized basis set has a number of advantages over plane waves: \textit{1}) 0D (molecular), 1D, 2D, and 3D systems can be treated on an equal computational footing, \textit{2}) Calculations on low-density systems with large unit cells can be more efficient\cite{Vandevondele2005,Kuhne2020,Dovesi2020} \textit{3}) Hartree-Fock exchange can be more efficiently computed in the smaller, local-orbital basis\cite{Guidon2009,Guidon2010,Dovesi2020,Kuhne2020} and \textit{4}) The local-orbital representations can lower the computational cost of correlation corrections with a more compact representation of the virtual space.
(\textit{1}) - (\textit{3}) have spurred the development of local-orbital DFT and Hartree-Fock methods with Gaussian orbitals \cite{Pisani1988,Dovesi2020,Kuhne2020,Guidon2009} and numerical atomic orbitals \cite{Blum2009}, while (\textit{4}) has been behind recent work to apply correlated electronic structure theory to periodic solids \cite{Pisani2003,Pisani2005,DelBen2012,DelBen2013,Booth2013,Booth2016,Mcclain2017}. In the following subsection we describe the symmetry-adapted periodic sum of Gaussian-type orbitals used in this work.

\subsection{Basis functions and matrix elements}\label{sec:electronic_structureA}
A local basis function, $\tilde{\chi}_p$, can be adapted to the translational symmetry of a lattice to form a periodized function
\begin{align}
\chi_{p,\K}(\R) = \sum_{\vv{T}}
e^{i\K\cdot \vv{T}}\tilde{\chi}_{p}(\R - \vv{T}),
\end{align}
where $\vv{T}$ represents a lattice translation vector and $\K$ is a crystal momentum vector lying in the first Brillouin zone. The lattice momentum $\K$ labels an irreducible representation of the group of translations defined by the translational symmetry of the material.  Functions of this form are easily verified to be Bloch functions in that
\begin{equation}
\chi_{p,\K}(\R) = e^{i\K \cdot \R}u_{p,\K}(\R)
\end{equation}
where $u_{p,\K}(\R)$ has the same periodicity as the lattice.

Orbitals are constructed from a linear combination of the underlying Bloch orbitals,
\begin{align}
\phi_{i\K}(\R) = N_k^{-1/2}\sum_{p}c_{p,i}(\K)\chi_{p,\K}(\R),
\end{align}
where $N_k$ is the total number of $k$ points. The expansion coefficients, $c_{p,i}(\K)$ are determined from the appropriate periodic self-consistent field procedure, usually Hartree-Fock or Kohn-Sham DFT.  The resulting orbitals are normally constrained to be orthogonal by convention and can serve as a basis for representing the second-quantized Hamiltonian.

The matrix elements of a one-electron operator,
\begin{align}
T_{p\vv{k}_{p},q\vv{k}_{q}} =  \int dr \, \phi^{*}_{p \vv{k}_{p}}(r) \mathcal{O}_{1} \phi_{q\vv{k}_{q}}(\R)
\end{align}
are non-zero only when $\vv{k}_{p} = \vv{k}_{q}$ as long as $\mathcal{O}_1$ has the translational symmetry of the lattice. We can use a similar strategy to derive the structure of the two-electron integrals which are given by
\begin{align}\label{eq:block_tei}
V_{p\K_{p},q\K_{q},r\K_{r},s\K_{s}}
&= \int \int dr_{1}\, dr_{2}\, \phi_{p\K_{p}}^{*}(\R_{1}) \phi_{q\K_{q}}(\R_{1})\mathcal{O}_{2} \phi_{r\K_{r}}^{*}(\R_{2})\phi_{s\K_{s}}(\R_{2}).
\end{align}
The translational symmetry of the Bloch orbitals implies the 2-electron operator $\mathcal{O}_2$ matrix elements can only be nonzero when $\left(\K_{p} + \K_{r} - \K_{q} - \K_{s}\right) = \mathbf{G}$ where $\mathbf{G}$ is a reciprocal lattice vector. We note that this expression for nonzero matrix elements by symmetry is a specific instance of the more general expression.  More generally, given a group $\mathfrak{g}$ with its irreducible representations labeled by $\{\Gamma_{i}\}$, the two-electron integral is nonzero by symmetry whenever
$\Gamma_{p} \otimes \Gamma_{q} \otimes \Gamma_{r} \otimes \Gamma_{s}$
contains the complete symmetric representation~\cite{cotton1991chemical}.  For periodic systems $\mathfrak{g}$ is the set of translational symmetries.  
Despite this sparsity, the evaluation of the nonzero matrix elements for all basis functions is often a major computational bottleneck whenever local basis sets are used.

Local orbitals provide a more compact representation than plane waves, so fewer basis functions are needed. Unfortunately, there are $O(N_k^3N^4)$ generally nonzero two-electron matrix elements for $N_k$ $k$-points and $N$ basis functions in the primitive cell. For very large calculations locality can be exploited to yield asymptotically linear-scaling DFT methods \cite{Goedecker1999,Bowler2012}. Linear scaling Hartree-Fock is also possible for insulators \cite{Wu2009}. This linear regime is almost never reached in practice, and it is usually advantageous to instead reduce the cost by tensor factorization.

Though our discussion has been thus far general with regard to the choice of local basis functions, Gaussian basis functions are by far the most popular choice in molecular calculations, and crystalline Gaussian orbitals are also a popular choice for periodic calculations. This popularity is due to the existence of analytic formulas which allow for fast, numerically exact evaluation of the matrix elements of most common operators. Despite the existence of efficient numerical techniques, the large number of two-electron integrals that must be evaluated in periodic calculations requires a more efficient procedure. Traditionally, this is accomplished with the Gaussian plane wave (GPW) method \cite{lippert1997hybrid,Vandevondele2005} which only requires storage of $O(N_{k}^2N^2n_{\mathrm{pw}})$ integrals where $n_{\mathrm{pw}}$ is the number of plane waves used to evaluate the integrals. In molecular calculations, the most common decomposition is called the resolution of the identity (RI) or sometimes density fitting (DF) \cite{Whitten1973,Mintmire1982,Dunlap2000a,Weigend2002}. This procedure requires the storage of $O(N_{k}^2N^2n_{\mathrm{aux}})$ integrals where $n_{\mathrm{aux}}$ is the size of the auxiliary basis set. Both the GPW and the RI method can be considered as density fitting approaches where the former uses a plane-wave fitting basis and the latter uses a Gaussian fitting basis. For this reason, the RI approach is often called ``Gaussian density fitting" (GDF) in the context of periodic calculations \cite{Varga2005,Maschio2008,Burow2009,Wang2020,Ye2021}.

The two-electron integral tensor can be further factorized into a product of five two-index tensors as was done in the tensor hypercontraction (THC) method of Martinez and coworkers \cite{Hohenstein2012,Parrish2012,Hohenstein2012a}. Factorizations of this form are most useful for correlated methods where they have the potential to lower the computational scaling. In this work we present a translational symmetry-adapted form of the tensor hypercontraction for the two-electron integral tensors of periodic systems.  

\subsection{The second-quantized Hamiltonian}\label{sec:electronic_structureB}
We can express the second-quantized electronic structure Hamiltonian as
\begin{align}
H &= H_{1} + H_{2} \, ,\\
H_{1} &= \sum_{\sigma} \sum_{\K }\sum_{pq} h_{p\K,q\K} a_{p\K\sigma}^\dagger a_{q\K\sigma} \, , \\
h_{p\K,q\K} &= T_{p\K,q\K} - \frac{1}{2}\sum_{r, \Q }V_{p\K, r\Q, r\Q, q\K} \, , \label{eq:one_body_coeff} \\
H_{2} &= \frac{1}{2}  \sum_{\sigma, \tau}  \sum_{\Q,\K, \Kp}\sum_{pqrs}V_{p\K, q(\K\modmin\Q), r(\Kp\modmin\Q), s\Kp} a_{p\K\sigma}^\dagger a_{q(\K\modmin\Q)\sigma} a_{r(\Kp\modmin\Q)\tau}^\dagger a_{s\Kp\tau} \, .
\label{eq:V}
\end{align} 
We first introduce summation limits for each symbol as we will commonly use short hand summation formulas to indicate multiple sums.  For each variable $\{p,q,r,s\}$ summation is performed over the range $[0, N/2-1]$ indexing the spatial orbital or band, $\{\Q, \K, \Kp\}$ summation is performed over the Brillouin zone ($\bz$) at a set number of $k$-points of which there are $N_{k}$, and $\{\sigma, \tau\}$ are electron spin variables and summed over $\{\uparrow, \downarrow\}$. Non-modular differences of $\K$, $\Q$, and $\Kp$ span twice the Brillouin zone.
Because $V$ needs to be indexed by values in the Brillouin zone, we use modular subtraction indicated by $\modmin$.
That is, if the number of points in each dimension is $N_x,N_y,N_z$, we perform subtraction modulo $N_x,N_y,N_z$ in each direction, respectively.

The Hamiltonian is generally complex Hermitian with four-fold symmetry of the two-electron integrals.\footnote{We note that the following generic complex Coulomb integral symmetries are present
\begin{equation}
V_{p\K_{p},q\K_{q},r\K_{r},s\K_{s}} = V_{r\K_{r},s\K_{s},p\K_{p},q\K_{q}} = V_{q\K_{q},p\K_{p},s\K_{s},r\K_{r}}^{*} = V_{s\K_{s},r\K_{r},q\K_{q},p\K_{p}}^{*}
\end{equation}
from integration index relabeling and complex conjugation.} In the following sections we demonstrate how the sparse structure of the two-electron integral tensor affects the scaling of block encoding the Hamiltonian for implementation of qubitized quantum walk oracles.  The cost of qubitization is greatly affected by the representational freedom of the the underlying Hamiltonian expressed as a linear combination of unitaries.  We demonstrate how to construct the sparse, single-factorization (SF), double-factorization (DF), and tensor-hypercontraction (THC) integral decompositions of Bloch orbital Hamiltonians and cost out simulations for a variety of materials.

For all algorithms, we will make a comparison to the case of a $\Gamma$-point calculation on a supercell composed of $N_{k}$ primitive cells in the geometry described by the $k$-point sampling. This allows us to directly observe the proposed speedup due to symmetry-adapting. To demonstrate the scaling of symmetry-adapted block encoding, we estimate quantum simulation resource requirements for the series of systems listed in Table~\ref{tab:benchmark_systems}.  Range-separated density fitting~\cite{Ye2021} is used to construct integrals with Dunning type correlation-consistent basis sets~\cite{ye2022correlation} and the Goedecker-Teter-Hutter (GTH) family of pseudopotentials for Hartree-Fock~\cite{PhysRevB.58.3641}.  For each Hamiltonian, cutoffs for the factorization are selected so that the M{\o}ller-Plesset second order perturbation theory (MP2) error in the total energy is below one milliHartree per cell or formula unit depending on the system.  While prior works used coupled-cluster theory, MP2 is used here for computational efficiency.
\begin{table}[H]
\centering
\begin{minipage}{0.85\textwidth}
\begin{ruledtabular}
\begin{tabular}{llcccc}
System & Structure & Atoms in Cell & Lattice Parameters & spin-orbitals cc-pVDZ & spin-orbitals cc-pVTZ \\ \hline
C & diamond & 2 & 3.567 \cite{heyd2005energy} & 52 & 116 \\
Si & diamond & 2 & 5.43 \cite{heyd2005energy} & 52 & 116\\
BN & zinc blende & 2 & 3.616 \cite{heyd2005energy} & 52 & 116\\
LiCl & rocksalt & 2 & 5.106 \cite{gruneis2010second} & 52 & 98\\
AlN & wurzite & 4 & (a) 3.11 (c) 4.981 \cite{heyd2005energy} & 104 & 220 \\
Li & bcc& 2 & 3.51 \cite{nadler1959crystallographic} & 52 & 80 \\
Al & fcc& 2 & 4.0479 \cite{tang2009surface} & 52 & 104 \\
%H & bcc & 2 & 2.0 & 20 & 56\\
\end{tabular}
\end{ruledtabular}
\caption{Crystal structures lattice parameters used for the systems studied in this work. The lattice parameters were chosen to be at or near their experimental equilibrium values.\label{tab:benchmark_systems}}
\end{minipage}
\end{table}

\section{Qubitization of materials Hamiltonians}\label{sec:qubitization_section}
Similar to fault tolerant resource estimates for molecular systems represented in second quantization~\cite{vonBurg2020, Lee2020, BabbushSpectraB, Berry2019B, Goints_pnas.2203533119}, we compare the number of logical qubits and number of Toffoli gates required to implement phase estimation on unitaries that use block encoding~\cite{gilyen2019quantum} and qubitization~\cite{low2019hamiltonian} to encode the Hamiltonian spectrum in a Szegedy walk operator~\cite{Szegedy2004} for various linear combination of unitaries (LCU)~\cite{child_wiebe_LCU_2012} representations of the Hamiltonian.  All LCUs represent the Hamiltonian as
\begin{align}\label{eq:def_LCU}
H = \sum_{\ell=1}^{L}\omega_{\ell}U_{\ell}
\end{align}
where $\omega_{\ell} \in \mathbb{R}$, $\omega_{\ell} \geq 0$, and $U_{\ell}$ is a unitary operator.
One can then construct the operators
\begin{align}
&\mathrm{PREPARE}\vert 0\rangle^{\otimes \log(L)} \mapsto \sum_{\ell=1}^{L}\sqrt{\frac{\omega_{l}}{\lambda}}\vert \ell \rangle  \equiv \vert \mathcal{L}\rangle \label{eq:prepare_def} \\
&\mathrm{SELECT}\vert \ell \rangle \vert \psi \rangle \mapsto \vert \ell \rangle U_{\ell}\vert \psi \rangle \label{eq:select_def} \\
&\lambda = \sum_{\ell=1}^{L} \omega_{\ell} \label{eq:def_lambda}
\end{align}
where $\vert \psi \rangle$ is the system register, and $\vert \ell\rangle$ is an ancilla register used to index each term in the LCU.  The walk operator constructed from {\sel} and a reflection operator built from {\prep}, $R = 2 \vert \mathcal{L}\rangle\langle \mathcal{L}\vert \otimes \mathbb{1}  - \mathbb{1}$, has eigenvalues proportional $e^{\pm i \arccos{E_{n}/\lambda}}$ where $E_{n}$ is an eigenvalue of the Hamiltonian in Eq.~\eqref{eq:def_LCU}.

It was shown in References~\cite{BabbushSpectraB} and~\cite{Lee2020} when ensuring that {\sel} is self-inverse, only the reflection operator $R$ needs to be controlled on the ancilla for phase estimation and not {\sel}. Therefore, the Toffoli cost of phase estimating the walk operator scales as
\begin{align}
\ceil{\frac{\pi \lambda}{2 \epsilon_{\mathrm{PEA}}}} \left(C_{S} + C_{P} + C_{P^{\dagger}} + \log(L)\right)
\end{align}
where $C_{S}$ is the cost for implementing the {\sel} oracle and $C_{P}$ is the cost for implementing the {\prep} oracle, $C_{P^{\dagger}}$ is the cost for the inverse {\prep} oracle, and $\epsilon_{\mathrm{PEA}}$ is the target precision for phase estimation. Thus the main costs for sampling from the eigenspectrum of a second quantized operator are the costs to implement {\sel}, {\prep}, and {\prep}$^{\dagger}$. 
These costs need to be multiplied by a factor proportional to $\lambda/\epsilon_{\rm PEA}$ for the number of walk steps needed for phase estimation.
Note that when computing intensive quantities, such as the the energy per cell, the $\lambda$ factor is scaled by $1/N_k$.

The particular choice of LCU changes all of these costs. Prior works have investigated the resource requirements for simulating molecules with four different LCUs.  While all these methods can be used without modification in supercell calculations at the $\Gamma$-point, the construction of molecular {\sel} and {\prep} do not exploit any symmetries and are not applicable away from the $\Gamma$-point -- \textit{e.g.}\ at the Baldereschi point~\cite{PhysRevB.7.5212}. 

The leading costs in constructing {\sel} and {\prep} for second quantized Hamiltonians is the circuit primitive that functions similar to a read-only-memory (ROM) called QROM. The QROM primitive is a gadget that takes a memory address, potentially in superposition, and outputs data, also potentially in superposition. There are currently two variations of QROM that have different costs; traditional QROM that has linear Toffoli complexity when outputting $L$ items with any amount of data associated with each item, and advanced QROM (called QROAM) with reduced non-Clifford complexity~\cite{low2018trading}.
It uses a select-swap circuit construction with Toffoli cost
\begin{align}
\ceil{\frac{L}{k}} + m (k - 1)
\end{align}
where $k$ is a power of $2$ for outputting $L$ items of data where each item of data is $m$ bits long.
The notation $k$ here for an integer should not be confused with $\K$ for the crystal momentum vector.
It needs $m (k - 1)$ ancillas, so increases the logical ancilla count in exchange for reduced Toffoli complexity.
When $L > m$ this function is minimized by selecting $k \approx \sqrt{L/m}$ and thus the Toffoli and ancilla cost generically go as $\mathcal{O}(\sqrt{Lm})$. 
It is also possible to adjust $k$ to reduce the ancilla count while increasing the Toffoli count.
Having QROAM output the minimal amount of information to represent the Hamiltonian is at the core of the $\sqrt{N_{k}}$ improvements we derive in many of the block encodings.  We will also demonstrate that for all LCUs the lowest scaling can be linear in the Bloch orbital basis size $\mathcal{O}(N_{k}N)$ due to the requirement to perform unary iteration at least once over the entire basis.

Another primitive that becomes the dominant cost in constructing symmetry-adapted {\sel} is the multiplexed-controlled swap between two registers.  The controlled swap between two registers uses unary iteration~\cite{BabbushSpectraB} on $L$ items to swap $M$ elements between two registers at the cost of $\mathcal{O}(LM)$ Toffolis.  For simulating materials this primitive is commonly encountered when swapping all band indices with a particular irreducible representation label, or $k$-point, into a working register at a cost of $\mathcal{O}(N_{k}N)$.  The necessity of coherently moving data thus puts a limit on the total savings one can achieve by leveraging Abelian symmetries. The cost of moving data must be weighed against the benefits, which we describe in each section below.
In Table~\ref{tab:qubitization_factorized_cost_table} we summarize the space complexity, in terms of logical qubits, and time complexity, in terms of Toffolis of the four LCUs when considering translational symmetry on the primitive cell and without (denoted as SC for supercell).

\begin{table}[tbh]
    \centering
    \begin{minipage}{0.98\textwidth}
        \caption{Generically, qubitized quantum walks scale as ${\mathcal{O}}(\sqrt{\Gamma})$ in space and ${\mathcal{O}}(\lambda \sqrt{\Gamma}/\epsilon)$ in time where $\Gamma$ is the amount of information required to specify the Hamiltonian within a particular representation.
        For double factorization, $\Xi$ is the sum of the average rank of the second factorization which is expected to scale as $\mathcal{O}(N_{k}N)$, which is the number of orbitals in the primitive cell or bands. $\tilde{\Xi}$ is the average rank of the second factorization in the supercell calculation and is also expected to scale as $\mathcal{O}(N_{k}N)$.
        The tilde on $\mathcal{O}$ is used to account for logarithmic factors, and can include variables not explicitly given in the scaling. $\lambda$ for each LCU is different and is denoted as a subscript indicating the LCU type and if it $\lambda$ for the supercell version.}
    \label{tab:qubitization_factorized_cost_table}
    \begin{ruledtabular}
    \begin{tabular}{ccccc}
    Representation  & Qubits &  Toffoli Complexity  & SC Qubits & SC Toffoli\\
    \hline
    sparse & $\widetilde{\mathcal{O}}(N_{k}^{3/2}N^{2})$   &  $\widetilde{\mathcal{O}}(N_{k}^{3/2}N^{2}\lambda_{\mathrm{sparse}} / \epsilon)$ & $\widetilde{\mathcal{O}}(N_{k}^{2}N^{2})$ & $\widetilde{\mathcal{O}}(N_{k}^{2}N^{2}\lambda_{\mathrm{sparse}, \mathrm{SC}} / \epsilon)$ \\
     SF  & $\widetilde{\mathcal{O}}(N_{k}N^{3/2})$    & $\widetilde{\mathcal{O}}(N_{k}N^{3/2}\lambda_{\mathrm{SF}} / \epsilon)$ &  $\widetilde{\mathcal{O}}(N_{k}^{3/2}N^{3/2})$ & $\widetilde{\mathcal{O}}(N_{k}^{3/2}N^{3/2} \lambda_{\mathrm{SF},\mathrm{SC}}/\epsilon)$ \\
     DF & $\widetilde{\mathcal{O}}(\sqrt{N_{k}}N\sqrt{\Xi})$ & $\widetilde{\mathcal{O}}(\sqrt{N_{k}}N\sqrt{\Xi} \lambda_{\mathrm{DF}} / \epsilon)$ &$\widetilde{\mathcal{O}}(N_{k}N\sqrt{\tilde{\Xi}})$  & $\widetilde{\mathcal{O}}(N_{k}N\sqrt{\tilde{\Xi}} \lambda_{\mathrm{DF},\mathrm{SC}}/\epsilon)$ \\
     THC & $\widetilde{\mathcal{O}}(N_{k}N)$          &  $\widetilde{\mathcal{O}}(N_{k}N \lambda_{\mathrm{THC}} / \epsilon)$ & $\widetilde{\mathcal{O}}(N_{k}N)$          &  $\widetilde{\mathcal{O}}(N_{k}N \lambda_{\mathrm{THC},\mathrm{SC}} / \epsilon)$ \\
    \end{tabular}
    \end{ruledtabular}
    \end{minipage}
\end{table}
For the sparse LCU, exploiting primitive cell translational symmetry reduces the amount of symmetry unique information in the Hamiltonian by a factor of $N_{k}$, which translates to a reduction of $\sqrt{N_{k}}$ savings in Toffoli complexity and ancilla complexity.  For sparse {\prep}, the square root savings originates from the QROAM cost of outputting ``alt'' and ``keep'' values for the coherent alias sampling component of the state preparation.  For sparse {\sel} controlled application of all Pauli terms has linear cost in the basis size $\mathcal{O}(N_{k}N)$ and is not the dominant cost. The supercell calculation does not exploit the $k$-point symmetry unique non-zero coefficients of the Hamiltonian and thus has worse scaling.

The single factorization LCU leverages the fact that the Coulomb integral tensor is positive semidefinite and can be written in a quadratic form. For molecular systems without symmetry--\textit{i.e.}\ $C1$ symmetry, the factorization results in a three-tensor where there are two orbital indices and one auxiliary index that scales as the number of orbitals in the system~\cite{werner2003fast}.  For simulations where orbitals now have point group symmetry labels, such as $k$-points, each three-tensor factor can now be arranged into a five-tensor; two symmetry labels (irreducible representation labels), two-band labels (orbital labels), and one auxiliary index which still scales with the number of bands due to density fitting of the cell periodic part of the density~\cite{PhysRevB.78.073102, Ye2021}.  Thus the origin of the $\sqrt{N_{k}}$ improvement for the symmetry-adapted block encoding with a single-factorization LCU lies in the fact that the auxiliary index has $N_{k}$ lower scaling in comparison to a supercell variation where the Cholesky factorization or density fitting is performed on the entire supercell two-electron integral tensor. The single factorization algorithm is also dominated by the QROM cost of {\prep}--of which there are two state preparations.  The inner state preparation~\cite{Lee2020} for the $k$-point symmetry-adapted algorithm requires outputting $\mathcal{O}(N_{k}^{2}N^{3})$ to be used in the state preparation leading to $\mathcal{O}(N_{k}N^{3/2})$ Toffoli and qubit complexity. Contrasting this to the supercell calculation, we see a $\sqrt{N_{k}}$ savings due to the fact that the inner state preparation requires only $N_{k}^{2}$ information and not $N_{k}^{3}$ information.  We elaborate on this point further in Section~\ref{sec_sub:sf_derivation}.  {\sel} is implemented in a similar fashion to sparse, scaling as $\mathcal{O}(N_{k}N)$, and is not a dominant cost.

The double factorization LCU represents the Hamiltonian in a series of non-orthogonal bases and leverages that a linear combination of ladder operators can be constructed by a similarity transform of a single fermionic ladder operator, or Majorana operator, by a unitary generated by a quadratic fermionic Hamiltonian.  In the molecular case, the dominant cost for these algorithms is the QROM to output the rotations for the similarity transform and implementing the basis rotations with the programmable gate array circuit primitive~\cite{vonBurg2020, Lee2020} for {\sel}.  When taking advantage of primitive cell symmetry we reduce the amount of data needed to be output by QROM by $N_{k}$, which results in a $\sqrt{N_{k}}$ savings in the Toffoli complexity.  Because we are using advanced QROM, this output size advantage is also observed in the logical qubit requirements.  As mentioned previously, computing the total number of Toffolis requires scaling the walk operator cost by a linear function of $\lambda$.  We find that using a canonical orbtial basis set, the total Toffoli cost is higher than the commensurate supercell total Toffoli cost because $\lambda$ for the symmetry-adapted case increases. The origin of the increase is related to a reduced variational freedom when selecting non-orthogonal bases and is further discussed in Section~\ref{sec:df_lcu}.

Finally, in the THC LCU there is no asymptotic speedup because the molecular algorithm had the lowest possible scaling for second quantized algorithms.  This stems from the fact that even iterating over the basis once with unary iteration to apply an operator indexed by basis element has a Toffoli cost of $\mathcal{O}(N_{k}N)$.  As we will discuss in Section~\ref{sec_sub:thc_qubitization} our symmetry-adapted algorithm offers other benefits such as enabling the classical precomputation of the THC factors by exploiting symmetry and lowering the number of controlled rotations.

We now describe the Hamiltonian factorization used in each LCU, the calculation of $\lambda$ associated with each Hamiltonian factorization, and outline the construction of the qubitization oracles. Detailed compilations are provided for each LCU in the Appendices. In each section we provide numerical evidence that the symmetry-adapted oracles have the reported scaling by plotting the Toffoli requirements to synthesize {\sel} $+$ {\prep} $+$ {\prep}$^{-1}$ compared against the number of $k$-points sampled ($N_{k}$).
\subsection{The sparse Hamiltonian representation}
In the ``sparse'' method the Hamiltonian described in Eq.~\eqref{eq:one_body_coeff} and Eq.~\eqref{eq:V} is directly translated to Pauli operators which form the LCU. Under the Jordan-Wigner transformation, we take
\begin{align}
a_{p\K\sigma} & \mapsto \vec Z (X_{p\K\sigma} + iY_{p\K\sigma})/2,  \\
a^\dagger_{p\K\sigma} & \mapsto \vec Z (X_{p\K\sigma} - iY_{p\K\sigma})/2,
\end{align}
where the notation $\vec Z$ is being used to indicate that there is a string of $Z$ operators on qubits up to (not including) that on which $X_{p\K\sigma}$ or $Y_{p\K\sigma}$ acts upon.
This requires a choice of ordering for the qubits indexed by $p$, $\K$, and $\sigma$.
We need only apply the string of $Z$ operators for the same value of $\sigma$, because we always have matching annihilation and creation operators for the same spin $\sigma$ (so any $Z$ gates on the other spin would cancel).
We also adopt a convention that the ordering of qubits for the Jordan-Wigner transformation takes $\K$ as the more significant bits, with qubits for all $p$ with a given $\K$ grouped together.
For most of the discussion we will not need to explicitly consider this ordering.

With the Jordan-Wigner transform the one-body component of the Hamiltonian takes on the form
\begin{align}\label{eq:simpleham}
H_{1} &= \frac{i}{4} \sum_{\sigma\in\{\uparrow,\downarrow\}}\sum_{\K}\sum_{p,q=1} \Rea(h_{p\K, q\K}) \left\{\vec Z X_{p\K\sigma} \vec Z Y_{q\K\sigma} - \vec Z Y_{p\K\sigma} \vec Z X_{q\K\sigma} \right\} \nn &\quad
+ \frac{i}{4} \sum_{\sigma\in\{\uparrow,\downarrow\}}\sum_{\K}\sum_{p,q=1}{\rm Im}(h_{p\K,q\K}) \left\{\vec Z X_{p\K\sigma} \vec Z X_{q\K\sigma}  + \vec Z Y_{p\K\sigma}\vec Z Y_{q\K\sigma}\right\}  +\sum_{\K}\sum_{p=1} h_{p\K,p\K} \openone .
\end{align}
We provide the full derivation for this expression in Appendix~\ref{app:onebodytermderv}. To derive the two-body operator LCU we use only complex conjugation symmetry in contrast to the molecular derivation that used eight-fold symmetry. The two-body Hamiltonian can be written as
\begin{align}
H_{2} = \frac{1}{4}  \sum_{\sigma, \tau \in \{\uparrow, \downarrow\}}  \sum_{\Q,\K, \Kp}^{N_{k}}\sum_{p,q,r,s=1}^{N/2} &\left[ V_{p\K,q(\K\modmin\Q),r(\Kp\modmin\Q),s\Kp} a_{p\K\sigma}^\dagger a_{q(\K\modmin\Q)\sigma} a_{r(\Kp\modmin\Q)\tau}^\dagger a_{s\Kp\tau} \right. \nn
& \left. + V^*_{p\K,q(\K\modmin\Q),r(\Kp\modmin\Q),s\Kp}a_{q(\K\modmin\Q)\sigma}^{\dagger}a_{p\K\sigma}a_{s\Kp\tau}^{\dagger}a_{r(\Kp\modmin\Q)\tau} \right],\label{eq:H2genform}
\end{align}
where $\modmin$ indicates modular subtraction as defined above.
In the case where $\Q\ne 0$ or $p\ne q$ and $r\ne s$, we can move the creation and annihilation operators using the fermionic anticommutation relations to give the term on the second line as
\begin{equation}\label{eq:conjterm}
V^*_{p\K,q(\K\modmin\Q),r(\Kp\modmin\Q),s\Kp}a_{p\K\sigma}a_{q(\K\modmin\Q)\sigma}^{\dagger}a_{r(\Kp\modmin\Q)\tau}a_{s\Kp\tau}^{\dagger} \, .
\end{equation}
The Jordan-Wigner representation then gives the expression in square brackets in \eq{H2genform} as 
\begin{align}\label{eq:sparse_two_body_pauli}
&\frac 1{16} \left\{V_{p\K,q(\K\modmin\Q),r(\Kp\modmin\Q),s\Kp} [\vec Z (X_{p\K\sigma}-iY_{p\K\sigma})] [\vec Z(X_{q(\K\modmin\Q)\sigma}+iY_{q(\K\modmin\Q)\sigma})] [\vec Z (X_{r(\Kp\modmin\Q)\tau}-iY_{r(\Kp\modmin\Q)\tau})] [\vec Z (X_{s\Kp\tau}+iY_{s\Kp\tau})] \right. \nn
& \left. + V^*_{p\K,q(\K\modmin\Q),r(\Kp\modmin\Q),s\Kp}[\vec Z (X_{p\K\sigma}+iY_{p\K\sigma})] [\vec Z(X_{q(\K\modmin\Q)\sigma}-iY_{q(\K\modmin\Q)\sigma})] [\vec Z (X_{r(\Kp\modmin\Q)\tau}+iY_{r(\Kp\modmin\Q)\tau})] [\vec Z (X_{s\Kp\tau}-iY_{s\Kp\tau})] \right\}.
\end{align}
Then we can separate Eq.~\eqref{eq:sparse_two_body_pauli} into real and imaginary components as
\begin{align}\label{eq:twobody}
&\frac 1{8} \left\{ {\rm Re}(V_{p\K,q(\K\modmin\Q),r(\Kp\modmin\Q),s\Kp}) [\vec Z X_{p\K\sigma}\vec Z X_{q(\K\modmin\Q)\sigma}+\vec Z Y_{p,\K, \sigma}\vec Z Y_{q,\K\modmin\Q, \sigma})] [\vec Z X_{r(\Kp\modmin\Q)\tau}\vec Z X_{s\Kp\tau}+\vec Z Y_{r(\Kp\modmin\Q)\tau}\vec Z Y_{s\Kp\tau}] \right. \nn
&-{\rm Re}(V_{p\K,q(\K\modmin\Q),r(\Kp\modmin\Q),s\Kp}) [\vec Z Y_{p,\K, \sigma}\vec Z X_{q(\K\modmin\Q)\sigma}-\vec Z X_{p\K\sigma}\vec Z Y_{q(\K\modmin\Q)\sigma})] [\vec Z Y_{r(\Kp\modmin\Q)\tau}\vec Z X_{s\Kp\tau}-\vec Z X_{r(\Kp\modmin\Q)\tau}\vec Z Y_{s\Kp\tau}] \nn
&+{\rm Im}(V_{p\K,q(\K\modmin\Q),r(\Kp\modmin\Q),s\Kp}) [\vec Z Y_{p\K\sigma}\vec Z X_{q(\K\modmin\Q)\sigma}-\vec Z X_{p\K\sigma}\vec Z Y_{q(\K\modmin\Q)\sigma})] [\vec Z X_{r(\Kp\modmin\Q)\tau}\vec Z X_{s\Kp\tau}+\vec Z Y_{r(\Kp\modmin\Q)\tau}\vec Z Y_{s\Kp\tau}] \nn
& \left.+{\rm Im}(V_{p\K,q(\K\modmin\Q),r(\Kp\modmin\Q),s\Kp}) [\vec Z X_{p\K\sigma}\vec Z X_{q(\K\modmin\Q)\sigma}+\vec Z Y_{p\K\sigma}\vec Z Y_{q(\K\modmin\Q)\sigma})] [\vec Z Y_{r(\Kp\modmin\Q)\tau}\vec Z X_{s\Kp\tau}-\vec Z X_{r(\Kp\modmin\Q)\tau}\vec Z Y_{s\Kp\tau}] \right\} .
\end{align}

In accounting for cases where $\Q=0$ with $p=q$ or $r=s$, the same expression is obtained, but there are also one-body terms obtained.
These result in a total one-body operator
\begin{align}
\tilde{H}_{1} = H_{1} + \sum_{\sigma \in \{\uparrow, \downarrow\}} \sum_{\K}^{N_{k}}\sum_{p,q=1}^{N/2} \left(\sum_{r=1}^{N/2}\sum_{\Kp}^{N_{k}} V_{p\K,q\K,r\Kp,r\Kp} \right)  a_{p\K\sigma}^{\dagger} a_{q\K\sigma}  \,  .
\label{eq:general_complex_1body}
\end{align}
A full derivation of this expression can be found in Appendix~\ref{app:sparse_one_body_derivation}.

Using the representation of the one-body and two-body operators as Pauli operators, we have a linear combination of unitaries form.  The $\lambda$ associated with this LCU is 
\begin{align}\label{eq:sparse_lambda}
\lambda &= \lambda_{\tilde{H}_{1}} + \lambda_{H_{2}} \\
\lambda_{\tilde{H}_{1}} &= \sum_{\K}\sum_{pq}\left\{\left|\mathrm{Re}[h_{p\K, q\K}] + \mathrm{Re}\left[\sum_{\Kp,r}V_{p\K,q\K,r\Kp,r\Kp}\right]\right| + \left|\mathrm{Im}[h_{p\K,q\K}] + \mathrm{Im}\left[\sum_{\Kp,r}V_{p\K,q\K,r\Kp,r\Kp}\right]\right|\right\} \\
\lambda_{H_{2}} &= \sum_{\K,\Kp,\Q}\sum_{pqrs}\left\{\left|{\rm Re}(V_{p\K,q(\K\modmin\Q),r(\Kp\modmin\Q),s\Kp})\right| + \left|{\rm Im}(V_{p\K,q(\K\modmin\Q),r(\Kp\modmin\Q),s\Kp})\right|\right\} .
\end{align}
In determining $\lambda_{\tilde{H}_{1}}$ there is a factor of $2$ due to the summation over spin $\sigma$ and then a factor of $2$ accounting for the fact that each expression in braces in \eq{simpleham} is the sum of two different Pauli strings.
As a result these factors have cancelled the original $1/4$ prefactor.
In the expression for $\lambda_{\tilde{H}_{1}}$ we have also summed over the native one-body terms and the contributions from the two-body terms.
For $\lambda_{H_{2}}$ we had a factor of $1/8$ in \eq{twobody}, which is multiplied by the factor of $1/4$ in \eq{H2genform}.
The two sums over spin $\sigma$ and $\tau$ give a factor of $4$.
Then for each of the real and imaginary parts in \eq{twobody} there were sums over 8 Pauli strings, giving a factor of 8.
As a result these factors have also cancelled in the expression for $\lambda_{H_{2}}$. Note that there is a factor of 2 between this expression and that in \cite{Lee2020}, even when we just consider $V$ that is real.
The reason is that in Ref.~\cite{Lee2020} there was eight-fold symmetry, where here we only have four-fold symmetry.
That is, here we have symmetry when simultaneously swapping the pairs $p,q$ and $r,s$, whereas in Ref.~\cite{Lee2020} there are two symmetries from swapping $p,q$ or $r,s$ on their own.
That meant it was possible to express the Hamiltonian as in Eq.~(A2) of that work, then in Eq.~(A3) of that work the Jordan-Wigner mapping was used in the form
\begin{equation}
a_{p\sigma}^\dagger a_{q\sigma} +a_{q\sigma}^\dagger a_{p\sigma} \mapsto \frac{X_{p\sigma}\vec Z X_{q\sigma} + Y_{p\sigma} \vec Z Y_{q\sigma}}2.
\end{equation}
In this mapping there has been a cancellation of half the Pauli strings, which results in $\lambda$ being reduced by a factor of 2.
Here we only have four-fold symmetry, so the value of $\lambda$ for the two-body term is a factor of 2 larger than that in \cite{Lee2020}.

In order to implement the Hamiltonian as a linear combination of unitaries, the first step is to perform a state preparation on $\Q,\K,\Kp,p,q,r,s$.
This state preparation corresponds to the sum, then we will perform controlled operations for each of the operators in \eq{twobody}.
The state preparation is applied using coherent alias sampling as described in \cite{Berry2019B}.
Because there are multiple variables that the state needs to be prepared over, it is convenient to use the QROM to output ``ind'' values as well as ``alt'' and ``keep'' values.
Both ``ind'' and ``alt'' give values of all variables $\Q,\K,\Kp,p,q,r,s$.
Then an inequality test is performed between keep and an equal superposition state, and the result is used to control a swap between ind and alt.

There are both real and imaginary values of $V_{p\K,q(\K\modmin\Q),r(\Kp\modmin\Q),s\Kp}$, so we also include a qubit to distinguish between these values in the state preparation.
We also do not prepare all values of $\{p\K,q(\K\modmin\Q),r(\Kp\modmin\Q),s\Kp\}$.
There is symmetry in swapping $p\K,q(\K\modmin\Q)$ with $r(\Kp\modmin\Q),s\Kp$, or simultaneously $p\K$ with $q(\K\modmin\Q)$ and $r(\Kp\modmin\Q)$ with $s\Kp$.
Only those values of $\{p\K,q(\K\modmin\Q),r(\Kp\modmin\Q),s\Kp\}$ that give unique values of $V$ will be prepared.
Then the full range can be obtained by using qubits to control these swaps.
There is also a complex conjugate needed in the symmetry, which can be applied with a Clifford gate.

The dominant complexity in the preparation comes from the QROM.
The number of items of data is $\mathcal{O}(N_k^3 N^4)$, and by using the advanced form of QROM the complexity can be made approximately the square root of the total amount of data (number of items of data times the size of each).
The size of each item of data is logarithmic in $N_k$ and $N$ as well as the allowable error.
Therefore, the scaling of the complexity can be given ignoring these logarithmic parts as $\widetilde{\mathcal{O}}(N_k^{3/2} N^2)$.

To describe the controlled operations needed in order to implement the operation as in \eq{twobody}, we need to account for the fact that there are two lines for each of the real and imaginary components of $V$.
In addition, for each line in \eq{twobody} there is a product of two factors, each of which is a sum of two terms.
To describe the linear combination of unitaries we therefore introduce three more qubits.
\begin{itemize}
    \item The first is used to distinguish between the two lines for each of the real and imaginary parts in \eq{twobody}.
    \item The second distinguishes between the two terms in the first set of square brackets.
    \item The third distinguishes between the two terms in the second set of square brackets.
\end{itemize}

When implementing the controlled operations, we perform four operations of the form of $\vec ZX$ or $\vec Z Y$, with $X$ or $Y$ being applied on target qubits indexed by $p\K$, $q(\K\modmin\Q)$ and so forth.
These Pauli strings are applied using the approach of \cite{Lee2020}, but in this case there is the additional complication that we need to select between $X$ or $Y$.
This selection can be performed simply by performing the controlled Pauli string twice, once for $X$ and once for $Y$.
The complexity is proportional to $N_kN$, which is trivial compared to the complexity of the state preparation.
The choice of whether $X$ or $Y$ is performed depends on the value of the three qubits selecting between the terms, as well as the qubit selecting between the real and imaginary parts.
The processing of these qubits to determine the appropriate choice of $X$ or $Y$ can be performed with a trivial number of gates.

The last part to consider is how the implementation of the one-body part of the Hamiltonian is integrated with the implementation of the two-body part.
In the state preparation, amplitudes corresponding to the real and imaginary parts of $h_{p\K,q\K}$ will be produced, as well as a qubit selecting between the one- and two-body parts.
That qubit will be used to also select between the choice of $X$ and $Y$.
For the one-body part there is a product of only two of the Pauli strings, so the other two will not be applied at all for the one-body part.
See Appendix~\ref{app:sparseimp} for a more detailed description of the implementation.

\begin{figure}[tbh]
    \centering
    \begin{picture}(400,185)
    \put(-45,0){\includegraphics[width=8.5cm]{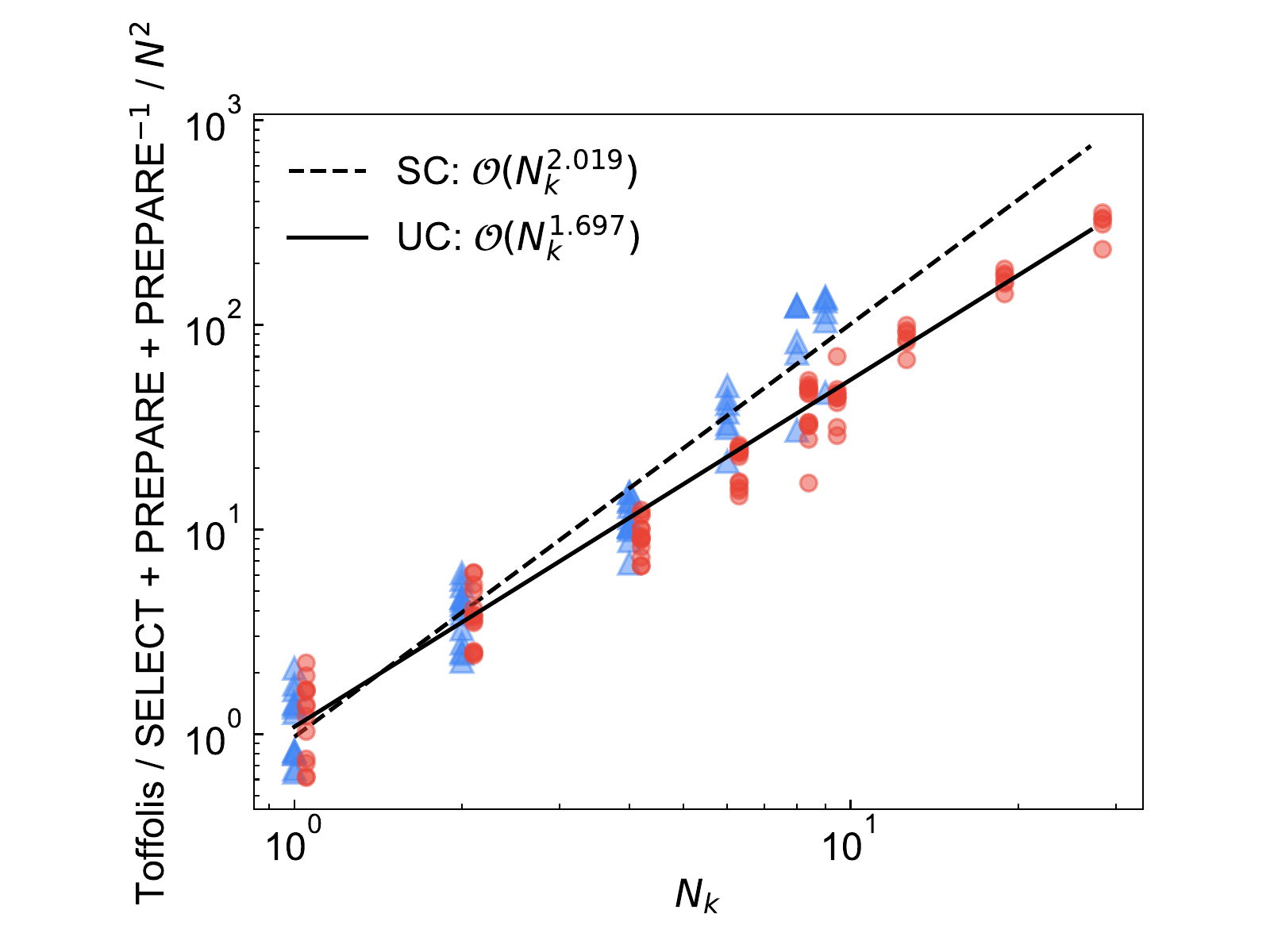}}
    \put(195,0){\includegraphics[width=8.5cm]{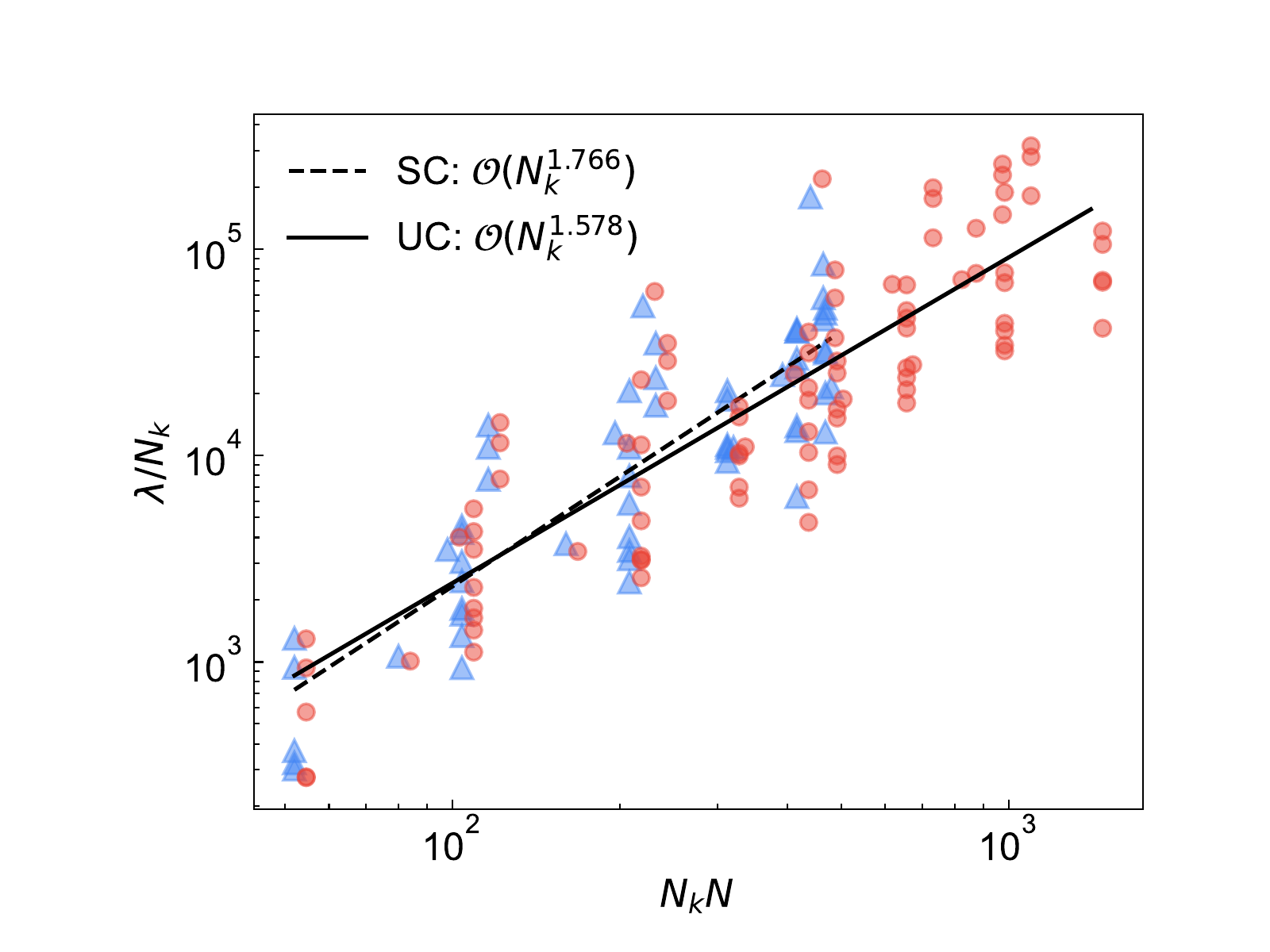}}
    \put(5,170){(a)}
    \put(245,170){(b)}
    \end{picture}
    \caption{(a) Sparse Toffoli step complexity versus the number of $k$-points for systems in Table~\ref{tab:benchmark_systems} using the cc-pVDZ and cc-pVTZ basis set $\Gamma$-centered Monkhorst-Pack grids of size [1, 1, 1] to [3, 3, 3]. Each point is a single system described at a particular basis set and $k$-mesh where the threshold for zeroing each nonzero two-electron integral coefficient is determined by MP2 as described earlier. The scaling for implementing the block encoding is shown in the legend. To isolate the $N_{k}$ scaling behavior we divide the Toffoli step complexity by the square of the number of basis functions $(N^{2})$. For supercell we expect a scaling going as $\mathcal{O}(N_{k}^{2})$ and for symmetry-adapted block encodings we expect a scaling going as $\mathcal{O}(N_{k}^{1.5})$. The ideal symmetry-adapted scaling is not reached due to finite size effects which are further discussed in Figure~\ref{fig:carbon_block_dependence}.  (b) Total $\lambda$ for the symmetry-adapted version (denoted UC) and the supercell calculation without explicit primitive cell symmetry (denoted SC) demonstrating no deterioration of $\lambda$ by symmetry-adapting. Similar to (a) all points are a particular system from the benchmark set in a fixed basis and $k$-mesh. }
    \label{fig:sparse_toff_step_complexity}
\end{figure}

In Figure~\ref{fig:sparse_toff_step_complexity} we plot the Toffoli complexity to implement {\sel} $+$ {\prep} $+$ {\prep}$^{-1}$ for simulating the aforementioned sample systems using a symmetry-adapted {\sel} and {\prep} at different Monkhorst-Pack grids and different number of bands (cc-pVDZ and cc-pVTZ).  We compare the symmetry-adapted calculations to supercell calculations using the same {\sel} and {\prep}.  The supercell calculations do not explicitly take into account the symmetry of the primitive cell in the full simulation cell.  To demonstrate the symmetry-adapted Hartree-Fock orbitals do not appreciably change the overall scaling with respect to a supercell calculation we plot the total $\lambda$ for the supercell calculation (which reruns Hartree-Fock on the supercell) and the symmetry-adpated version.

\begin{figure}[tbh]
    \centering
    \includegraphics[width=8.5cm]{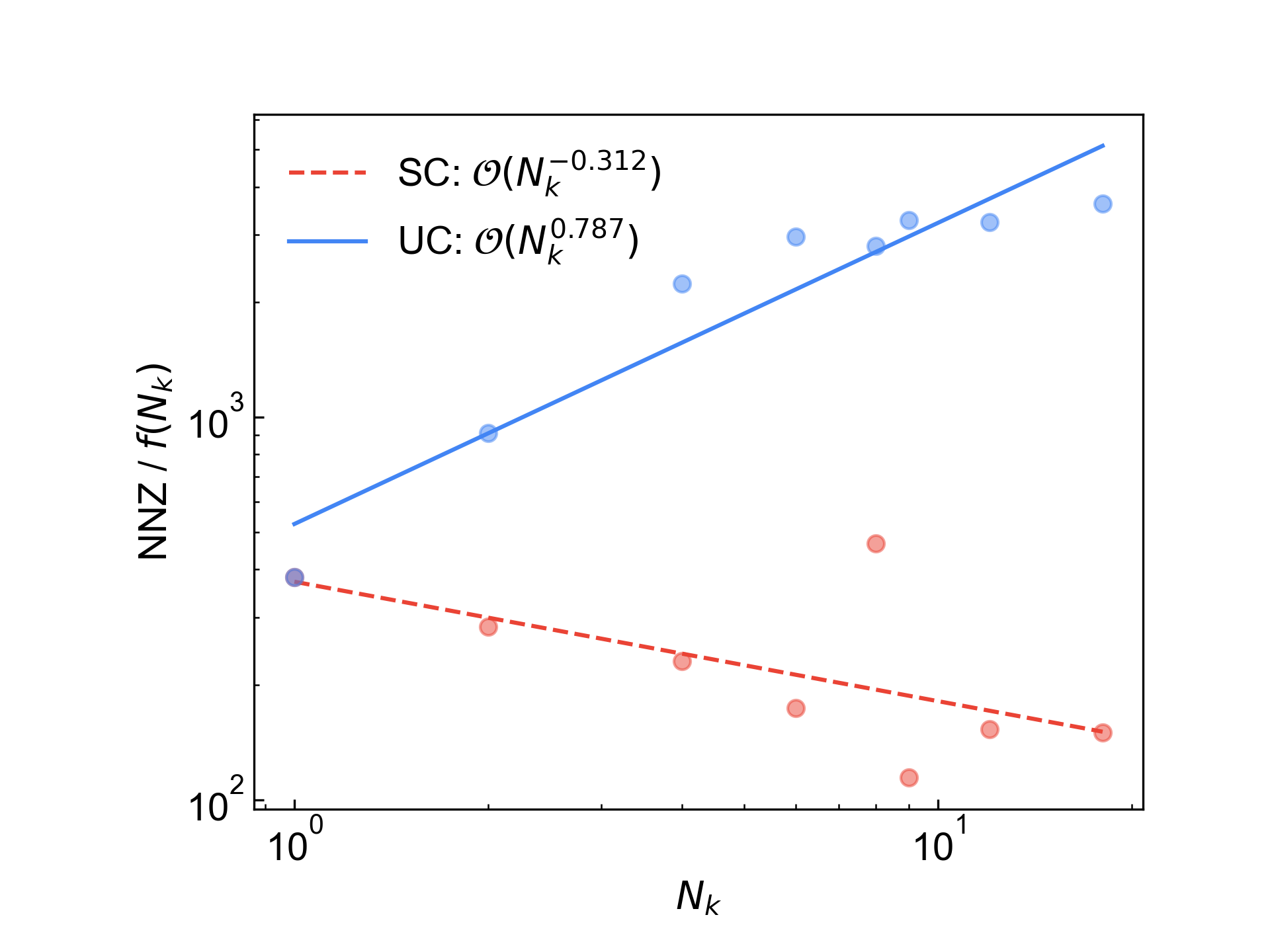}
    \caption{$N_{k}$ dependence on the number of non-zero elements in each two-electron integral block indexed by irrep.\ labels for Diamond in a single-zeta-valence basis with a $k$-mesh shifted to (1/8, 1/8, 1/8) of the simulation cell. In the thermodynamic limit this value should be independent of $N_{k}$ and thus both supercell (SC) and symmetry-adapted (UC) should have no correlation with $N_{k}$--\textit{i.e.}\ a slope of zero.  The $N_{k}$ dependence for small $N_{k}$ makes it difficult to observe the asymptotic improvements from symmetry-adapting the sparse qubitization oracles.}
    \label{fig:carbon_block_dependence}
\end{figure}

Though Figure~\ref{fig:sparse_toff_step_complexity} indicates some computational advantage for the symmetry-adapted case the expected $\sqrt{N_{k}}$ improvement over the supercell case is not easily observed.  The cost of the sparse method largely depends on the number of nonzero elements of $V$, which is generically expected to go as $\mathcal{O}(N_{k}^{3}N^{4})$ for the symmetry-adapted case and $\mathcal{O}(N_{k}^{4}N^{4})$ for the supercell case.  The scaling ultimately depends on the number of nonzero elements in each block of integrals (indexed by three-momentum indices), which we expect to be independent of supercell size $N_{k}$.  For Diamond we plot this dependence in Figure~\ref{fig:carbon_block_dependence} and demonstrate that convergence is slow and there is a strong $N_{k}$ dependence in the number of nonzero elements in each two-electron integral block. This dependence makes observing the improvement in Toffoli cost for symmetry-adapted oracles difficult in the low $N_{k}$ regime.

\subsection{The single-factorization Hamiltonian representation}\label{sec_sub:sf_derivation}
For the ``single-factorization'' method, the Cholesky decomposition of the 2-electron integral tensor can be applied iteratively or the factorized forms can be directly recovered from a density fitted representation of the atomic orbital integral. The quadratic representation of the two-electron integral tensor is
\begin{equation}
V_{p\K_{p},q \K_{q}, r \K_{r}, s \K_{s}} = \sum_n L_{p \K_p q \K_q, n} L_{s \K_s r \K_r, n}^*
\end{equation}
where $\K_p + \K_r = \K_q + \K_s$ modulo a reciprocal lattice vector $\mathbf{G}$, or $\K_p -\K_q - (\K_s - \K_r) = \mathbf{G}$. We can identify $\K_p - \K_q = \mathbf{Q} + \mathbf{G} = \K_s -\K_r$. Thus the two-body interaction operator can be written as
\begin{align}
\hat{H}_2' &= \frac{1}{2} \sum_{\Q}^{N_{k}} \sum_{n}^{M} \left(\sum_{\sigma \in \{\uparrow, \downarrow\}}\sum_{\K}^{N_{k}} \sum_{pq}^{N/2} L_{p \K q (\K\modmin\Q), n} a_{p\K\sigma}^\dagger a_{q(\K\modmin\Q)\sigma}\right) \left( \sum_{\tau \in \{\uparrow, \downarrow\}}\sum_{\Kp}^{N_{k}} \sum_{rs}^{N/2} L^{*}_{s \Kp r (\Kp\modmin\Q), n} a_{r(\Kp\modmin\Q)\tau}^\dagger a_{s\Kp\tau} \right).  \label{eq:hfac_mod_press0}
\end{align}
Due to the reduced symmetry of the complex valued two-electron integral tensor we take additional steps to form Hermitian operators which can be expressed as Pauli operators under the Jordan-Wigner transform.  We express each one-body operator in the product of particle-conserving one-body operators forming the two-electron operator as
\begin{equation}
\hat{\rho}_{n}(\Q) = \left(\sum_{\sigma \in \{\uparrow, \downarrow\}}\sum_{\K}^{N_{k}} \sum_{pq}^{N/2}L_{p \K q (\K\modmin\Q), n} a_{p\K\sigma}^\dagger a_{q(\K\modmin\Q)\sigma}\right)  ,\qquad \hat{\rho}^\dagger_{n}(\Q) = \left( \sum_{\sigma \in \{\uparrow, \downarrow\}}\sum_{\K}^{N_{k}} \sum_{pq}^{N/2}L^{*}_{p \K q (\K\modmin\Q), n} a_{q(\K\modmin\Q)\sigma}^\dagger a_{p\K\sigma}\right) .
\end{equation}
We now take a linear combination of $\hat{\rho}_{n}(\Q)$ to form Hermitian operators and represent our two-electron integral operator as a sum of squares of Hermitian operators that are amenable to the approach for the qubitization of one-body sparse operators via a linear combination of unitaries. These operators are denoted $\hat{A}_{n}(\Q)$ and $\hat{B}_{n}(\Q)$ and are defined as
\begin{align}
\hat{A}_{n}(\Q) =\frac{1}{2}(\hat{\rho}_{n}(\Q) + \hat{\rho}^\dagger_{n}(\Q)),\\
\hat{B}_{n}(\Q) = \frac{i}{2}(\hat{\rho}_{n}(\Q) - \hat{\rho}^\dagger_{n}(\Q)),
\end{align}
to give
\begin{equation}\label{eq:sos_form_sf}
\hat{H}'_2 = \frac{1}{2}  \sum_{\Q}^{N_{k}}\sum_{n}^{M} \left(\hat{A}^2_{n}(\Q) + \hat{B}^2_{n}(\Q)\right).
\end{equation}
We have taken advantage of the translational symmetry by performing the sum over $\Q$ outside the squares of $\hat{A}$ and $\hat{B}$, which reduces the amount of information needed in the representation. In the case $\Q \neq 0$ we can write $\hat{A}_{n}$ as
\begin{align}
\hat{A}_{n}(\Q \neq 0) &= \frac{1}{2}\sum_{\sigma \in \{\uparrow,\downarrow\}}\sum_{\K}^{N_{k}}\sum_{pq}^{N/2}\left(L_{p\K q(\K\modmin\Q),n}a_{p\K\sigma}^{\dagger}a_{q(\K\modmin\Q)\sigma} + L_{p\K q(\K\modmin\Q),n}^{*}a_{q(\K\modmin\Q)\sigma}^{\dagger}a_{p\K\sigma} \right) \nonumber \\
&= \frac{1}{2}\sum_{\sigma \in \{\uparrow,\downarrow\}}\sum_{\K}^{N_{k}}\sum_{pq}^{N/2}\Rea[L_{p\K q(\K\modmin\Q),n}]\left(a_{p\K\sigma}^{\dagger}a_{q(\K\modmin\Q)\sigma} + a_{q(\K\modmin\Q)\sigma}^{\dagger}a_{p\K\sigma}\right) \nonumber \\
&\quad +\frac{i}{2}\sum_{\sigma \in \{\uparrow,\downarrow\}}\sum_{\K}^{N_{k}}\sum_{pq}^{N/2} \Ima[L_{p\K q(\K\modmin\Q),n}]\left(a_{p\K\sigma}^{\dagger}a_{q(\K\modmin\Q)\sigma} - a_{q(\K\modmin\Q)\sigma}^{\dagger}a_{p\K\sigma}\right) . 
\label{eq:AQzero}
\end{align}
Applying the Jordan-Wigner representation then gives
\begin{align}
\hat{A}_{n}(\Q \neq 0) = \sum_{\sigma \in \{\uparrow,\downarrow\}}\sum_{\K}^{N_{k}}\sum_{pq}^{N/2}&\left(\frac{i \Rea[L_{p\K q(\K\modmin\Q),n}]}{4}\left(\vec{Z}X_{p\K\sigma}\vec{Z}Y_{q(\K\modmin\Q)\sigma} - \vec{Z}Y_{p\K \sigma}\vec{Z}X_{q(\K\modmin\Q)\sigma}\right) \right. \nonumber \\
&+ \left. \frac{i\Ima[L_{p\K q(\K\modmin\Q),n}]}{4}\left(\vec{Z}X_{p\K\sigma}\vec{Z}X_{q(\K\modmin\Q)\sigma} + \vec{Z}Y_{p\K \sigma}\vec{Z}Y_{q(\K\modmin\Q)\sigma}\right)  \right).
\end{align}
The same reasoning can be performed for $\hat{B}_{n}(\Q \neq 0)$, which gives the plus and minus signs between $a_{p\K\sigma}^{\dagger}a_{q(\K\modmin\Q)\sigma}$ and $a_{q(\K\modmin\Q)\sigma}^{\dagger}a_{p\K}$ in \eq{AQzero} reversed, so the roles of the real and imaginary parts are reversed.
As a result we obtain
\begin{align}\label{eq:Bform}
\hat{B}_{n}(\Q \neq 0) =\sum_{\sigma \in \{\uparrow,\downarrow\}}\sum_{\K}^{N_{k}}\sum_{pq}^{N/2}&\left(\frac{i \Ima[L_{p\K q(\K\modmin\Q),n}]}{4}\left(\vec{Z}X_{p\K\sigma}\vec{Z}Y_{q(\K\modmin\Q)\sigma} - \vec{Z}Y_{p\K \sigma}\vec{Z}X_{q(\K\modmin\Q)\sigma}\right) \right. \nonumber \\
&+ \left. \frac{i\Rea[L_{p\K q(\K\modmin\Q),n}]}{4}\left(\vec{Z}X_{p\K\sigma}\vec{Z}X_{q(\K\modmin\Q)\sigma} + \vec{Z}Y_{p\K \sigma}\vec{Z}Y_{q(\K\modmin\Q)\sigma}\right)  \right).
\end{align}

Accounting for the cases with $\Q=0$, we may use the same expressions with an extra identity, which yields a one-body correction when squaring.
We show in \app{singleonecor} that this results in the total one-body operator
\begin{align}
\tilde{H}_1 &= \sum_{\sigma \in \{\uparrow, \downarrow\}}\sum_{\K}^{N_{k}} \sum_{p,q=1}^{N/2} \left(h_{p\K, q\K} + \sum_{r=1}^{N/2}\sum_{\Kp}^{N_{k}}V_{p\K, q\K, r\Kp, r\Kp}\right)a_{p\K\sigma}^{\dagger}a_{q\K\sigma}
\label{eq:H1Vp}
\end{align}
as before.
Therefore the associated $\lambda$ is again $\lambda_{\tilde{H}_{1}}$ as given in \eq{sparse_lambda}.
The $\lambda$ for the two-body term is then
\begin{align}
\lambda_{V} = \frac 12 \sum_{\Q}\sum_{n}^{M}\left(\sum_{\K,pq}(|\Rea[L_{p\K q(\K\modmin\Q),n}]|+|\Ima[L_{p\K q(\K\modmin\Q),n}]|)\right)^{2} .
\end{align}
This expression can be obtained by first summing the absolute values of the weights in the linear combination of unitaries for $\hat{A}$ and $\hat{B}$ to give
\begin{equation}
\sum_{\K,pq}(|\Rea[L_{p\K q(\K\modmin\Q),n}]|+|\Ima[L_{p\K q(\K\modmin\Q),n}]|).
\end{equation}
This is obtained by noting that the sum over the spin gives a factor of 2, and there are two unitary operators for each of the real and imaginary parts; together these cancel the factor of 4.
Then this expression is squared for each of $\hat{A}$ and $\hat{B}$, and there is a sum over $\Q$ and $n$ in \eq{sos_form_sf}.
A further factor of $1/2$ is obtained because we use amplitude amplification on each operator as described in~\cite{Lee2020}, thus giving our expression for $\lambda_{V}$.  

Next we describe the method to block encode the Hamiltonian in this single-factorized representation.
The key idea is to perform a state preparation over $\Q$ and $n$, then block encode the squares of $\hat{A}_{n}(\Q)$ and $\hat{B}_{n}(\Q)$ using a single step of oblivious amplitude amplification (which saves a factor of 2 for the value of $\lambda$). 
That is, we perform block encodings of $\hat{A}_{n}(\Q)$ and $\hat{B}_{n}(\Q)$, reflect on an ancilla register, then apply the block encodings again.

For the initial state preparation on $\Q,n$, the number of items of data is $MN_k+1$, where the $+1$ is for the one-body part of the Hamiltonian.
This state preparation is via coherent alias sampling, so the dominant cost is from the QROM needed to output ind and alt values.
That has complexity scaling as $\widetilde{\mathcal{O}}(\sqrt{MN_k})$, where the tilde accounts for the size of the items of data.

For both $\hat{A}_{n}(\Q)$ and $\hat{B}_{n}(\Q)$ we have weightings according to the real and imaginary parts of $L_{p\K q(\K\modmin\Q),n}$, but the difference is in what operations are performed in the sum.
Therefore, for an LCU block encoding, the state preparation step may be identical between $\hat{A}_{n}(\Q)$ and $\hat{B}_{n}(\Q)$.
For each value of $\Q,n$, the number of unique values of $\K,p,q$ to consider is $N_k N^2/4$.
Unlike the supercell case we cannot take advantage of symmetry between $p$ and $q$, because we have $p\K$ and $q(\K\modmin\Q)$. The relation between $\K$ and $\K\modmin\Q$ is governed by the value of $\Q$ which is given in the outer sum, and so we cannot exchange $p$ and $q$.
 %accounting for the symmetry between $p$ and $q$.
There is a further factor of $2$ for the number of items of data, because both real and imaginary parts are needed.

Accounting for the values of $\Q,n$, the total number of items of data that must be output by the QROM used in the state preparation is $(MN_k+1)N_kN^2/2=\mathcal{O}(N_k^2 N^3)$, given that $M$ scales as $\mathcal{O}(N)$.
Again because the size of the items of data is logarithmic, this gives a complexity $\widetilde{\mathcal{O}}(N_k N^{3/2})$.
In contrast, in the supercell calculation, each $\hat{A}_{n}$ and $\hat{B}_{n}$ would have $\mathcal{O}(N_k^2N^2)$ entries, and the rank would be $\mathcal{O}(N_kN)$, for a total number of items of data $\mathcal{O}(N_k^3N^3)$.
That would give a complexity $\widetilde{\mathcal{O}}(N_k^{3/2} N^{3/2})$, so there is a factor of $\sqrt{N_{k}}$ improvement obtained by taking advantage of the symmetry.

In the state preparation we only prepare $p,q$ for $p\le q$, and the full range of values should be produced using a swap controlled by an ancilla register.
A further subtlety in the implementation as compared to prior work is that the complex conjugate is needed as well.
This may be implemented using a sign flip on the qubit indicating the imaginary part, so it is just a Clifford gate.

A major difference is in the selection of operations for $\hat{A}_{n}(\Q)$ and $\hat{B}_{n}(\Q)$.
We see that there are two steps where we need to apply an operation of the form $\vec Z X$ or $\vec Z Y$, and the choice of $X$ or $Y$.
The selection of where the $X$ or $Y$ is applied (indicated by the subscript) can be implemented in the standard way.
The choice of whether $X$ or $Y$ is applied depends on four qubits.
\begin{enumerate}
\item The qubit selecting between the one- and two-body parts.
\item A qubit selecting between $A$ and $B$, which can simply be prepared in an equal superposition using a Hadamard because there are equal weightings between these operators.
\item A qubit selecting between the real and imaginary parts of $L_{p\K q(\K\modmin\Q),n}$, which was prepared in the state preparation.
\item A qubit selecting between the two terms shown above in each line of the expressions for $\hat{A}_{n}(\Q)$ and $\hat{B}_{n}(\Q)$. This qubit can also be prepared using a Hadamard.
\end{enumerate}
Using a trivial number of operations on these qubits we can determine whether it is $X$ or $Y$ that needs to be performed.
The cost of the controlled unitary is doubled because we apply a controlled $\vec Z X$ and a controlled $\vec Z Y$, but this cost is trivial compared to the state preparation cost so has little effect on the overall complexity.

A further subtlety in the implementation is that in the second implementation of $\hat{A}_{n}(\Q)$ and $\hat{B}_{n}(\Q)$, we simply use the qubit flagging the one-body part to control whether the Pauli string $\vec Z X$ or $\vec Z Y$ is applied at all.
This ensures that the square is not obtained for the one-body part.
For a more in-depth explanation of the implementation, see the circuit diagram in Figure~\ref{fig:selectsf} and the explanation in Appendix~\ref{app:single}.

\begin{figure}[tbh]
    \centering
\includegraphics[scale=0.97]{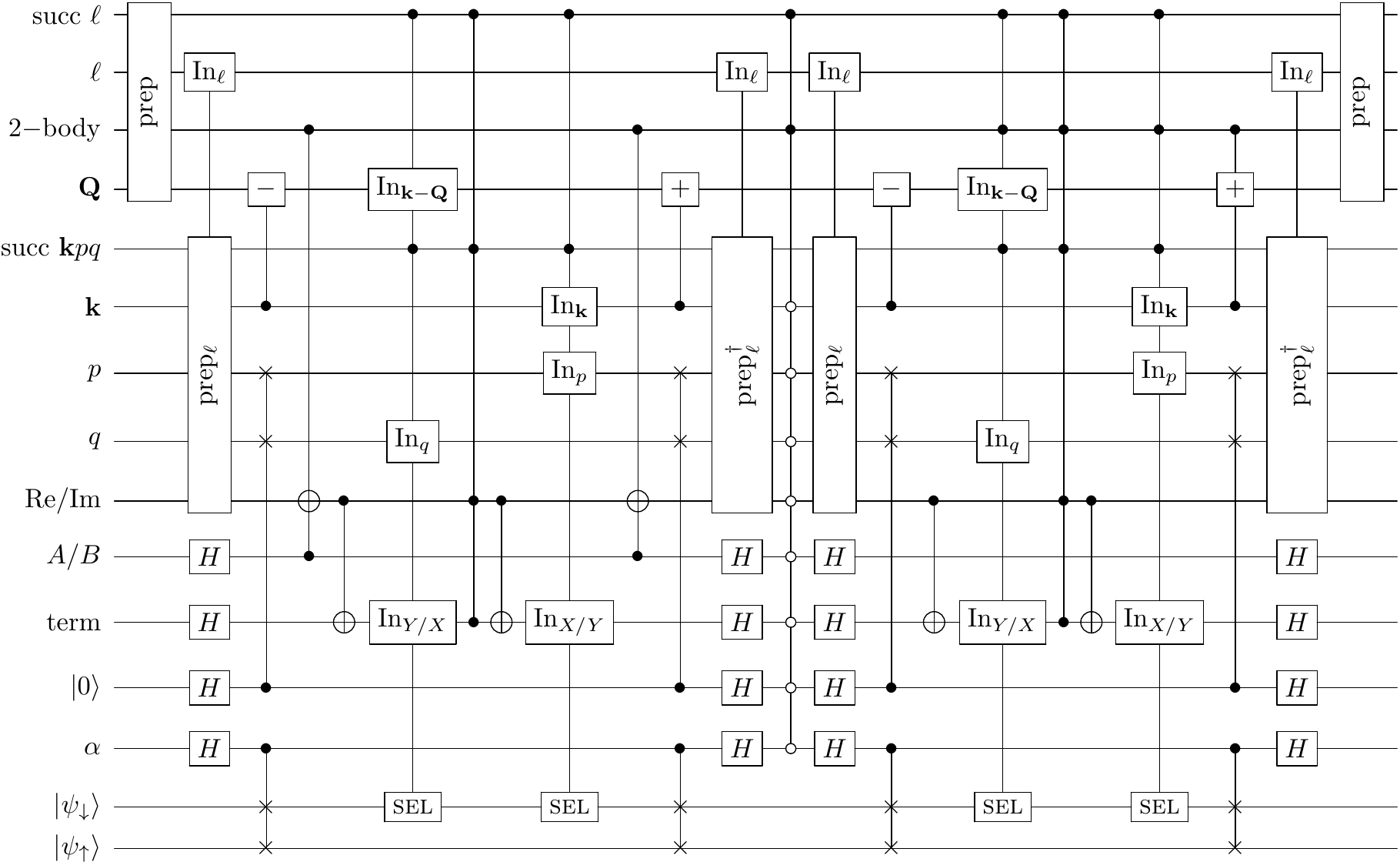}
\caption{The circuit for performing the state preparation and controlled operations for the single factorization approach.
The register labelled $\ell$ is a contiguous register for $\Q,n$, with $\Q$ also output in the state preparation.
The inner state preparation uses $\ell$ as a control.
Then the minus on $\Q$ controlled by $\K$ is to compute $\K\modmin\Q$, with that register reset to $\Q$ later with a controlled addition.
A qubit is used to swap $p$ and $q$ to generate that symmetry, and another is used to swap the spin up and spin down components of the system so the selection only need act on the spin down component.
The qubits labelled Re/Im, $A/B$, and ``term'' are the qubits selecting the real versus imaginary parts, $A$ versus $B$, and the two terms in each line of the Hamiltonian.
These correspond to $b_1,b_2,b_3$ in Appendix~\ref{app:single}, and the Toffoli and CNOT gates are used so that the ``term'' qubit can be used to select whether the Paul string with $X$ or $Y$ is applied.
The selection is performed twice, once for each of the Pauli strings and so is controlled by $\K\modmin\Q$ and $q$ the first time, then $\K$ and $p$ the second time.
A controlled phase between Re/Im and ``term'' (also controlled by the success flag qubits) is used to generate the correct sign for the term.
The block encoding of $A/B$ is performed twice, with the reflection on the ancilla qubits in the middle generating the step of oblivious amplitude amplification.
The third register flags that we have the two-body part of the Hamiltonian, and is used to control the block encoding of $A/B$ the second time to ensure it is not performed for the one-body part.
}
    \label{fig:selectsf}
\end{figure}

\begin{figure}[htb]
    \centering
    \begin{picture}(400,185)
    \put(-45,0){\includegraphics[width=8.5cm]{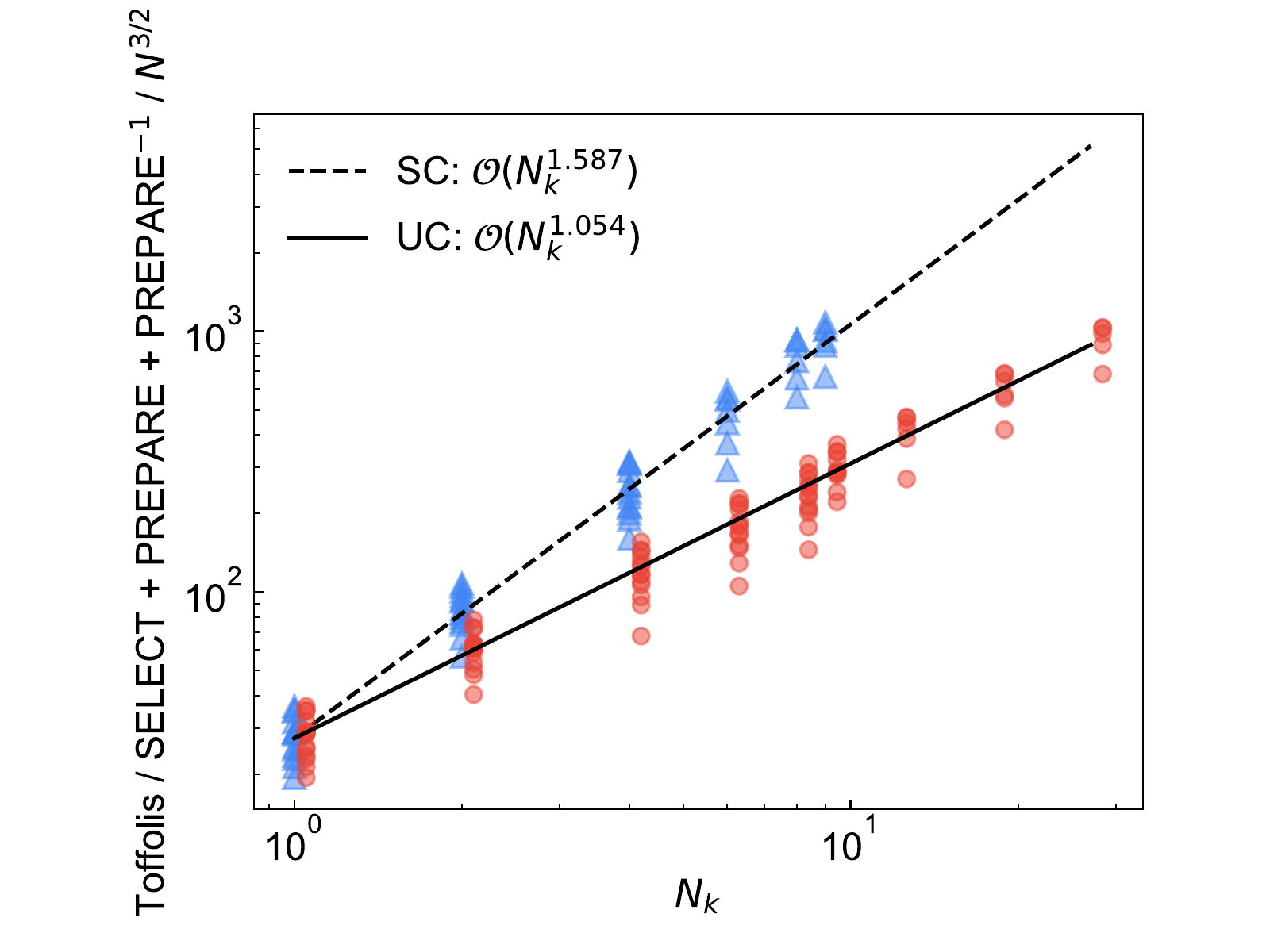}}
    \put(195,0){\includegraphics[width=8.5cm]{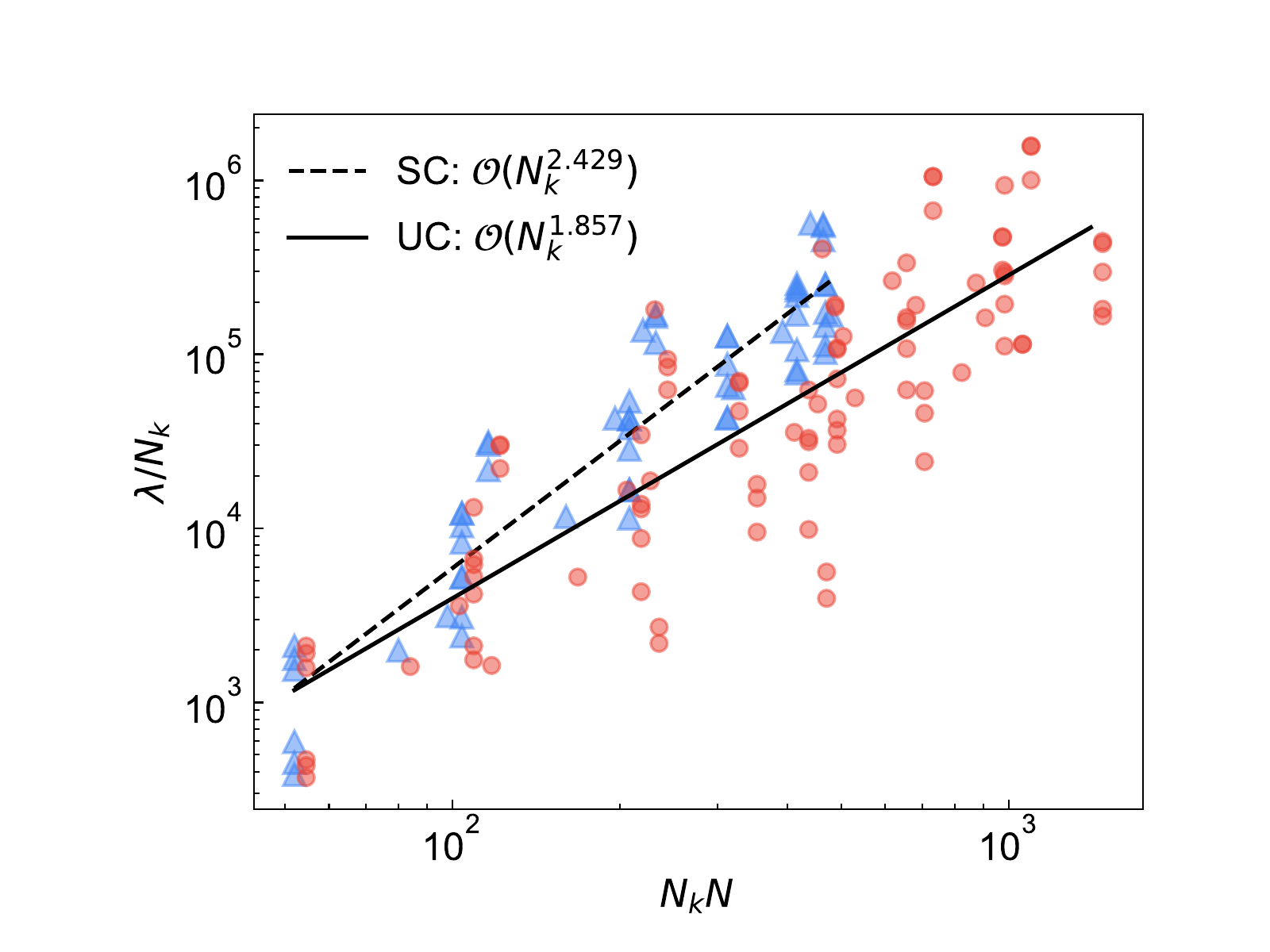}}
    \put(5,170){(a)}
    \put(245,170){(b)}
    \end{picture}
    \caption{(a) Number of $k$-points verses Toffoli cost to implement the block encoding for the single factorization LCU evaluated for the benchmark systems listed in Table~\ref{tab:benchmark_systems} described using the cc-pVDZ and cc-pVTZ basis sets and $\Gamma$-centered Monkhorst-Pack grids of size [1, 1, 1] to [3, 3, 3]. Each point is a single system described at a particular basis set and $k$-mesh where the range of the auxiliary index of the Cholesky factorization is selected to produce two-electron integrals corresponding to an MP2 error of one 1 milliHartree per unit cell with respect to an untruncated auxiliary index range. We divide the Toffoli complexity for implementing {\sel} + {\prep} + {\prep}$^{-1}$ by $N^{3/2}$, which is the shared scaling in the number of bands.  The different scaling in number of $k$-points becomes clear: $N_{k}$ for symmetry-adapted block encodings and $N_{k}^{3/2}$ for supercell non-symmetry-adapted block encodings. We observe similar behavior for qubit count and for plotting oracle Toffoli complexity versus the number of bands.
    (b) The value of $\lambda$ per unit cell ($\lambda / N_{k}$) as a function of the total system size $NN_{k}$ for the same systems described with the same cutoffs used in (a).}
    \label{fig:sf_kpoint_vs_toff_step}
\end{figure}
Figure~\ref{fig:sf_kpoint_vs_toff_step} demonstrates the $\sqrt{N_{k}}$ improvement in constructing the walk operator by symmetry-adapting.  Even for small $N_{k}$ there is a clear separation between the cost of supercell (SC) and symmetry-adapted oracles that agrees with the theoretical scalings of 1.5 and 1.0, respectively.

\subsection{The double-factorization Hamiltonian representation}\label{sec:df_lcu}
In the sparse and SF LCU approaches we have found that there is a factor of $\sqrt{N_{k}}$ savings in Toffoli costs and logical qubit costs for symmetry-adapted block encoding constructions over their non-symmetry-adapted counterparts (supercell calculations).
The double-factorization (DF) representation continues this trend, though the origin of the speedup is different.
In the double-factorization circuits each unitary of the LCU is a rank-one one-body operator that can be thought of as the outer product of two vectors of ladder operators, where each vector of ladder operators is obtained by a Givens rotation with multiqubit control based on other indices.
First notice that for SF there is $\mathcal{O}(N_{k}^{2}N^{3})$ data to output to specify the Hamiltonian via the Cholesky factors.
The factors come from two momentum indices $\K,\Q$, two band indices $p,q$, and one auxiliary index $n$.  In this section we demonstrate that by using a workspace register to apply Givens rotations to pairs of band indices, $\{\K, \K\modmin\Q\}$, the complexity of the DF LCU can also be improved by a factor of $\sqrt{N_{k}}$ over supercell calculations.

To construct the DF LCU, we will separate $\hat{A}_n(\Q)$ and $\hat{B}_n(\Q)$ out into sums over $\K$.
To express this, instead of having $\rho_{n}(\Q)$, we define $\rho_{n}(\Q, \K)$
\begin{equation}
\hat{\rho}_{n}(\Q, \K) = \left(\sum_{\sigma \in \{\uparrow, \downarrow\}}\sum_{pq}^{N/2}L_{p \K q (\K\modmin\Q), n} a_{p\K\sigma}^\dagger a_{q(\K\modmin\Q)\sigma}  \right)  ,\qquad \hat{\rho}^\dagger_{n}(\Q, \K) = \left( \sum_{\sigma \in \{\uparrow, \downarrow\}} \sum_{pq}^{N/2}L^{*}_{p \K q (\K\modmin\Q), n} a_{q(\K\modmin\Q)\sigma}^\dagger a_{p\K\sigma}\right) 
\end{equation}
so then the Hermitian one-body operators that are squared to form the two body part of the Hamiltonian are
\begin{align}
\hat{A}_{n}(\Q) = \sum_{\K}\frac{1}{2}(\hat{\rho}_{n}(\Q, \K) + \hat{\rho}^\dagger_{n}(\Q, \K))\, ,\\
\hat{B}_{n}(\Q) = \sum_{\K}\frac{i}{2}(\hat{\rho}_{n}(\Q, \K) - \hat{\rho}^\dagger_{n}(\Q, \K))\, .
\end{align}
Just as in the single factorization case, we have the two-body part of the Hamiltonian
\begin{equation}
\hat{H}'_2 = \frac{1}{2}  \sum_{\Q}^{N_{k}}\sum_{n}^{M} \left(\hat{A}^2_{n}(\Q) + \hat{B}^2_{n}(\Q)\right).
\end{equation}
We can write $\hat{A}_{n}(\Q)$ as
\begin{align}
\hat{A}_{n}(\Q) = \sum_{\K} \left[ U^A_{n}(\Q, \K) \left( \sum_{\sigma}\sum_{p}^{\Xi_{\Q, n,\K,A}}f^A_{p}(\Q, n, \K)n_{p\K\sigma} \right) U^A_{n}(\Q, \K)^{\dagger} \right]
\end{align}
where the basis rotation unitary $U_{n}(\Q, \K)$ acts on orbitals indexed by $\K$ and $\K\modmin\Q$, $\Xi_{\Q, n,\K,A}$ corresponds to a rank cutoff for $A$, and $f_{p}^{A}(\Q, n, \K)$ is the eigenvalue of the one body operator that is diagonalized by $U_{n}(\Q, \K)$.
The expression for $\hat{B}_{n}(\Q)$ is similar, and we use $\Xi_{\Q, n,\K,B}$ to denote the rank cutoff.

In practice, for the implementation we would apply a different basis rotation for each individual value of $p$.
As explained by \cite{Lee2020}, when doing that the number of Givens rotations needed only corresponds to the number of orbitals it is acting upon, instead of the square.
Here we have two momentum modes $\{\K, \K\modmin\Q\}$ with $N$ orbitals for each, suggesting there should be $2N$.
However, there is no mixture between the different spin states indexed by $\sigma$, so that gives the number of orbitals as $N$.

To quantify the amount of information needed to specify the rotations for the Hamiltonian, there is a $\Q$ summation, $n$ summation, $\K$ summation, $p$ summation, and we need to specify $N$ Givens rotations for each.
In turn, each Givens rotation needs two angles.
The total data here therefore scales as
\begin{equation}
    \widetilde{\mathcal{O}}\left(N\sum_{\Q,n,\K} (\Xi_{\Q,n,k,A}+\Xi_{\Q,n,k,B})\right) ,
\end{equation}
where a factor of $N$ comes from the number of Givens rotations, and the tilde accounts for the bits of precision given for the rotations.
By analogy with the supercell case, it is convenient to define an average rank
\begin{equation}\label{eq:Xidef}
    \ranktwo \coloneqq \frac 1{2 N_k M} \sum_{\Q,n,\K} (\Xi_{\Q,n,k,A}+\Xi_{\Q,n,k,B}) .
\end{equation}
This has division by $N_k$ for the $\Q$ sum and $M$ for the $n$ sum, but no division by a factor accounting for $\K$.
That is, it is the average rank for each value of $\Q$ and $n$, with the sum over $\K$ regarded as part of the rank.
Then it is most closely analogous to the rank in the supercell case, and it is found that it similarly scales as $\mathcal{O}(N_k N)$.
In terms of $\Xi$, the scaling of the amount of data can be given as $\widetilde{\mathcal{O}}(N_{k}N^{2}\ranktwo)$, using $M = \mathcal{O}(N)$.

Next we describe in general terms how to perform the block encoding of the linear combination of unitaries, with the full explanation in Appendix~\ref{app:df_lambda_one_larger_derivation}.
As in the case of single factorisation, the general principle is to perform state preparation over $\Q$ and $n$, then block encode the squares of $\hat{A}_{n}(\Q)$ and $\hat{B}_{n}(\Q)$ using oblivious amplitude amplification.
The difference is that $\hat{A}_{n}(\Q)$ and $\hat{B}_{n}(\Q)$ are now block encoded in a factorized form.
In more detail, the key parts are as follows.
\begin{enumerate}
    \item Perform a state preparation over $\Q$ and $n$, as well as a qubit distinguishing between $\hat{A}_{n}(\Q)$ and $\hat{B}_{n}(\Q)$.
    Using the advanced QROM, the complexity of this state preparation scales approximately as the square root of the number of items of data, so as $\widetilde{O}(\sqrt{N_k N})$.
    The tilde accounts for logarithmic factors from the size of the output.
    For convenience here we use a contiguous register for combined values of $\Q$ and $n$.
    \item Apply a QROM which outputs the value of $\Q$, as well as an offset needed for the contiguous register needed in the state preparation for $\hat{A}_{n}(\Q)$ and $\hat{B}_{n}(\Q)$.
    \item Perform the inner state preparation over $\K$ and $p$.  Here the number of items of data is ${\mathcal{O}}(N_{k}N\ranktwo)$, accounting for the sums over $\Q,n,\K,p$.
    The complexity via advanced QROM is approximately the square root of this quantity, $\widetilde{\mathcal{O}}(\sqrt{N_{k}N\ranktwo})$.
    \item Apply the QROM again to output the rotation angles for the Givens rotations needed for the basis rotation.  This time the size of the output scales as $\mathcal{O}(N)$,
    so the complexity of the QROM scales as $\widetilde{\mathcal{O}}(N\sqrt{N_{k}\ranktwo})$.
    This is the dominating term in the complexity.
    \item Use control qubits to swap the system registers into the correct location.
    This is done first controlled by a qubit labelling the spin, $\sigma$, which is similar to what was done in prior work.
    The new feature here is that registers containing $\K$ and $\K\modmin\Q$ are also used to swap system registers into $N$ target qubits.
    \item Apply the Givens rotations on these $N$ target qubits. The complexity here only scales as $\widetilde{\mathcal{O}}(N)$, so is smaller than in the other steps.
    \item Apply a controlled $Z$ for part of the number operator.
    This comes from representing the number operator as $(\openone-Z)/2$ and combining the identity with the one-body part of the Hamiltonian.
    \item Invert the Givens rotations, controlled swaps, QROM for the Givens rotations, and state preparation over $\K,p$.
    The complexities here are similar to those in the previous steps, but the complexities for QROM erasure are reduced.
    This completes the block encoding of $\hat{A}_{n}(\Q)$ and $\hat{B}_{n}(\Q)$.
    \item Perform a reflection on the ancilla qubits used for the state preparation on $\K,p$.
    This is needed for the oblivious amplitude amplification.
    \item Perform steps 3 to 8 again for a second block encoding of $\hat{A}_{n}(\Q)$ and $\hat{B}_{n}(\Q)$.
    This together with the reflection gives a step of oblivious amplitude amplification, and therefore the squares of $\hat{A}_{n}(\Q)$ and $\hat{B}_{n}(\Q)$.
    \item Invert the QROM from step 2. This has reduced complexity because it is an erasure.
    \item Invert the state preparation from step 1.
\end{enumerate}

\begin{figure}[]
    \centering
\includegraphics[angle=270,origin=c]{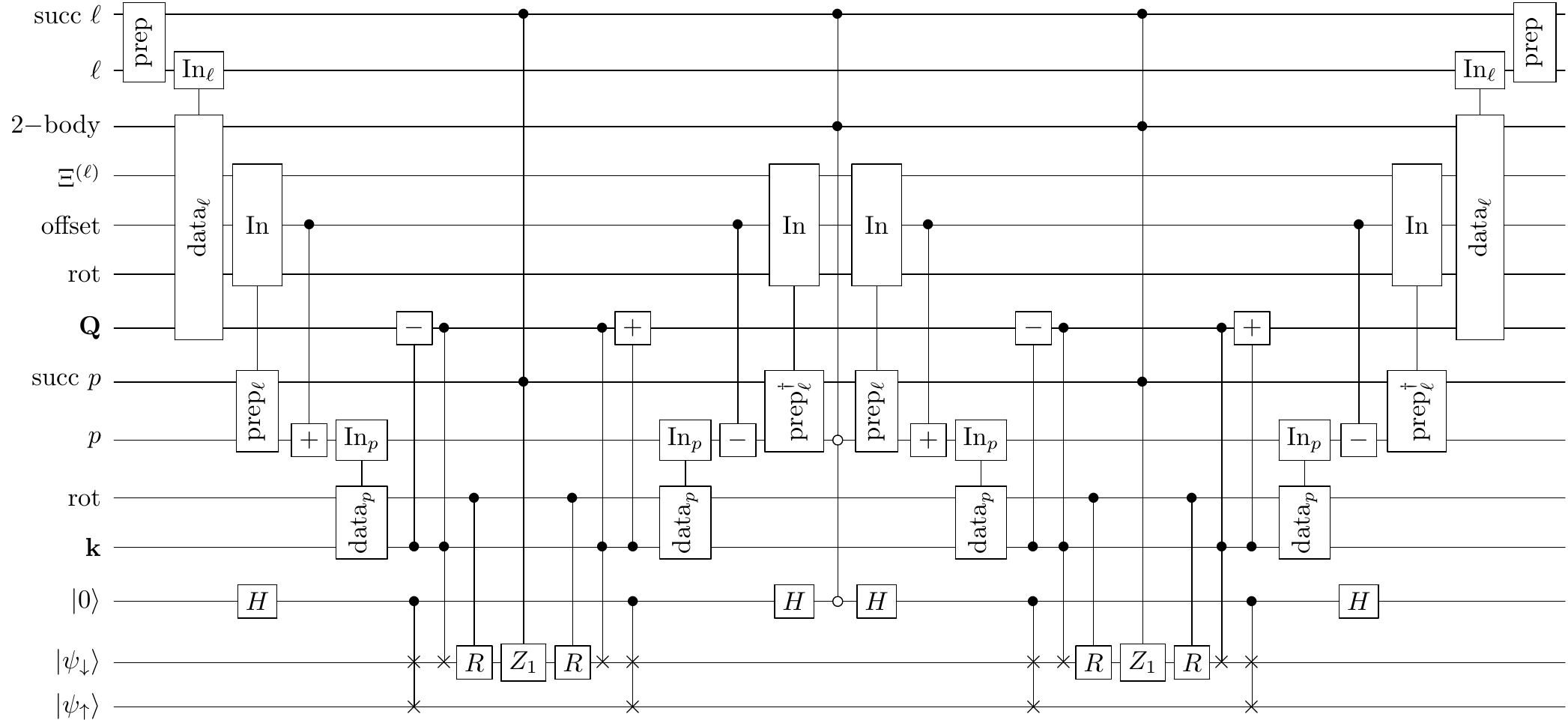}
\caption{The circuit for performing the state preparation and controlled operations for the double factorization approach.
The register labelled $\ell$ is a contiguous register for preparing $\Q,n$, with $\Q$ output next in the QROM.
The register labelled $p$ is actually a contiguous register for preparing both $\K$ and $p$.
The value of $\K$ is output in the next step together with the rotations.
Then the minus on $\Q$ controlled by $\K$ is to compute $\K\modmin\Q$.
The ``$\times$'' on $\ket{\psi_\downarrow}$ controlled by the $\Q$ and $\K$ registers indicates that these registers are used to swap the required part of the system register into $N$ target qubits that the Givens rotations $R$ act upon.
}
    \label{fig:selectdf}
\end{figure}

A quantum circuit for the procedure is shown in Figure~\ref{fig:selectdf}.
This is similar to Figure~16 in \cite{Lee2020}, except it is including the extra parts needed in order to account for the momentum $\K$ used here.
In particular, $\ell$ shown here is a contiguous register for $\Q,n$, and $p$ shown in the diagram is actually a contiguous register for $\K,p$.
The values of $\Q$ and $\K$ need to be output via QROM after the state preparations.
Then $\K\modmin\Q$ is computed and $\K$ and $\K\modmin\Q$ are used to swap the required part of the system register into $N$ target qubits where we apply the Givens rotations.

The lambda value for the Hamiltonian can be calculated by determining the total L1-norm of the coefficients of the unitaries used to represent the Hamiltonian.
To determine this norm, note first that the number operator is replaced with $(\openone-Z)/2$, and the identity is combined with the one-body part of the Hamiltonian.
For $\hat{A}_{n}(\Q)$, what is implemented therefore corresponds to
\begin{align}
 -\frac 12 \sum_{\K} \left[ U^A_{n}(\Q, \K) \left( \sum_{\sigma}\sum_{p}^{\ranktwo_{\Q, n,\K,A}}f^A_{p}(\Q, n, \K)Z_{p\K\sigma} \right) U^A_{n}(\Q, \K)^{\dagger} \right] .
\end{align}
Summing the absolute values of coefficients here gives
\begin{align}
\sum_{\K} \sum_{p}^{\Xi_{\Q, n,\K,A}}|f^A_{p}(\Q, n, \K)| ,
\end{align}
where the sum over the spin $\sigma$ has given a factor of 2 which canceled the factor of $1/2$.
In implementing the square of $\hat{A}_{n}(\Q)$ we use oblivious amplitude amplification, which provides a factor of $1/2$ to $\lambda$.
Combining this with the $1/2$ in the definition of $\hat{H}'_2$ gives $1/4$, and combining with the contribution from $\hat{B}_{n}(\Q)$ then gives
\begin{align}
\lambda_{\mathrm{DF},2} = \frac{1}{4}\sum_{\Q,n}\left[ \left(\sum_{\K,p}^{N_{k}\ranktwo_{\Q,n,\K,A}}|f^A_{n}(p, \Q,\K)|\right)^{2} + \left(\sum_{\K,p}^{N_{k}\ranktwo_{\Q,n,\K,B}}|f^B_{n}(p, \Q,\K)|\right)^{2} \right],
\end{align}
where the superscript $B$ on $f$ indicates the corresponding quantity for $\hat{B}_{n}(\Q)$.

The one-body Hamiltonian is adjusted by the one-body term arising from the identity in the representation of the number operator in the two-body Hamiltonian.
This yields an effective one-body Hamiltonian (see Appendix~\ref{app:df_lambda_one_larger_derivation})
\begin{align}\label{eq:df_one_body_finalb}
H_{1}' = \sum_{\K,p,q,\sigma}\left(h_{p\K, q\K} + \sum_{\Kp,r}V_{r\Kp,r\Kp,q\K,p\K}\right)a_{p\K\sigma}^{\dagger}a_{q\K\sigma}.
\end{align}
We can rewrite this as
\begin{align}
H_{1}' = \sum_{\K,\sigma} \left[ U^C(\K) \left( \sum_{p}^{N/2}\lambda_{\K,p} n_{p\K\sigma} \right) U^C(\K)^{\dagger} \right]
\end{align}
where $\lambda_{\K,p}$ are eigenvalues of the matrix indexed by $p,q$ in the the brackets in \eq{df_one_body_finalb}.
Thus the L1-norm of $H_{1}'$ is the sum
\begin{align}
\lambda_{\mathrm{DF},1} = \sum_{\K}\sum_{p}|\lambda_{\K,p}| .
\end{align}
\begin{figure}[H]
    \centering
    \begin{picture}(400,185)
    \put(-45,0){\includegraphics[width=8.5cm]{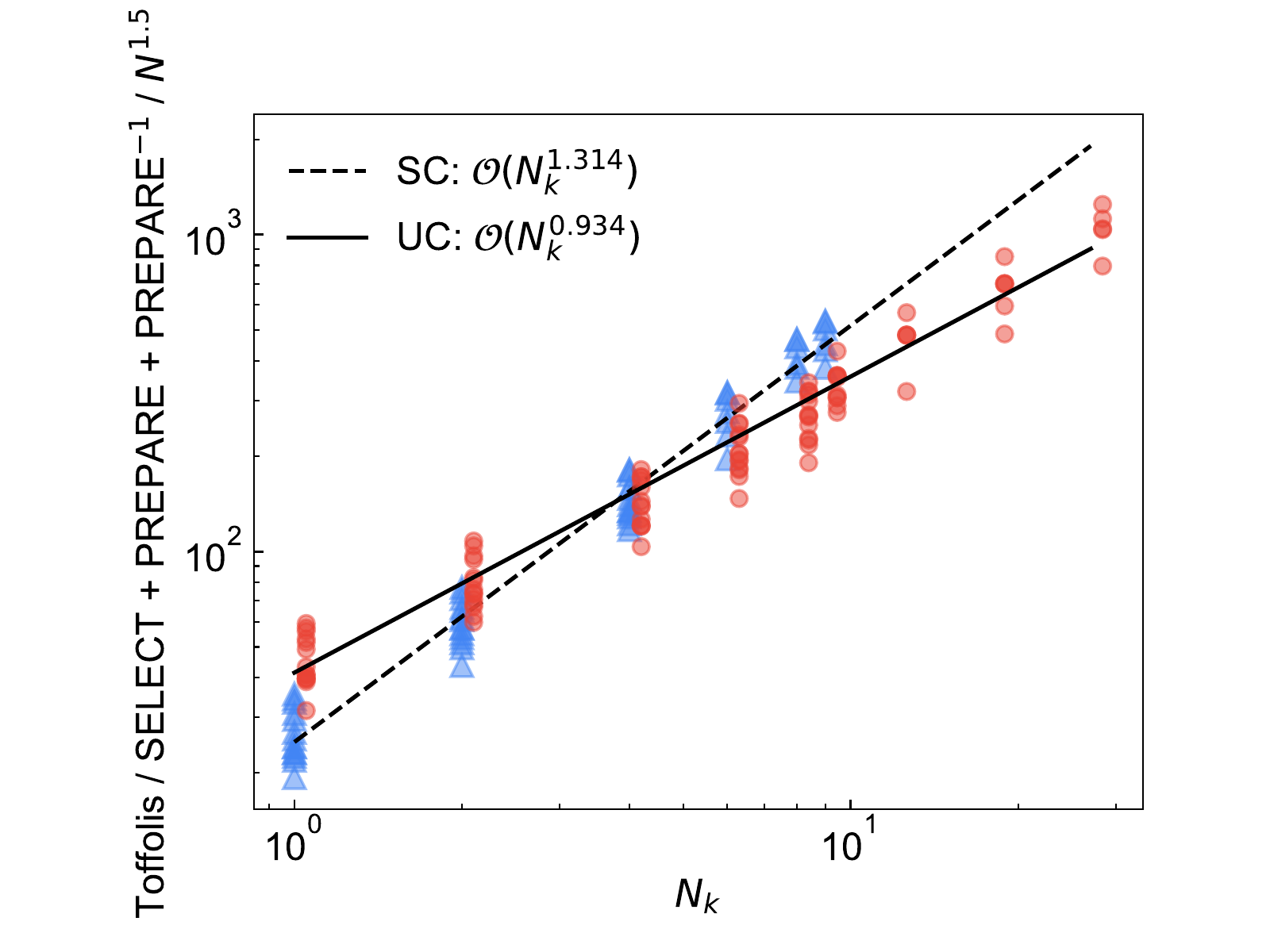}}
    \put(195,0){\includegraphics[width=8.5cm]{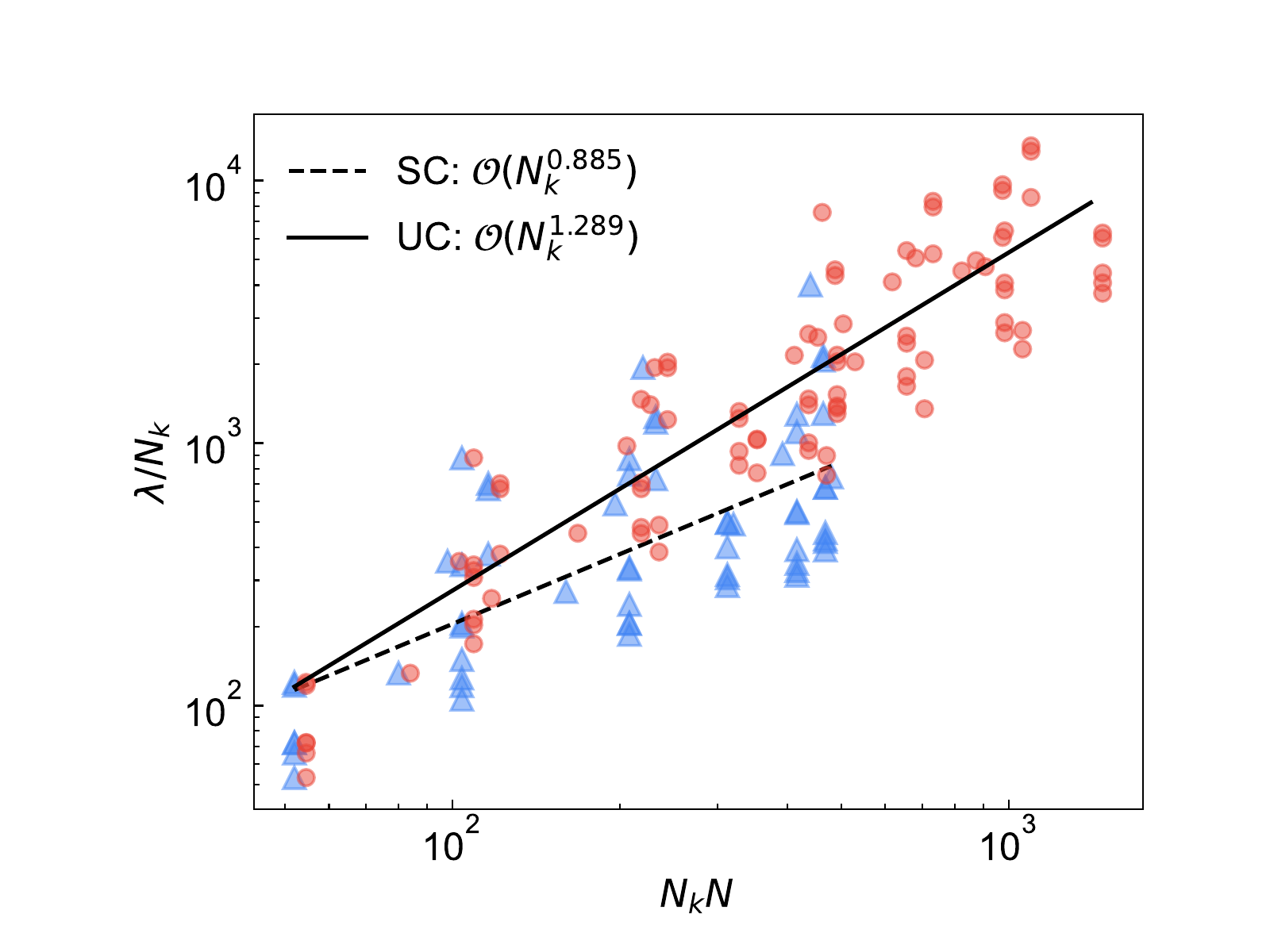}}
    \put(5,170){(a)}
    \put(245,170){(b)}
    \end{picture}    
    \caption{(a) The number of $k$-points verses Toffoli cost to implement the block encoding for the double factorization LCU evaluated for the benchmark systems listed in Table~\ref{tab:benchmark_systems} described using the cc-pVDZ and cc-pVTZ basis sets and $\Gamma$-centered Monkhorst-Pack grids of size [1, 1, 1] to [3, 3, 3]. Each point is a single system described at a particular basis set and $k$-mesh where the threshold to keep eigenvalues and vectors of the second factorization is selected to produce two-electron integrals corresponding to an MP2 error of one 1 milliHartree with respect to an untruncated double factorization. On average this corresponds to a threshold value of $1 \times 10^{-4}$ for the benchmark systems.  The expected $\mathcal{O}(\sqrt{N_{k}})$ scaling improvement for symmetry-adapted walk operators is demonstrated.  (b) The value of $\lambda$ per unit cell as a function of the total system size $NN_{k}$ for the same systems described with the same cutoffs used in (a). The reduced variational freedom in compression of the two-electron integral tensors for the symmetry-adapted walk operator construction translates to an increased value of $\lambda$ at all system sizes.}
    \label{fig:df_kmesh_vs_toff}
\end{figure}
Figure~\ref{fig:df_kmesh_vs_toff} demonstrates the improved $\sqrt{N_{k}}$ scaling of the block encodings coming from reducing the number of controlled rotations by $N_{k}$. Unlike the SF case $\lambda$ for DF has worse scaling in the symmetry-adapted setting compared to the supercell case.  This is rationalized by the fact that there is a larger degree of variational freedom in the second factorization for supercell calculations (and thus more compression) compared to the symmetry-adapted case. The $\lambda$ value is basis set dependent and can potentially be reduced by orbital optimization~\cite{oumarou2022accelerating}.  

\subsection{The tensor hypercontraction Hamiltonian representation}\label{sec_sub:thc_qubitization}
In the tensor hypercontraction (THC) LCU representation the fact that the two-electron integrals can be represented in a symmetric Canonical Polyadic like decomposition is used to define a set of non-orthogonal basis function in which to represent the Hamiltonian, and we use a similar infrastructure to the DF algorithm to implement each term in the factorization (which is in a different non-orthogonal basis) sequentially. In the following section we describe the Bloch orbital version (symmetry-adapted) of the THC decomposition and the resulting LCU, $\lambda$ calculation, and qubitization complexities. 
First we review the salient features of tensor hypercontraction for the molecular case before introducing symmetry labels.  Recall that in the molecular THC approach we expand density like terms over a grid of $M$ points (labeled $\mu$) and weight each grid point with a function $\xi_{\mu}(r)$
\begin{equation}
\phi_p(\mathbf{r}) \phi_q(\mathbf{r}) \approx \sum_\mu \xi_\mu(\mathbf{r}) \phi_p(\mathbf{r}_\mu) \phi_q(\mathbf{r}_{\mu})
\end{equation}
which allows us to write the two-electron integral tensor as
\begin{equation}
V_{pqrs} = \sum_{\mu\nu} \chi_p^{(\mu)}\chi_q^{(\mu)}\zeta_{\mu\nu} \chi_r^{(\nu)}\chi_s^{(\nu)}
\end{equation}
where the central tensor is defined as
\begin{equation}
\zeta_{\mu\nu} = \int d\mathbf{r} \, \int d\mathbf{r}' \, \frac{\xi_{\mu}(\mathbf{r})\xi_{\nu}(\mathbf{r}')}{|\mathbf{r}-\mathbf{r}'|}.
\end{equation}
In order to incorporate translational symmetry into the THC factorization the decomposition of the density is performed on the \emph{cell periodic} part of the Bloch orbitals as \cite{Lu2016,wu2021low}
\begin{equation}
u_{p\K_p}^*(\mathbf{r}) u_{q \K_q}(\mathbf{r}) \approx \sum_\mu  \xi_\mu(\mathbf{r})u_{p\K_p}^*(\mathbf{r}_\mu) u_{q \K_q}(\mathbf{r}_\mu), \label{eq:thc_cell_periodic}
\end{equation}
where  $u_{p\K_p}(\mathbf{r}) = e^{-i \K_p \mathbf{r}} \phi_{p\K_p}(\mathbf{r})$.
Then the two-electron integral tensor has the form
\begin{align}
V_{p\K_p,q\K_q,r\K_r, s\K_s} &= 
\int d \R \int d\R' \phi_{p \K_p}^*\phi_{q\K_q}V(\R,\R')\phi_{r \K_r}^*\phi_{s\K_s}\nn
&= \sum_{\mu\nu} u_{p\K_p}^*(\mathbf{r}_\mu) u_{q \K_q}(\mathbf{r}_\mu) 
\zeta_{\mu\nu}^{\K_p,\K_q,\K_r,\K_s}
u_{r\K_r}^*(\mathbf{r}_\nu) u_{s \K_s}(\mathbf{r}_\nu) \nn
&= \sum_{\mu\nu} \chi_{p\K_p}^{(\mu)*} \chi_{q \K_q}^{(\mu)}
\zeta_{\mu\nu}^{\K_p,\K_q,\K_r,\K_s}
\chi_{r\K_r}^{(\nu)*} \chi_{s\K_s}^{(\nu)} \label{eq:thc_eri_expressions_4index}
\end{align}
where $\chi_{q \K_q}^{(\mu)}=u_{q \K_q}(\mathbf{r}_\mu) $, $V(\R, \R') = |\R - \R'|^{-1}$, and
\begin{align}\label{eq:kpoint_thc_zeta_q}
\zeta_{\mu\nu}^{\K_p,\K_q,\K_r,\K_s}
&=   \int d \R \int d\R' e^{-i(\K_p-\K_q)\cdot\R} \xi_\mu(\R) V(\R,\R') \xi_\nu(\R')e^{i(\K_s-\K_r)\cdot\R'}. 
\end{align}
Some care needs to be taken when bringing this into a form similar to \cref{eq:block_tei}.
First recall that we have $\K_p - \K_q + \K_r - \K_s = \G_{pqrs}$, where $\G_{pqrs}$ is a reciprocal lattice vector, and we are working with a uniform $\Gamma$-point centered momentum grid with dimensions $\mathbf{N} = [N_{x}, N_{y}, N_{z}]$ and $N_{k} = N_{x}N_{y}N_{z}$ 
.
To eliminate one of the four momentum modes, we identify $\Q = \K_p \modmin \K_q$, and $\Q = \K_s \modmin \K_r$, and set $\K_p = \K$, $\K_q = \K \modmin \Q$, $\K_s = \K'$ and $\K_r = (\Kp \modmin \Q) $.
To evaluate the $\zeta$ tensor we still need to know the values of $\K_p-\K_q$ in absolute terms given a value for $\Q$ and $\K$.
We note that mapping the difference $\K_p-\K_q$ back into our $k$-point mesh amounts to adding a specific reciprocal lattice vector $\G^\Q_{pq} = (\K_p-\K_q) - \Q = (\K - (\K \modmin \Q)) - \Q \equiv \G_{\K,\K-\Q}$, with a similar expression for $\K'$
(the subtraction here is not modular).
Thus, given a $\Q$ and $\K$ we can determine $\K-\Q$ and $\G_{\K,\K-\Q}$. 
With these replacements we can write
\begin{align}\label{eq:thc_eri_expression_q}
    V_{p\K_p,q\K_q,r\K_r, s\K_s} 
    \rightarrow V_{p\K,q(\K\modmin\Q),r(\Kp\modmin\Q),s\K'}
    &= \sum_{\mu\nu} \chi_{p\K}^{(\mu)*} \chi_{q \K\modmin\Q}^{(\mu)}
       \zeta_{\mu\nu}^{\Q,\K,\K'}
        \chi_{r(\Kp\modmin\Q)}^{(\nu)*} \chi_{s\K'}^{(\nu)}  \nn
     &= \sum_{\mu\nu} \chi_{p\K}^{(\mu)*} \chi_{q \K\modmin\Q}^{(\mu)}
        \zeta_{\mu\nu}^{\Q,\G_{\K,\K-\Q},\G_{\K',\K'-\Q}}
         \chi_{r(\Kp\modmin\Q)}^{(\nu)*} \chi_{s\K'}^{(\nu)}, 
\end{align}
where we have used
\begin{align}\label{eq:kpoint_thc_zeta_q}
\zeta_{\mu\nu}^{\K_p,\K_q,\K_r,\K_s} \rightarrow \zeta_{\mu\nu}^{\Q,\K,\K'}
&=   \int d \R \int d\R' e^{-i(\Q + \G_{\K,\K-\Q})\cdot\R} \xi_\mu(\R) V(\R,\R') \xi_\nu(\R')e^{i(\Q+\G_{\K'\K'-\Q})\cdot\R')} \nn 
&= \zeta_{\mu\nu}^{\Q,\G_{\K,\K-\Q},\G_{\K',\K'-\Q}}.
\end{align}
In practice there are at most 8 values of $\G$, so we only need to classically determine at most $8^2N_k$ values of $\zeta$, as opposed to $N_k^3$.

We can then write
\begin{align}
H_{2} &= \frac{1}{2}\sum_{\Q,\K,\Kp}\sum_{pqrs}\sum_{\sigma\tau}V_{p\K,q(\K\modmin\Q),r(\Kp\modmin\Q),s\Kp}a_{p\K\sigma}^{\dagger}a_{q(\K\modmin\Q)\sigma}a_{r(\Kp\modmin\Q)\tau}^{\dagger}a_{s\Kp\tau} \nn
&= \frac{1}{2}\sum_{\Q,\K,\Kp}\sum_{pqrs}\sum_{\sigma\tau}\sum_{\mu\nu}\chi_{p\K,\mu}^{*}\chi_{q(\K\modmin\Q),\mu} \zeta_{\mu\nu}^{\Q,\G_{\K,\K-\Q}, \G_{\Kp,\Kp-\Q}}\chi_{r(\Kp\modmin\Q),\nu}^{*}\chi_{s\Kp,\nu}a_{p\K\sigma}^{\dagger}a_{q(\K\modmin\Q)\sigma}a_{r(\Kp\modmin\Q)\tau}^{\dagger}a_{s\Kp\tau} \nn 
&= \frac{1}{2}\sum_{\Q,\G_1,\G_2}\sum_{\mu\nu}\sum_{\sigma\tau}\zeta_{\mu\nu}^{\Q,\G_{1},\G_{2}} \nn
& \quad \times\left(\sum_{\K|\G_{\K,\K-\Q} = \G_1} \sum_{pq}\chi_{p\K,\mu}^{*}\chi_{q(\K\modmin\Q),\mu}
a_{p\K\sigma}^{\dagger}a_{q(\K\modmin\Q)\sigma}\right)
\left(\sum_{\K'|\G_{\K',\K'-\Q} = \G_2} \sum_{rs}
\chi_{r(\Kp\modmin\Q),\nu}^{*}\chi_{s\Kp,\nu}a_{r(\Kp\modmin\Q)\tau}^{\dagger}a_{s\Kp\tau}\right), \label{eq:thc_ham_restricted_sum}
\end{align}
where in going from the second to the third line of \cref{eq:thc_ham_restricted_sum} we have rewritten the sum over $\K$ and $\K'$ as a double sum over all $8^2$ values of $\G_1$ and $\G_2$, and a restricted sum on $\K$ such that for a given $\G_1$ and $\Q$ we only sum over those $\K$ which satisfy $\G_{\K,\K-\Q} = \G_1$.
Here the notation $\G_{\K_p,\K_q}$ is used as equivalent to $\G_{pq}$ above.
The fourfold symmetry of the two-electron integrals carries over to analogous symmetries in $\zeta$, which are listed in \cref{app:THCcomplex}. 

We will then define $\tilde\chi$ which are individually normalized for each $\K$ and $\mu$ so $\sum_{p}\tilde{\chi}_{p\K,\mu}^{*}\tilde{\chi}_{p\K,\mu} = 1$ and
\begin{align}
\nrm_{\K,\mu} \tilde{\chi}_{p\K,\mu} = \chi_{p\K,\mu}
\end{align}
with $\nrm_{\K, \mu}:=\sqrt{\sum_{p}|\chi_{p\K,\mu}|^2}$.
We then use these normalized $\tilde\chi$ to give transformed annihilation and creation operators
\begin{align}
c_{\mu\K\sigma} = \sum_{p}\tilde{\chi}_{p\K,\mu}a_{p\K\sigma},  \qquad
c_{\mu\K\sigma}^{\dagger} = \sum_{p}\tilde{\chi}_{p\K,\mu}^{*}a_{p\K\sigma}^{\dagger} \, .
\end{align}
We can then write the two-body Hamiltonian as
\begin{align}\label{eq:THCcform}
\hat{H}_2 &=\frac{1}{2}\sum_{\Q,\G_1,\G_2}\sum_{\mu,\nu}\sum_{\sigma\tau}\zeta_{\mu\nu}^{\Q,\G_{1},\G_{2}} \nn
& \quad \times\left(\sum_{\K|\G_{\K,\K-\Q} = \G_1} \nrm_{\K,\mu}\nrm_{\K\modmin\Q,\mu} c_{\mu\K\sigma}^\dagger c_{\mu(\K\modmin\Q)\sigma}
\right)
\left(\sum_{\K'|\G_{\Kp,\Kp-\Q} = \G_2}\nrm_{\Kp\modmin\Q,\nu}\nrm_{\Kp,\nu} c_{\nu(\Kp\modmin\Q)\tau}^{\dagger}c_{\nu\Kp\tau}\right).
\end{align}

A complication for the implementation is that we would like to be able to choose the relative weighting between $\zeta$ and $\chi$ such that
\begin{equation}
    \sum_{\K|\G_{\K,\K-\Q} = \G}\nrm_{\K,\mu}\nrm_{\K\modmin\Q,\mu} = 1.
\end{equation}
The difficulty here is that the values of $\nrm_{\K,\mu}$ only depend on $\K,\mu$, because they are based on $\chi_{p\K,\mu}$.
This sum is also dependent on $\Q$ and $\G$, so for this normalization condition to hold it would mean we need to have $\chi_{p\K,\mu}$ also dependent on $\Q$ and $\G$ in a multiplicative factor (so a non-$\mu$-dependent way).
That will leave the normalized $\tilde{\chi}_{p\K,\mu}$ unaffected, but means that the values of $\nrm_{\K,\mu}$ need to have dependence on $\Q,\G$, which will need to be taken account of in the state preparation.

The form in \eq{THCcform} then gives us a recipe for block encoding the Hamiltonian as a linear combination of unitaries.
\begin{enumerate}
    \item First prepare a superposition state proportional to
    \begin{equation}
        \sum_{\Q,\G_1,\G_2,\mu,\nu}\sqrt{|\zeta_{\mu\nu}^{\Q,\G_1,\G_2}|}\ket{\Q,\G_1,\G_2,\mu,\nu}.
    \end{equation}
    This state may be prepared via the coherent alias sampling approach with a complexity dominated by the complexity of the QROM.
    Accounting for symmetry the dimension is about $32N_k M^2$ and the size of the QROM output is approximately the log of that plus the number of bits for the keep probability.
    That gives a Toffoli complexity scaling as
    \begin{equation}
        \sqrt{32N_kM^2[\log(32N_kM^2)+\zetabits]} \, .\label{eq:thc_qrom}
    \end{equation}
    \item For each of the two expressions in brackets in Eq.~\eqref{eq:THCcform}, a preparation over $\K$ or $\Kp$ is needed to give a state of the form
    \begin{equation}
        \sum_{\K|\G_{\K,\K-\Q} = \G}\sqrt{\nrm_{\K,\mu}\nrm_{\K\modmin\Q,\mu}}\ket{\K}.
    \end{equation}
    As explained above, the values of $\nrm_{\K,\mu}$ need to be chosen with (implicit) dependence on $\Q,\G$ for this to be a normalised state.
    This means that the amplitudes here need to be indexed by $\K$, $\Q$, $\G_1$ and $\mu$.
    The restricted range of values in the sum over $\K$ means that the indexing over $\K,\Q,\G_1$ gives $N_k^2$ items of data, which is multiplied by $M$ for the indexing over $\mu$.
    So there are $N_k^2 M$ items of data needed, which is smaller than that in the first step, because it is missing the factor of 32 and typically $N_k<M$.
    Given that the output size is approximately $\log(N_k)+\zetabits$, the Toffoli complexity is approximately
    \begin{equation}
        \sqrt{N_k^2M[\log(N_k)+\zetabits]}\label{eq:thc_norm_toff}.
    \end{equation}
    This cost is incurred twice, once for each of the factors in brackets in \eq{THCcform}.
    \item For each of the $c$ annihilation and creation operators we perform a rotation of the basis from $a$.
    This is done in the following way.
    \begin{enumerate}
        \item First use the spin $\sigma$ or $\tau$ to control a swap of the system registers.
        This is done once and inverted for each of the two $c^\dagger c$ factors.
        Each of these 4 swaps has cost $N_kN/2$.
        \item
        Then use $\K$ or $\K\modmin\Q$ to control the swap of the registers we wish to act on into working registers.
        The value of $\K\modmin\Q$ is used for
        $c_{\mu(\K\modmin\Q)\sigma}$, and needs to be computed to use as a control.
        Each of these eight swaps may be done with a Toffoli complexity approximately as half the number of system registers $N_kN/2$.
        \item Next $\K$ (or $\K\modmin\Q$) and $\mu$ (or $\nu$) are used as a control for a QROM to output the angles for Givens rotations.
        There are two angles for each of $N/2$ Givens rotations, so if they have $\rotbits$ each the size of the output is $N\rotbits$.
        Then the QROM complexity is about
        \begin{equation}
            \sqrt{N_kN^2\rotbits}.
        \end{equation}
        This must be done 4 times (and has a smaller erasure cost).
        \item The sequence of $N/2$ Givens rotations is performed, each with 4 individual rotations on $\rotbits$, for a cost of $2N\rotbits$.
        This cost is incurred 8 times, twice for each of the annihilation and creation operators.
    \end{enumerate}
    \item After the rotation of the basis, we simply need to perform the linear combination of $\vec ZX$ and $\vec ZY$ for $c^\dagger$ and $c$.
    The $X$ or $Y$ is applied in a fixed location, but the $\vec Z$ needs to be applied on a range of qubits chosen by $\K$ or $\K\modmin\Q$.
    We therefore have approximately $N_k$ for the unary iteration for each $\vec Z$ for a total cost of about $4N_k$.
\end{enumerate}
Lastly we would perform reflections on control ancillas as usual to construct a qubitised quantum walk from the block encoding.
This cost is trivial compared to that in the other steps.
For a more detailed explanation, see the circuit diagram in Figure \ref{fig:thcirc} and the discussion in Appendix \ref{app:THCcomplex}.

\begin{figure}[tbh]
    \centering
    \includegraphics[scale=0.8]{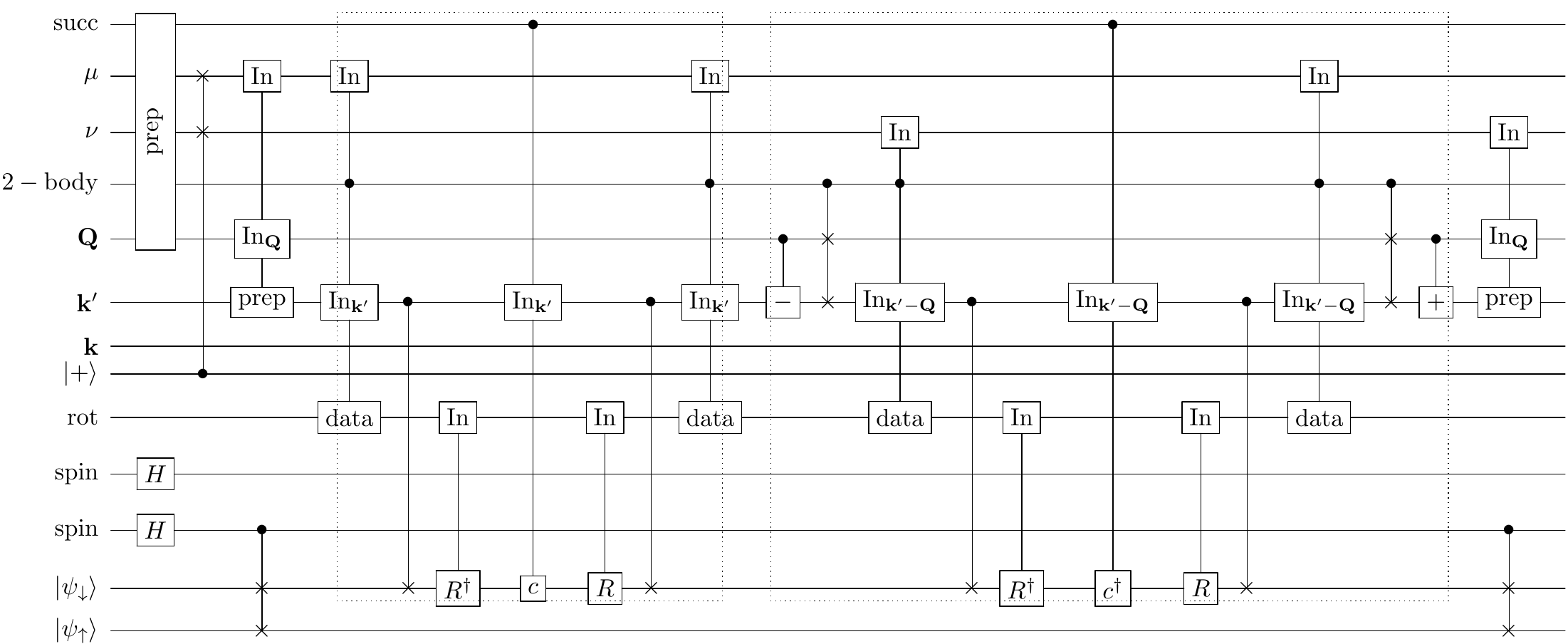}
    \includegraphics[scale=0.8]{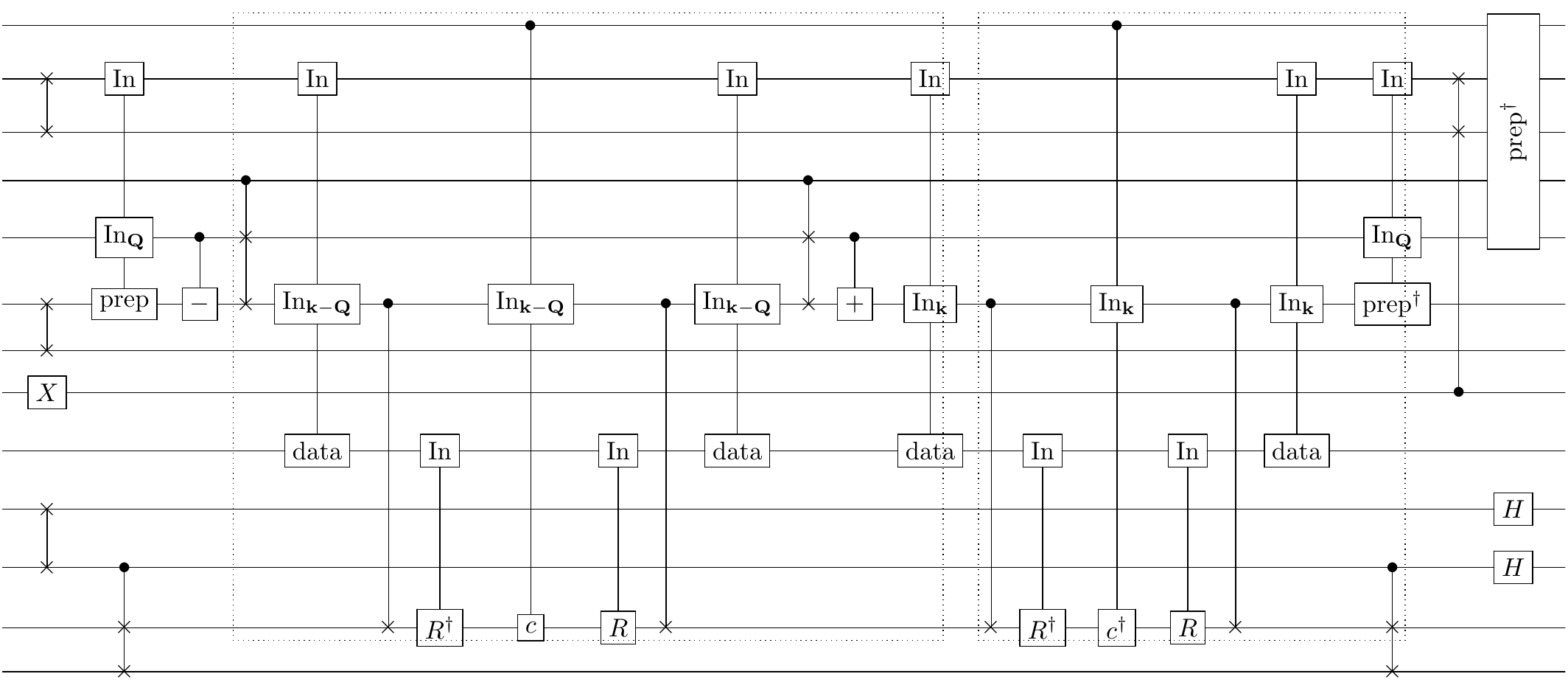}
    \caption{The quantum circuit for the block encoding of the THC representation, split into two parts with the right half at the bottom.
    The top shows the portion of the circuit for the first part controlled by $\Kp$ and $\nu$, and the bottom shows the (right) part of the circuit where it is controlled by $\K$ and $\mu$.
    The dotted rectangles show the regions for implementing the $c$ and $c^\dagger$ operators together with the Givens rotations needed to change the basis.
    The swaps controlled by the $\Kp$ and $\K$ registers are to move the appropriate qubits into target registers in order to apply the Givens rotations.
    The $c$ and $c^\dagger$ are applied using a superposition of $X$ and $iY$ applied using an ancilla qubit (not shown for simplicity), together with a string of $Z$ gates for the Jordan-Wigner representation.
    The preparation at the beginning includes an inequality test between $\mu$ and $\nu$ to give a qubit flagging whether the real or imaginary part is produced.
    To make the implementation self-inverse, the $\mu,\nu$ and $\K,\Kp$ pairs of registers are swapped in the middle (the left of the lower half).
    Also, an $X$ gate is applied to the qubit that controls the swaps at the beginning and end.}
    \label{fig:thcirc}
\end{figure}

\begin{figure}[H]
    \centering
    \includegraphics[width=8.5cm]{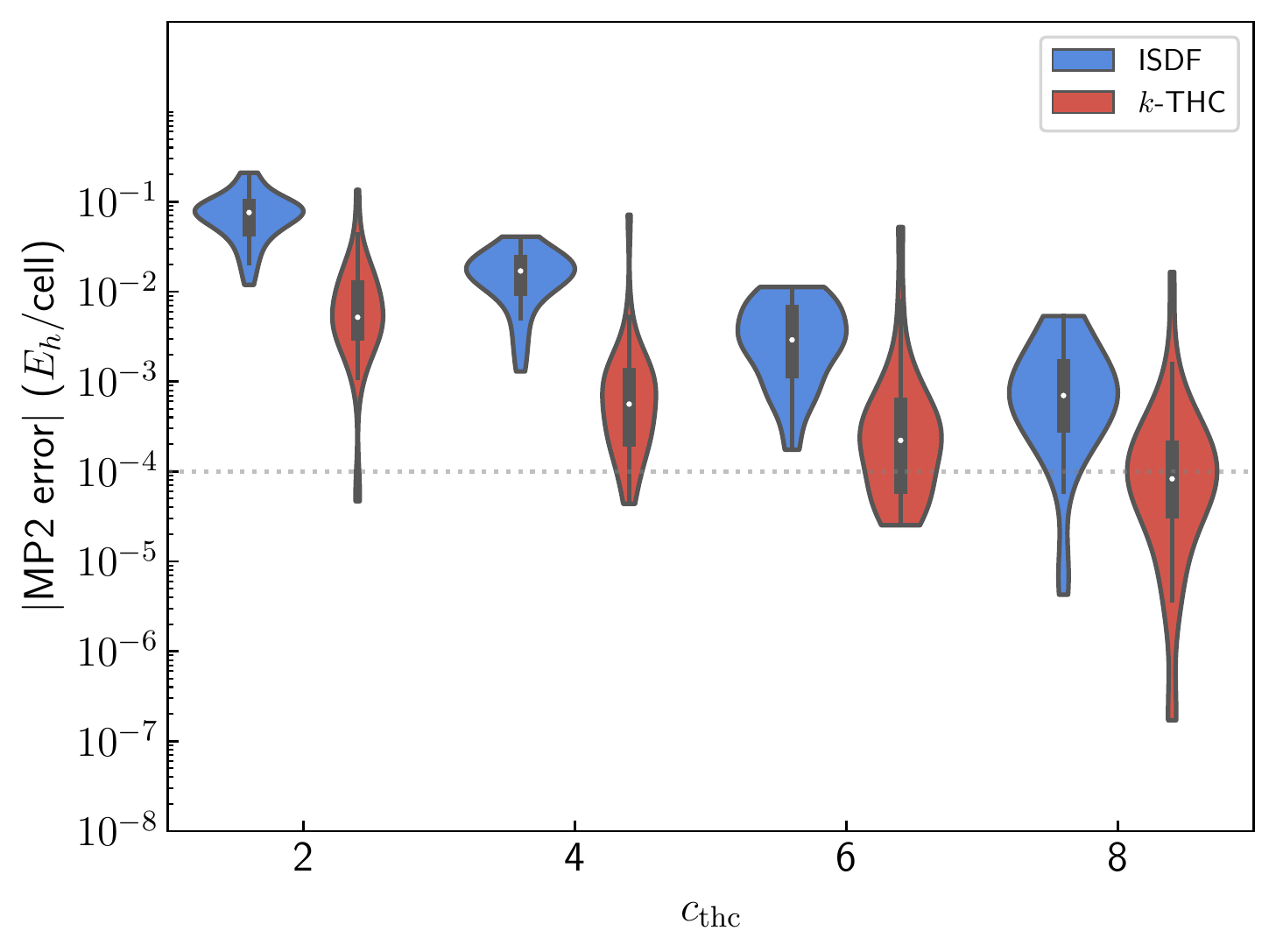}
    \caption{Violin plot of absolute errors in the $k$-THC-MP2 energy per cell for the benchmark set in \cref{tab:benchmark_systems}. Here we compare the MP2 errors as a function of the THC rank parameter $c_{\mathrm{THC}}$ using ISDF or subsequent reoptimization to generate the THC factors.}
    \label{fig:kthc_mp2_error}
\end{figure}

\begin{figure}[H]
    \centering
    \begin{picture}(400,185)
    \put(-50,0){\includegraphics[width=8.5cm]{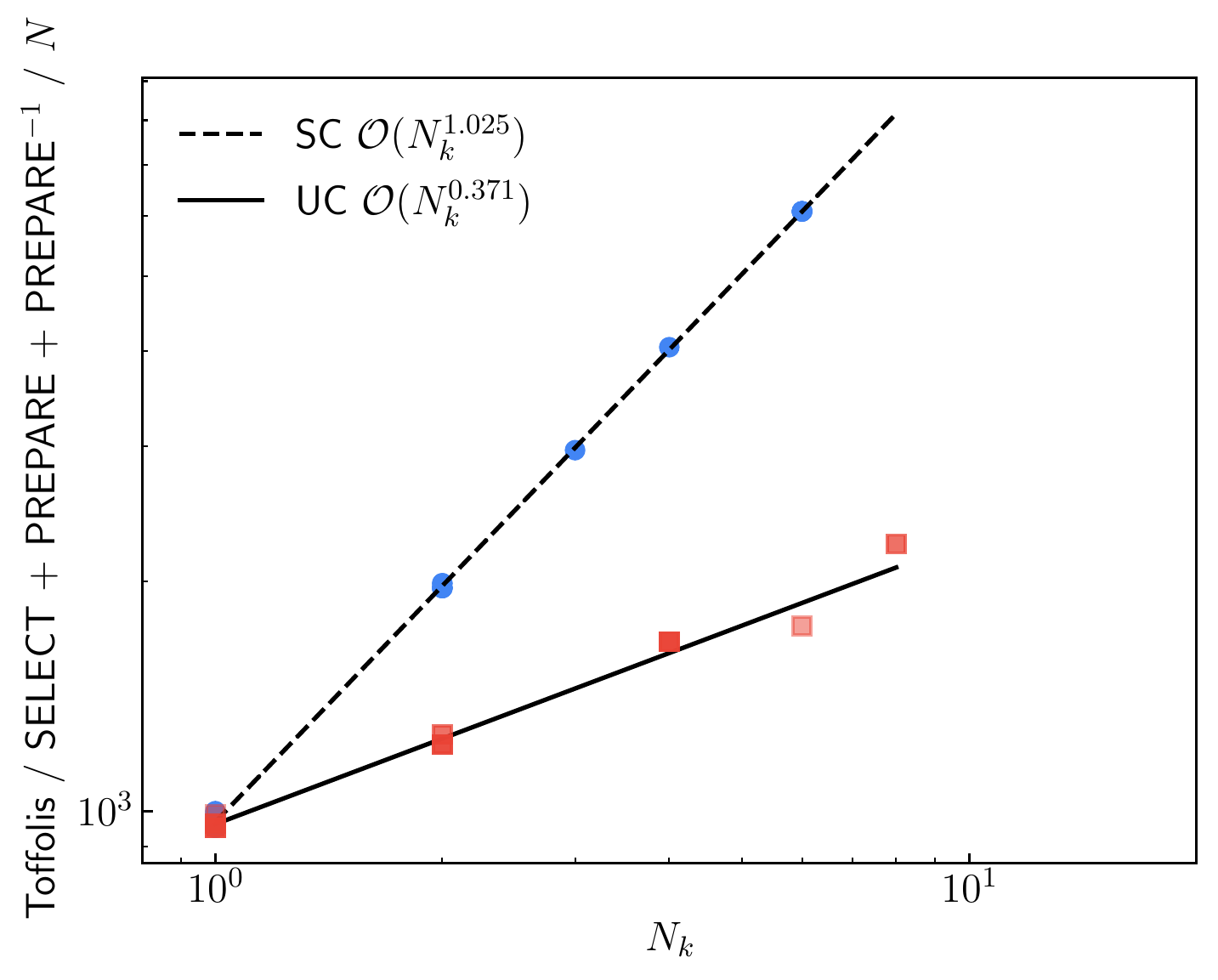}}
    \put(200,0){\includegraphics[width=8.5cm]{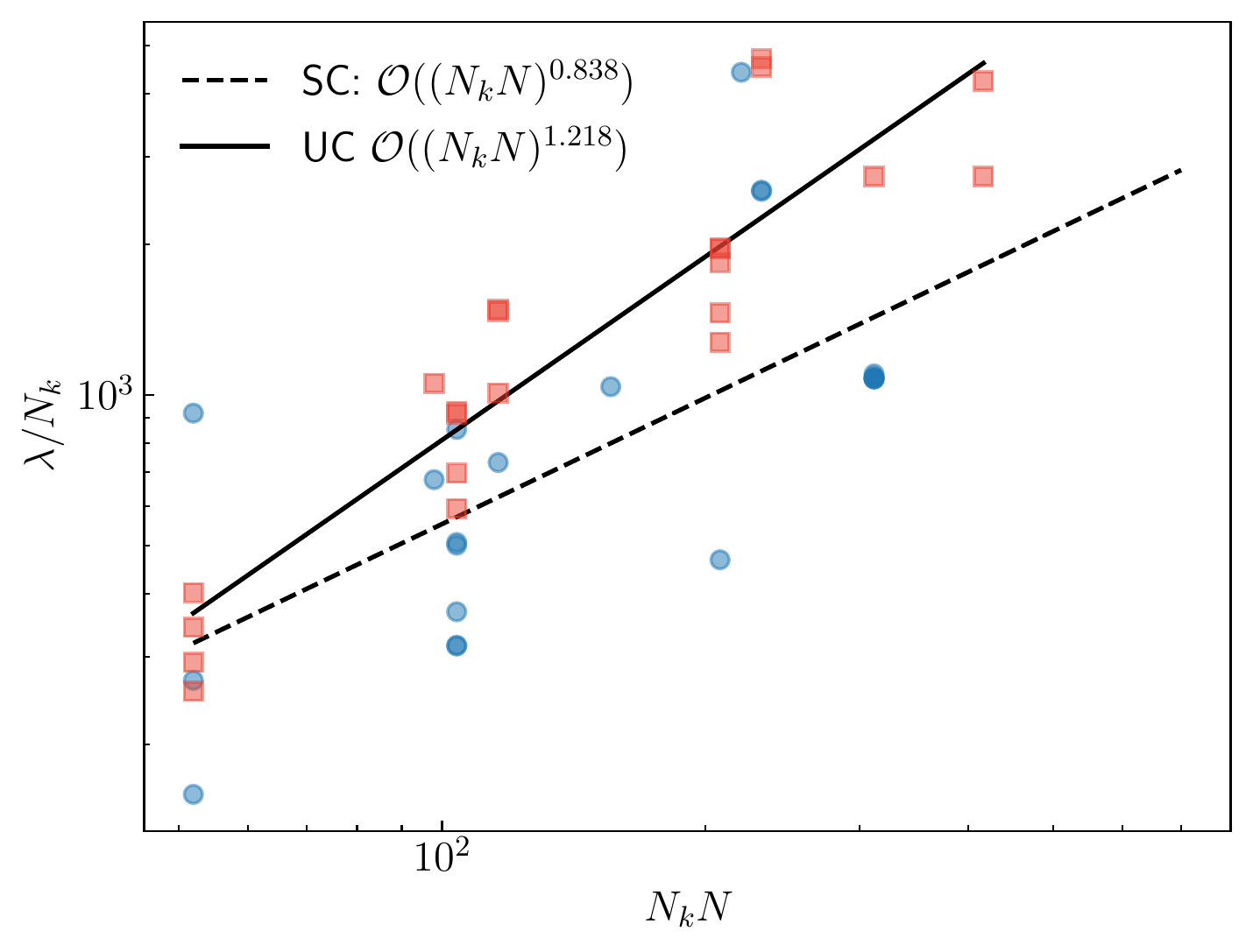}}
    \put(-10,185){(a)}
    \put(240,185){(b)}
    \end{picture}  
    \caption{(a) The number of $k$-points verses Toffoli cost to implement the block encoding for the THC factorization LCU evaluated
for the benchmark systems listed in Table I described using the cc-pVDZ and cc-pVTZ basis sets and $\Gamma$-centered Monkhorst-
Pack grids of size [1, 1, 1] to [3, 3, 3]. Each point is a single system described at a particular basis set and k-mesh where the
range of the auxiliary index of the THC factorization is selected to produce two-electron integrals corresponding to an MP2
error of one 1 milliHartree with respect to an untruncated auxiliary index range. This corresponds to an auxiliary
index that is eight times the number of orbitals in the primitive cell for symmetry adapted THC, and eight times the total number of orbitals ($N_kN$) for supercell THC. We divide the Toffoli complexity for implementing SELECT + PREPARE +
PREPARE$^{-1}$ by N, which is the shared scaling in the number of bands. While we observe a $\sqrt{N_k}$ scaling improvement for symmetry-adapted walk operations, we believe this is a finite size effect and both methods should scale linear with $N_k$ for sufficiently large $N_k$ The value of $\lambda$ as a
function of the total system size $N N_k$ for the same systems described with the same cutoffs used in (a). The reduced variational freedom in compression of the two-electron
integral tensors for the symmetry-adapted walk operator construction translates to an increased value of $\lambda$ at all system sizes.\label{fig:thc_resource_comparison}}
\end{figure}

\begin{figure}[H]
    \centering
    \includegraphics[width=8.5cm]{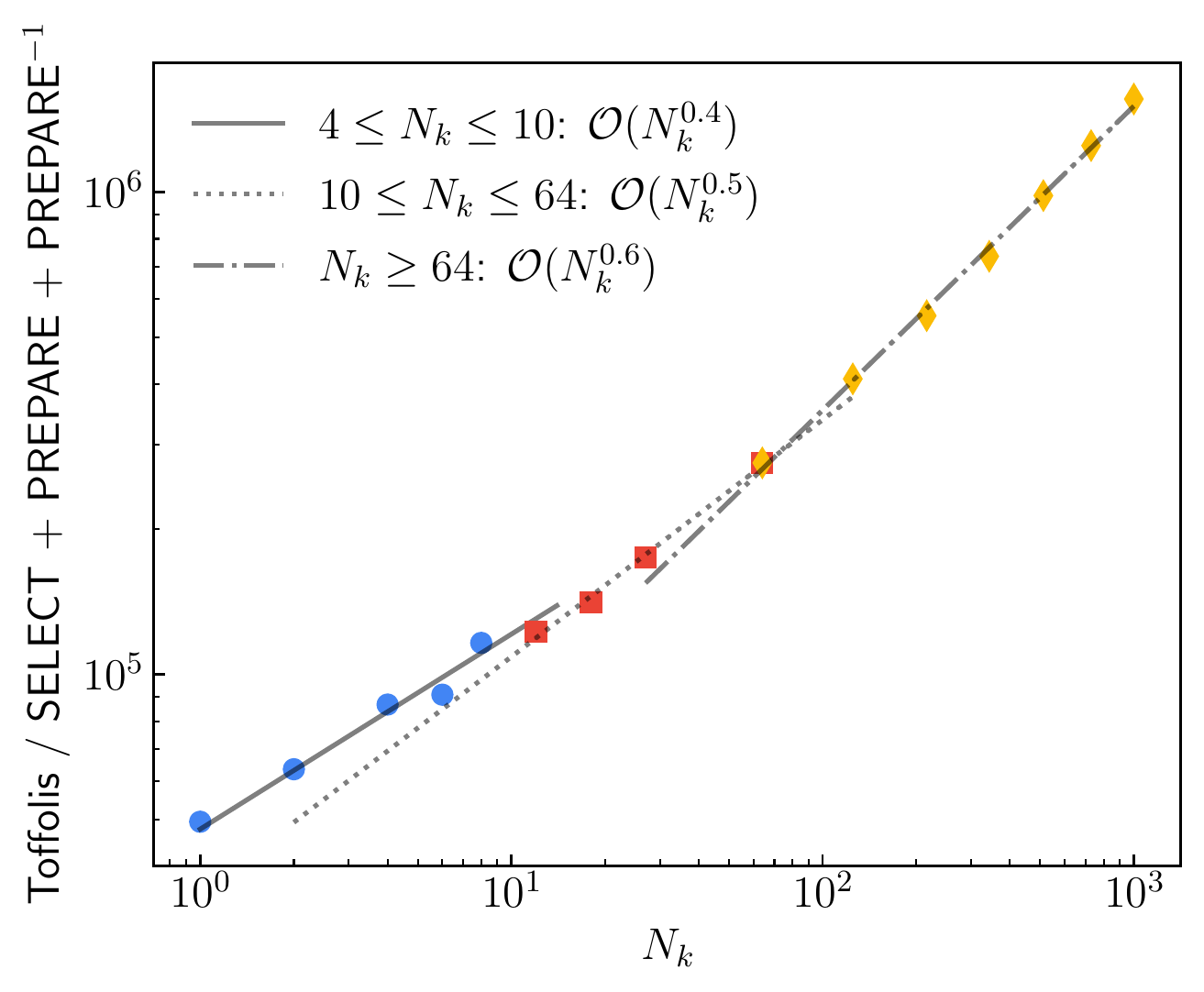} 
    \caption{ Synthetic data for the number of Toffolis required to implement the qubitization oracles  with the $k$-THC factorization demonstrating the challenge of extracting the correct asymptotic scaling with limited finite size data. To generate the data we used the system parameters of carbon diamond in the cc-pVDZ basis set ($N=52$, $M=208$) \label{fig:thc_asymptotics}}
\end{figure}

The $\lambda_{\rm THC}$ value has a one-body component and two-body component.  Unlike molecular THC where the two-body component is reduced because we evolve by number operators in the non-orthogonal basis, in this version of the THC algorithm we will evolve by ladder operators in a non-orthogonal basis, and thus there is no one-body part to remove. The one-body contribution to $\lambda_{\rm THC}$, $\lambda_{{\rm THC},1}$, is computed in a similar way as for the double factorization algorithm but noting that the extra factor of $1/2$ coming from the $Z$ operator is no-longer present to cancel the factor of two from spin summing. The one-body contribution to $\lambda_{\rm THC}$ is
\begin{align}
\lambda_{{\rm THC},1} &= 2\sum_{\K}\sum_{p}|\lambda_{p,\K}|.
\end{align}
The two-body contribution to $\lambda_{\rm THC}$, $\lambda_{{\rm THC},2}$, is determined by summing over all unitaries in the LCU. This summation can be rewritten in the form
\begin{align}
\lambda_{{\rm THC},2}
&= 2 \sum_{\Q}\sum_{\mu,\nu}\sum_{\G_1,\G_2}\left(|\mathrm{Re}[\zeta_{\mu\nu}^{\Q, \G_1, \G_2}]| + |\mathrm{Im}[\zeta_{\mu\nu}^{\Q, \G_1, \G_2}]|\right) \nn
& \quad \times\left(\sum_{\K|\G_{\K,\K\modmin\Q} = \G_1} \nrm_{\K,\mu}\nrm_{\K\modmin\Q,\mu}\right) 
\left(\sum_{\K'|\G_{\K',\K'-\Q} = \G_2} \nrm_{\Kp\modmin\Q,\nu}\nrm_{\Kp,\nu}\right)  \label{eq:lambda_kthc}
\end{align}
using the expression for $\zeta$ described in Eq.~\eqref{eq:kpoint_thc_zeta_q}.

To obtain resource estimates for THC with $k$-points we follow a similar procedure to previous molecular work \cite{Lee2020} and first compress the rank of the THC factors ($M = c_{\mathrm{THC}} N/2$, where $c_\mathrm{THC}$ is the THC rank parameter).
In particular, we use the interpolative separable density fitting (ISDF) approach \cite{Lu2016,Hu2017,Dong2018} as a starting point before subsequently reoptimizing these factors in order to compress the THC rank while regularizing $\lambda$ \cite{Lee2020,Goints_pnas.2203533119}, which we will call $k$-THC.
Further details of this procedure are provided in \cref{app:kthc}.
In \cref{fig:kthc_mp2_error} we demonstrate that a $c_\mathrm{THC}=8$ is sufficient to obtain MP2 correlation energies within approximately $0.1$ mHa/Cell for a subset of the systems considered in the benchmark set. 
We note that the equivalent ISDF rank may be on the order of 10-15 for comparable accuracy, which would correspond to a much larger value for $\lambda$.

\cref{fig:thc_resource_comparison} (a) demonstrates a $\sqrt{N_k}$ scaling improvement of the block encodings in the symmetry adapted case.
Note that this $\sqrt{N_k}$ speedup for the block encodings is partially a finite size effect. In \cref{fig:thc_asymptotics} we plot the Toffoli complexity per step as a function of $N_k$ using artificially generated data to explore the large $N_k$ behavior. 
We see that depending on the fitting range employed the extracted asymptotic scaling trends towards linear.
While ultimately both the symmetry-adapted and supercell encodings should scale linearly with the system size due to the cost of unary iteration over all basis states, there are several factors that yield a $\sqrt{N_k}$ saving in the symmetry-adapted case, and the relative size of the prefactors becomes important.
Similar to DF, we find from \cref{fig:thc_resource_comparison} (b) that $\lambda$ in the symmetry-adapted setting exhibits slightly worse scaling than for supercell calculations. 
This worsening of $\lambda$ in the symmetry-adapted case can be understood again as a reduction in variational freedom in the symmetry adapated case, leading to smaller compression.
Note that while \cref{eq:lambda_kthc} nominally scales cubicly with $N_k$, we expect each individual matrix element to decay like $N_k^{-1}$, which yields the expected quadratic dependence of $\lambda$, or a linear dependence of $\lambda$ when targeting the total energy per cell. 
In the supercell case, there are simply $M^2 = (N_k N)^2$ elements in the central tensor, which in turn controls the scaling of $\lambda$.
From \cref{tab:qubitization_factorized_cost_table} and \cref{fig:thc_resource_comparison}  we can conclude that there is asymptotically no advantage to incorporating symmetry in the THC factorization for the Toffoli complexity, with both the supercell and symmetry-adapted methods exhibiting approximately quadratic scaling with system size for a fixed target accuracy of the total energy per cell.

%%%%%%%%%%%%%%%%%%%%%%%%%%%%%%%%%%%%%%%%%%%%%%%%%%%%%%%%%%%%%%%%%%%%%%%%%%%%%%
\section{Scaling comparison and runtimes for diamond}\label{sec:scaling_comparison}
\begin{figure}[H]
    \centering
    \begin{picture}(400,175)
    \put(-50,0){\includegraphics[width=8.5cm]{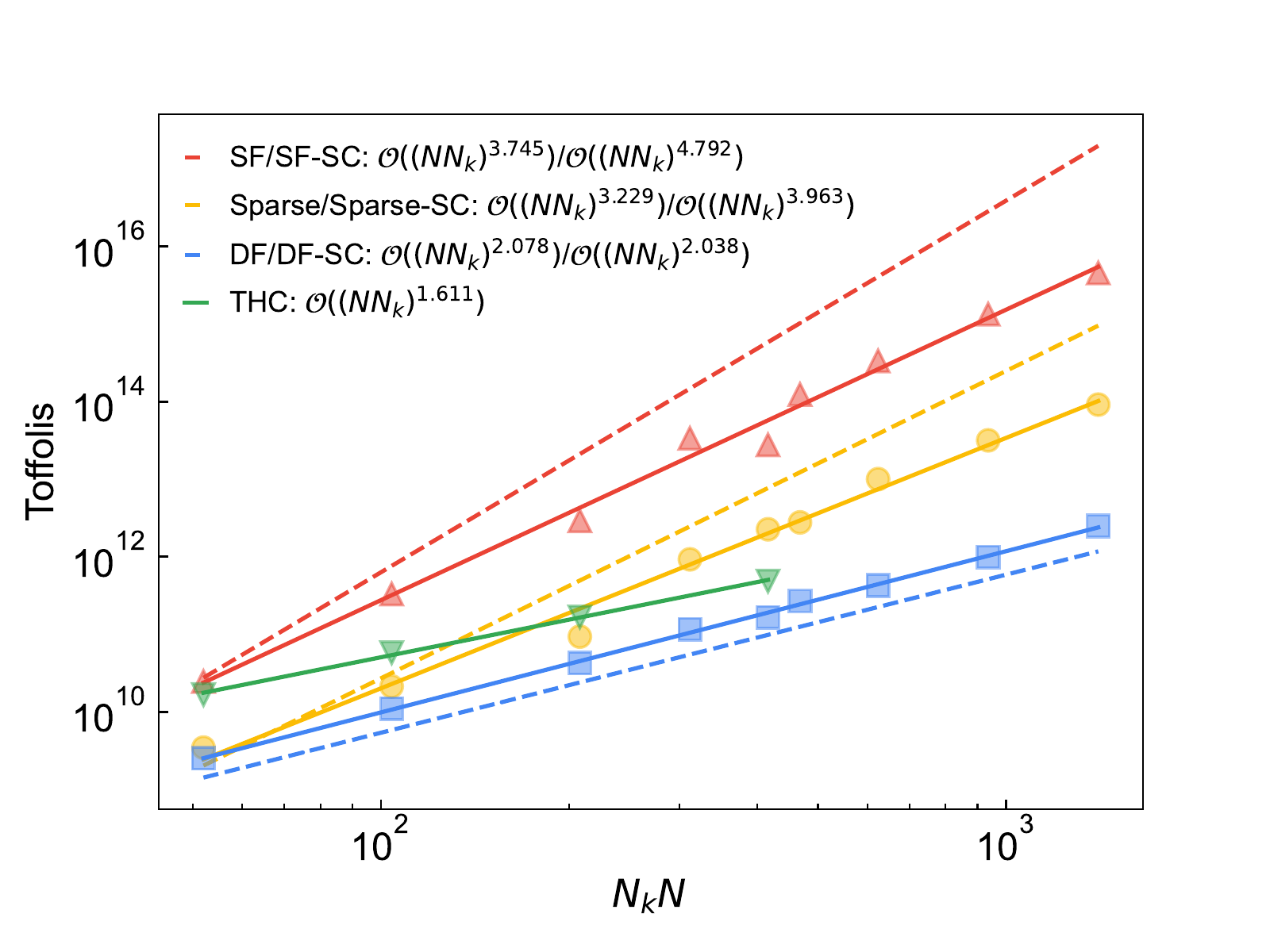}}
    \put(200,0){\includegraphics[width=8.5cm]{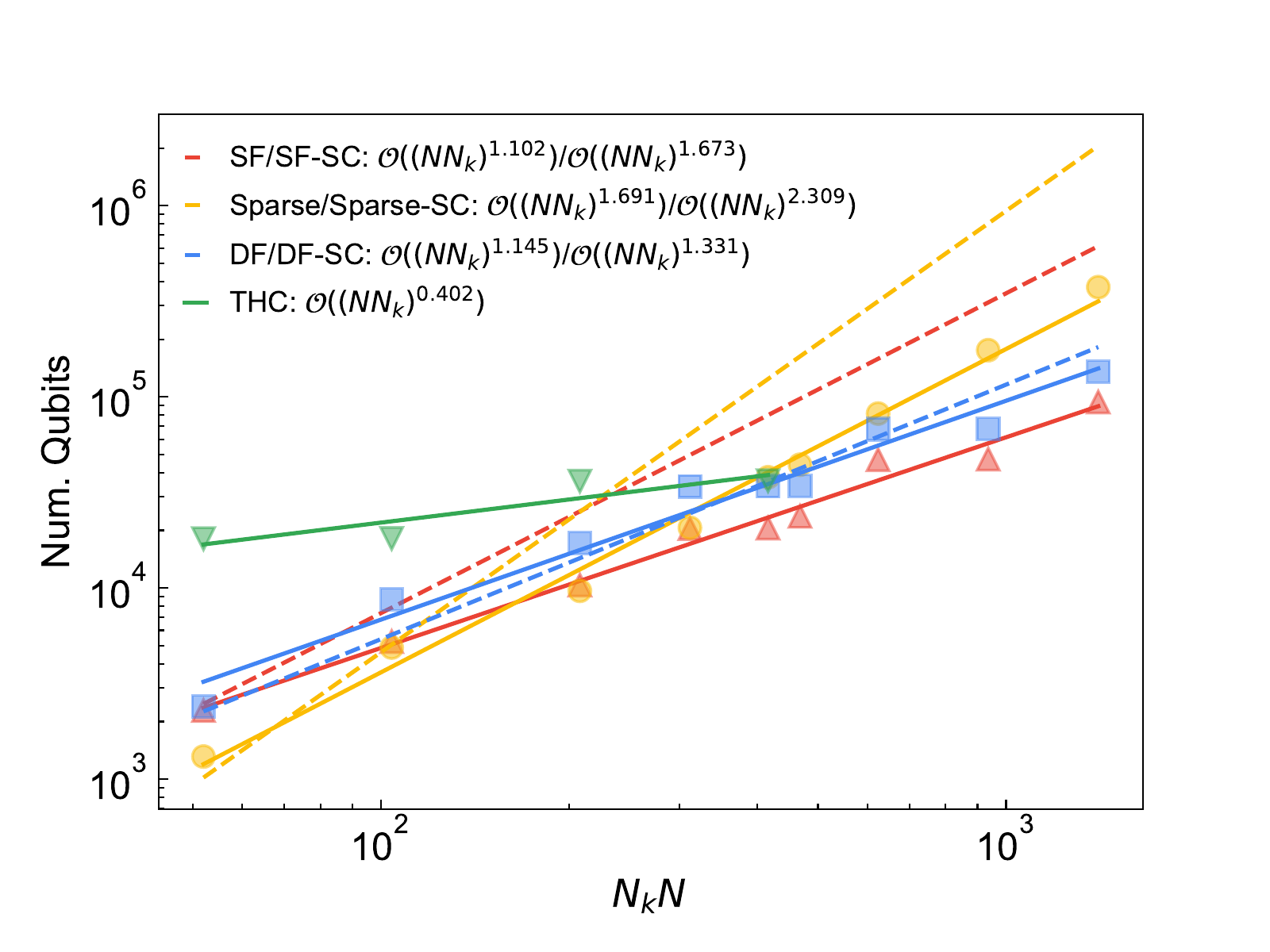}}
    \put(-10,165){(a)}
    \put(240,165){(b)}
    \end{picture} 
    \caption{(a) Total Toffoli requirements for Diamond in a cc-pVDZ basis at various Monkhorst-Pack samplings of the Brillouin zone with $\Gamma$-point centered grids of size [1,1,1] to [3, 3, 3]. Dashed lines are fits to the supercell data that is not plotted. Solid lines are fits to the symmetry-adapted data shown as data points.  (b) Total logical qubits for symmetry-adapted oracles and supercell (dotted lines). All values are estimated from 0.1 mHa per unit cell thresholds on the MP2 energy.
    In the case of THC we only plot the symmetry adapted data  due limited THC data arising from difficulty in optimizing the supercell THC factors.
    \label{fig:diamond_all_methods}}
\end{figure}

We now compare runtimes and estimate total physical requirements to simulate Diamond as a representative material.  In Figure~\ref{fig:diamond_all_methods} we plot the total Toffoli complexity for the sparse, SF, DF, and THC LCUs using symmetry-adapted block encodings and supercell calculations for Diamond with cc-pVDZ and cc-pVTZ basis sets at various Monkhorst-Pack samplings.  In sparse and SF there is a clear asymptotic separation between supercell and symmetry-adapted Toffoli counts. This is expected from the fact that both block encoding constructions are asymptotically improved and $\lambda$ does not increase.  For the DF case, total Toffoli complexity for supercell and symmetry-adapted cases is similar due to the larger $\lambda$ for the symmetry-adapted algorithm.
For THC, the total Toffoli complexity is similar in the supercell and symmetry adapted case, but the asymptotic scaling is identical for the supercell and symmetry-adapted algorithms.
This is due to the increase in $\lambda$ for the symmetry-adapted algorithm.

In Table~\ref{tab:diamond_physical_costs} we tabulate the quantum resource requirements and estimated runtimes after compiling into a surface code using physical qubits with error rates of $0.01$\% and a \SI{1}{\micro\second} cycle time.  We assume four Toffoli factories similar to References~\cite{Lee2020} and~\cite{Goints_pnas.2203533119} and observe that for systems with 52-1404 spin-orbitals the quantum resource estimates are roughly in line with extrapolated estimates from the molecular algorithms.

It is important to note that while the THC resource requirements look competitive for these small systems, in its current form it is not a practical way to simulate materials at scale.
This is due to the prohibitive cost of reoptimizing the THC factors which significantly limits the system sizes that can be simulated.
Moreover, as discussed in \cref{sec_sub:thc_qubitization}, we caution that the THC trend lines are only valid within the fitting range, and we expect that asymptotic THC Toffoli count will trend more towards $\mathcal{O}(N_k^2)$ in the thermodynamic limit.
\begin{table}[H]
    \centering
    \begin{tabular}{ccccD{.}{.}{5}c}
    \hline \hline 
    LCU & $k$-mesh & Toffolis & Logical Qubits & \multicolumn{1}{c}{Physical Qubits[M]} & \multicolumn{1}{c}{Surface Code Runtime [days]} \\
    \hline
    sparse & $[1, 1, 1]$ & $4.84 \times  10^{9}$ &  2478 &  2.20 & $9.10 \times  10^{-1}$ \\
& $[2, 2, 2]$ & $2.66 \times  10^{12}$ &  75287 &  90.57 & $5.77 \times  10^{2}$ \\
& $[3, 3, 3]$ & $1.06 \times  10^{14}$ &  374274 &  543.76 & $2.61 \times  10^{4}$ \\
SF & $[1, 1, 1]$ & $3.20 \times  10^{9}$ &  2283 &  2.05 & $6.02 \times  10^{-1}$ \\
 & $[2, 2, 2]$ & $3.27 \times  10^{12}$ &  20567 &  24.91 & $7.11 \times  10^{2}$ \\
 & $[3, 3, 3]$ & $1.13 \times  10^{15}$ &  47665 &  69.52 & $3.10 \times  10^{5}$ \\
DF & $[1, 1, 1]$ & $9.61 \times  10^{8}$ &  2396 &  1.55 & $1.81 \times  10^{-1}$ \\
 & $[2, 2, 2]$ & $6.74 \times  10^{10}$ &  18693 &  18.47 & $1.27 \times  10^{1}$ \\
 & $[3, 3, 3]$ & $1.09 \times  10^{12}$ &  68470 &  82.39 & $2.37 \times  10^{2}$ \\
    THC & $[1, 1, 1]$ & $1.67 \times  10^{10}$ &  18095 &  14.20 & $3.14$ \\
    & $[2, 2, 2]$ & $4.85 \times  10^{11}$ &  36393 &  35.60 & $1.05 \times 10^{2}$ \\
      \hline \hline
    \end{tabular}
    \caption{Diamond represented in a cc-pVDZ basis (52 spin-orbitals in the primitive cell) at various $k$-mesh sizes and the associated quantum resource requirements to compute the total energy per cell to within 1 kcal / mol. The surface code runtime is estimated using four T-factories, a physical error rate of $0.01$\%, and a cycle time of \SI{1}{\micro\second}. The physical qubit count is given in millions. 
    \label{tab:diamond_physical_costs}}
\end{table}

\section{Classical and quantum simulations of LNO}\label{sec:LNOcompare}
In this section, we compare modern classical computational methods with quantum resource estimates in the context of a challenging problem of industrial interest: the ground state of LiNiO$_2$.

\subsection{LNO background}\label{sec_sub:context_LNO}
Layered oxides have been the most popular cathode active materials for Li-ion batteries since their commercialization in the early `90s.
While LiCoO$_2$ is still the material of choice in the electronics industry, the increasing human, environmental and financial cost of cobalt spells out the need for cobalt-free cathode active materials, especially for automotive applications\cite{basf,Olivetti2017}.

The isostructural compound LiNiO$_2$ (LNO) had been identified as an ideal replacement for LiCoO$_2$ already in the `90s, due to its comparably high theoretical capacity at a lower cost \cite{Dahn1990,Ohzuku}.
Despite its numerous drawbacks, LNO still serves as the perfect model system for many derivative compounds such as lithium nickel-cobalt-manganese (NCM) and lithium nickel-cobalt-aluminum oxides (NCA) that are nowadays the gold standard in the automotive industry  \cite{Bianchini}.
Moreover, the constant demand for better performing materials pushes the amount of substituted Ni to the dilute regime and the research trend is approaching the asymptotic LiNiO$_2$ limit, making LiNiO$_2$ a system of interest in battery research \cite{Bianchini}.

\begin{figure}[ht]
\centering
\includegraphics[width=0.7\textwidth]{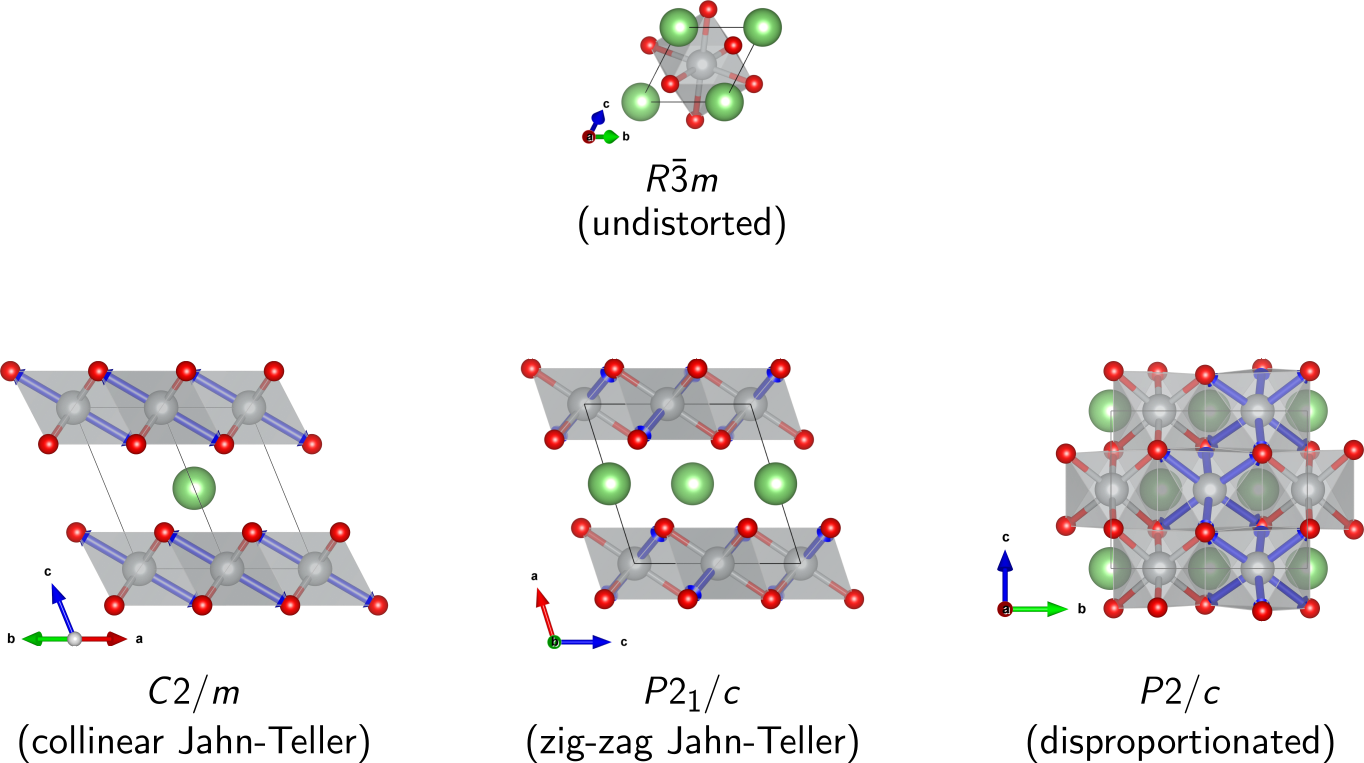}
\caption{\label{fig:gs} The four known LiNiO$_2$ polymorphs: high-symmetry $\mathrm{R\bar{3m}}$, collinear JT-distorted $\mathrm{C2/m}$, zig-zag JT-distorted $\mathrm{P2{_1}/c}$, and
disproportionated $\mathrm{P2/c}$. Green spheres represent Li, gray polyhedra are NiO$_6$ octahedra, and elongated Ni-O bonds are depicted as bold blue arrows.
}
\end{figure}

Even the nature of the ground state of LNO is still under debate.
The universally observed rhombohedral $\mathrm{R\bar3m}$ symmetry \cite{Bianchini}, with Ni being octahedrally coordinated to six oxygen atoms through six equivalent Ni-O bonds conflicts with the renowned Jahn-Teller (JT) activity of low-spin trivalent Ni, which has been experimentally proven on a local scale \cite{Bianchini}.
In a recent DFT study \cite{acs.chemmater.0c03442}, we argued that this apparent discrepancy might be resolved by the dynamics and low spatial correlation of Jahn-Teller distortions.
In that work, the energy distance between Jahn-Teller distorted and non-distorted candidates (Figure \ref{fig:gs}) compared to zero-point vibrational energies makes a strong argument in favor of the dynamic Jahn-Teller effect.
A non-JT distorted structure resulting from the disproportionation of Ni$^{3+}$ has also been reported as a ground state candidate \cite{PhysRevB.84.085108} despite the 1:1 ratio between long and short Ni-O bonds, which conflicts with the experimentally determined 2:1 ratio. 
In the original study, the stability of this structure has been found to depend heavily on the value of the on-site Hubbard correction applied to the PBE functional. 
With the SCAN-rVV10 functional (with and without on-site Hubbard correction) \cite{acs.chemmater.0c03442}, this candidate is consistently less stable than the JT-distorted models; it is also worth mentioning that the on-site Hubbard correction considerably increases the stability of the JT-distorted models.
The dependence of Jahn-Teller stabilization energies on the functional had already been observed by Radin \cite{PhysRevMaterials.4.043601} and is ascribed to the difficulty to adequately describe the doubly degenerate high-symmetry, undistorted state.

In light of previous studies, we will focus on four candidate structures for the LNO ground state. These structures are shown in Figure \ref{fig:gs}. We will furthermore focus only on the energetics of the problem. The goal is to compute the relative energies of these different crystal structures without the uncertainty of DFT.

\subsection{Correlated $k$-point calculations}\label{sec:LNOCor}
\begin{figure}[ht]
\centering
\includegraphics{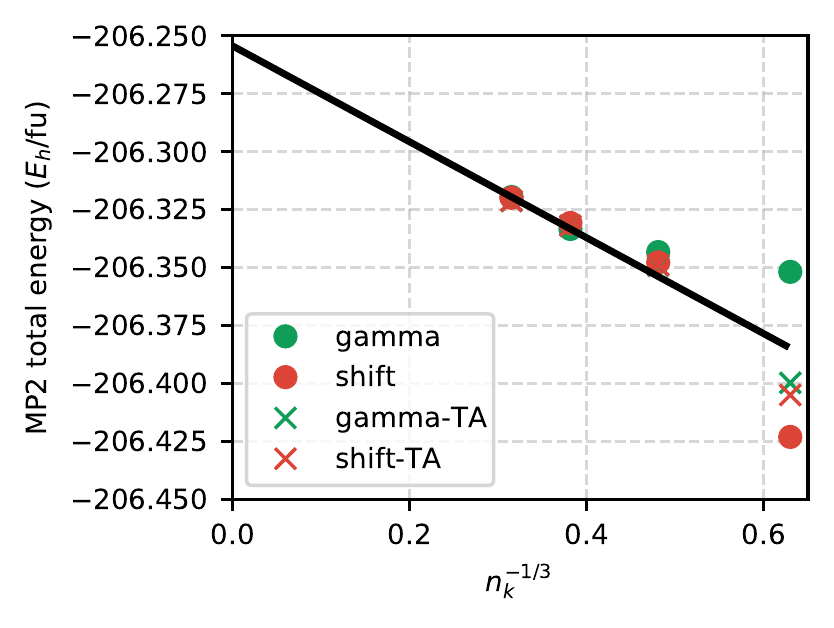}
\caption{\label{fig:MP2_szv} Convergence of the total MP2 energy for the C2/m structure using the minimal basis set as a function of $n_k^{-1/3}$ where $n_k$ is the number of $k$-points. $\Gamma$-centered and shifted $k$-point grids are compared with and without twist averaging (TA).}
\end{figure}

Local-basis quantum chemistry methods for electron correlation have been increasingly applied to periodic systems  as an alternative to DFT with more controllable accuracy. Here we apply two such methods, second order M{\o}ller-Plesset perturbation theory (MP2)\cite{Moller1934,Cremer2011} and coupled cluster singles and doubles (CCSD)\cite{Purvis1982,Bartlett2007}, to the three distorted structures of LNO (Figure \ref{fig:gs}). Local basis methods like these can be directly compared to quantum algorithms described in this work, since both are formulated within the same framework of a crystalline Gaussian one-particle basis. While these methods cannot be easily applied to the symmetric structure, which is metallic at the mean-field level, they should provide accurate results for the distorted structures provided that the finite-size and finite-basis errors can be controlled. All mean field, MP2 and CCSD calculations were performed with the PySCF program package \cite{Sun2018,Sun2020}. QMCPACK \cite{kim2018qmcpack,Kent2020May} was used to perform the ph-AFQMC calculations, where we used at least 600 walkers and a timestep of 0.005 Ha$^{-1}$. The population control bias was found to be negligible. In all calculations, we use separable, norm-conserving GTH pseudopotentials \cite{Goedecker1996,PhysRevB.58.3641} that have been recently optimized for Hartree-Fock \cite{Hutter}. In all calculations on LNO we use the GTH basis sets \cite{Vandevondele2005,VandeVondele2007} (GTH-SZV and GTH-DZVP specifically) that are distributed with the CP2K \cite{Kuhne2020} and PySCF \cite{Sun2020} packages. In Figure~\ref{fig:MP2_szv} we show the convergence of the minimal-basis MP2 energy as a function of effective cell size for increasingly large $k$-point calculations. This demonstrates the essential difficulty in converging to the bulk limit for correlated calculations: the finite-size error will converge with  $n_k^{-1/3}$. Shifting the $k$-point grid to (1/8, 1/8, 1/8) and/or twist averaging (TA) does not change the asymptotic behavior of the energy. In all other LNO calculations, we use $\Gamma$-centered $k$-point grids. In all calculations, the density of $k$-points along each reciprocal lattice vector was chosen so that the density of $k$-points is as close to constant as possible.

While a minimal basis is useful for a qualitative understanding of the finite-size error, it is does not provide sufficient accuracy to resolve the different LNO structures examined in this work. The double-zeta basis set (GTH-DZVP) is large enough to provide qualitative accuracy, but converging the result to the bulk limit is prohibitively expensive for the systems considered here. We can nonetheless provide some estimates of the ground-state CCSD and MP2 (DZVP) energies as shown in Table~\ref{tab:dzvp_results} and~\ref{tab:dzvp_results2}. Since there is no evidence of particularly "strong correlation" in any of these systems (see Appendix~\ref{app:LNOCorrelation} for a more detailed discussion), MP2 and CCSD should provide qualitatively correct estimates of the ground state energy. The unusually large MP2 correlation energy for the P2/c structure suggests it may not be as reliable for this structure, and this suspicion is confirmed by the CCSD and ph-AFQMC calculations. For CCSD and ph-AFQMC, the P2$_1$/c structure is lowest in energy which agrees qualitatively with the DFT calculations in Ref. \cite{acs.chemmater.0c03442}. However, this prediction carries with it a great deal of uncertainty due to the small simulation size, small one-particle basis set, and error in the MP2/CCSD/ph-AFQMC approximations.
\begin{table}
\centering
\begin{minipage}{0.85\textwidth}
\begin{ruledtabular}
\begin{tabular}{cccccc}
structure& $k$-points& ROHF& MP2& CCSD& ph-AFQMC \\ \hline
C2/m& 2x2x1& -206.557491& -0.750524& -0.767350& -0.7997(5)\\
P2$_1$/c& 1x2x1& -206.567049& -0.747717& -0.765445& -0.7982(5)\\
P2/c& 1x1x1& -206.551767& -0.811386& -0.780580& -0.8078(4)\\
\end{tabular}
\end{ruledtabular}
\caption{\label{tab:dzvp_results} ROHF total energy and MP2, CCSD, and ph-AFQMC correlation energies computed with a double-zeta basis set and a small k-mesh, the equivalent of 4 primitive formula units (16 atoms total) for each distorted structure. All units are Hartrees per formula unit.}
\end{minipage}
\end{table}

\begin{table}
\centering
\begin{minipage}{0.85\textwidth}
\begin{ruledtabular}
\begin{tabular}{ccccc}
structure& ROHF& MP2& CCSD& ph-AFQMC\\ \hline
C2/m& 260&	184&	208&	218(18)\\
P2/c&  416& -1317& 4&  155(17)\\
%structure & MP2& CCSD& TA-MP2& TA-CCSD \\ \hline
\end{tabular}
\end{ruledtabular}
\caption{\label{tab:dzvp_results2} Energies in meV relative to the energy of the P2$_1$/c structure for each method. The P2$_1$/c structure is lowest in energy for all methods except MP2.}
\end{minipage}
\end{table}

\subsection{Single shot density matrix embedding theory}
Another way to apply high-level correlated methods to periodic solids is through quantum embedding methods in which a local {\it impurity} is treated with a high-level method and the remainder of the system, the {\it bath}, is treated at a lower level of theory. For periodic solids, dynamical mean-field theory (DMFT) is perhaps the most widely successful such method \cite{Georges1996,Kotliar2006,Held2007,Vollhardt2020,Zhu2020}. Density matrix embedding theory (DMET) is an efficient quantum embedding method for the ground state of quantum systems \cite{Knizia2012,Knizia2013}, and it has recently been applied to periodic solids with a fully {\it ab initio} Hamiltonian \cite{Zhu2020}.

Though very large impurities are necessary to converge to the bulk limit of the correlated method used for the impurity, a fixed impurity size provides a local, systematically improvable approximation to the correlation energy. This is particularly useful in cases where a local treatment of correlation is sufficient for a qualitatively correct solution. Here we apply DMET with a CCSD impurity solver to the distorted structures in a minimal basis set (GTH-SZV). The libdmet code \cite{Cui2020} was used for the DMET calculations with the PySCF program package \cite{Sun2018,Sun2020} used in the impurity solver.

\begin{figure}[ht]
\centering
\includegraphics{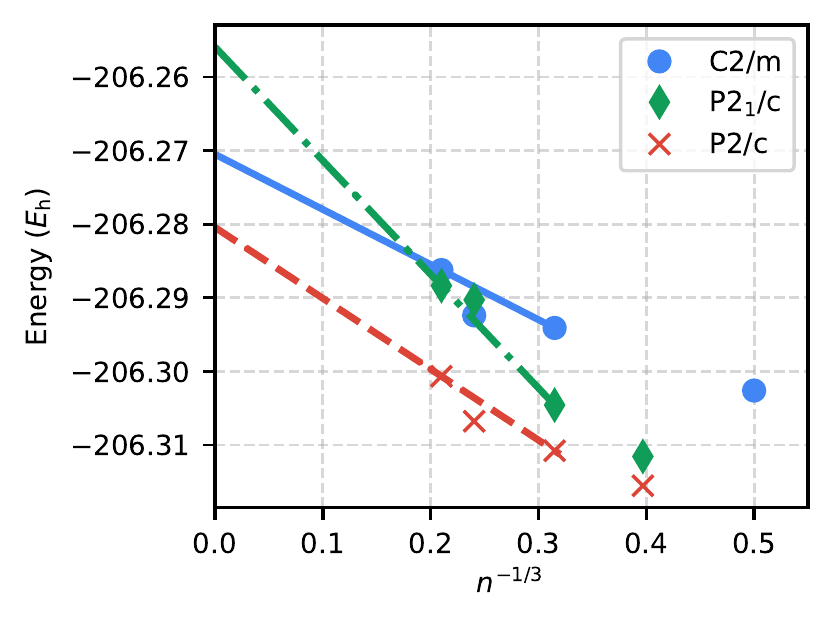}
\caption{\label{fig:DMET} Convergence of the total DMET energy for a 4 formula unit (16 atom) impurity with respect to effective size of the mean-field calculation for the different distorted structures.}
\end{figure}

Figure~\ref{fig:DMET} shows the minimal-basis DMET results with for each of the three distorted structures.
In the context of DMET, we can effectively converge the mean-field part of the problem. Unfortunately, the small one-particle basis set and modest impurity size make it unable to meaningfully resolve these three structure of LiNiO$_2$. Quantum simulation can potentially overcome some of these limits by acting as a lower scaling, unbiased impurity solver~\cite{liu2023bootstrap}.

\subsection{Quantum resource estimates for LNO}
Quantum resource estimates for LNO using the SF and DF LCUs are reported in Table~\ref{tab:LNO_quantum_resource_estimates_V2}.  THC is not reported due to the difficulty of re-optimizing the THC tensors to have low L1-norm as discussed in References~\cite{Lee2020, Goints_pnas.2203533119, oumarou2022accelerating}.  For the sparse LCU, a threshold of $1 \times 10^{-4}$ was determined by averaging the thresholds for the systems in Table~\ref{tab:benchmark_systems} required to achieve 1 m$E_{\mathrm{H}}$ per unit cell. For the SF LCU the truncation of the auxiliary basis was set to eight times the number of molecular orbitals which was determined by requiring the error in the MP2 energy for the smallest C2/m system to be less than 1 m$E_{\mathrm{h}}$ per formula unit. For DF, the same requirement was used to determine a cutoff for the second factorization of $1 \times 10^{-3}$. The trends are consistent with what was observed in Section~\ref{sec:scaling_comparison}: DF is consistently more efficient than either sparse or SF LCUs.

For the smaller systems, these calculations are anticipated to be useful for benchmarking faster classical methods. For the larger systems, the estimated run times are daunting, but we are optimistic that further algorithmic improvements can make calculations like these feasible in the future.

\begin{table}[ht]
    \centering
    \begin{tabular}{cccD{.}{.}{5}ccrD{.}{.}{5}c}
    \hline \hline
       System  & LCU & $k$-mesh & \multicolumn{1}{c}{$\lambda$} & Num. Spin-Orbs. & Toffolis & Logical Qubits & \multicolumn{1}{l}{Physical Qubits [M]} & run time [days]\\
       \hline
       $\mathrm{R\bar3m}$  & Sparse & $[2, 2, 2]$ & 120382.037 &   116 &  6.16$\times 10^{13}$ &   166946 &   242.72 &  1.51$\times 10^{4}$ \\
                           &        & $[3, 3, 3]$ & 718377.133 &   116 &  3.57$\times 10^{15}$ &  1625295 &  2808.82 &  9.82$\times 10^{5}$ \\
                           & SF     & $[2, 2, 2]$ & 183778.821 &   116 &  7.86$\times 10^{13}$ &    89162 &   129.77 &  1.93$\times 10^{4}$ \\
                           &        & $[3, 3, 3]$ & 2966279.293 &  116 &  4.60$\times 10^{15}$ &   404723 &   699.68 &  1.27$\times 10^{6}$ \\
                           & DF     & $[2, 2, 2]$ &   10730.422 &  116 &  4.97$\times 10^{12}$ &   149939 &   180.16 &  1.08$\times 10^{3}$ \\
                           &        & $[3, 3, 3]$ &   44794.803 &  116 &  7.28$\times 10^{13}$ &   598286 &   869.02 &  1.79$\times 10^{4}$ \\                          
       $\mathrm{C2/m}$     & Sparse & $[2, 2, 1]$ &  58422.522 &  116 &  1.03$\times 10^{13}$ &    83532 &   100.47 &  2.53$\times 10^{3}$ \\
                           &        & $[4, 4, 2]$ & 893339.394 &  116 &  5.37$\times 10^{15}$ &  3051285 &  5272.93 &  1.48$\times 10^{6}$ \\
                           & SF     & $[2, 2, 1]$ &  95803.204 &  116 &  2.05$\times 10^{13}$ &    44657 &    53.90 &  5.05$\times 10^{3}$ \\
                           &        & $[4, 4, 2]$ &2899609.300 &  116 &  5.23$\times 10^{15}$ &   405310 &   700.69 &  1.44$\times 10^{6}$ \\
                           & DF     & $[2, 2, 1]$ &   4873.648 &  116 &  1.18$\times 10^{12}$ &    75178 &    90.44 &  2.56$\times 10^{2}$ \\
                           &        & $[4, 4, 2]$ &  51416.281 &  116 &  9.82$\times 10^{13}$ &   598736 &   869.68 &  2.41$\times 10^{4}$ \\
       $\mathrm{P2/c}$     & Sparse & $[1, 1, 1]$ &   84977.359 &  464 &  2.06$\times 10^{13}$ &    99918 &   120.21 &  5.07$\times 10^{3}$ \\
                           &        & $[2, 2, 2]$ & 1627121.892 &  464 &  1.67$\times 10^{16}$ &  3182362 &  6454.14 &  4.59$\times 10^{6}$ \\
                           & SF     & $[1, 1, 1]$ &  201894.726 &  464 &  8.74$\times 10^{13}$ &    92786 &   135.04 &  2.15$\times 10^{4}$ \\
                           &        & $[2, 2, 2]$ & 5666363.179 &  464 &  2.07$\times 10^{16}$ &   839487 &  1450.95 &  5.68$\times 10^{6}$ \\
                           & DF     & $[1, 1, 1]$ &    2753.901 &  464 &  9.72$\times 10^{11}$ &    75834 &    91.23 &  2.11$\times 10^{2}$ \\
                           &        & $[2, 2, 2]$ &   40788.113 &  464 &  1.40$\times 10^{14}$ &  1192900 &  1732.40 &  3.44$\times 10^{4}$ \\
       $\mathrm{P2_{1}/c}$ & Sparse & $[1, 2, 1]$ &  105584.297 &  232 &  3.39$\times 10^{13}$ &   182864 &   265.83 &  8.34$\times 10^{3}$ \\
                           &        & $[2, 4, 2]$ & 1714723.913 &  232 &  1.50$\times 10^{16}$ &  3116825 &  6321.24 &  4.12$\times 10^{6}$ \\
                           & SF     & $[1, 2, 1]$ &  271178.934 &  232 &  8.92$\times 10^{13}$ &    96882 &   140.98 &  2.19$\times 10^{4}$ \\
                           &        & $[2, 4, 2]$ & 7798992.981 &  232 &  2.13$\times 10^{16}$ &   438080 &   757.32 &  5.85$\times 10^{6}$ \\
                           & DF     & $[1, 2, 1]$ &    3958.111 &  232 &  1.27$\times 10^{12}$ &    75383 &    90.69 &  2.76$\times 10^{2}$ \\
                           &        & $[2, 4, 2]$ &   46189.645 &  232 &  1.23$\times 10^{14}$ &  1192758 &  1732.20 &  3.02$\times 10^{4}$ \\
       \hline \hline
    \end{tabular}
    \caption{Quantum Resource estimates for all four LNO structures normalized by the number of formula units represented in each simulation cell. $\mathrm{R\bar3m}$ and $\mathrm{C2/m}$ are both one formula unit while $\mathrm{P2/c}$ is four formula units and $\mathrm{P2_{1}/c}$ is two formula units.  The sparse threshold is selected to be $1.0 \times 10^{-4}$, the SF the auxiliary index is truncated at eight times the number of molecular orbitals, and the DF the second factorization is truncated at $1.0 \times 10^{-4}$. }
    \label{tab:LNO_quantum_resource_estimates_V2}
\end{table}

\section{Conclusion}

In this work we developed the theory of symmetry-adapted block encodings for extended system simulation using four different representations of the Hamiltonian as LCUs in order to improve quantum resource costs for reaching the thermodynamic limit when simulating solids.  In order to realize an asymptotic speedup due to symmetry, we substantially modify the block encodings compared with their molecular counterparts.  To demonstrate these asymptotic improvements we compiled constant factors for all four LCUs and compared their performance on a suite of benchmark systems and a realistic problem in materials simulation. We find that despite a clear asymptotic speedup for walk operator construction there are competing factors (such as lower compression in Hamiltonian tensor factorizations) that make it difficult to observe a large speedup using symmetry. It was recently shown that variationally constructing tensor compressions for Hamiltonian simulation can improve quantum resource requirements~\cite{rubin2022compressing, oumarou2022accelerating, Goints_pnas.2203533119} and thus we believe the compressions can be improved to ultimately demonstrate a speedup for these types of simulations.  

For the sparse and SF LCUs we derive a $\mathcal{O}(\sqrt{N_{k}})$ speedup in constructing {\sel} and {\prep} by ensuring only the minimal amount of symmetry unique information is accessed by the quantum circuit through QROM.  In both cases a speedup is observable, though it is much clearer in the SF case.  Observing the sparse LCU speedup is more challenging due to the difficulty of converging the $N_{k}$ and $N$ dependence of the two-electron integrals. Compared with the molecular case where sparse was competitive with the DF and THC algorithms~\cite{Berry2019B,Lee2020}, largely due to the simplicity of {\sel}, we find that sparse is not viable for converging to the thermodynamic limit of solids.

The DF and THC tensor factorizations yield LCUs as unitaries in non-orthogonal bases and lead to much higher compression than sparse and SF LCUs.  In the DF case we derive an asymptotic $\mathcal{O}(\sqrt{N_{k}})$ improvement in Toffoli complexity and qubit cost when constructing the qubitization walk operator.  Unfortunately, $\lambda$ is increased in these cases.  The increase is attributed to the lower variational freedom in constructing non-orthogonal bases when representing the two-electron integral tensor in factorized form compared with the non-symmetry adapted setting.  For the THC case, no asymptotic speedup is formally possible.  This stems from the linear cost of unary iteration over all basis states.  Nevertheless, due to competing prefactors between unary iteration and state preparation, we do observe a $\sqrt{N_k}$ scaling improvement in the Toffoli per step and logical qubit cost for the range of systems studied.
This is likely a finite size effect, but may be a practically important when considering which algorithm to chose in the future.
Thus, improving the $\lambda$ value of THC through more sophisticated and affordable means is worth further investigation. 

Reaching the thermodynamic and complete basis set limit is very challenging, even for classical wavefunction methods like CCSD and ph-AFQMC.
Previous ph-AFQMC results for simple insulating solids with two-atom unit cells suggest that at least a $3\times3\times3$ and $4\times4\times4$ sampling of the Brillouin zone is required to extrapolate correlation energies to the thermodynamic limit \cite{malone2020accelerating}.
Similarly, it has been found that quadruple-zeta quality basis sets are required to converge the cohesive energy to less than 0.1~eV / atom, while a triple-zeta quality basis is likely sufficient for quantities such as the lattice constant and bulk modulus \cite{morales2020accelerating}. 
Similar system sizes and basis sets were found to be required for CCSD simulations of metallic systems \cite{neufeld2022ccsdmetals}. 
Although the theory of finite size corrections \cite{chiesa2006finite,drummond2008finite,azadi2015systematic,holzmann2016theory} is still an area of active research \cite{dagrada2016exact,mihm2021shortcut}, the simulation of bulk systems even with these corrections typically requires on the order of 50 atoms, which in turn corresponds to hundreds of electrons and thousands of orbitals. 
For excited state properties, particularly those concerning charged excitations, even larger system sizes may be required without the use of sophisticated finite size correction schemes \cite{yang2020electronic}. 
Thus, we suspect that simulating large system sizes will continue to be necessary in order to obtain high accuracy for condensed phase simulations.
It is important to note that high accuracy classical wavefunction methods are often considered too expensive for practical materials simulation, and DFT is still the workhorse of the field.
\cref{app:ccsd_timing} shows that simulating even simple solids with coarse $k$-meshes can take on the order of  hours, which would otherwise take seconds for a modern DFT code.
From the quantum resource estimates it is clear that several orders of magnitude of improvement are necessary before practical materials simulation is possible.
Despite this, the fairly low scaling of phase estimation as a function of system size serves as encouragement to pursue quantum simulation for materials further. 

The aforementioned convergence difficulties are demonstrated in our classical calculations on the LNO system when attempting to resolve the discrepancy between band-theory predictions and experimental observations of the ground state geometry. Furthermore, the variance in energy between CCSD, MP2, ph-AFQMC, and DMET (and their expenses) make it difficult to select an efficient method for determining Hamiltonian parameter cutoffs to use in quantum resource estimation. If anything, this highlights the need for high accuracy classical computation when performing quantum resource estimates and ultimately picking an algorithm for quantum simulation.  The quantum resource estimates for LNO simulations are exorbitantly expensive even at small $k$-mesh; estimated to run in $\mathcal{O}(10^{2})-\mathcal{O}(10^{3})$ days using the DF LCU. Just as resource estimates for chemistry fell drastically with algorithmic developments clearly further algorithmic improvements are needed to make a LNO sized problem feasible on a quantum computer. 

Qubitization is a general tool for Hamiltonian simulation and there may be other simulation scenarios when the improved walk operators yield faster simulations.  There are also areas to further improve the quantum algorithms by taking advantage of space group symmetry along with translational symmetry.  In classical calculations this can lead to substantial computational savings even at the mean-field level.  Just as in the case of quantum algorithms for molecular simulations we expect the quantum resource costs to fall with further exploration of algorithmic improvements.

\section*{Acknowledgements}

The authors thank Yuan Su for helpful conversations. FDM thanks Miguel Morales for helpful discussions on the form of the $k$-point THC factorization. DWB worked on this project under a sponsored research agreement with Google Quantum AI. DWB is also supported by Australian Research Council Discovery Projects DP190102633 and DP210101367.

\bibliographystyle{apsrev4-1}
\bibliography{extra,biblo,addedrefs,Mendeley}

\begin{thebibliography}{122}%
\makeatletter
\providecommand \@ifxundefined [1]{%
 \@ifx{#1\undefined}
}%
\providecommand \@ifnum [1]{%
 \ifnum #1\expandafter \@firstoftwo
 \else \expandafter \@secondoftwo
 \fi
}%
\providecommand \@ifx [1]{%
 \ifx #1\expandafter \@firstoftwo
 \else \expandafter \@secondoftwo
 \fi
}%
\providecommand \natexlab [1]{#1}%
\providecommand \enquote  [1]{``#1''}%
\providecommand \bibnamefont  [1]{#1}%
\providecommand \bibfnamefont [1]{#1}%
\providecommand \citenamefont [1]{#1}%
\providecommand \href@noop [0]{\@secondoftwo}%
\providecommand \href [0]{\begingroup \@sanitize@url \@href}%
\providecommand \@href[1]{\@@startlink{#1}\@@href}%
\providecommand \@@href[1]{\endgroup#1\@@endlink}%
\providecommand \@sanitize@url [0]{\catcode `\\12\catcode `\$12\catcode
  `\&12\catcode `\#12\catcode `\^12\catcode `\_12\catcode `\%12\relax}%
\providecommand \@@startlink[1]{}%
\providecommand \@@endlink[0]{}%
\providecommand \url  [0]{\begingroup\@sanitize@url \@url }%
\providecommand \@url [1]{\endgroup\@href {#1}{\urlprefix }}%
\providecommand \urlprefix  [0]{URL }%
\providecommand \Eprint [0]{\href }%
\providecommand \doibase [0]{http://dx.doi.org/}%
\providecommand \selectlanguage [0]{\@gobble}%
\providecommand \bibinfo  [0]{\@secondoftwo}%
\providecommand \bibfield  [0]{\@secondoftwo}%
\providecommand \translation [1]{[#1]}%
\providecommand \BibitemOpen [0]{}%
\providecommand \bibitemStop [0]{}%
\providecommand \bibitemNoStop [0]{.\EOS\space}%
\providecommand \EOS [0]{\spacefactor3000\relax}%
\providecommand \BibitemShut  [1]{\csname bibitem#1\endcsname}%
\let\auto@bib@innerbib\@empty
%</preamble>
\bibitem [{\citenamefont {Kassal}\ \emph {et~al.}(2008)\citenamefont {Kassal},
  \citenamefont {Jordan}, \citenamefont {Love}, \citenamefont {Mohseni},\ and\
  \citenamefont {Aspuru-Guzik}}]{kassal2008polynomial}%
  \BibitemOpen
  \bibfield  {author} {\bibinfo {author} {\bibfnamefont {I.}~\bibnamefont
  {Kassal}}, \bibinfo {author} {\bibfnamefont {S.~P.}\ \bibnamefont {Jordan}},
  \bibinfo {author} {\bibfnamefont {P.~J.}\ \bibnamefont {Love}}, \bibinfo
  {author} {\bibfnamefont {M.}~\bibnamefont {Mohseni}}, \ and\ \bibinfo
  {author} {\bibfnamefont {A.}~\bibnamefont {Aspuru-Guzik}},\ }\href
  {https://www.pnas.org/doi/abs/10.1073/pnas.0808245105} {\bibfield  {journal}
  {\bibinfo  {journal} {Proceedings of the National Academy of Sciences}\
  }\textbf {\bibinfo {volume} {105}},\ \bibinfo {pages} {18681} (\bibinfo
  {year} {2008})}\BibitemShut {NoStop}%
\bibitem [{\citenamefont {Babbush}\ \emph
  {et~al.}(2018{\natexlab{a}})\citenamefont {Babbush}, \citenamefont {Gidney},
  \citenamefont {Berry}, \citenamefont {Wiebe}, \citenamefont {McClean},
  \citenamefont {Paler}, \citenamefont {Fowler},\ and\ \citenamefont
  {Neven}}]{BabbushSpectra}%
  \BibitemOpen
  \bibfield  {author} {\bibinfo {author} {\bibfnamefont {R.}~\bibnamefont
  {Babbush}}, \bibinfo {author} {\bibfnamefont {C.}~\bibnamefont {Gidney}},
  \bibinfo {author} {\bibfnamefont {D.}~\bibnamefont {Berry}}, \bibinfo
  {author} {\bibfnamefont {N.}~\bibnamefont {Wiebe}}, \bibinfo {author}
  {\bibfnamefont {J.}~\bibnamefont {McClean}}, \bibinfo {author} {\bibfnamefont
  {A.}~\bibnamefont {Paler}}, \bibinfo {author} {\bibfnamefont
  {A.}~\bibnamefont {Fowler}}, \ and\ \bibinfo {author} {\bibfnamefont
  {H.}~\bibnamefont {Neven}},\ }\href
  {https://journals.aps.org/prx/abstract/10.1103/PhysRevX.8.041015} {\bibfield
  {journal} {\bibinfo  {journal} {Physical Review X}\ }\textbf {\bibinfo
  {volume} {8}},\ \bibinfo {pages} {041015} (\bibinfo {year}
  {2018}{\natexlab{a}})}\BibitemShut {NoStop}%
\bibitem [{\citenamefont {Su}\ \emph {et~al.}(2021)\citenamefont {Su},
  \citenamefont {Berry}, \citenamefont {Wiebe}, \citenamefont {Rubin},\ and\
  \citenamefont {Babbush}}]{PRXQuantum.2.040332}%
  \BibitemOpen
  \bibfield  {author} {\bibinfo {author} {\bibfnamefont {Y.}~\bibnamefont
  {Su}}, \bibinfo {author} {\bibfnamefont {D.~W.}\ \bibnamefont {Berry}},
  \bibinfo {author} {\bibfnamefont {N.}~\bibnamefont {Wiebe}}, \bibinfo
  {author} {\bibfnamefont {N.}~\bibnamefont {Rubin}}, \ and\ \bibinfo {author}
  {\bibfnamefont {R.}~\bibnamefont {Babbush}},\ }\href {\doibase
  10.1103/PRXQuantum.2.040332} {\bibfield  {journal} {\bibinfo  {journal} {PRX
  Quantum}\ }\textbf {\bibinfo {volume} {2}},\ \bibinfo {pages} {040332}
  (\bibinfo {year} {2021})}\BibitemShut {NoStop}%
\bibitem [{\citenamefont {Babbush}\ \emph
  {et~al.}(2018{\natexlab{b}})\citenamefont {Babbush}, \citenamefont {Wiebe},
  \citenamefont {McClean}, \citenamefont {McClain}, \citenamefont {Neven},\
  and\ \citenamefont {Chan}}]{PhysRevX.8.011044}%
  \BibitemOpen
  \bibfield  {author} {\bibinfo {author} {\bibfnamefont {R.}~\bibnamefont
  {Babbush}}, \bibinfo {author} {\bibfnamefont {N.}~\bibnamefont {Wiebe}},
  \bibinfo {author} {\bibfnamefont {J.}~\bibnamefont {McClean}}, \bibinfo
  {author} {\bibfnamefont {J.}~\bibnamefont {McClain}}, \bibinfo {author}
  {\bibfnamefont {H.}~\bibnamefont {Neven}}, \ and\ \bibinfo {author}
  {\bibfnamefont {G.~K.-L.}\ \bibnamefont {Chan}},\ }\href {\doibase
  10.1103/PhysRevX.8.011044} {\bibfield  {journal} {\bibinfo  {journal}
  {Physical Review X}\ }\textbf {\bibinfo {volume} {8}},\ \bibinfo {pages}
  {011044} (\bibinfo {year} {2018}{\natexlab{b}})}\BibitemShut {NoStop}%
\bibitem [{\citenamefont {Gr{\"{u}}neis}\ \emph {et~al.}(2013)\citenamefont
  {Gr{\"{u}}neis}, \citenamefont {Shepherd}, \citenamefont {Alavi},
  \citenamefont {Tew},\ and\ \citenamefont {Booth}}]{gruneis2013explicitly}%
  \BibitemOpen
  \bibfield  {author} {\bibinfo {author} {\bibfnamefont {A.}~\bibnamefont
  {Gr{\"{u}}neis}}, \bibinfo {author} {\bibfnamefont {J.~J.}\ \bibnamefont
  {Shepherd}}, \bibinfo {author} {\bibfnamefont {A.}~\bibnamefont {Alavi}},
  \bibinfo {author} {\bibfnamefont {D.~P.}\ \bibnamefont {Tew}}, \ and\
  \bibinfo {author} {\bibfnamefont {G.~H.}\ \bibnamefont {Booth}},\ }\href
  {http://aip.scitation.org/doi/abs/10.1063/1.4818753} {\bibfield  {journal}
  {\bibinfo  {journal} {The Journal of Chemical Physics}\ }\textbf {\bibinfo
  {volume} {139}},\ \bibinfo {pages} {84112} (\bibinfo {year}
  {2013})}\BibitemShut {NoStop}%
\bibitem [{\citenamefont {Kato}(1957)}]{kato1957eigenfunctions}%
  \BibitemOpen
  \bibfield  {author} {\bibinfo {author} {\bibfnamefont {T.}~\bibnamefont
  {Kato}},\ }\href {https://onlinelibrary.wiley.com/doi/10.1002/cpa.3160100201}
  {\bibfield  {journal} {\bibinfo  {journal} {Communications on Pure and
  Applied Mathematics}\ }\textbf {\bibinfo {volume} {10}},\ \bibinfo {pages}
  {151} (\bibinfo {year} {1957})}\BibitemShut {NoStop}%
\bibitem [{\citenamefont {Imada}\ \emph {et~al.}(1998)\citenamefont {Imada},
  \citenamefont {Fujimori},\ and\ \citenamefont {Tokura}}]{RevModPhys.70.1039}%
  \BibitemOpen
  \bibfield  {author} {\bibinfo {author} {\bibfnamefont {M.}~\bibnamefont
  {Imada}}, \bibinfo {author} {\bibfnamefont {A.}~\bibnamefont {Fujimori}}, \
  and\ \bibinfo {author} {\bibfnamefont {Y.}~\bibnamefont {Tokura}},\ }\href
  {\doibase 10.1103/RevModPhys.70.1039} {\bibfield  {journal} {\bibinfo
  {journal} {Reviews of Modern Physics}\ }\textbf {\bibinfo {volume} {70}},\
  \bibinfo {pages} {1039} (\bibinfo {year} {1998})}\BibitemShut {NoStop}%
\bibitem [{\citenamefont {Dagotto}(1994)}]{RevModPhys.66.763}%
  \BibitemOpen
  \bibfield  {author} {\bibinfo {author} {\bibfnamefont {E.}~\bibnamefont
  {Dagotto}},\ }\href {\doibase 10.1103/RevModPhys.66.763} {\bibfield
  {journal} {\bibinfo  {journal} {Reviews of Modern Physics}\ }\textbf
  {\bibinfo {volume} {66}},\ \bibinfo {pages} {763} (\bibinfo {year}
  {1994})}\BibitemShut {NoStop}%
\bibitem [{\citenamefont {Sachdev}(2003)}]{RevModPhys.75.913}%
  \BibitemOpen
  \bibfield  {author} {\bibinfo {author} {\bibfnamefont {S.}~\bibnamefont
  {Sachdev}},\ }\href {\doibase 10.1103/RevModPhys.75.913} {\bibfield
  {journal} {\bibinfo  {journal} {Reviews of Modern Physics}\ }\textbf
  {\bibinfo {volume} {75}},\ \bibinfo {pages} {913} (\bibinfo {year}
  {2003})}\BibitemShut {NoStop}%
\bibitem [{\citenamefont {Pisani}(2003)}]{Pisani2003}%
  \BibitemOpen
  \bibfield  {author} {\bibinfo {author} {\bibfnamefont {C.}~\bibnamefont
  {Pisani}},\ }\href {www.elsevier.com/locate/theochem} {\bibfield  {journal}
  {\bibinfo  {journal} {Journal of Molecular Structure (Theochem)}\ }\textbf
  {\bibinfo {volume} {621}},\ \bibinfo {pages} {141} (\bibinfo {year}
  {2003})}\BibitemShut {NoStop}%
\bibitem [{\citenamefont {Pisani}\ \emph {et~al.}(2005)\citenamefont {Pisani},
  \citenamefont {Busso}, \citenamefont {Capecchi}, \citenamefont {Casassa},
  \citenamefont {Dovesi}, \citenamefont {Maschio}, \citenamefont
  {Zicovich-Wilson},\ and\ \citenamefont {Schütz}}]{Pisani2005}%
  \BibitemOpen
  \bibfield  {author} {\bibinfo {author} {\bibfnamefont {C.}~\bibnamefont
  {Pisani}}, \bibinfo {author} {\bibfnamefont {M.}~\bibnamefont {Busso}},
  \bibinfo {author} {\bibfnamefont {G.}~\bibnamefont {Capecchi}}, \bibinfo
  {author} {\bibfnamefont {S.}~\bibnamefont {Casassa}}, \bibinfo {author}
  {\bibfnamefont {R.}~\bibnamefont {Dovesi}}, \bibinfo {author} {\bibfnamefont
  {L.}~\bibnamefont {Maschio}}, \bibinfo {author} {\bibfnamefont
  {C.}~\bibnamefont {Zicovich-Wilson}}, \ and\ \bibinfo {author} {\bibfnamefont
  {M.}~\bibnamefont {Schütz}},\ }\href {\doibase 10.1063/1.1857479} {\bibfield
   {journal} {\bibinfo  {journal} {Journal of Chemical Physics}\ }\textbf
  {\bibinfo {volume} {122}} (\bibinfo {year} {2005}),\
  10.1063/1.1857479}\BibitemShut {NoStop}%
\bibitem [{\citenamefont {Grüneis}\ \emph {et~al.}(2011)\citenamefont
  {Grüneis}, \citenamefont {Booth}, \citenamefont {Marsman}, \citenamefont
  {Spencer}, \citenamefont {Alavi},\ and\ \citenamefont
  {Kresse}}]{Gruneis2011}%
  \BibitemOpen
  \bibfield  {author} {\bibinfo {author} {\bibfnamefont {A.}~\bibnamefont
  {Grüneis}}, \bibinfo {author} {\bibfnamefont {G.~H.}\ \bibnamefont {Booth}},
  \bibinfo {author} {\bibfnamefont {M.}~\bibnamefont {Marsman}}, \bibinfo
  {author} {\bibfnamefont {J.}~\bibnamefont {Spencer}}, \bibinfo {author}
  {\bibfnamefont {A.}~\bibnamefont {Alavi}}, \ and\ \bibinfo {author}
  {\bibfnamefont {G.}~\bibnamefont {Kresse}},\ }\href {\doibase
  10.1021/ct200263g} {\bibfield  {journal} {\bibinfo  {journal} {Journal of
  Chemical Theory and Computation}\ }\textbf {\bibinfo {volume} {7}},\ \bibinfo
  {pages} {2780} (\bibinfo {year} {2011})}\BibitemShut {NoStop}%
\bibitem [{\citenamefont {Ben}\ \emph {et~al.}(2012)\citenamefont {Ben},
  \citenamefont {Hutter},\ and\ \citenamefont {Vandevondele}}]{DelBen2012}%
  \BibitemOpen
  \bibfield  {author} {\bibinfo {author} {\bibfnamefont {M.~D.}\ \bibnamefont
  {Ben}}, \bibinfo {author} {\bibfnamefont {J.}~\bibnamefont {Hutter}}, \ and\
  \bibinfo {author} {\bibfnamefont {J.}~\bibnamefont {Vandevondele}},\ }\href
  {\doibase 10.1021/CT300531W/SUPPL_FILE/CT300531W_SI_001.PDF} {\bibfield
  {journal} {\bibinfo  {journal} {Journal of Chemical Theory and Computation}\
  }\textbf {\bibinfo {volume} {8}},\ \bibinfo {pages} {4177} (\bibinfo {year}
  {2012})}\BibitemShut {NoStop}%
\bibitem [{\citenamefont {Ben}\ \emph {et~al.}(2013)\citenamefont {Ben},
  \citenamefont {Hutter},\ and\ \citenamefont {Vandevondele}}]{DelBen2013}%
  \BibitemOpen
  \bibfield  {author} {\bibinfo {author} {\bibfnamefont {M.~D.}\ \bibnamefont
  {Ben}}, \bibinfo {author} {\bibfnamefont {J.}~\bibnamefont {Hutter}}, \ and\
  \bibinfo {author} {\bibfnamefont {J.}~\bibnamefont {Vandevondele}},\ }\href
  {\doibase 10.1021/ct4002202} {\bibfield  {journal} {\bibinfo  {journal}
  {Journal of Chemical Theory and Computation}\ }\textbf {\bibinfo {volume}
  {9}},\ \bibinfo {pages} {2654} (\bibinfo {year} {2013})}\BibitemShut
  {NoStop}%
\bibitem [{\citenamefont {Booth}\ \emph {et~al.}(2013)\citenamefont {Booth},
  \citenamefont {Grüneis}, \citenamefont {Kresse},\ and\ \citenamefont
  {Alavi}}]{Booth2013}%
  \BibitemOpen
  \bibfield  {author} {\bibinfo {author} {\bibfnamefont {G.~H.}\ \bibnamefont
  {Booth}}, \bibinfo {author} {\bibfnamefont {A.}~\bibnamefont {Grüneis}},
  \bibinfo {author} {\bibfnamefont {G.}~\bibnamefont {Kresse}}, \ and\ \bibinfo
  {author} {\bibfnamefont {A.}~\bibnamefont {Alavi}},\ }\href {\doibase
  10.1038/nature11770} {\bibfield  {journal} {\bibinfo  {journal} {Nature}\
  }\textbf {\bibinfo {volume} {493}},\ \bibinfo {pages} {365} (\bibinfo {year}
  {2013})}\BibitemShut {NoStop}%
\bibitem [{\citenamefont {Booth}\ \emph {et~al.}(2016)\citenamefont {Booth},
  \citenamefont {Tsatsoulis}, \citenamefont {Chan},\ and\ \citenamefont
  {Gr{\"u}neis}}]{Booth2016}%
  \BibitemOpen
  \bibfield  {author} {\bibinfo {author} {\bibfnamefont {G.~H.}\ \bibnamefont
  {Booth}}, \bibinfo {author} {\bibfnamefont {T.}~\bibnamefont {Tsatsoulis}},
  \bibinfo {author} {\bibfnamefont {G.~K.~L.}\ \bibnamefont {Chan}}, \ and\
  \bibinfo {author} {\bibfnamefont {A.}~\bibnamefont {Gr{\"u}neis}},\ }\href
  {\doibase 10.1063/1.4961301} {\bibfield  {journal} {\bibinfo  {journal}
  {Journal of Chemical Physics}\ }\textbf {\bibinfo {volume} {145}} (\bibinfo
  {year} {2016}),\ 10.1063/1.4961301}\BibitemShut {NoStop}%
\bibitem [{\citenamefont {McClain}\ \emph {et~al.}(2017)\citenamefont
  {McClain}, \citenamefont {Sun}, \citenamefont {Chan},\ and\ \citenamefont
  {Berkelbach}}]{Mcclain2017}%
  \BibitemOpen
  \bibfield  {author} {\bibinfo {author} {\bibfnamefont {J.}~\bibnamefont
  {McClain}}, \bibinfo {author} {\bibfnamefont {Q.}~\bibnamefont {Sun}},
  \bibinfo {author} {\bibfnamefont {G.~K.~L.}\ \bibnamefont {Chan}}, \ and\
  \bibinfo {author} {\bibfnamefont {T.~C.}\ \bibnamefont {Berkelbach}},\ }\href
  {\doibase 10.1021/acs.jctc.7b00049} {\bibfield  {journal} {\bibinfo
  {journal} {Journal of Chemical Theory and Computation}\ }\textbf {\bibinfo
  {volume} {13}},\ \bibinfo {pages} {1209} (\bibinfo {year}
  {2017})}\BibitemShut {NoStop}%
\bibitem [{\citenamefont {Neufeld}\ \emph {et~al.}(2022)\citenamefont
  {Neufeld}, \citenamefont {Ye},\ and\ \citenamefont
  {Berkelbach}}]{neufeld2022ccsdmetals}%
  \BibitemOpen
  \bibfield  {author} {\bibinfo {author} {\bibfnamefont {V.~A.}\ \bibnamefont
  {Neufeld}}, \bibinfo {author} {\bibfnamefont {H.-Z.}\ \bibnamefont {Ye}}, \
  and\ \bibinfo {author} {\bibfnamefont {T.~C.}\ \bibnamefont {Berkelbach}},\
  }\href {https://doi.org/10.1021/acs.jpclett.2c01828} {\bibfield  {journal}
  {\bibinfo  {journal} {The Journal of Physical Chemistry Letters}\ }\textbf
  {\bibinfo {volume} {13}},\ \bibinfo {pages} {7497} (\bibinfo {year}
  {2022})}\BibitemShut {NoStop}%
\bibitem [{\citenamefont {Cui}\ \emph {et~al.}(2020)\citenamefont {Cui},
  \citenamefont {Zhu},\ and\ \citenamefont {Chan}}]{Cui2020}%
  \BibitemOpen
  \bibfield  {author} {\bibinfo {author} {\bibfnamefont {Z.~H.}\ \bibnamefont
  {Cui}}, \bibinfo {author} {\bibfnamefont {T.}~\bibnamefont {Zhu}}, \ and\
  \bibinfo {author} {\bibfnamefont {G.~K.~L.}\ \bibnamefont {Chan}},\ }\href
  {\doibase 10.1021/acs.jctc.9b00934} {\bibfield  {journal} {\bibinfo
  {journal} {Journal of Chemical Theory and Computation}\ }\textbf {\bibinfo
  {volume} {16}},\ \bibinfo {pages} {119–129} (\bibinfo {year}
  {2020})}\BibitemShut {NoStop}%
\bibitem [{\citenamefont {Zhu}\ \emph {et~al.}(2020)\citenamefont {Zhu},
  \citenamefont {Cui},\ and\ \citenamefont {Chan}}]{Zhu2020}%
  \BibitemOpen
  \bibfield  {author} {\bibinfo {author} {\bibfnamefont {T.}~\bibnamefont
  {Zhu}}, \bibinfo {author} {\bibfnamefont {Z.~H.}\ \bibnamefont {Cui}}, \ and\
  \bibinfo {author} {\bibfnamefont {G.~K.~L.}\ \bibnamefont {Chan}},\ }\href
  {\doibase 10.1021/acs.jctc.9b00934} {\bibfield  {journal} {\bibinfo
  {journal} {Journal of Chemical Theory and Computation}\ }\textbf {\bibinfo
  {volume} {16}},\ \bibinfo {pages} {141} (\bibinfo {year} {2020})}\BibitemShut
  {NoStop}%
\bibitem [{\citenamefont {Cui}\ \emph {et~al.}(2022)\citenamefont {Cui},
  \citenamefont {Zhai}, \citenamefont {Zhang},\ and\ \citenamefont
  {Chan}}]{cui2022systematic}%
  \BibitemOpen
  \bibfield  {author} {\bibinfo {author} {\bibfnamefont {Z.-H.}\ \bibnamefont
  {Cui}}, \bibinfo {author} {\bibfnamefont {H.}~\bibnamefont {Zhai}}, \bibinfo
  {author} {\bibfnamefont {X.}~\bibnamefont {Zhang}}, \ and\ \bibinfo {author}
  {\bibfnamefont {G.~K.-L.}\ \bibnamefont {Chan}},\ }\href {\doibase
  10.1126/science.abm2295} {\bibfield  {journal} {\bibinfo  {journal}
  {Science}\ }\textbf {\bibinfo {volume} {377}},\ \bibinfo {pages} {1192}
  (\bibinfo {year} {2022})}\BibitemShut {NoStop}%
\bibitem [{\citenamefont {Motta}\ \emph {et~al.}(2019)\citenamefont {Motta},
  \citenamefont {Zhang},\ and\ \citenamefont {Chan}}]{PhysRevB.100.045127}%
  \BibitemOpen
  \bibfield  {author} {\bibinfo {author} {\bibfnamefont {M.}~\bibnamefont
  {Motta}}, \bibinfo {author} {\bibfnamefont {S.}~\bibnamefont {Zhang}}, \ and\
  \bibinfo {author} {\bibfnamefont {G.~K.-L.}\ \bibnamefont {Chan}},\ }\href
  {\doibase 10.1103/PhysRevB.100.045127} {\bibfield  {journal} {\bibinfo
  {journal} {Physical Review B}\ }\textbf {\bibinfo {volume} {100}},\ \bibinfo
  {pages} {045127} (\bibinfo {year} {2019})}\BibitemShut {NoStop}%
\bibitem [{\citenamefont {Cotton}(1991)}]{cotton1991chemical}%
  \BibitemOpen
  \bibfield  {author} {\bibinfo {author} {\bibfnamefont {F.~A.}\ \bibnamefont
  {Cotton}},\ }\href@noop {} {\emph {\bibinfo {title} {Chemical applications of
  group theory}}}\ (\bibinfo  {publisher} {John Wiley \& Sons},\ \bibinfo
  {year} {1991})\BibitemShut {NoStop}%
\bibitem [{\citenamefont {Crawford}\ and\ \citenamefont
  {Di~Remigio}(2019)}]{crawford2019tensor}%
  \BibitemOpen
  \bibfield  {author} {\bibinfo {author} {\bibfnamefont {T.~D.}\ \bibnamefont
  {Crawford}}\ and\ \bibinfo {author} {\bibfnamefont {R.}~\bibnamefont
  {Di~Remigio}},\ }in\ \href
  {https://www.sciencedirect.com/science/article/abs/pii/S1574140019300052}
  {\emph {\bibinfo {booktitle} {Annual Reports in Computational Chemistry}}},\
  Vol.~\bibinfo {volume} {15}\ (\bibinfo  {publisher} {Elsevier},\ \bibinfo
  {year} {2019})\ pp.\ \bibinfo {pages} {79--101}\BibitemShut {NoStop}%
\bibitem [{\citenamefont {Georges}\ \emph
  {et~al.}(1996{\natexlab{a}})\citenamefont {Georges}, \citenamefont {Kotliar},
  \citenamefont {Krauth},\ and\ \citenamefont {Rozenberg}}]{RevModPhys.68.13}%
  \BibitemOpen
  \bibfield  {author} {\bibinfo {author} {\bibfnamefont {A.}~\bibnamefont
  {Georges}}, \bibinfo {author} {\bibfnamefont {G.}~\bibnamefont {Kotliar}},
  \bibinfo {author} {\bibfnamefont {W.}~\bibnamefont {Krauth}}, \ and\ \bibinfo
  {author} {\bibfnamefont {M.~J.}\ \bibnamefont {Rozenberg}},\ }\href {\doibase
  10.1103/RevModPhys.68.13} {\bibfield  {journal} {\bibinfo  {journal} {Rev.
  Mod. Phys.}\ }\textbf {\bibinfo {volume} {68}},\ \bibinfo {pages} {13}
  (\bibinfo {year} {1996}{\natexlab{a}})}\BibitemShut {NoStop}%
\bibitem [{\citenamefont {Cortona}(1991)}]{PhysRevB.44.8454}%
  \BibitemOpen
  \bibfield  {author} {\bibinfo {author} {\bibfnamefont {P.}~\bibnamefont
  {Cortona}},\ }\href {\doibase 10.1103/PhysRevB.44.8454} {\bibfield  {journal}
  {\bibinfo  {journal} {Phys. Rev. B}\ }\textbf {\bibinfo {volume} {44}},\
  \bibinfo {pages} {8454} (\bibinfo {year} {1991})}\BibitemShut {NoStop}%
\bibitem [{\citenamefont {Inglesfield}(1981)}]{inglesfield1981method}%
  \BibitemOpen
  \bibfield  {author} {\bibinfo {author} {\bibfnamefont {J.}~\bibnamefont
  {Inglesfield}},\ }\href@noop {} {\bibfield  {journal} {\bibinfo  {journal}
  {Journal of Physics C: Solid State Physics}\ }\textbf {\bibinfo {volume}
  {14}},\ \bibinfo {pages} {3795} (\bibinfo {year} {1981})}\BibitemShut
  {NoStop}%
\bibitem [{\citenamefont {Knizia}\ and\ \citenamefont
  {Chan}(2012{\natexlab{a}})}]{PhysRevLett.109.186404}%
  \BibitemOpen
  \bibfield  {author} {\bibinfo {author} {\bibfnamefont {G.}~\bibnamefont
  {Knizia}}\ and\ \bibinfo {author} {\bibfnamefont {G.~K.-L.}\ \bibnamefont
  {Chan}},\ }\href {\doibase 10.1103/PhysRevLett.109.186404} {\bibfield
  {journal} {\bibinfo  {journal} {Phys. Rev. Lett.}\ }\textbf {\bibinfo
  {volume} {109}},\ \bibinfo {pages} {186404} (\bibinfo {year}
  {2012}{\natexlab{a}})}\BibitemShut {NoStop}%
\bibitem [{\citenamefont {Cui}\ \emph {et~al.}(2019)\citenamefont {Cui},
  \citenamefont {Zhu},\ and\ \citenamefont {Chan}}]{cui2019efficient}%
  \BibitemOpen
  \bibfield  {author} {\bibinfo {author} {\bibfnamefont {Z.-H.}\ \bibnamefont
  {Cui}}, \bibinfo {author} {\bibfnamefont {T.}~\bibnamefont {Zhu}}, \ and\
  \bibinfo {author} {\bibfnamefont {G.~K.-L.}\ \bibnamefont {Chan}},\ }\href
  {https://pubs.acs.org/doi/10.1021/acs.jctc.9b00933} {\bibfield  {journal}
  {\bibinfo  {journal} {Journal of Chemical Theory and Computation}\ }\textbf
  {\bibinfo {volume} {16}},\ \bibinfo {pages} {119} (\bibinfo {year}
  {2019})}\BibitemShut {NoStop}%
\bibitem [{\citenamefont {Pham}\ \emph {et~al.}(2019)\citenamefont {Pham},
  \citenamefont {Hermes},\ and\ \citenamefont {Gagliardi}}]{pham2019periodic}%
  \BibitemOpen
  \bibfield  {author} {\bibinfo {author} {\bibfnamefont {H.~Q.}\ \bibnamefont
  {Pham}}, \bibinfo {author} {\bibfnamefont {M.~R.}\ \bibnamefont {Hermes}}, \
  and\ \bibinfo {author} {\bibfnamefont {L.}~\bibnamefont {Gagliardi}},\ }\href
  {https://pubs.acs.org/doi/abs/10.1021/acs.jctc.9b00939} {\bibfield  {journal}
  {\bibinfo  {journal} {Journal of Chemical Theory and Computation}\ }\textbf
  {\bibinfo {volume} {16}},\ \bibinfo {pages} {130} (\bibinfo {year}
  {2019})}\BibitemShut {NoStop}%
\bibitem [{\citenamefont {Zheng}\ \emph {et~al.}(2018)\citenamefont {Zheng},
  \citenamefont {Changlani}, \citenamefont {Williams}, \citenamefont
  {Busemeyer},\ and\ \citenamefont {Wagner}}]{zheng2018real}%
  \BibitemOpen
  \bibfield  {author} {\bibinfo {author} {\bibfnamefont {H.}~\bibnamefont
  {Zheng}}, \bibinfo {author} {\bibfnamefont {H.~J.}\ \bibnamefont
  {Changlani}}, \bibinfo {author} {\bibfnamefont {K.~T.}\ \bibnamefont
  {Williams}}, \bibinfo {author} {\bibfnamefont {B.}~\bibnamefont {Busemeyer}},
  \ and\ \bibinfo {author} {\bibfnamefont {L.~K.}\ \bibnamefont {Wagner}},\
  }\href {https://www.frontiersin.org/articles/10.3389/fphy.2018.00043/full}
  {\bibfield  {journal} {\bibinfo  {journal} {Frontiers in Physics}\ }\textbf
  {\bibinfo {volume} {6}},\ \bibinfo {pages} {43} (\bibinfo {year}
  {2018})}\BibitemShut {NoStop}%
\bibitem [{\citenamefont {Lee}\ \emph {et~al.}(2021)\citenamefont {Lee},
  \citenamefont {Berry}, \citenamefont {Gidney}, \citenamefont {Huggins},
  \citenamefont {McClean}, \citenamefont {Wiebe},\ and\ \citenamefont
  {Babbush}}]{Lee2020}%
  \BibitemOpen
  \bibfield  {author} {\bibinfo {author} {\bibfnamefont {J.}~\bibnamefont
  {Lee}}, \bibinfo {author} {\bibfnamefont {D.~W.}\ \bibnamefont {Berry}},
  \bibinfo {author} {\bibfnamefont {C.}~\bibnamefont {Gidney}}, \bibinfo
  {author} {\bibfnamefont {W.~J.}\ \bibnamefont {Huggins}}, \bibinfo {author}
  {\bibfnamefont {J.~R.}\ \bibnamefont {McClean}}, \bibinfo {author}
  {\bibfnamefont {N.}~\bibnamefont {Wiebe}}, \ and\ \bibinfo {author}
  {\bibfnamefont {R.}~\bibnamefont {Babbush}},\ }\href {\doibase
  10.1103/PRXQuantum.2.030305} {\bibfield  {journal} {\bibinfo  {journal} {PRX
  Quantum}\ }\textbf {\bibinfo {volume} {2}},\ \bibinfo {pages} {030305}
  (\bibinfo {year} {2021})}\BibitemShut {NoStop}%
\bibitem [{\citenamefont {Low}\ and\ \citenamefont
  {Chuang}(2019)}]{low2019hamiltonian}%
  \BibitemOpen
  \bibfield  {author} {\bibinfo {author} {\bibfnamefont {G.~H.}\ \bibnamefont
  {Low}}\ and\ \bibinfo {author} {\bibfnamefont {I.~L.}\ \bibnamefont
  {Chuang}},\ }\href {\doibase 10.22331/q-2019-07-12-163} {\bibfield  {journal}
  {\bibinfo  {journal} {Quantum}\ }\textbf {\bibinfo {volume} {3}},\ \bibinfo
  {pages} {163} (\bibinfo {year} {2019})}\BibitemShut {NoStop}%
\bibitem [{\citenamefont {Ivanov}\ \emph {et~al.}(2022)\citenamefont {Ivanov},
  \citenamefont {S{\"u}nderhauf}, \citenamefont {Holzmann}, \citenamefont
  {Ellaby}, \citenamefont {Kerber}, \citenamefont {Jones},\ and\ \citenamefont
  {Camps}}]{ivanov2022quantum}%
  \BibitemOpen
  \bibfield  {author} {\bibinfo {author} {\bibfnamefont {A.~V.}\ \bibnamefont
  {Ivanov}}, \bibinfo {author} {\bibfnamefont {C.}~\bibnamefont
  {S{\"u}nderhauf}}, \bibinfo {author} {\bibfnamefont {N.}~\bibnamefont
  {Holzmann}}, \bibinfo {author} {\bibfnamefont {T.}~\bibnamefont {Ellaby}},
  \bibinfo {author} {\bibfnamefont {R.~N.}\ \bibnamefont {Kerber}}, \bibinfo
  {author} {\bibfnamefont {G.}~\bibnamefont {Jones}}, \ and\ \bibinfo {author}
  {\bibfnamefont {J.}~\bibnamefont {Camps}},\ }\href
  {https://arxiv.org/abs/2210.02403} {\bibfield  {journal} {\bibinfo  {journal}
  {arXiv:2210.02403}\ } (\bibinfo {year} {2022})}\BibitemShut {NoStop}%
\bibitem [{\citenamefont {Reiher}\ \emph {et~al.}(2017)\citenamefont {Reiher},
  \citenamefont {Wiebe}, \citenamefont {Svore}, \citenamefont {Wecker},\ and\
  \citenamefont {Troyer}}]{Reiher2017}%
  \BibitemOpen
  \bibfield  {author} {\bibinfo {author} {\bibfnamefont {M.}~\bibnamefont
  {Reiher}}, \bibinfo {author} {\bibfnamefont {N.}~\bibnamefont {Wiebe}},
  \bibinfo {author} {\bibfnamefont {K.~M.}\ \bibnamefont {Svore}}, \bibinfo
  {author} {\bibfnamefont {D.}~\bibnamefont {Wecker}}, \ and\ \bibinfo {author}
  {\bibfnamefont {M.}~\bibnamefont {Troyer}},\ }\href
  {http://www.pnas.org/content/114/29/7555.abstract} {\bibfield  {journal}
  {\bibinfo  {journal} {Proceedings of the National Academy of Sciences}\
  }\textbf {\bibinfo {volume} {114}},\ \bibinfo {pages} {7555} (\bibinfo {year}
  {2017})}\BibitemShut {NoStop}%
\bibitem [{\citenamefont {von Burg}\ \emph {et~al.}(2021)\citenamefont {von
  Burg}, \citenamefont {Low}, \citenamefont {H{\"{a}}ner}, \citenamefont
  {Steiger}, \citenamefont {Reiher}, \citenamefont {Roetteler},\ and\
  \citenamefont {Troyer}}]{vonBurg2020}%
  \BibitemOpen
  \bibfield  {author} {\bibinfo {author} {\bibfnamefont {V.}~\bibnamefont {von
  Burg}}, \bibinfo {author} {\bibfnamefont {G.~H.}\ \bibnamefont {Low}},
  \bibinfo {author} {\bibfnamefont {T.}~\bibnamefont {H{\"{a}}ner}}, \bibinfo
  {author} {\bibfnamefont {D.~S.}\ \bibnamefont {Steiger}}, \bibinfo {author}
  {\bibfnamefont {M.}~\bibnamefont {Reiher}}, \bibinfo {author} {\bibfnamefont
  {M.}~\bibnamefont {Roetteler}}, \ and\ \bibinfo {author} {\bibfnamefont
  {M.}~\bibnamefont {Troyer}},\ }\href
  {https://journals.aps.org/prresearch/abstract/10.1103/PhysRevResearch.3.033055}
  {\bibfield  {journal} {\bibinfo  {journal} {Physical Review Research}\
  }\textbf {\bibinfo {volume} {3}},\ \bibinfo {pages} {033055} (\bibinfo {year}
  {2021})}\BibitemShut {NoStop}%
\bibitem [{\citenamefont {Goings}\ \emph {et~al.}(2022)\citenamefont {Goings},
  \citenamefont {White}, \citenamefont {Lee}, \citenamefont {Tautermann},
  \citenamefont {Degroote}, \citenamefont {Gidney}, \citenamefont {Shiozaki},
  \citenamefont {Babbush},\ and\ \citenamefont
  {Rubin}}]{Goints_pnas.2203533119}%
  \BibitemOpen
  \bibfield  {author} {\bibinfo {author} {\bibfnamefont {J.~J.}\ \bibnamefont
  {Goings}}, \bibinfo {author} {\bibfnamefont {A.}~\bibnamefont {White}},
  \bibinfo {author} {\bibfnamefont {J.}~\bibnamefont {Lee}}, \bibinfo {author}
  {\bibfnamefont {C.~S.}\ \bibnamefont {Tautermann}}, \bibinfo {author}
  {\bibfnamefont {M.}~\bibnamefont {Degroote}}, \bibinfo {author}
  {\bibfnamefont {C.}~\bibnamefont {Gidney}}, \bibinfo {author} {\bibfnamefont
  {T.}~\bibnamefont {Shiozaki}}, \bibinfo {author} {\bibfnamefont
  {R.}~\bibnamefont {Babbush}}, \ and\ \bibinfo {author} {\bibfnamefont
  {N.~C.}\ \bibnamefont {Rubin}},\ }\href {\doibase 10.1073/pnas.2203533119}
  {\bibfield  {journal} {\bibinfo  {journal} {Proceedings of the National
  Academy of Sciences}\ }\textbf {\bibinfo {volume} {119}},\ \bibinfo {pages}
  {e2203533119} (\bibinfo {year} {2022})}\BibitemShut {NoStop}%
\bibitem [{\citenamefont {Bianchini}\ \emph {et~al.}(2019)\citenamefont
  {Bianchini}, \citenamefont {Roca-Ayats}, \citenamefont {Hartmann},
  \citenamefont {Brezesinski},\ and\ \citenamefont {Janek}}]{Bianchini}%
  \BibitemOpen
  \bibfield  {author} {\bibinfo {author} {\bibfnamefont {M.}~\bibnamefont
  {Bianchini}}, \bibinfo {author} {\bibfnamefont {M.}~\bibnamefont
  {Roca-Ayats}}, \bibinfo {author} {\bibfnamefont {P.}~\bibnamefont
  {Hartmann}}, \bibinfo {author} {\bibfnamefont {T.}~\bibnamefont
  {Brezesinski}}, \ and\ \bibinfo {author} {\bibfnamefont {J.}~\bibnamefont
  {Janek}},\ }\href {\doibase 10.1002/anie.201812472} {\bibfield  {journal}
  {\bibinfo  {journal} {Angewandte Chemie Int.\,Ed.}\ }\textbf {\bibinfo
  {volume} {58}},\ \bibinfo {pages} {2} (\bibinfo {year} {2019})}\BibitemShut
  {NoStop}%
\bibitem [{\citenamefont {Sicolo}\ \emph {et~al.}(2020)\citenamefont {Sicolo},
  \citenamefont {Mock}, \citenamefont {Bianchini},\ and\ \citenamefont
  {Albe}}]{acs.chemmater.0c03442}%
  \BibitemOpen
  \bibfield  {author} {\bibinfo {author} {\bibfnamefont {S.}~\bibnamefont
  {Sicolo}}, \bibinfo {author} {\bibfnamefont {M.}~\bibnamefont {Mock}},
  \bibinfo {author} {\bibfnamefont {M.}~\bibnamefont {Bianchini}}, \ and\
  \bibinfo {author} {\bibfnamefont {K.}~\bibnamefont {Albe}},\ }\href {\doibase
  https://doi.org/10.1021/acs.chemmater.0c03442} {\bibfield  {journal}
  {\bibinfo  {journal} {Chemistry of Materials}\ }\textbf {\bibinfo {volume}
  {32}},\ \bibinfo {pages} {10096} (\bibinfo {year} {2020})}\BibitemShut
  {NoStop}%
\bibitem [{\citenamefont {Chen}\ \emph {et~al.}(2011)\citenamefont {Chen},
  \citenamefont {Freeman},\ and\ \citenamefont {Harding}}]{PhysRevB.84.085108}%
  \BibitemOpen
  \bibfield  {author} {\bibinfo {author} {\bibfnamefont {H.}~\bibnamefont
  {Chen}}, \bibinfo {author} {\bibfnamefont {C.~L.}\ \bibnamefont {Freeman}}, \
  and\ \bibinfo {author} {\bibfnamefont {J.~H.}\ \bibnamefont {Harding}},\
  }\href {\doibase 10.1103/PhysRevB.84.085108} {\bibfield  {journal} {\bibinfo
  {journal} {Physical Review B}\ }\textbf {\bibinfo {volume} {84}},\ \bibinfo
  {pages} {085108} (\bibinfo {year} {2011})}\BibitemShut {NoStop}%
\bibitem [{\citenamefont {Vandevondele}\ \emph {et~al.}(2005)\citenamefont
  {Vandevondele}, \citenamefont {Krack}, \citenamefont {Mohamed}, \citenamefont
  {Parrinello}, \citenamefont {Chassaing},\ and\ \citenamefont
  {Hutter}}]{Vandevondele2005}%
  \BibitemOpen
  \bibfield  {author} {\bibinfo {author} {\bibfnamefont {J.}~\bibnamefont
  {Vandevondele}}, \bibinfo {author} {\bibfnamefont {M.}~\bibnamefont {Krack}},
  \bibinfo {author} {\bibfnamefont {F.}~\bibnamefont {Mohamed}}, \bibinfo
  {author} {\bibfnamefont {M.}~\bibnamefont {Parrinello}}, \bibinfo {author}
  {\bibfnamefont {T.}~\bibnamefont {Chassaing}}, \ and\ \bibinfo {author}
  {\bibfnamefont {J.}~\bibnamefont {Hutter}},\ }\href {\doibase
  10.1016/j.cpc.2004.12.014} {\bibfield  {journal} {\bibinfo  {journal}
  {Computer Physics Communications}\ }\textbf {\bibinfo {volume} {167}},\
  \bibinfo {pages} {103} (\bibinfo {year} {2005})}\BibitemShut {NoStop}%
\bibitem [{\citenamefont {Kühne}\ \emph {et~al.}(2020)\citenamefont {Kühne},
  \citenamefont {Iannuzzi}, \citenamefont {Ben}, \citenamefont {Rybkin},
  \citenamefont {Seewald}, \citenamefont {Stein}, \citenamefont {Laino},
  \citenamefont {Khaliullin}, \citenamefont {Schütt}, \citenamefont
  {Schiffmann}, \citenamefont {Golze}, \citenamefont {Wilhelm}, \citenamefont
  {Chulkov}, \citenamefont {Bani-Hashemian}, \citenamefont {Weber},
  \citenamefont {Borštnik}, \citenamefont {Taillefumier}, \citenamefont
  {Jakobovits}, \citenamefont {Lazzaro}, \citenamefont {Pabst}, \citenamefont
  {Müller}, \citenamefont {Schade}, \citenamefont {Guidon}, \citenamefont
  {Andermatt}, \citenamefont {Holmberg}, \citenamefont {Schenter},
  \citenamefont {Hehn}, \citenamefont {Bussy}, \citenamefont {Belleflamme},
  \citenamefont {Tabacchi}, \citenamefont {Glöß}, \citenamefont {Lass},
  \citenamefont {Bethune}, \citenamefont {Mundy}, \citenamefont {Plessl},
  \citenamefont {Watkins}, \citenamefont {VandeVondele}, \citenamefont
  {Krack},\ and\ \citenamefont {Hutter}}]{Kuhne2020}%
  \BibitemOpen
  \bibfield  {author} {\bibinfo {author} {\bibfnamefont {T.~D.}\ \bibnamefont
  {Kühne}}, \bibinfo {author} {\bibfnamefont {M.}~\bibnamefont {Iannuzzi}},
  \bibinfo {author} {\bibfnamefont {M.~D.}\ \bibnamefont {Ben}}, \bibinfo
  {author} {\bibfnamefont {V.~V.}\ \bibnamefont {Rybkin}}, \bibinfo {author}
  {\bibfnamefont {P.}~\bibnamefont {Seewald}}, \bibinfo {author} {\bibfnamefont
  {F.}~\bibnamefont {Stein}}, \bibinfo {author} {\bibfnamefont
  {T.}~\bibnamefont {Laino}}, \bibinfo {author} {\bibfnamefont {R.~Z.}\
  \bibnamefont {Khaliullin}}, \bibinfo {author} {\bibfnamefont
  {O.}~\bibnamefont {Schütt}}, \bibinfo {author} {\bibfnamefont
  {F.}~\bibnamefont {Schiffmann}}, \bibinfo {author} {\bibfnamefont
  {D.}~\bibnamefont {Golze}}, \bibinfo {author} {\bibfnamefont
  {J.}~\bibnamefont {Wilhelm}}, \bibinfo {author} {\bibfnamefont
  {S.}~\bibnamefont {Chulkov}}, \bibinfo {author} {\bibfnamefont {M.~H.}\
  \bibnamefont {Bani-Hashemian}}, \bibinfo {author} {\bibfnamefont
  {V.}~\bibnamefont {Weber}}, \bibinfo {author} {\bibfnamefont
  {U.}~\bibnamefont {Borštnik}}, \bibinfo {author} {\bibfnamefont
  {M.}~\bibnamefont {Taillefumier}}, \bibinfo {author} {\bibfnamefont {A.~S.}\
  \bibnamefont {Jakobovits}}, \bibinfo {author} {\bibfnamefont
  {A.}~\bibnamefont {Lazzaro}}, \bibinfo {author} {\bibfnamefont
  {H.}~\bibnamefont {Pabst}}, \bibinfo {author} {\bibfnamefont
  {T.}~\bibnamefont {Müller}}, \bibinfo {author} {\bibfnamefont
  {R.}~\bibnamefont {Schade}}, \bibinfo {author} {\bibfnamefont
  {M.}~\bibnamefont {Guidon}}, \bibinfo {author} {\bibfnamefont
  {S.}~\bibnamefont {Andermatt}}, \bibinfo {author} {\bibfnamefont
  {N.}~\bibnamefont {Holmberg}}, \bibinfo {author} {\bibfnamefont {G.~K.}\
  \bibnamefont {Schenter}}, \bibinfo {author} {\bibfnamefont {A.}~\bibnamefont
  {Hehn}}, \bibinfo {author} {\bibfnamefont {A.}~\bibnamefont {Bussy}},
  \bibinfo {author} {\bibfnamefont {F.}~\bibnamefont {Belleflamme}}, \bibinfo
  {author} {\bibfnamefont {G.}~\bibnamefont {Tabacchi}}, \bibinfo {author}
  {\bibfnamefont {A.}~\bibnamefont {Glöß}}, \bibinfo {author} {\bibfnamefont
  {M.}~\bibnamefont {Lass}}, \bibinfo {author} {\bibfnamefont {I.}~\bibnamefont
  {Bethune}}, \bibinfo {author} {\bibfnamefont {C.~J.}\ \bibnamefont {Mundy}},
  \bibinfo {author} {\bibfnamefont {C.}~\bibnamefont {Plessl}}, \bibinfo
  {author} {\bibfnamefont {M.}~\bibnamefont {Watkins}}, \bibinfo {author}
  {\bibfnamefont {J.}~\bibnamefont {VandeVondele}}, \bibinfo {author}
  {\bibfnamefont {M.}~\bibnamefont {Krack}}, \ and\ \bibinfo {author}
  {\bibfnamefont {J.}~\bibnamefont {Hutter}},\ }\href {\doibase
  10.1063/5.0007045} {\bibfield  {journal} {\bibinfo  {journal} {Journal of
  Chemical Physics}\ }\textbf {\bibinfo {volume} {152}},\ \bibinfo {pages}
  {194103} (\bibinfo {year} {2020})}\BibitemShut {NoStop}%
\bibitem [{\citenamefont {Dovesi}\ \emph {et~al.}(2020)\citenamefont {Dovesi},
  \citenamefont {Pascale}, \citenamefont {Civalleri}, \citenamefont {Doll},
  \citenamefont {Harrison}, \citenamefont {Bush}, \citenamefont {D'Arco},
  \citenamefont {Noel}, \citenamefont {Rera}, \citenamefont {Carbonniere},
  \citenamefont {Causa}, \citenamefont {Salustro}, \citenamefont {Lacivita},
  \citenamefont {Kirtman}, \citenamefont {Ferrari}, \citenamefont {Gentile},
  \citenamefont {Baima}, \citenamefont {Ferrero}, \citenamefont {Demichelis},\
  and\ \citenamefont {Pierre}}]{Dovesi2020}%
  \BibitemOpen
  \bibfield  {author} {\bibinfo {author} {\bibfnamefont {R.}~\bibnamefont
  {Dovesi}}, \bibinfo {author} {\bibfnamefont {F.}~\bibnamefont {Pascale}},
  \bibinfo {author} {\bibfnamefont {B.}~\bibnamefont {Civalleri}}, \bibinfo
  {author} {\bibfnamefont {K.}~\bibnamefont {Doll}}, \bibinfo {author}
  {\bibfnamefont {N.~M.}\ \bibnamefont {Harrison}}, \bibinfo {author}
  {\bibfnamefont {I.}~\bibnamefont {Bush}}, \bibinfo {author} {\bibfnamefont
  {P.}~\bibnamefont {D'Arco}}, \bibinfo {author} {\bibfnamefont
  {Y.}~\bibnamefont {Noel}}, \bibinfo {author} {\bibfnamefont {M.}~\bibnamefont
  {Rera}}, \bibinfo {author} {\bibfnamefont {P.}~\bibnamefont {Carbonniere}},
  \bibinfo {author} {\bibfnamefont {M.}~\bibnamefont {Causa}}, \bibinfo
  {author} {\bibfnamefont {S.}~\bibnamefont {Salustro}}, \bibinfo {author}
  {\bibfnamefont {V.}~\bibnamefont {Lacivita}}, \bibinfo {author}
  {\bibfnamefont {B.}~\bibnamefont {Kirtman}}, \bibinfo {author} {\bibfnamefont
  {A.~M.}\ \bibnamefont {Ferrari}}, \bibinfo {author} {\bibfnamefont {F.~S.}\
  \bibnamefont {Gentile}}, \bibinfo {author} {\bibfnamefont {J.}~\bibnamefont
  {Baima}}, \bibinfo {author} {\bibfnamefont {M.}~\bibnamefont {Ferrero}},
  \bibinfo {author} {\bibfnamefont {R.}~\bibnamefont {Demichelis}}, \ and\
  \bibinfo {author} {\bibfnamefont {M.~D.~L.}\ \bibnamefont {Pierre}},\ }\href
  {\doibase 10.1063/5.0004892} {\bibfield  {journal} {\bibinfo  {journal}
  {Journal of Chemical Physics}\ }\textbf {\bibinfo {volume} {152}},\ \bibinfo
  {pages} {204111 (2020);} (\bibinfo {year} {2020})}\BibitemShut {NoStop}%
\bibitem [{\citenamefont {Guidon}\ \emph {et~al.}(2009)\citenamefont {Guidon},
  \citenamefont {Hutter},\ and\ \citenamefont {VandeVondele}}]{Guidon2009}%
  \BibitemOpen
  \bibfield  {author} {\bibinfo {author} {\bibfnamefont {M.}~\bibnamefont
  {Guidon}}, \bibinfo {author} {\bibfnamefont {J.}~\bibnamefont {Hutter}}, \
  and\ \bibinfo {author} {\bibfnamefont {J.}~\bibnamefont {VandeVondele}},\
  }\href {\doibase 10.1021/ct900494g} {\bibfield  {journal} {\bibinfo
  {journal} {Journal of Chemical Theory and Computation}\ }\textbf {\bibinfo
  {volume} {5}},\ \bibinfo {pages} {3010} (\bibinfo {year} {2009})}\BibitemShut
  {NoStop}%
\bibitem [{\citenamefont {Guidon}\ \emph {et~al.}(2010)\citenamefont {Guidon},
  \citenamefont {Hutter},\ and\ \citenamefont {VandeVondele}}]{Guidon2010}%
  \BibitemOpen
  \bibfield  {author} {\bibinfo {author} {\bibfnamefont {M.}~\bibnamefont
  {Guidon}}, \bibinfo {author} {\bibfnamefont {J.}~\bibnamefont {Hutter}}, \
  and\ \bibinfo {author} {\bibfnamefont {J.}~\bibnamefont {VandeVondele}},\
  }\href@noop {} {\bibfield  {journal} {\bibinfo  {journal} {Journal of
  Chemical Theory and Computation}\ }\textbf {\bibinfo {volume} {6}},\ \bibinfo
  {pages} {2348–2364} (\bibinfo {year} {2010})}\BibitemShut {NoStop}%
\bibitem [{\citenamefont {Pisani}\ \emph {et~al.}(1988)\citenamefont {Pisani},
  \citenamefont {Dovesi},\ and\ \citenamefont {Roetti}}]{Pisani1988}%
  \BibitemOpen
  \bibfield  {author} {\bibinfo {author} {\bibfnamefont {C.}~\bibnamefont
  {Pisani}}, \bibinfo {author} {\bibfnamefont {R.}~\bibnamefont {Dovesi}}, \
  and\ \bibinfo {author} {\bibfnamefont {C.}~\bibnamefont {Roetti}},\
  }\href@noop {} {\enquote {\bibinfo {title} {Hartree-fock ab initio treatment
  of crystalline systems},}\ } (\bibinfo {year} {1988})\BibitemShut {NoStop}%
\bibitem [{\citenamefont {Blum}\ \emph {et~al.}(2009)\citenamefont {Blum},
  \citenamefont {Gehrke}, \citenamefont {Hanke}, \citenamefont {Havu},
  \citenamefont {Havu}, \citenamefont {Ren}, \citenamefont {Reuter},\ and\
  \citenamefont {Scheffler}}]{Blum2009}%
  \BibitemOpen
  \bibfield  {author} {\bibinfo {author} {\bibfnamefont {V.}~\bibnamefont
  {Blum}}, \bibinfo {author} {\bibfnamefont {R.}~\bibnamefont {Gehrke}},
  \bibinfo {author} {\bibfnamefont {F.}~\bibnamefont {Hanke}}, \bibinfo
  {author} {\bibfnamefont {P.}~\bibnamefont {Havu}}, \bibinfo {author}
  {\bibfnamefont {V.}~\bibnamefont {Havu}}, \bibinfo {author} {\bibfnamefont
  {X.}~\bibnamefont {Ren}}, \bibinfo {author} {\bibfnamefont {K.}~\bibnamefont
  {Reuter}}, \ and\ \bibinfo {author} {\bibfnamefont {M.}~\bibnamefont
  {Scheffler}},\ }\href {\doibase 10.1016/j.cpc.2009.06.022} {\bibfield
  {journal} {\bibinfo  {journal} {Computer Physics Communications}\ }\textbf
  {\bibinfo {volume} {180}},\ \bibinfo {pages} {2175} (\bibinfo {year}
  {2009})}\BibitemShut {NoStop}%
\bibitem [{\citenamefont {Goedecker}(1999)}]{Goedecker1999}%
  \BibitemOpen
  \bibfield  {author} {\bibinfo {author} {\bibfnamefont {S.}~\bibnamefont
  {Goedecker}},\ }\href {\doibase 10.1103/revmodphys.71.1085} {\bibfield
  {journal} {\bibinfo  {journal} {Reviews of Modern Physics}\ }\textbf
  {\bibinfo {volume} {71}},\ \bibinfo {pages} {1085} (\bibinfo {year}
  {1999})}\BibitemShut {NoStop}%
\bibitem [{\citenamefont {Bowler}\ and\ \citenamefont
  {Miyazaki}(2012)}]{Bowler2012}%
  \BibitemOpen
  \bibfield  {author} {\bibinfo {author} {\bibfnamefont {D.~R.}\ \bibnamefont
  {Bowler}}\ and\ \bibinfo {author} {\bibfnamefont {T.}~\bibnamefont
  {Miyazaki}},\ }\href {\doibase 10.1088/0034-4885/75/3/036503} {\bibfield
  {journal} {\bibinfo  {journal} {Reports on Progress in Physics}\ }\textbf
  {\bibinfo {volume} {75}},\ \bibinfo {pages} {036503} (\bibinfo {year}
  {2012})}\BibitemShut {NoStop}%
\bibitem [{\citenamefont {Wu}\ \emph {et~al.}(2009)\citenamefont {Wu},
  \citenamefont {Selloni},\ and\ \citenamefont {Car}}]{Wu2009}%
  \BibitemOpen
  \bibfield  {author} {\bibinfo {author} {\bibfnamefont {X.}~\bibnamefont
  {Wu}}, \bibinfo {author} {\bibfnamefont {A.}~\bibnamefont {Selloni}}, \ and\
  \bibinfo {author} {\bibfnamefont {R.}~\bibnamefont {Car}},\ }\href {\doibase
  10.1103/PhysRevB.79.085102} {\bibfield  {journal} {\bibinfo  {journal}
  {Physical Review B - Condensed Matter and Materials Physics}\ }\textbf
  {\bibinfo {volume} {79}},\ \bibinfo {pages} {085102} (\bibinfo {year}
  {2009})}\BibitemShut {NoStop}%
\bibitem [{\citenamefont {Lippert}\ \emph {et~al.}(1997)\citenamefont
  {Lippert}, \citenamefont {Hutter},\ and\ \citenamefont
  {Parinello}}]{lippert1997hybrid}%
  \BibitemOpen
  \bibfield  {author} {\bibinfo {author} {\bibfnamefont {G.}~\bibnamefont
  {Lippert}}, \bibinfo {author} {\bibfnamefont {J.}~\bibnamefont {Hutter}}, \
  and\ \bibinfo {author} {\bibfnamefont {M.}~\bibnamefont {Parinello}},\ }\href
  {\doibase 10.1080/002689797170220} {\bibfield  {journal} {\bibinfo  {journal}
  {Molecular Physics}\ }\textbf {\bibinfo {volume} {92}},\ \bibinfo {pages}
  {477} (\bibinfo {year} {1997})}\BibitemShut {NoStop}%
\bibitem [{\citenamefont {Whitten}(1973)}]{Whitten1973}%
  \BibitemOpen
  \bibfield  {author} {\bibinfo {author} {\bibfnamefont {J.~L.}\ \bibnamefont
  {Whitten}},\ }\href {\doibase 10.1063/1.1679012} {\bibfield  {journal}
  {\bibinfo  {journal} {The Journal of Chemical Physics}\ }\textbf {\bibinfo
  {volume} {58}},\ \bibinfo {pages} {4496} (\bibinfo {year}
  {1973})}\BibitemShut {NoStop}%
\bibitem [{\citenamefont {Mintmire}\ and\ \citenamefont
  {Dunlap}(1982)}]{Mintmire1982}%
  \BibitemOpen
  \bibfield  {author} {\bibinfo {author} {\bibfnamefont {J.~W.}\ \bibnamefont
  {Mintmire}}\ and\ \bibinfo {author} {\bibfnamefont {B.~I.}\ \bibnamefont
  {Dunlap}},\ }\href@noop {} {\bibfield  {journal} {\bibinfo  {journal}
  {Physical Review A}\ }\textbf {\bibinfo {volume} {25}},\ \bibinfo {pages}
  {88} (\bibinfo {year} {1982})}\BibitemShut {NoStop}%
\bibitem [{\citenamefont {Dunlap}(2000)}]{Dunlap2000a}%
  \BibitemOpen
  \bibfield  {author} {\bibinfo {author} {\bibfnamefont {B.~I.}\ \bibnamefont
  {Dunlap}},\ }\href {\doibase 10.1016/S0166-1280(00)00528-5} {\bibfield
  {journal} {\bibinfo  {journal} {Journal of Molecular Structure: THEOCHEM}\
  }\textbf {\bibinfo {volume} {529}},\ \bibinfo {pages} {37} (\bibinfo {year}
  {2000})}\BibitemShut {NoStop}%
\bibitem [{\citenamefont {Weigend}(2002)}]{Weigend2002}%
  \BibitemOpen
  \bibfield  {author} {\bibinfo {author} {\bibfnamefont {F.}~\bibnamefont
  {Weigend}},\ }\href@noop {} {\bibfield  {journal} {\bibinfo  {journal}
  {Physical Chemistry Chemical Physics}\ }\textbf {\bibinfo {volume} {4}},\
  \bibinfo {pages} {4285} (\bibinfo {year} {2002})}\BibitemShut {NoStop}%
\bibitem [{\citenamefont {Varga}(2005)}]{Varga2005}%
  \BibitemOpen
  \bibfield  {author} {\bibinfo {author} {\bibfnamefont {S.}~\bibnamefont
  {Varga}},\ }\href@noop {} {\bibfield  {journal} {\bibinfo  {journal}
  {Physical Review B}\ }\textbf {\bibinfo {volume} {71}},\ \bibinfo {pages}
  {073103} (\bibinfo {year} {2005})}\BibitemShut {NoStop}%
\bibitem [{\citenamefont {Maschio}\ and\ \citenamefont
  {Usvyat}(2008{\natexlab{a}})}]{Maschio2008}%
  \BibitemOpen
  \bibfield  {author} {\bibinfo {author} {\bibfnamefont {L.}~\bibnamefont
  {Maschio}}\ and\ \bibinfo {author} {\bibfnamefont {D.}~\bibnamefont
  {Usvyat}},\ }\href@noop {} {\bibfield  {journal} {\bibinfo  {journal}
  {Physical Review B1}\ }\textbf {\bibinfo {volume} {78}},\ \bibinfo {pages}
  {073102} (\bibinfo {year} {2008}{\natexlab{a}})}\BibitemShut {NoStop}%
\bibitem [{\citenamefont {Burow}\ \emph {et~al.}(2009)\citenamefont {Burow},
  \citenamefont {Sierka},\ and\ \citenamefont {Mohamed}}]{Burow2009}%
  \BibitemOpen
  \bibfield  {author} {\bibinfo {author} {\bibfnamefont {A.~M.}\ \bibnamefont
  {Burow}}, \bibinfo {author} {\bibfnamefont {M.}~\bibnamefont {Sierka}}, \
  and\ \bibinfo {author} {\bibfnamefont {F.}~\bibnamefont {Mohamed}},\ }\href
  {\doibase 10.1063/1.3267858} {\bibfield  {journal} {\bibinfo  {journal}
  {Journal of Chemical Physics}\ }\textbf {\bibinfo {volume} {131}} (\bibinfo
  {year} {2009}),\ 10.1063/1.3267858}\BibitemShut {NoStop}%
\bibitem [{\citenamefont {Wang}\ \emph {et~al.}(2020)\citenamefont {Wang},
  \citenamefont {Lewis},\ and\ \citenamefont {Valeev}}]{Wang2020}%
  \BibitemOpen
  \bibfield  {author} {\bibinfo {author} {\bibfnamefont {X.}~\bibnamefont
  {Wang}}, \bibinfo {author} {\bibfnamefont {C.~A.}\ \bibnamefont {Lewis}}, \
  and\ \bibinfo {author} {\bibfnamefont {E.~F.}\ \bibnamefont {Valeev}},\
  }\href@noop {} {\bibfield  {journal} {\bibinfo  {journal} {The Journal of
  Chemical Physics}\ }\textbf {\bibinfo {volume} {153}},\ \bibinfo {pages}
  {124116} (\bibinfo {year} {2020})}\BibitemShut {NoStop}%
\bibitem [{\citenamefont {Ye}\ and\ \citenamefont {Berkelbach}(2021)}]{Ye2021}%
  \BibitemOpen
  \bibfield  {author} {\bibinfo {author} {\bibfnamefont {H.~Z.}\ \bibnamefont
  {Ye}}\ and\ \bibinfo {author} {\bibfnamefont {T.~C.}\ \bibnamefont
  {Berkelbach}},\ }\href {\doibase 10.1063/5.0046617} {\bibfield  {journal}
  {\bibinfo  {journal} {Journal of Chemical Physics}\ }\textbf {\bibinfo
  {volume} {154}} (\bibinfo {year} {2021}),\ 10.1063/5.0046617}\BibitemShut
  {NoStop}%
\bibitem [{\citenamefont {Hohenstein}\ \emph
  {et~al.}(2012{\natexlab{a}})\citenamefont {Hohenstein}, \citenamefont
  {Parrish},\ and\ \citenamefont {Mart{\'{i}}nez}}]{Hohenstein2012}%
  \BibitemOpen
  \bibfield  {author} {\bibinfo {author} {\bibfnamefont {E.~G.}\ \bibnamefont
  {Hohenstein}}, \bibinfo {author} {\bibfnamefont {R.~M.}\ \bibnamefont
  {Parrish}}, \ and\ \bibinfo {author} {\bibfnamefont {T.~J.}\ \bibnamefont
  {Mart{\'{i}}nez}},\ }\href {\doibase 10.1063/1.4732310} {\bibfield  {journal}
  {\bibinfo  {journal} {The Journal of Chemical Physics}\ }\textbf {\bibinfo
  {volume} {137}},\ \bibinfo {pages} {1085} (\bibinfo {year}
  {2012}{\natexlab{a}})}\BibitemShut {NoStop}%
\bibitem [{\citenamefont {Parrish}\ \emph {et~al.}(2012)\citenamefont
  {Parrish}, \citenamefont {Hohenstein}, \citenamefont {Mart{\'{i}}nez},\ and\
  \citenamefont {Sherrill}}]{Parrish2012}%
  \BibitemOpen
  \bibfield  {author} {\bibinfo {author} {\bibfnamefont {R.~M.}\ \bibnamefont
  {Parrish}}, \bibinfo {author} {\bibfnamefont {E.~G.}\ \bibnamefont
  {Hohenstein}}, \bibinfo {author} {\bibfnamefont {T.~J.}\ \bibnamefont
  {Mart{\'{i}}nez}}, \ and\ \bibinfo {author} {\bibfnamefont {C.~D.}\
  \bibnamefont {Sherrill}},\ }\href {\doibase 10.1063/1.4768233} {\bibfield
  {journal} {\bibinfo  {journal} {The Journal of Chemical Physics}\ }\textbf
  {\bibinfo {volume} {137}},\ \bibinfo {pages} {224106} (\bibinfo {year}
  {2012})}\BibitemShut {NoStop}%
\bibitem [{\citenamefont {Hohenstein}\ \emph
  {et~al.}(2012{\natexlab{b}})\citenamefont {Hohenstein}, \citenamefont
  {Parrish}, \citenamefont {Sherrill},\ and\ \citenamefont
  {Mart{\'{i}}nez}}]{Hohenstein2012a}%
  \BibitemOpen
  \bibfield  {author} {\bibinfo {author} {\bibfnamefont {E.~G.}\ \bibnamefont
  {Hohenstein}}, \bibinfo {author} {\bibfnamefont {R.~M.}\ \bibnamefont
  {Parrish}}, \bibinfo {author} {\bibfnamefont {C.~D.}\ \bibnamefont
  {Sherrill}}, \ and\ \bibinfo {author} {\bibfnamefont {T.~J.}\ \bibnamefont
  {Mart{\'{i}}nez}},\ }\bibfield  {booktitle} {\emph {\bibinfo {booktitle} {The
  Journal of Chemical Physics}},\ }\href {\doibase 10.1063/1.4768241}
  {\bibfield  {journal} {\bibinfo  {journal} {The Journal of Chemical Physics}\
  }\textbf {\bibinfo {volume} {137}},\ \bibinfo {pages} {221101} (\bibinfo
  {year} {2012}{\natexlab{b}})}\BibitemShut {NoStop}%
\bibitem [{\citenamefont {Ye}\ and\ \citenamefont
  {Berkelbach}(2022)}]{ye2022correlation}%
  \BibitemOpen
  \bibfield  {author} {\bibinfo {author} {\bibfnamefont {H.-Z.}\ \bibnamefont
  {Ye}}\ and\ \bibinfo {author} {\bibfnamefont {T.~C.}\ \bibnamefont
  {Berkelbach}},\ }\href
  {https://pubs.acs.org/doi/pdf/10.1021/acs.jctc.1c01245} {\bibfield  {journal}
  {\bibinfo  {journal} {Journal of Chemical Theory and Computation}\ }\textbf
  {\bibinfo {volume} {18}},\ \bibinfo {pages} {1595} (\bibinfo {year}
  {2022})}\BibitemShut {NoStop}%
\bibitem [{\citenamefont {Hartwigsen}\ \emph {et~al.}(1998)\citenamefont
  {Hartwigsen}, \citenamefont {Goedecker},\ and\ \citenamefont
  {Hutter}}]{PhysRevB.58.3641}%
  \BibitemOpen
  \bibfield  {author} {\bibinfo {author} {\bibfnamefont {C.}~\bibnamefont
  {Hartwigsen}}, \bibinfo {author} {\bibfnamefont {S.}~\bibnamefont
  {Goedecker}}, \ and\ \bibinfo {author} {\bibfnamefont {J.}~\bibnamefont
  {Hutter}},\ }\href {\doibase 10.1103/PhysRevB.58.3641} {\bibfield  {journal}
  {\bibinfo  {journal} {Physical Review B}\ }\textbf {\bibinfo {volume} {58}},\
  \bibinfo {pages} {3641} (\bibinfo {year} {1998})}\BibitemShut {NoStop}%
\bibitem [{\citenamefont {Heyd}\ \emph {et~al.}(2005)\citenamefont {Heyd},
  \citenamefont {Peralta}, \citenamefont {Scuseria},\ and\ \citenamefont
  {Martin}}]{heyd2005energy}%
  \BibitemOpen
  \bibfield  {author} {\bibinfo {author} {\bibfnamefont {J.}~\bibnamefont
  {Heyd}}, \bibinfo {author} {\bibfnamefont {J.~E.}\ \bibnamefont {Peralta}},
  \bibinfo {author} {\bibfnamefont {G.~E.}\ \bibnamefont {Scuseria}}, \ and\
  \bibinfo {author} {\bibfnamefont {R.~L.}\ \bibnamefont {Martin}},\ }\href
  {\doibase 10.1063/1.2085170} {\bibfield  {journal} {\bibinfo  {journal} {The
  Journal of Chemical Physics}\ }\textbf {\bibinfo {volume} {123}},\ \bibinfo
  {pages} {174101} (\bibinfo {year} {2005})}\BibitemShut {NoStop}%
\bibitem [{\citenamefont {Gr{\"u}neis}\ \emph {et~al.}(2010)\citenamefont
  {Gr{\"u}neis}, \citenamefont {Marsman},\ and\ \citenamefont
  {Kresse}}]{gruneis2010second}%
  \BibitemOpen
  \bibfield  {author} {\bibinfo {author} {\bibfnamefont {A.}~\bibnamefont
  {Gr{\"u}neis}}, \bibinfo {author} {\bibfnamefont {M.}~\bibnamefont
  {Marsman}}, \ and\ \bibinfo {author} {\bibfnamefont {G.}~\bibnamefont
  {Kresse}},\ }\href {\doibase 10.1063/1.3466765} {\bibfield  {journal}
  {\bibinfo  {journal} {The Journal of Chemical Physics}\ }\textbf {\bibinfo
  {volume} {133}},\ \bibinfo {pages} {074107} (\bibinfo {year}
  {2010})}\BibitemShut {NoStop}%
\bibitem [{\citenamefont {Nadler}\ and\ \citenamefont
  {Kempier}(1959)}]{nadler1959crystallographic}%
  \BibitemOpen
  \bibfield  {author} {\bibinfo {author} {\bibfnamefont {M.~R.}\ \bibnamefont
  {Nadler}}\ and\ \bibinfo {author} {\bibfnamefont {C.}~\bibnamefont
  {Kempier}},\ }\href {\doibase 10.1021/ac60156a007} {\bibfield  {journal}
  {\bibinfo  {journal} {Analytical Chemistry}\ }\textbf {\bibinfo {volume}
  {31}},\ \bibinfo {pages} {2109} (\bibinfo {year} {1959})}\BibitemShut
  {NoStop}%
\bibitem [{\citenamefont {Tang}\ \emph {et~al.}(2009)\citenamefont {Tang},
  \citenamefont {Che}, \citenamefont {Lu}, \citenamefont {Chen}, \citenamefont
  {Xie},\ and\ \citenamefont {Yu}}]{tang2009surface}%
  \BibitemOpen
  \bibfield  {author} {\bibinfo {author} {\bibfnamefont {F.~L.}\ \bibnamefont
  {Tang}}, \bibinfo {author} {\bibfnamefont {X.~X.}\ \bibnamefont {Che}},
  \bibinfo {author} {\bibfnamefont {W.~J.}\ \bibnamefont {Lu}}, \bibinfo
  {author} {\bibfnamefont {G.~B.}\ \bibnamefont {Chen}}, \bibinfo {author}
  {\bibfnamefont {Y.}~\bibnamefont {Xie}}, \ and\ \bibinfo {author}
  {\bibfnamefont {W.~Y.}\ \bibnamefont {Yu}},\ }\href
  {https://www.sciencedirect.com/science/article/abs/pii/S0921452609002890}
  {\bibfield  {journal} {\bibinfo  {journal} {Physica B: Condensed Matter}\
  }\textbf {\bibinfo {volume} {404}},\ \bibinfo {pages} {2489} (\bibinfo {year}
  {2009})}\BibitemShut {NoStop}%
\bibitem [{\citenamefont {Babbush}\ \emph
  {et~al.}(2018{\natexlab{c}})\citenamefont {Babbush}, \citenamefont {Gidney},
  \citenamefont {Berry}, \citenamefont {Wiebe}, \citenamefont {McClean},
  \citenamefont {Paler}, \citenamefont {Fowler},\ and\ \citenamefont
  {Neven}}]{BabbushSpectraB}%
  \BibitemOpen
  \bibfield  {author} {\bibinfo {author} {\bibfnamefont {R.}~\bibnamefont
  {Babbush}}, \bibinfo {author} {\bibfnamefont {C.}~\bibnamefont {Gidney}},
  \bibinfo {author} {\bibfnamefont {D.~W.}\ \bibnamefont {Berry}}, \bibinfo
  {author} {\bibfnamefont {N.}~\bibnamefont {Wiebe}}, \bibinfo {author}
  {\bibfnamefont {J.}~\bibnamefont {McClean}}, \bibinfo {author} {\bibfnamefont
  {A.}~\bibnamefont {Paler}}, \bibinfo {author} {\bibfnamefont
  {A.}~\bibnamefont {Fowler}}, \ and\ \bibinfo {author} {\bibfnamefont
  {H.}~\bibnamefont {Neven}},\ }\href
  {https://journals.aps.org/prx/abstract/10.1103/PhysRevX.8.041015} {\bibfield
  {journal} {\bibinfo  {journal} {Physical Review X}\ }\textbf {\bibinfo
  {volume} {8}},\ \bibinfo {pages} {041015} (\bibinfo {year}
  {2018}{\natexlab{c}})}\BibitemShut {NoStop}%
\bibitem [{\citenamefont {Berry}\ \emph {et~al.}(2019)\citenamefont {Berry},
  \citenamefont {Gidney}, \citenamefont {Motta}, \citenamefont {McClean},\ and\
  \citenamefont {Babbush}}]{Berry2019B}%
  \BibitemOpen
  \bibfield  {author} {\bibinfo {author} {\bibfnamefont {D.~W.}\ \bibnamefont
  {Berry}}, \bibinfo {author} {\bibfnamefont {C.}~\bibnamefont {Gidney}},
  \bibinfo {author} {\bibfnamefont {M.}~\bibnamefont {Motta}}, \bibinfo
  {author} {\bibfnamefont {J.}~\bibnamefont {McClean}}, \ and\ \bibinfo
  {author} {\bibfnamefont {R.}~\bibnamefont {Babbush}},\ }\href
  {https://quantum-journal.org/papers/q-2019-12-02-208/} {\bibfield  {journal}
  {\bibinfo  {journal} {Quantum}\ }\textbf {\bibinfo {volume} {3}},\ \bibinfo
  {pages} {208} (\bibinfo {year} {2019})}\BibitemShut {NoStop}%
\bibitem [{\citenamefont {Gily{\'e}n}\ \emph {et~al.}(2019)\citenamefont
  {Gily{\'e}n}, \citenamefont {Su}, \citenamefont {Low},\ and\ \citenamefont
  {Wiebe}}]{gilyen2019quantum}%
  \BibitemOpen
  \bibfield  {author} {\bibinfo {author} {\bibfnamefont {A.}~\bibnamefont
  {Gily{\'e}n}}, \bibinfo {author} {\bibfnamefont {Y.}~\bibnamefont {Su}},
  \bibinfo {author} {\bibfnamefont {G.~H.}\ \bibnamefont {Low}}, \ and\
  \bibinfo {author} {\bibfnamefont {N.}~\bibnamefont {Wiebe}},\ }in\ \href
  {https://dl.acm.org/doi/10.1145/3313276.3316366} {\emph {\bibinfo {booktitle}
  {Proceedings of the 51st Annual ACM SIGACT Symposium on Theory of
  Computing}}}\ (\bibinfo {year} {2019})\ pp.\ \bibinfo {pages}
  {193--204}\BibitemShut {NoStop}%
\bibitem [{\citenamefont {Szegedy}(2004)}]{Szegedy2004}%
  \BibitemOpen
  \bibfield  {author} {\bibinfo {author} {\bibfnamefont {M.}~\bibnamefont
  {Szegedy}},\ }in\ \href {\doibase 10.1109/FOCS.2004.53} {\emph {\bibinfo
  {booktitle} {45th Annual IEEE Symposium on Foundations of Computer
  Science}}}\ (\bibinfo  {publisher} {IEEE},\ \bibinfo {year} {2004})\ pp.\
  \bibinfo {pages} {32--41}\BibitemShut {NoStop}%
\bibitem [{\citenamefont {Childs}\ and\ \citenamefont
  {Wiebe}(2012)}]{child_wiebe_LCU_2012}%
  \BibitemOpen
  \bibfield  {author} {\bibinfo {author} {\bibfnamefont {A.~M.}\ \bibnamefont
  {Childs}}\ and\ \bibinfo {author} {\bibfnamefont {N.}~\bibnamefont {Wiebe}},\
  }\href {https://dl.acm.org/doi/10.5555/2481569.2481570} {\bibfield  {journal}
  {\bibinfo  {journal} {Quantum Information and Computation}\ }\textbf
  {\bibinfo {volume} {12}},\ \bibinfo {pages} {901–924} (\bibinfo {year}
  {2012})}\BibitemShut {NoStop}%
\bibitem [{\citenamefont {Baldereschi}(1973)}]{PhysRevB.7.5212}%
  \BibitemOpen
  \bibfield  {author} {\bibinfo {author} {\bibfnamefont {A.}~\bibnamefont
  {Baldereschi}},\ }\href {\doibase 10.1103/PhysRevB.7.5212} {\bibfield
  {journal} {\bibinfo  {journal} {Physical Review B}\ }\textbf {\bibinfo
  {volume} {7}},\ \bibinfo {pages} {5212} (\bibinfo {year} {1973})}\BibitemShut
  {NoStop}%
\bibitem [{\citenamefont {Low}\ \emph {et~al.}(2018)\citenamefont {Low},
  \citenamefont {Kliuchnikov},\ and\ \citenamefont
  {Schaeffer}}]{low2018trading}%
  \BibitemOpen
  \bibfield  {author} {\bibinfo {author} {\bibfnamefont {G.~H.}\ \bibnamefont
  {Low}}, \bibinfo {author} {\bibfnamefont {V.}~\bibnamefont {Kliuchnikov}}, \
  and\ \bibinfo {author} {\bibfnamefont {L.}~\bibnamefont {Schaeffer}},\ }\href
  {https://arxiv.org/abs/1812.00954} {\bibfield  {journal} {\bibinfo  {journal}
  {arXiv: 1812.00954}\ } (\bibinfo {year} {2018})}\BibitemShut {NoStop}%
\bibitem [{\citenamefont {Werner}\ \emph {et~al.}(2003)\citenamefont {Werner},
  \citenamefont {Manby},\ and\ \citenamefont {Knowles}}]{werner2003fast}%
  \BibitemOpen
  \bibfield  {author} {\bibinfo {author} {\bibfnamefont {H.-J.}\ \bibnamefont
  {Werner}}, \bibinfo {author} {\bibfnamefont {F.~R.}\ \bibnamefont {Manby}}, \
  and\ \bibinfo {author} {\bibfnamefont {P.~J.}\ \bibnamefont {Knowles}},\
  }\href {https://aip.scitation.org/doi/10.1063/1.1564816} {\bibfield
  {journal} {\bibinfo  {journal} {The Journal of chemical physics}\ }\textbf
  {\bibinfo {volume} {118}},\ \bibinfo {pages} {8149} (\bibinfo {year}
  {2003})}\BibitemShut {NoStop}%
\bibitem [{\citenamefont {Maschio}\ and\ \citenamefont
  {Usvyat}(2008{\natexlab{b}})}]{PhysRevB.78.073102}%
  \BibitemOpen
  \bibfield  {author} {\bibinfo {author} {\bibfnamefont {L.}~\bibnamefont
  {Maschio}}\ and\ \bibinfo {author} {\bibfnamefont {D.}~\bibnamefont
  {Usvyat}},\ }\href {\doibase 10.1103/PhysRevB.78.073102} {\bibfield
  {journal} {\bibinfo  {journal} {Physical Review B}\ }\textbf {\bibinfo
  {volume} {78}},\ \bibinfo {pages} {073102} (\bibinfo {year}
  {2008}{\natexlab{b}})}\BibitemShut {NoStop}%
\bibitem [{\citenamefont {Oumarou}\ \emph {et~al.}(2022)\citenamefont
  {Oumarou}, \citenamefont {Scheurer}, \citenamefont {Parrish}, \citenamefont
  {Hohenstein},\ and\ \citenamefont {Gogolin}}]{oumarou2022accelerating}%
  \BibitemOpen
  \bibfield  {author} {\bibinfo {author} {\bibfnamefont {O.}~\bibnamefont
  {Oumarou}}, \bibinfo {author} {\bibfnamefont {M.}~\bibnamefont {Scheurer}},
  \bibinfo {author} {\bibfnamefont {R.~M.}\ \bibnamefont {Parrish}}, \bibinfo
  {author} {\bibfnamefont {E.~G.}\ \bibnamefont {Hohenstein}}, \ and\ \bibinfo
  {author} {\bibfnamefont {C.}~\bibnamefont {Gogolin}},\ }\href
  {https://arxiv.org/abs/2212.07957} {\bibfield  {journal} {\bibinfo  {journal}
  {arXiv:2212.07957}\ } (\bibinfo {year} {2022})}\BibitemShut {NoStop}%
\bibitem [{\citenamefont {Lu}\ and\ \citenamefont {Ying}(2016)}]{Lu2016}%
  \BibitemOpen
  \bibfield  {author} {\bibinfo {author} {\bibfnamefont {J.}~\bibnamefont
  {Lu}}\ and\ \bibinfo {author} {\bibfnamefont {L.}~\bibnamefont {Ying}},\
  }\href {\doibase 10.4310/AMSA.2016.v1.n2.a3} {\bibfield  {journal} {\bibinfo
  {journal} {Annals of Mathematical Sciences and Applications}\ }\textbf
  {\bibinfo {volume} {1}},\ \bibinfo {pages} {321} (\bibinfo {year}
  {2016})}\BibitemShut {NoStop}%
\bibitem [{\citenamefont {Wu}\ \emph {et~al.}(2021)\citenamefont {Wu},
  \citenamefont {Qin}, \citenamefont {Hu},\ and\ \citenamefont
  {Yang}}]{wu2021low}%
  \BibitemOpen
  \bibfield  {author} {\bibinfo {author} {\bibfnamefont {K.}~\bibnamefont
  {Wu}}, \bibinfo {author} {\bibfnamefont {X.}~\bibnamefont {Qin}}, \bibinfo
  {author} {\bibfnamefont {W.}~\bibnamefont {Hu}}, \ and\ \bibinfo {author}
  {\bibfnamefont {J.}~\bibnamefont {Yang}},\ }\href
  {https://pubs.acs.org/doi/abs/10.1021/acs.jctc.1c00874} {\bibfield  {journal}
  {\bibinfo  {journal} {Journal of Chemical Theory and Computation}\ }\textbf
  {\bibinfo {volume} {18}},\ \bibinfo {pages} {206} (\bibinfo {year}
  {2021})}\BibitemShut {NoStop}%
\bibitem [{\citenamefont {Hu}\ \emph {et~al.}(2017)\citenamefont {Hu},
  \citenamefont {Lin},\ and\ \citenamefont {Yang}}]{Hu2017}%
  \BibitemOpen
  \bibfield  {author} {\bibinfo {author} {\bibfnamefont {W.}~\bibnamefont
  {Hu}}, \bibinfo {author} {\bibfnamefont {L.}~\bibnamefont {Lin}}, \ and\
  \bibinfo {author} {\bibfnamefont {C.}~\bibnamefont {Yang}},\ }\href {\doibase
  10.1021/acs.jctc.7b00807} {\bibfield  {journal} {\bibinfo  {journal} {Journal
  of Chemical Theory and Computation}\ }\textbf {\bibinfo {volume} {13}},\
  \bibinfo {pages} {5420} (\bibinfo {year} {2017})}\BibitemShut {NoStop}%
\bibitem [{\citenamefont {Dong}\ \emph {et~al.}(2018)\citenamefont {Dong},
  \citenamefont {Hu},\ and\ \citenamefont {Lin}}]{Dong2018}%
  \BibitemOpen
  \bibfield  {author} {\bibinfo {author} {\bibfnamefont {K.}~\bibnamefont
  {Dong}}, \bibinfo {author} {\bibfnamefont {W.}~\bibnamefont {Hu}}, \ and\
  \bibinfo {author} {\bibfnamefont {L.}~\bibnamefont {Lin}},\ }\href {\doibase
  10.1021/acs.jctc.7b01113} {\bibfield  {journal} {\bibinfo  {journal} {Journal
  of Chemical Theory and Computation}\ }\textbf {\bibinfo {volume} {14}},\
  \bibinfo {pages} {1311} (\bibinfo {year} {2018})}\BibitemShut {NoStop}%
\bibitem [{bas()}]{basf}%
  \BibitemOpen
  \href@noop {} {\enquote {\bibinfo {title} {Project by {BMW, BASF, Samsung SDI
  and Samsung Electronics} to enhance sustainable cobalt mining},}\ }\bibinfo
  {note}
  {\url{https://www.basf.com/global/en/who-we-are/sustainability/responsible-partnering/cobalt-initiative.html}
  (accessed October 26, 2020)}\BibitemShut {NoStop}%
\bibitem [{\citenamefont {Olivetti}\ \emph {et~al.}(2017)\citenamefont
  {Olivetti}, \citenamefont {Ceder}, \citenamefont {Gaustad},\ and\
  \citenamefont {Fu}}]{Olivetti2017}%
  \BibitemOpen
  \bibfield  {author} {\bibinfo {author} {\bibfnamefont {E.~A.}\ \bibnamefont
  {Olivetti}}, \bibinfo {author} {\bibfnamefont {G.}~\bibnamefont {Ceder}},
  \bibinfo {author} {\bibfnamefont {G.~G.}\ \bibnamefont {Gaustad}}, \ and\
  \bibinfo {author} {\bibfnamefont {X.}~\bibnamefont {Fu}},\ }\href {\doibase
  https://doi.org/10.1016/j.joule.2017.08.019} {\bibfield  {journal} {\bibinfo
  {journal} {Joule}\ }\textbf {\bibinfo {volume} {1}},\ \bibinfo {pages} {229}
  (\bibinfo {year} {2017})}\BibitemShut {NoStop}%
\bibitem [{\citenamefont {Dahn}\ \emph {et~al.}(1990)\citenamefont {Dahn},
  \citenamefont {Vonsacken},\ and\ \citenamefont {Michal}}]{Dahn1990}%
  \BibitemOpen
  \bibfield  {author} {\bibinfo {author} {\bibfnamefont {J.~R.}\ \bibnamefont
  {Dahn}}, \bibinfo {author} {\bibfnamefont {U.}~\bibnamefont {Vonsacken}}, \
  and\ \bibinfo {author} {\bibfnamefont {C.~A.}\ \bibnamefont {Michal}},\
  }\href {\doibase Doi 10.1016/0167-2738(90)90049-W} {\bibfield  {journal}
  {\bibinfo  {journal} {Solid State Ionics}\ }\textbf {\bibinfo {volume}
  {44}},\ \bibinfo {pages} {87} (\bibinfo {year} {1990})}\BibitemShut {NoStop}%
\bibitem [{\citenamefont {Ohzuku}\ \emph {et~al.}(1993)\citenamefont {Ohzuku},
  \citenamefont {Ueda},\ and\ \citenamefont {Nagayama}}]{Ohzuku}%
  \BibitemOpen
  \bibfield  {author} {\bibinfo {author} {\bibfnamefont {T.}~\bibnamefont
  {Ohzuku}}, \bibinfo {author} {\bibfnamefont {A.}~\bibnamefont {Ueda}}, \ and\
  \bibinfo {author} {\bibfnamefont {M.}~\bibnamefont {Nagayama}},\ }\href
  {\doibase 10.1149/1.2220730} {\bibfield  {journal} {\bibinfo  {journal}
  {Journal of The Electrochemical Society}\ }\textbf {\bibinfo {volume}
  {140}},\ \bibinfo {pages} {1862} (\bibinfo {year} {1993})}\BibitemShut
  {NoStop}%
\bibitem [{\citenamefont {Radin}\ \emph {et~al.}(2020)\citenamefont {Radin},
  \citenamefont {Thomas},\ and\ \citenamefont {Van~der
  Ven}}]{PhysRevMaterials.4.043601}%
  \BibitemOpen
  \bibfield  {author} {\bibinfo {author} {\bibfnamefont {M.~D.}\ \bibnamefont
  {Radin}}, \bibinfo {author} {\bibfnamefont {J.~C.}\ \bibnamefont {Thomas}}, \
  and\ \bibinfo {author} {\bibfnamefont {A.}~\bibnamefont {Van~der Ven}},\
  }\href {\doibase 10.1103/PhysRevMaterials.4.043601} {\bibfield  {journal}
  {\bibinfo  {journal} {Physical Review Materials}\ }\textbf {\bibinfo {volume}
  {4}},\ \bibinfo {pages} {043601} (\bibinfo {year} {2020})}\BibitemShut
  {NoStop}%
\bibitem [{\citenamefont {M{\o}ller}\ and\ \citenamefont
  {Plesset}(1934)}]{Moller1934}%
  \BibitemOpen
  \bibfield  {author} {\bibinfo {author} {\bibfnamefont {C.}~\bibnamefont
  {M{\o}ller}}\ and\ \bibinfo {author} {\bibfnamefont {M.~S.}\ \bibnamefont
  {Plesset}},\ }\href {\doibase 10.1103/PhysRev.46.618} {\bibfield  {journal}
  {\bibinfo  {journal} {Physical Review}\ }\textbf {\bibinfo {volume} {46}},\
  \bibinfo {pages} {618} (\bibinfo {year} {1934})}\BibitemShut {NoStop}%
\bibitem [{\citenamefont {Cremer}(2011)}]{Cremer2011}%
  \BibitemOpen
  \bibfield  {author} {\bibinfo {author} {\bibfnamefont {D.}~\bibnamefont
  {Cremer}},\ }\href {\doibase 10.1002/wcms.58} {\bibfield  {journal} {\bibinfo
   {journal} {Ltd. WIREs Comput Mol Sci}\ }\textbf {\bibinfo {volume} {1}},\
  \bibinfo {pages} {509} (\bibinfo {year} {2011})}\BibitemShut {NoStop}%
\bibitem [{\citenamefont {Purvis}\ and\ \citenamefont
  {Bartlett}(1982)}]{Purvis1982}%
  \BibitemOpen
  \bibfield  {author} {\bibinfo {author} {\bibfnamefont {G.~D.}\ \bibnamefont
  {Purvis}}\ and\ \bibinfo {author} {\bibfnamefont {R.~J.}\ \bibnamefont
  {Bartlett}},\ }\href {\doibase 10.1063/1.443164} {\bibfield  {journal}
  {\bibinfo  {journal} {The Journal of Chemical Physics}\ }\textbf {\bibinfo
  {volume} {76}},\ \bibinfo {pages} {1910} (\bibinfo {year}
  {1982})}\BibitemShut {NoStop}%
\bibitem [{\citenamefont {Bartlett}\ and\ \citenamefont
  {Musiał}(2007)}]{Bartlett2007}%
  \BibitemOpen
  \bibfield  {author} {\bibinfo {author} {\bibfnamefont {R.~J.}\ \bibnamefont
  {Bartlett}}\ and\ \bibinfo {author} {\bibfnamefont {M.}~\bibnamefont
  {Musiał}},\ }\href {\doibase 10.1103/RevModPhys.79.291} {\bibfield
  {journal} {\bibinfo  {journal} {Reviews of Modern Physics}\ }\textbf
  {\bibinfo {volume} {79}},\ \bibinfo {pages} {291} (\bibinfo {year}
  {2007})}\BibitemShut {NoStop}%
\bibitem [{\citenamefont {Sun}\ \emph {et~al.}(2018)\citenamefont {Sun},
  \citenamefont {Berkelbach}, \citenamefont {Blunt}, \citenamefont {Booth},
  \citenamefont {Guo}, \citenamefont {Li}, \citenamefont {Liu}, \citenamefont
  {McClain}, \citenamefont {Sayfutyarova}, \citenamefont {Sharma},
  \citenamefont {Wouters},\ and\ \citenamefont {Chan}}]{Sun2018}%
  \BibitemOpen
  \bibfield  {author} {\bibinfo {author} {\bibfnamefont {Q.}~\bibnamefont
  {Sun}}, \bibinfo {author} {\bibfnamefont {T.~C.}\ \bibnamefont {Berkelbach}},
  \bibinfo {author} {\bibfnamefont {N.~S.}\ \bibnamefont {Blunt}}, \bibinfo
  {author} {\bibfnamefont {G.~H.}\ \bibnamefont {Booth}}, \bibinfo {author}
  {\bibfnamefont {S.}~\bibnamefont {Guo}}, \bibinfo {author} {\bibfnamefont
  {Z.}~\bibnamefont {Li}}, \bibinfo {author} {\bibfnamefont {J.}~\bibnamefont
  {Liu}}, \bibinfo {author} {\bibfnamefont {J.~D.}\ \bibnamefont {McClain}},
  \bibinfo {author} {\bibfnamefont {E.~R.}\ \bibnamefont {Sayfutyarova}},
  \bibinfo {author} {\bibfnamefont {S.}~\bibnamefont {Sharma}}, \bibinfo
  {author} {\bibfnamefont {S.}~\bibnamefont {Wouters}}, \ and\ \bibinfo
  {author} {\bibfnamefont {G.~K.~L.}\ \bibnamefont {Chan}},\ }\href {\doibase
  10.1002/wcms.1340} {\bibfield  {journal} {\bibinfo  {journal} {Wiley
  Interdisciplinary Reviews: Computational Molecular Science}\ }\textbf
  {\bibinfo {volume} {8}},\ \bibinfo {pages} {e1340} (\bibinfo {year}
  {2018})}\BibitemShut {NoStop}%
\bibitem [{\citenamefont {Sun}\ \emph {et~al.}(2020)\citenamefont {Sun},
  \citenamefont {Zhang}, \citenamefont {Banerjee}, \citenamefont {Bao},
  \citenamefont {Barbry}, \citenamefont {Blunt}, \citenamefont {Bogdanov},
  \citenamefont {Booth}, \citenamefont {Chen}, \citenamefont {Cui},
  \citenamefont {Eriksen}, \citenamefont {Gao}, \citenamefont {Guo},
  \citenamefont {Hermann}, \citenamefont {Hermes}, \citenamefont {Koh},
  \citenamefont {Koval}, \citenamefont {Lehtola}, \citenamefont {Li},
  \citenamefont {Liu}, \citenamefont {Mardirossian}, \citenamefont {McClain},
  \citenamefont {Motta}, \citenamefont {Mussard}, \citenamefont {Pham},
  \citenamefont {Pulkin}, \citenamefont {Purwanto}, \citenamefont {Robinson},
  \citenamefont {Ronca}, \citenamefont {Sayfutyarova}, \citenamefont
  {Scheurer}, \citenamefont {Schurkus}, \citenamefont {Smith}, \citenamefont
  {Sun}, \citenamefont {Sun}, \citenamefont {Upadhyay}, \citenamefont {Wagner},
  \citenamefont {Wang}, \citenamefont {White}, \citenamefont {Whitfield},
  \citenamefont {Williamson}, \citenamefont {Wouters}, \citenamefont {Yang},
  \citenamefont {Yu}, \citenamefont {Zhu}, \citenamefont {Berkelbach},
  \citenamefont {Sharma}, \citenamefont {Sokolov},\ and\ \citenamefont
  {Chan}}]{Sun2020}%
  \BibitemOpen
  \bibfield  {author} {\bibinfo {author} {\bibfnamefont {Q.}~\bibnamefont
  {Sun}}, \bibinfo {author} {\bibfnamefont {X.}~\bibnamefont {Zhang}}, \bibinfo
  {author} {\bibfnamefont {S.}~\bibnamefont {Banerjee}}, \bibinfo {author}
  {\bibfnamefont {P.}~\bibnamefont {Bao}}, \bibinfo {author} {\bibfnamefont
  {M.}~\bibnamefont {Barbry}}, \bibinfo {author} {\bibfnamefont {N.~S.}\
  \bibnamefont {Blunt}}, \bibinfo {author} {\bibfnamefont {N.~A.}\ \bibnamefont
  {Bogdanov}}, \bibinfo {author} {\bibfnamefont {G.~H.}\ \bibnamefont {Booth}},
  \bibinfo {author} {\bibfnamefont {J.}~\bibnamefont {Chen}}, \bibinfo {author}
  {\bibfnamefont {Z.~H.}\ \bibnamefont {Cui}}, \bibinfo {author} {\bibfnamefont
  {J.~J.}\ \bibnamefont {Eriksen}}, \bibinfo {author} {\bibfnamefont
  {Y.}~\bibnamefont {Gao}}, \bibinfo {author} {\bibfnamefont {S.}~\bibnamefont
  {Guo}}, \bibinfo {author} {\bibfnamefont {J.}~\bibnamefont {Hermann}},
  \bibinfo {author} {\bibfnamefont {M.~R.}\ \bibnamefont {Hermes}}, \bibinfo
  {author} {\bibfnamefont {K.}~\bibnamefont {Koh}}, \bibinfo {author}
  {\bibfnamefont {P.}~\bibnamefont {Koval}}, \bibinfo {author} {\bibfnamefont
  {S.}~\bibnamefont {Lehtola}}, \bibinfo {author} {\bibfnamefont
  {Z.}~\bibnamefont {Li}}, \bibinfo {author} {\bibfnamefont {J.}~\bibnamefont
  {Liu}}, \bibinfo {author} {\bibfnamefont {N.}~\bibnamefont {Mardirossian}},
  \bibinfo {author} {\bibfnamefont {J.~D.}\ \bibnamefont {McClain}}, \bibinfo
  {author} {\bibfnamefont {M.}~\bibnamefont {Motta}}, \bibinfo {author}
  {\bibfnamefont {B.}~\bibnamefont {Mussard}}, \bibinfo {author} {\bibfnamefont
  {H.~Q.}\ \bibnamefont {Pham}}, \bibinfo {author} {\bibfnamefont
  {A.}~\bibnamefont {Pulkin}}, \bibinfo {author} {\bibfnamefont
  {W.}~\bibnamefont {Purwanto}}, \bibinfo {author} {\bibfnamefont {P.~J.}\
  \bibnamefont {Robinson}}, \bibinfo {author} {\bibfnamefont {E.}~\bibnamefont
  {Ronca}}, \bibinfo {author} {\bibfnamefont {E.~R.}\ \bibnamefont
  {Sayfutyarova}}, \bibinfo {author} {\bibfnamefont {M.}~\bibnamefont
  {Scheurer}}, \bibinfo {author} {\bibfnamefont {H.~F.}\ \bibnamefont
  {Schurkus}}, \bibinfo {author} {\bibfnamefont {J.~E.}\ \bibnamefont {Smith}},
  \bibinfo {author} {\bibfnamefont {C.}~\bibnamefont {Sun}}, \bibinfo {author}
  {\bibfnamefont {S.~N.}\ \bibnamefont {Sun}}, \bibinfo {author} {\bibfnamefont
  {S.}~\bibnamefont {Upadhyay}}, \bibinfo {author} {\bibfnamefont {L.~K.}\
  \bibnamefont {Wagner}}, \bibinfo {author} {\bibfnamefont {X.}~\bibnamefont
  {Wang}}, \bibinfo {author} {\bibfnamefont {A.}~\bibnamefont {White}},
  \bibinfo {author} {\bibfnamefont {J.~D.}\ \bibnamefont {Whitfield}}, \bibinfo
  {author} {\bibfnamefont {M.~J.}\ \bibnamefont {Williamson}}, \bibinfo
  {author} {\bibfnamefont {S.}~\bibnamefont {Wouters}}, \bibinfo {author}
  {\bibfnamefont {J.}~\bibnamefont {Yang}}, \bibinfo {author} {\bibfnamefont
  {J.~M.}\ \bibnamefont {Yu}}, \bibinfo {author} {\bibfnamefont
  {T.}~\bibnamefont {Zhu}}, \bibinfo {author} {\bibfnamefont {T.~C.}\
  \bibnamefont {Berkelbach}}, \bibinfo {author} {\bibfnamefont
  {S.}~\bibnamefont {Sharma}}, \bibinfo {author} {\bibfnamefont {A.~Y.}\
  \bibnamefont {Sokolov}}, \ and\ \bibinfo {author} {\bibfnamefont {G.~K.~L.}\
  \bibnamefont {Chan}},\ }\href {\doibase 10.1063/5.0006074} {\bibfield
  {journal} {\bibinfo  {journal} {The Journal of chemical physics}\ }\textbf
  {\bibinfo {volume} {153}},\ \bibinfo {pages} {024109} (\bibinfo {year}
  {2020})}\BibitemShut {NoStop}%
\bibitem [{\citenamefont {Kim}\ \emph {et~al.}(2018)\citenamefont {Kim},
  \citenamefont {Baczewski}, \citenamefont {Beaudet}, \citenamefont {Benali},
  \citenamefont {Bennett}, \citenamefont {Berrill}, \citenamefont {Blunt},
  \citenamefont {Borda}, \citenamefont {Casula}, \citenamefont {Ceperley} \emph
  {et~al.}}]{kim2018qmcpack}%
  \BibitemOpen
  \bibfield  {author} {\bibinfo {author} {\bibfnamefont {J.}~\bibnamefont
  {Kim}}, \bibinfo {author} {\bibfnamefont {A.~D.}\ \bibnamefont {Baczewski}},
  \bibinfo {author} {\bibfnamefont {T.~D.}\ \bibnamefont {Beaudet}}, \bibinfo
  {author} {\bibfnamefont {A.}~\bibnamefont {Benali}}, \bibinfo {author}
  {\bibfnamefont {M.~C.}\ \bibnamefont {Bennett}}, \bibinfo {author}
  {\bibfnamefont {M.~A.}\ \bibnamefont {Berrill}}, \bibinfo {author}
  {\bibfnamefont {N.~S.}\ \bibnamefont {Blunt}}, \bibinfo {author}
  {\bibfnamefont {E.~J.~L.}\ \bibnamefont {Borda}}, \bibinfo {author}
  {\bibfnamefont {M.}~\bibnamefont {Casula}}, \bibinfo {author} {\bibfnamefont
  {D.~M.}\ \bibnamefont {Ceperley}},  \emph {et~al.},\ }\href {\doibase
  10.1088/1361-648X/aab9c3} {\bibfield  {journal} {\bibinfo  {journal} {Journal
  of Physics: Condensed Matter}\ }\textbf {\bibinfo {volume} {30}},\ \bibinfo
  {pages} {195901} (\bibinfo {year} {2018})}\BibitemShut {NoStop}%
\bibitem [{\citenamefont {Kent}\ \emph {et~al.}(2020)\citenamefont {Kent},
  \citenamefont {Annaberdiyev}, \citenamefont {Benali}, \citenamefont
  {Bennett}, \citenamefont {Landinez~Borda}, \citenamefont {Doak},
  \citenamefont {Hao}, \citenamefont {Jordan}, \citenamefont {Krogel},
  \citenamefont
  {Kyl{\ifmmode\ddot{a}\else\"{a}\fi}np{\ifmmode\ddot{a}\else\"{a}\fi}{\ifmmode\ddot{a}\else\"{a}\fi}},
  \citenamefont {Lee}, \citenamefont {Luo}, \citenamefont {Malone},
  \citenamefont {Melton}, \citenamefont {Mitas}, \citenamefont {Morales},
  \citenamefont {Neuscamman}, \citenamefont {Reboredo}, \citenamefont
  {Rubenstein}, \citenamefont {Saritas}, \citenamefont {Upadhyay},
  \citenamefont {Wang}, \citenamefont {Zhang},\ and\ \citenamefont
  {Zhao}}]{Kent2020May}%
  \BibitemOpen
  \bibfield  {author} {\bibinfo {author} {\bibfnamefont {P.~R.~C.}\
  \bibnamefont {Kent}}, \bibinfo {author} {\bibfnamefont {A.}~\bibnamefont
  {Annaberdiyev}}, \bibinfo {author} {\bibfnamefont {A.}~\bibnamefont
  {Benali}}, \bibinfo {author} {\bibfnamefont {M.~C.}\ \bibnamefont {Bennett}},
  \bibinfo {author} {\bibfnamefont {E.~J.}\ \bibnamefont {Landinez~Borda}},
  \bibinfo {author} {\bibfnamefont {P.}~\bibnamefont {Doak}}, \bibinfo {author}
  {\bibfnamefont {H.}~\bibnamefont {Hao}}, \bibinfo {author} {\bibfnamefont
  {K.~D.}\ \bibnamefont {Jordan}}, \bibinfo {author} {\bibfnamefont {J.~T.}\
  \bibnamefont {Krogel}}, \bibinfo {author} {\bibfnamefont {I.}~\bibnamefont
  {Kyl{\ifmmode\ddot{a}\else\"{a}\fi}np{\ifmmode\ddot{a}\else\"{a}\fi}{\ifmmode\ddot{a}\else\"{a}\fi}}},
  \bibinfo {author} {\bibfnamefont {J.}~\bibnamefont {Lee}}, \bibinfo {author}
  {\bibfnamefont {Y.}~\bibnamefont {Luo}}, \bibinfo {author} {\bibfnamefont
  {F.~D.}\ \bibnamefont {Malone}}, \bibinfo {author} {\bibfnamefont {C.~A.}\
  \bibnamefont {Melton}}, \bibinfo {author} {\bibfnamefont {L.}~\bibnamefont
  {Mitas}}, \bibinfo {author} {\bibfnamefont {M.~A.}\ \bibnamefont {Morales}},
  \bibinfo {author} {\bibfnamefont {E.}~\bibnamefont {Neuscamman}}, \bibinfo
  {author} {\bibfnamefont {F.~A.}\ \bibnamefont {Reboredo}}, \bibinfo {author}
  {\bibfnamefont {B.}~\bibnamefont {Rubenstein}}, \bibinfo {author}
  {\bibfnamefont {K.}~\bibnamefont {Saritas}}, \bibinfo {author} {\bibfnamefont
  {S.}~\bibnamefont {Upadhyay}}, \bibinfo {author} {\bibfnamefont
  {G.}~\bibnamefont {Wang}}, \bibinfo {author} {\bibfnamefont {S.}~\bibnamefont
  {Zhang}}, \ and\ \bibinfo {author} {\bibfnamefont {L.}~\bibnamefont {Zhao}},\
  }\href {\doibase 10.1063/5.0004860} {\bibfield  {journal} {\bibinfo
  {journal} {The Journal of Chemical Physics}\ }\textbf {\bibinfo {volume}
  {152}},\ \bibinfo {pages} {174105} (\bibinfo {year} {2020})}\BibitemShut
  {NoStop}%
\bibitem [{\citenamefont {Goedecker}\ \emph {et~al.}(1996)\citenamefont
  {Goedecker}, \citenamefont {Teter},\ and\ \citenamefont
  {Hutter}}]{Goedecker1996}%
  \BibitemOpen
  \bibfield  {author} {\bibinfo {author} {\bibfnamefont {S.}~\bibnamefont
  {Goedecker}}, \bibinfo {author} {\bibfnamefont {M.}~\bibnamefont {Teter}}, \
  and\ \bibinfo {author} {\bibfnamefont {J.}~\bibnamefont {Hutter}},\ }\href
  {\doibase 10.1103/PhysRevB.54.1703} {\bibfield  {journal} {\bibinfo
  {journal} {Physical Review B}\ }\textbf {\bibinfo {volume} {54}},\ \bibinfo
  {pages} {1703} (\bibinfo {year} {1996})}\BibitemShut {NoStop}%
\bibitem [{\citenamefont {Hutter}()}]{Hutter}%
  \BibitemOpen
  \bibfield  {author} {\bibinfo {author} {\bibfnamefont {J.}~\bibnamefont
  {Hutter}},\ }\href {https://github.com/juerghutter/GTH} {\enquote {\bibinfo
  {title} {New optimization of gth pseudopotentials for pbe, scan, pbe0
  functionals. gth pseudopotentials for hartree-fock. nlcc pseudopotentials for
  pbe},}\ }\BibitemShut {NoStop}%
\bibitem [{\citenamefont {VandeVondele}\ and\ \citenamefont
  {Hutter}(2007)}]{VandeVondele2007}%
  \BibitemOpen
  \bibfield  {author} {\bibinfo {author} {\bibfnamefont {J.}~\bibnamefont
  {VandeVondele}}\ and\ \bibinfo {author} {\bibfnamefont {J.}~\bibnamefont
  {Hutter}},\ }\href {\doibase 10.1063/1.2770708} {\bibfield  {journal}
  {\bibinfo  {journal} {Journal of Chemical Physics}\ }\textbf {\bibinfo
  {volume} {127}},\ \bibinfo {pages} {114105} (\bibinfo {year}
  {2007})}\BibitemShut {NoStop}%
\bibitem [{\citenamefont {Georges}\ \emph
  {et~al.}(1996{\natexlab{b}})\citenamefont {Georges}, \citenamefont {Kotliar},
  \citenamefont {Krauth},\ and\ \citenamefont {Rozenberg}}]{Georges1996}%
  \BibitemOpen
  \bibfield  {author} {\bibinfo {author} {\bibfnamefont {A.}~\bibnamefont
  {Georges}}, \bibinfo {author} {\bibfnamefont {G.}~\bibnamefont {Kotliar}},
  \bibinfo {author} {\bibfnamefont {W.}~\bibnamefont {Krauth}}, \ and\ \bibinfo
  {author} {\bibfnamefont {M.~J.}\ \bibnamefont {Rozenberg}},\ }\href {\doibase
  10.1103/RevModPhys.68.13} {\bibfield  {journal} {\bibinfo  {journal} {Reviews
  of Modern Physics}\ }\textbf {\bibinfo {volume} {68}},\ \bibinfo {pages} {13}
  (\bibinfo {year} {1996}{\natexlab{b}})}\BibitemShut {NoStop}%
\bibitem [{\citenamefont {Kotliar}\ \emph {et~al.}(2006)\citenamefont
  {Kotliar}, \citenamefont {Savrasov}, \citenamefont {Haule}, \citenamefont
  {Oudovenko}, \citenamefont {Parcollet},\ and\ \citenamefont
  {Marianetti}}]{Kotliar2006}%
  \BibitemOpen
  \bibfield  {author} {\bibinfo {author} {\bibfnamefont {G.}~\bibnamefont
  {Kotliar}}, \bibinfo {author} {\bibfnamefont {S.~Y.}\ \bibnamefont
  {Savrasov}}, \bibinfo {author} {\bibfnamefont {K.}~\bibnamefont {Haule}},
  \bibinfo {author} {\bibfnamefont {V.~S.}\ \bibnamefont {Oudovenko}}, \bibinfo
  {author} {\bibfnamefont {O.}~\bibnamefont {Parcollet}}, \ and\ \bibinfo
  {author} {\bibfnamefont {C.~A.}\ \bibnamefont {Marianetti}},\ }\href
  {\doibase 10.1103/RevModPhys.78.865} {\bibfield  {journal} {\bibinfo
  {journal} {Reviews of Modern Physics}\ }\textbf {\bibinfo {volume} {78}},\
  \bibinfo {pages} {865} (\bibinfo {year} {2006})}\BibitemShut {NoStop}%
\bibitem [{\citenamefont {Held}(2007)}]{Held2007}%
  \BibitemOpen
  \bibfield  {author} {\bibinfo {author} {\bibfnamefont {K.}~\bibnamefont
  {Held}},\ }\href {\doibase 10.1080/00018730701619647} {\bibfield  {journal}
  {\bibinfo  {journal} {Advances in Physics}\ }\textbf {\bibinfo {volume}
  {56}},\ \bibinfo {pages} {829} (\bibinfo {year} {2007})}\BibitemShut
  {NoStop}%
\bibitem [{\citenamefont {Vollhardt}(2020)}]{Vollhardt2020}%
  \BibitemOpen
  \bibfield  {author} {\bibinfo {author} {\bibfnamefont {D.}~\bibnamefont
  {Vollhardt}}\ }(\bibinfo {year} {2020})\ p.\ \bibinfo {pages}
  {11001}\BibitemShut {NoStop}%
\bibitem [{\citenamefont {Knizia}\ and\ \citenamefont
  {Chan}(2012{\natexlab{b}})}]{Knizia2012}%
  \BibitemOpen
  \bibfield  {author} {\bibinfo {author} {\bibfnamefont {G.}~\bibnamefont
  {Knizia}}\ and\ \bibinfo {author} {\bibfnamefont {G.~K.-L.}\ \bibnamefont
  {Chan}},\ }\href {\doibase 10.1103/PhysRevLett.109.186404} {\bibfield
  {journal} {\bibinfo  {journal} {Physical Review Letters}\ }\textbf {\bibinfo
  {volume} {109}},\ \bibinfo {pages} {186404} (\bibinfo {year}
  {2012}{\natexlab{b}})}\BibitemShut {NoStop}%
\bibitem [{\citenamefont {Knizia}\ and\ \citenamefont
  {Chan}(2013)}]{Knizia2013}%
  \BibitemOpen
  \bibfield  {author} {\bibinfo {author} {\bibfnamefont {G.}~\bibnamefont
  {Knizia}}\ and\ \bibinfo {author} {\bibfnamefont {G.~K.~L.}\ \bibnamefont
  {Chan}},\ }\href {\doibase 10.1021/ct301044e} {\bibfield  {journal} {\bibinfo
   {journal} {Journal of Chemical Theory and Computation}\ }\textbf {\bibinfo
  {volume} {9}},\ \bibinfo {pages} {1428} (\bibinfo {year} {2013})}\BibitemShut
  {NoStop}%
\bibitem [{\citenamefont {Liu}\ \emph {et~al.}(2023)\citenamefont {Liu},
  \citenamefont {Meitei}, \citenamefont {Chin}, \citenamefont {Dutt},
  \citenamefont {Tao}, \citenamefont {Van~Voorhis},\ and\ \citenamefont
  {Chuang}}]{liu2023bootstrap}%
  \BibitemOpen
  \bibfield  {author} {\bibinfo {author} {\bibfnamefont {Y.}~\bibnamefont
  {Liu}}, \bibinfo {author} {\bibfnamefont {O.~R.}\ \bibnamefont {Meitei}},
  \bibinfo {author} {\bibfnamefont {Z.~E.}\ \bibnamefont {Chin}}, \bibinfo
  {author} {\bibfnamefont {A.}~\bibnamefont {Dutt}}, \bibinfo {author}
  {\bibfnamefont {M.}~\bibnamefont {Tao}}, \bibinfo {author} {\bibfnamefont
  {T.}~\bibnamefont {Van~Voorhis}}, \ and\ \bibinfo {author} {\bibfnamefont
  {I.~L.}\ \bibnamefont {Chuang}},\ }\href {https://arxiv.org/abs/2301.01457}
  {\bibfield  {journal} {\bibinfo  {journal} {arXiv:2301.01457}\ } (\bibinfo
  {year} {2023})}\BibitemShut {NoStop}%
\bibitem [{\citenamefont {Rubin}\ \emph {et~al.}(2022)\citenamefont {Rubin},
  \citenamefont {Lee},\ and\ \citenamefont {Babbush}}]{rubin2022compressing}%
  \BibitemOpen
  \bibfield  {author} {\bibinfo {author} {\bibfnamefont {N.~C.}\ \bibnamefont
  {Rubin}}, \bibinfo {author} {\bibfnamefont {J.}~\bibnamefont {Lee}}, \ and\
  \bibinfo {author} {\bibfnamefont {R.}~\bibnamefont {Babbush}},\ }\href
  {https://pubs.acs.org/doi/abs/10.1021/acs.jctc.1c00912} {\bibfield  {journal}
  {\bibinfo  {journal} {Journal of Chemical Theory and Computation}\ }\textbf
  {\bibinfo {volume} {18}},\ \bibinfo {pages} {1480} (\bibinfo {year}
  {2022})}\BibitemShut {NoStop}%
\bibitem [{\citenamefont {Malone}\ \emph {et~al.}(2020)\citenamefont {Malone},
  \citenamefont {Zhang},\ and\ \citenamefont
  {Morales}}]{malone2020accelerating}%
  \BibitemOpen
  \bibfield  {author} {\bibinfo {author} {\bibfnamefont {F.~D.}\ \bibnamefont
  {Malone}}, \bibinfo {author} {\bibfnamefont {S.}~\bibnamefont {Zhang}}, \
  and\ \bibinfo {author} {\bibfnamefont {M.~A.}\ \bibnamefont {Morales}},\
  }\href {https://pubs.acs.org/doi/abs/10.1021/acs.jctc.0c00262} {\bibfield
  {journal} {\bibinfo  {journal} {Journal of Chemical Theory and Computation}\
  }\textbf {\bibinfo {volume} {16}},\ \bibinfo {pages} {4286} (\bibinfo {year}
  {2020})}\BibitemShut {NoStop}%
\bibitem [{\citenamefont {Morales}\ and\ \citenamefont
  {Malone}(2020)}]{morales2020accelerating}%
  \BibitemOpen
  \bibfield  {author} {\bibinfo {author} {\bibfnamefont {M.~A.}\ \bibnamefont
  {Morales}}\ and\ \bibinfo {author} {\bibfnamefont {F.~D.}\ \bibnamefont
  {Malone}},\ }\href {https://aip.scitation.org/doi/10.1063/5.0025390}
  {\bibfield  {journal} {\bibinfo  {journal} {The Journal of Chemical Physics}\
  }\textbf {\bibinfo {volume} {153}},\ \bibinfo {pages} {194111} (\bibinfo
  {year} {2020})}\BibitemShut {NoStop}%
\bibitem [{\citenamefont {Chiesa}\ \emph {et~al.}(2006)\citenamefont {Chiesa},
  \citenamefont {Ceperley}, \citenamefont {Martin},\ and\ \citenamefont
  {Holzmann}}]{chiesa2006finite}%
  \BibitemOpen
  \bibfield  {author} {\bibinfo {author} {\bibfnamefont {S.}~\bibnamefont
  {Chiesa}}, \bibinfo {author} {\bibfnamefont {D.~M.}\ \bibnamefont
  {Ceperley}}, \bibinfo {author} {\bibfnamefont {R.~M.}\ \bibnamefont
  {Martin}}, \ and\ \bibinfo {author} {\bibfnamefont {M.}~\bibnamefont
  {Holzmann}},\ }\href
  {https://journals.aps.org/prl/abstract/10.1103/PhysRevLett.97.076404}
  {\bibfield  {journal} {\bibinfo  {journal} {Physical Review Letters}\
  }\textbf {\bibinfo {volume} {97}},\ \bibinfo {pages} {076404} (\bibinfo
  {year} {2006})}\BibitemShut {NoStop}%
\bibitem [{\citenamefont {Drummond}\ \emph {et~al.}(2008)\citenamefont
  {Drummond}, \citenamefont {Needs}, \citenamefont {Sorouri},\ and\
  \citenamefont {Foulkes}}]{drummond2008finite}%
  \BibitemOpen
  \bibfield  {author} {\bibinfo {author} {\bibfnamefont {N.}~\bibnamefont
  {Drummond}}, \bibinfo {author} {\bibfnamefont {R.}~\bibnamefont {Needs}},
  \bibinfo {author} {\bibfnamefont {A.}~\bibnamefont {Sorouri}}, \ and\
  \bibinfo {author} {\bibfnamefont {W.}~\bibnamefont {Foulkes}},\ }\href
  {https://journals.aps.org/prb/abstract/10.1103/PhysRevB.78.125106} {\bibfield
   {journal} {\bibinfo  {journal} {Physical Review B}\ }\textbf {\bibinfo
  {volume} {78}},\ \bibinfo {pages} {125106} (\bibinfo {year}
  {2008})}\BibitemShut {NoStop}%
\bibitem [{\citenamefont {Azadi}\ and\ \citenamefont
  {Foulkes}(2015)}]{azadi2015systematic}%
  \BibitemOpen
  \bibfield  {author} {\bibinfo {author} {\bibfnamefont {S.}~\bibnamefont
  {Azadi}}\ and\ \bibinfo {author} {\bibfnamefont {W.}~\bibnamefont
  {Foulkes}},\ }\href {https://aip.scitation.org/doi/abs/10.1063/1.4922619}
  {\bibfield  {journal} {\bibinfo  {journal} {The Journal of Chemical Physics}\
  }\textbf {\bibinfo {volume} {143}},\ \bibinfo {pages} {102807} (\bibinfo
  {year} {2015})}\BibitemShut {NoStop}%
\bibitem [{\citenamefont {Holzmann}\ \emph {et~al.}(2016)\citenamefont
  {Holzmann}, \citenamefont {Clay~III}, \citenamefont {Morales}, \citenamefont
  {Tubman}, \citenamefont {Ceperley},\ and\ \citenamefont
  {Pierleoni}}]{holzmann2016theory}%
  \BibitemOpen
  \bibfield  {author} {\bibinfo {author} {\bibfnamefont {M.}~\bibnamefont
  {Holzmann}}, \bibinfo {author} {\bibfnamefont {R.~C.}\ \bibnamefont
  {Clay~III}}, \bibinfo {author} {\bibfnamefont {M.~A.}\ \bibnamefont
  {Morales}}, \bibinfo {author} {\bibfnamefont {N.~M.}\ \bibnamefont {Tubman}},
  \bibinfo {author} {\bibfnamefont {D.~M.}\ \bibnamefont {Ceperley}}, \ and\
  \bibinfo {author} {\bibfnamefont {C.}~\bibnamefont {Pierleoni}},\ }\href
  {https://journals.aps.org/prb/abstract/10.1103/PhysRevB.94.035126} {\bibfield
   {journal} {\bibinfo  {journal} {Physical Review B}\ }\textbf {\bibinfo
  {volume} {94}},\ \bibinfo {pages} {035126} (\bibinfo {year}
  {2016})}\BibitemShut {NoStop}%
\bibitem [{\citenamefont {Dagrada}\ \emph {et~al.}(2016)\citenamefont
  {Dagrada}, \citenamefont {Karakuzu}, \citenamefont {Vildosola}, \citenamefont
  {Casula},\ and\ \citenamefont {Sorella}}]{dagrada2016exact}%
  \BibitemOpen
  \bibfield  {author} {\bibinfo {author} {\bibfnamefont {M.}~\bibnamefont
  {Dagrada}}, \bibinfo {author} {\bibfnamefont {S.}~\bibnamefont {Karakuzu}},
  \bibinfo {author} {\bibfnamefont {V.~L.}\ \bibnamefont {Vildosola}}, \bibinfo
  {author} {\bibfnamefont {M.}~\bibnamefont {Casula}}, \ and\ \bibinfo {author}
  {\bibfnamefont {S.}~\bibnamefont {Sorella}},\ }\href
  {https://journals.aps.org/prb/abstract/10.1103/PhysRevB.94.245108} {\bibfield
   {journal} {\bibinfo  {journal} {Physical Review B}\ }\textbf {\bibinfo
  {volume} {94}},\ \bibinfo {pages} {245108} (\bibinfo {year}
  {2016})}\BibitemShut {NoStop}%
\bibitem [{\citenamefont {Mihm}\ \emph {et~al.}(2021)\citenamefont {Mihm},
  \citenamefont {Sch{\"a}fer}, \citenamefont {Ramadugu}, \citenamefont
  {Weiler}, \citenamefont {Gr{\"u}neis},\ and\ \citenamefont
  {Shepherd}}]{mihm2021shortcut}%
  \BibitemOpen
  \bibfield  {author} {\bibinfo {author} {\bibfnamefont {T.~N.}\ \bibnamefont
  {Mihm}}, \bibinfo {author} {\bibfnamefont {T.}~\bibnamefont {Sch{\"a}fer}},
  \bibinfo {author} {\bibfnamefont {S.~K.}\ \bibnamefont {Ramadugu}}, \bibinfo
  {author} {\bibfnamefont {L.}~\bibnamefont {Weiler}}, \bibinfo {author}
  {\bibfnamefont {A.}~\bibnamefont {Gr{\"u}neis}}, \ and\ \bibinfo {author}
  {\bibfnamefont {J.~J.}\ \bibnamefont {Shepherd}},\ }\href
  {https://www.nature.com/articles/s43588-021-00165-1} {\bibfield  {journal}
  {\bibinfo  {journal} {Nature Computational Science}\ }\textbf {\bibinfo
  {volume} {1}},\ \bibinfo {pages} {801} (\bibinfo {year} {2021})}\BibitemShut
  {NoStop}%
\bibitem [{\citenamefont {Yang}\ \emph {et~al.}(2020)\citenamefont {Yang},
  \citenamefont {Gorelov}, \citenamefont {Pierleoni}, \citenamefont {Ceperley},
  \citenamefont {Holzmann} \emph {et~al.}}]{yang2020electronic}%
  \BibitemOpen
  \bibfield  {author} {\bibinfo {author} {\bibfnamefont {Y.}~\bibnamefont
  {Yang}}, \bibinfo {author} {\bibfnamefont {V.}~\bibnamefont {Gorelov}},
  \bibinfo {author} {\bibfnamefont {C.}~\bibnamefont {Pierleoni}}, \bibinfo
  {author} {\bibfnamefont {D.~M.}\ \bibnamefont {Ceperley}}, \bibinfo {author}
  {\bibfnamefont {M.}~\bibnamefont {Holzmann}},  \emph {et~al.},\ }\href
  {https://journals.aps.org/prb/abstract/10.1103/PhysRevB.101.085115}
  {\bibfield  {journal} {\bibinfo  {journal} {Physical Review B}\ }\textbf
  {\bibinfo {volume} {101}},\ \bibinfo {pages} {085115} (\bibinfo {year}
  {2020})}\BibitemShut {NoStop}%
\bibitem [{\citenamefont {Sanders}\ \emph {et~al.}(2020)\citenamefont
  {Sanders}, \citenamefont {Berry}, \citenamefont {Costa}, \citenamefont
  {Tessler}, \citenamefont {Wiebe}, \citenamefont {Gidney}, \citenamefont
  {Neven},\ and\ \citenamefont {Babbush}}]{Sanders2020}%
  \BibitemOpen
  \bibfield  {author} {\bibinfo {author} {\bibfnamefont {Y.~R.}\ \bibnamefont
  {Sanders}}, \bibinfo {author} {\bibfnamefont {D.~W.}\ \bibnamefont {Berry}},
  \bibinfo {author} {\bibfnamefont {P.~C.~S.}\ \bibnamefont {Costa}}, \bibinfo
  {author} {\bibfnamefont {L.~W.}\ \bibnamefont {Tessler}}, \bibinfo {author}
  {\bibfnamefont {N.}~\bibnamefont {Wiebe}}, \bibinfo {author} {\bibfnamefont
  {C.}~\bibnamefont {Gidney}}, \bibinfo {author} {\bibfnamefont
  {H.}~\bibnamefont {Neven}}, \ and\ \bibinfo {author} {\bibfnamefont
  {R.}~\bibnamefont {Babbush}},\ }\href
  {https://journals.aps.org/prxquantum/abstract/10.1103/PRXQuantum.1.020312}
  {\bibfield  {journal} {\bibinfo  {journal} {PRX Quantum}\ }\textbf {\bibinfo
  {volume} {1}},\ \bibinfo {pages} {020312} (\bibinfo {year}
  {2020})}\BibitemShut {NoStop}%
\bibitem [{\citenamefont {DeYonker}\ \emph {et~al.}(2007)\citenamefont
  {DeYonker}, \citenamefont {Peterson}, \citenamefont {Steyl}, \citenamefont
  {Wilson},\ and\ \citenamefont {Cundari}}]{DeYonker2007}%
  \BibitemOpen
  \bibfield  {author} {\bibinfo {author} {\bibfnamefont {N.~J.}\ \bibnamefont
  {DeYonker}}, \bibinfo {author} {\bibfnamefont {K.~A.}\ \bibnamefont
  {Peterson}}, \bibinfo {author} {\bibfnamefont {G.}~\bibnamefont {Steyl}},
  \bibinfo {author} {\bibfnamefont {A.~K.}\ \bibnamefont {Wilson}}, \ and\
  \bibinfo {author} {\bibfnamefont {T.~R.}\ \bibnamefont {Cundari}},\ }\href
  {\doibase 10.1021/jp0715023} {\bibfield  {journal} {\bibinfo  {journal}
  {Journal of Physical Chemistry A}\ }\textbf {\bibinfo {volume} {111}},\
  \bibinfo {pages} {11269} (\bibinfo {year} {2007})}\BibitemShut {NoStop}%
\bibitem [{\citenamefont {Lee}\ and\ \citenamefont {Taylor}(1989)}]{Lee1989}%
  \BibitemOpen
  \bibfield  {author} {\bibinfo {author} {\bibfnamefont {T.~J.}\ \bibnamefont
  {Lee}}\ and\ \bibinfo {author} {\bibfnamefont {P.~R.}\ \bibnamefont
  {Taylor}},\ }\href {\doibase 10.1002/qua.560360824} {\bibfield  {journal}
  {\bibinfo  {journal} {International Journal of Quantum Chemistry}\ }\textbf
  {\bibinfo {volume} {36}},\ \bibinfo {pages} {199} (\bibinfo {year}
  {1989})}\BibitemShut {NoStop}%
\bibitem [{\citenamefont {Nielsen}\ and\ \citenamefont
  {Janssen}(1999)}]{Nielsen1999}%
  \BibitemOpen
  \bibfield  {author} {\bibinfo {author} {\bibfnamefont {I.~M.}\ \bibnamefont
  {Nielsen}}\ and\ \bibinfo {author} {\bibfnamefont {C.~L.}\ \bibnamefont
  {Janssen}},\ }\href {\doibase 10.1016/S0009-2614(99)00770-8} {\bibfield
  {journal} {\bibinfo  {journal} {Chemical Physics Letters}\ }\textbf {\bibinfo
  {volume} {310}},\ \bibinfo {pages} {568} (\bibinfo {year}
  {1999})}\BibitemShut {NoStop}%
\bibitem [{\citenamefont {Lee}(2003)}]{Lee2003}%
  \BibitemOpen
  \bibfield  {author} {\bibinfo {author} {\bibfnamefont {T.~J.}\ \bibnamefont
  {Lee}},\ }\href {\doibase 10.1016/S0009-2614(03)00435-4} {\bibfield
  {journal} {\bibinfo  {journal} {Chemical Physics Letters}\ }\textbf {\bibinfo
  {volume} {372}},\ \bibinfo {pages} {362} (\bibinfo {year}
  {2003})}\BibitemShut {NoStop}%
\bibitem [{\citenamefont {Lee}\ \emph {et~al.}(2022)\citenamefont {Lee},
  \citenamefont {Pham},\ and\ \citenamefont {Reichman}}]{lee2022twenty}%
  \BibitemOpen
  \bibfield  {author} {\bibinfo {author} {\bibfnamefont {J.}~\bibnamefont
  {Lee}}, \bibinfo {author} {\bibfnamefont {H.~Q.}\ \bibnamefont {Pham}}, \
  and\ \bibinfo {author} {\bibfnamefont {D.~R.}\ \bibnamefont {Reichman}},\
  }\href {\doibase 10.1021/acs.jctc.2c00802} {\bibfield  {journal} {\bibinfo
  {journal} {Journal of Chemical Theory and Computation}\ }\textbf {\bibinfo
  {volume} {18}},\ \bibinfo {pages} {7024} (\bibinfo {year}
  {2022})}\BibitemShut {NoStop}%
\end{thebibliography}%

\appendix

\section{Sparse representation derivations}
\subsection{The Pauli operator representation of the one-body term}
\label{app:onebodytermderv}
Here we derive the Pauli operator form of the one-body operator amenable to implementation as a Majorana select operation.
The one-body operator is rewritten as
\begin{align}
&\sum_{p,q=1}^{N/2} h_{p\K,q\K}a_{p\K\sigma}^\dagger a_{q\K\sigma} \mapsto \frac 14 \sum_{p,q=1}^{N/2} h_{p\K,q\K} [\vec Z (X_{p\K\sigma} - iY_{p\K\sigma})][\vec Z (X_{q\K\sigma} + iY_{q\K\sigma})] \nn
&= \frac 18 \sum_{p,q=1}^{N/2} h_{p\K,q\K} [\vec Z (X_{p\K\sigma} - iY_{p\K\sigma})][\vec Z (X_{q\K\sigma} + iY_{q\K\sigma})]  + \frac 18 \sum_{p,q=1}^{N/2} h_{q\K,p\K} [\vec Z (X_{q\K\sigma} - iY_{q\K\sigma})][\vec Z (X_{p\K\sigma}+ iY_{p\K\sigma})] \nn
&= \frac 18 \sum_{p\ne q=1}^{N/2} h_{p\K,q\K} [\vec Z (X_{p\K\sigma} - iY_{p\K\sigma})][\vec Z (X_{q\K\sigma} + iY_{q\K\sigma})]  - \frac 18 \sum_{p\ne q=1}^{N/2} h^{*}_{p\K,q\K} [\vec Z (X_{p\K\sigma} + iY_{p\K\sigma})][\vec Z (X_{q\K\sigma} - iY_{q\K\sigma})] \nn
&\quad +\frac 14 \sum_{p=1}^{N/2} h_{p\K,p\K} [\vec Z (X_{p\K\sigma} - iY_{p\K\sigma})][\vec Z (X_{p\K\sigma} + iY_{p\K\sigma})]\nn
&= \frac 18 \sum_{p\ne q=1}^{N/2} {\rm Re}(h_{p\K,q\K}) \left\{ [\vec Z (X_{p\K\sigma} - iY_{p\K\sigma})][\vec Z (X_{q\K\sigma} + iY_{q\K\sigma})] - [\vec Z (X_{p\K\sigma} + iY_{p\K\sigma})][\vec Z (X_{q\K\sigma} - iY_{q\K\sigma})]\right\} \nn & \quad 
+ \frac i8 \sum_{p,q=1}^{N/2}{\rm Im}(h_{p\K,q\K}) \left\{[\vec Z (X_{p\K\sigma} - iY_{p\K\sigma})][\vec Z (X_{q\K\sigma} + iY_{q\K\sigma})] + [\vec Z (X_{p\K\sigma} + iY_{p\K\sigma})][\vec Z (X_{q\K\sigma} - iY_{q\K\sigma})]\right\} \nn
&\quad +\frac 12 \sum_{p=1}^{N/2} h_{p\K,p\K} (\openone_{p\K\sigma} - Z_{p\K\sigma})\nn
&= \frac 14 \sum_{p\ne q=1}^{N/2} {\rm Re}(h_{p\K,q\K}) \left\{ -i\vec Z Y_{p\K\sigma}\vec Z X_{q\K\sigma}  + i\vec Z X_{p\K\sigma} \vec Z Y_{q\K\sigma}\right\} \nn & \quad 
+ \frac i4 \sum_{p,q=1}^{N/2}{\rm Im}(h_{p\K,q\K}) \left\{\vec Z X_{p\K\sigma} \vec Z X_{q\K\sigma}  + \vec Z Y_{p\K\sigma}\vec Z Y_{q\K\sigma}\right\}  +\frac 12 \sum_{p=1}^{N/2} h_{pp}(\K) (\openone_{p\K\sigma} - Z_{p\K\sigma})\nn
&= \frac i4 \sum_{p\ne q=1}^{N/2} {\rm Re}(h_{p\K, q\K}) \left\{ \vec Z X_{q\K\sigma}\vec Z Y_{p\K\sigma}  + \vec Z X_{p\K\sigma} \vec Z Y_{q\K\sigma}\right\} \nn & \quad 
+ \frac i4 \sum_{p,q=1}^{N/2}{\rm Im}(h_{p\K,q\K}) \left\{\vec Z X_{p\K\sigma} \vec Z X_{q\K\sigma}  + \vec Z Y_{p\K\sigma}\vec Z Y_{q\K\sigma}\right\} +\frac 12 \sum_{p=1}^{N/2} h_{pp}(\K) (\openone_{p\K\sigma} - Z_{p\K\sigma})\nn
&= \frac i2 \sum_{p,q=1}^{N/2} {\rm Re}(h_{p\K,q\K}) \vec Z X_{p\K\sigma} \vec Z Y_{q\K\sigma} 
+ \frac i4 \sum_{p,q=1}^{N/2}{\rm Im}(h_{p\K,q\K}) \left\{\vec Z X_{p\K\sigma} \vec Z X_{q\K\sigma}  + \vec Z Y_{p\K\sigma}\vec Z Y_{q\K\sigma}\right\}  +\frac 12 \sum_{p=1}^{N/2} h_{p\K,p\K} \openone .
\end{align}
In the last line we have used the symmetry of ${\rm Re}(h_{p\K,q\K})$ to combine $\vec Z X_{q\K\sigma}\vec Z Y_{p\K\sigma}$ and $\vec Z X_{p\K\sigma} \vec Z Y_{q\K\sigma}$, then used the fact that $iXY=-Z$ to combine the sum with $p\ne q$ with that for $p$.
The complete expression for the Hamiltonian has the sum over $\sigma$ and $\K$, which we have left out for simplicity here.
Including those gives the expression in \eq{simpleham}.

\subsection{One-body correction for sparse case}\label{app:sparse_one_body_derivation}
Next we derive the effective one-body term from the two-electron part of the Hamiltonian.
In the case $p=q$ and $\Q=0$, the second term in square brackets in \eq{H2genform} can be written as
\begin{align}
&- V^*_{p\K,q(\K\modmin\Q),r(\Kp\modmin\Q),s\Kp}a_{p\K\sigma}^{\dagger}a_{q(\K\modmin\Q)\sigma}a_{r(\Kp\modmin\Q)\tau}a_{s\Kp\tau}^{\dagger} \nn
&= V^*_{p\K,q(\K\modmin\Q),r(\Kp\modmin\Q),s\Kp}a_{p\K\sigma}a_{q(\K\modmin\Q)\sigma}^{\dagger}a_{r(\Kp\modmin\Q)\tau}a_{s\Kp\tau}^{\dagger} \nn
&\quad -V^*_{p\K,q(\K\modmin\Q),r(\Kp\modmin\Q),s\Kp}(a_{p\K\sigma}a_{q(\K\modmin\Q)\sigma}^{\dagger}+a_{p\K\sigma}^{\dagger}a_{q(\K\modmin\Q)\sigma})a_{r(\Kp\modmin\Q)\tau}a_{s\Kp\tau}^{\dagger} \nn
&= V^*_{p\K,q(\K\modmin\Q),r(\Kp\modmin\Q),s\Kp}a_{p\K\sigma}a_{q(\K\modmin\Q)\sigma}^{\dagger}a_{r(\Kp\modmin\Q)\tau}a_{s\Kp\tau}^{\dagger} \nn
&\quad -V^*_{p\K,q(\K\modmin\Q),r(\Kp\modmin\Q),s\Kp}a_{r(\Kp\modmin\Q)\tau}a_{s\Kp\tau}^{\dagger}.
\end{align}
In the last line we have used the fact that for $p=q$ and $\Q=0$, $a_{p\K\sigma}a_{q(\K\modmin\Q)\sigma}^{\dagger}+a_{p\K\sigma}^{\dagger}a_{q(\K\modmin\Q)\sigma}$ is just the identity, so this becomes a one-body operator.

Similarly, if $r=s$ and $\Q=0$ (but $p\ne q$), the second term in square brackets in \eq{H2genform} can be written as
\begin{align}
&- V^*_{p\K,q(\K\modmin\Q),r(\Kp\modmin\Q),s\Kp}a_{p\K\sigma}a_{q(\K\modmin\Q)\sigma}^{\dagger}a_{r(\Kp\modmin\Q)\tau}^{\dagger}a_{s\Kp\tau} \nn
&= V^*_{p\K,q(\K\modmin\Q),r(\Kp\modmin\Q),s\Kp}a_{p\K\sigma}a_{q(\K\modmin\Q)\sigma}^{\dagger}a_{r(\Kp\modmin\Q)\tau}a_{s\Kp\tau}^{\dagger} \nn
&\quad -V^*_{p\K,q(\K\modmin\Q),r(\Kp\modmin\Q),s\Kp}a_{p\K\sigma}a_{q(\K\modmin\Q)\sigma}^{\dagger}(a_{r(\Kp\modmin\Q)\tau}a_{s\Kp\tau}^{\dagger}+a_{r(\Kp\modmin\Q)\tau}^{\dagger}a_{s\Kp\tau}) \nn
&= V^*_{p\K,q(\K\modmin\Q),r(\Kp\modmin\Q),s\Kp}a_{p\K\sigma}a_{q(\K\modmin\Q)\sigma}^{\dagger}a_{r(\Kp\modmin\Q)\tau}a_{s\Kp\tau}^{\dagger} \nn
&\quad -V^*_{p\K,q(\K\modmin\Q),r(\Kp\modmin\Q),s\Kp}a_{p\K\sigma}a_{q(\K\modmin\Q)\sigma}^{\dagger}.
\end{align}
Thus we see that in either case ($p=q$ or $r=s$), we have the same expression as in Eq.~\eqref{eq:conjterm}, plus a one-body operator.
Moreover, because of the symmetry of $V$ (in swapping the $pq$ pair with the $rs$ pair), these corrections are equal.
Note also that we can relabel swapping $p$ with $q$ and $r$ with $s$ to replace $V^*_{p\K,q\K,r\Kp,s\Kp}a_{p\K\sigma}a_{q\K\sigma}^{\dagger}$ with (now explicitly taking $\Q=0$)
\begin{equation}
V^*_{q\K,p\K,s\Kp,r\Kp}a_{q\K\sigma}a_{p\K\sigma}^{\dagger} = -V_{p\K,q\K,r\Kp,s\Kp} a_{p\K\sigma}^{\dagger} a_{q\K\sigma} .
\end{equation}
This means that the contribution of these corrections is
\begin{equation}
\sum_{\sigma \in \{\uparrow, \downarrow\}} \sum_{\K}^{N_{k}}\sum_{p,q=1}^{N/2} \left(\sum_{r=1}^{N/2}\sum_{\Kp}^{N_{k}} V_{p\K,q\K,r\Kp,r\Kp} \right)  a_{p\K\sigma}^{\dagger} a_{q\K\sigma} .
\end{equation}
In this expression the constant factor is determined as follows.  There is a factor of $1/4$ in \eq{H2genform}.  Next, there is a factor of 2 because we have the contribution from $p=q$ as well as that from $r=s$.  Last, there is the factor of 2 from the summation over the spin $\tau$.  As a result, these factors cancel to give 1 above.
Therefore, for $p\ne q$, we can combine this one-body term with $\hpq$ as
\begin{equation}\label{eq:hhprime}
h'_{pq} = \hpq + \sum_{r=1}^{N/2}\sum_{\Kp}^{N_{k}} V_{p\K,q\K,r\Kp,r\Kp}.
\end{equation}

Next we consider the case where $p=q$, $r=s$, and $\Q=0$.
Then the second term in square brackets in \eq{H2genform} can be written as
\begin{align}
&V^*_{p\K,q(\K\modmin\Q),r(\Kp\modmin\Q),s\Kp}a_{p\K\sigma}^{\dagger}a_{q(\K\modmin\Q)\sigma}a_{r(\Kp\modmin\Q)\tau}^{\dagger}a_{s\Kp\tau} \nn
&= V^*_{p\K,q(\K\modmin\Q),r(\Kp\modmin\Q),s\Kp}a_{p\K\sigma}a_{q(\K\modmin\Q)\sigma}^{\dagger}a_{r(\Kp\modmin\Q)\tau}a_{s\Kp\tau}^{\dagger} \nn
&\quad +V^*_{p\K,q(\K\modmin\Q),r(\Kp\modmin\Q),s\Kp}(a_{p\K\sigma}^{\dagger}a_{q(\K\modmin\Q)\sigma}a_{r(\Kp\modmin\Q)\tau}^{\dagger}a_{s\Kp\tau}-a_{p\K\sigma}a_{q(\K\modmin\Q)\sigma}^{\dagger}a_{r(\Kp\modmin\Q)\tau}a_{s\Kp\tau}^{\dagger}) .\label{eq:rpeqrseq}
\end{align}
The operators in brackets in the final line can be written as, taking $p=q$, $r=s$, and $\Q=0$,
\begin{align}
& a_{p\K\sigma}^{\dagger}a_{p\K\sigma}a_{r\Kp\tau}^{\dagger}a_{r\Kp\tau}
+a_{p\K\sigma}a_{p\K\sigma}^{\dagger}a_{r\Kp\tau}^{\dagger}a_{r\Kp\tau}
-a_{p\K\sigma}a_{p\K\sigma}^{\dagger}a_{r\Kp\tau}^{\dagger}a_{r\Kp\tau}
-a_{p\K\sigma}a_{p\K\sigma}^{\dagger}a_{r\Kp\tau}a_{r\Kp\tau}^{\dagger} \nn
&=a_{r\Kp\tau}^{\dagger}a_{r\Kp\tau} - a_{p\K\sigma}a_{p\K\sigma}^{\dagger} \nn
& = a_{r\Kp\tau}^{\dagger}a_{r\Kp\tau} + a_{p\K\sigma}^{\dagger}a_{p\K\sigma} - \openone .
\end{align}
By symmetry of swapping $p$ and $q$, and swapping $r$ and $s$, we must be able to simplify the final line of \eqref{eq:rpeqrseq} to
\begin{equation}\label{eq:onebodycor}
V_{p\K,p\K,r\Kp,r\Kp} (a_{r\Kp\tau}^{\dagger} a_{r\Kp\tau} + a_{p\K\sigma}^{\dagger}a_{p\K\sigma} - \openone ).
\end{equation}
That is, this value of $V$ is real.
We can also relabel $r$ and $p$ and use symmetry to show the contribution from the first term in Eq.~\eqref{eq:onebodycor} is equivalent to
\begin{equation}
V_{r\Kp,r\Kp,p\K,p\K}a_{p\K\sigma}^{\dagger}a_{p\K\sigma} = V_{p\K,p\K,r\Kp,r\Kp} a_{p\K\sigma}^{\dagger}a_{p\K\sigma}.
\end{equation}
Hence the contribution of these corrections is
\begin{equation}
\sum_{\sigma \in \{\uparrow, \downarrow\}} \sum_{\K}^{N_{k}}\sum_{p=1}^{N/2} \left(\sum_{r=1}^{N/2}\sum_{\Kp}^{N_{k}} V_{p\K,p\K,r\Kp,r\Kp} \right)  (a_{p\K\sigma}^{\dagger}a_{p\K\sigma}-\openone/2).
\end{equation}
In this case, the constant factor comes from $1/4$ in \eq{H2genform}, and a factor of 2 from the sum over $\tau$.
As a result the expression in \eq{onebodycor} is divided by 2 here, and apart from the identity we have the same expression as that accounting for only one of the pairs $p,q$ and $r,s$ being equal.
The operator proportional to the identity can be ignored in the implementation of the Hamiltonian because it just gives a global shift in the eigenvalues.

\subsection{Complexity for sparse implementation}
\label{app:sparseimp}
The fundamental operator we are aiming to implement is in the form of \eq{simpleham} for the one-body term and \eq{twobody} for the two-body term.
In both we have a real part and an imaginary part; for the one-body term this is $\hpq$, and for the two-body term this is $V_{p\K,q(\K\modmin\Q),r(\Kp\modmin\Q),s\Kp}$.
We need to perform a state preparation that provides \emph{real} amplitudes for the real and imaginary parts of $h$ and $V$ on separate basis states (not just real and imaginary parts of an amplitude on each basis state).
This means the number of items of data to output is doubled in order to give the real and imaginary parts.
The state preparation is otherwise essentially unchanged from that in \cite{Berry2019B}, as described in Eq.~(48) of that work and the accompanying explanation.

Recall that in the sparse state preparation procedure, we use a register indexing the nonzero entries (see Eq.~(43) of \cite{Berry2019B}).
That is used to output ``ind'', ``alt'', and ``keep'' values via QROM (see Eq.~(44) of \cite{Berry2019B}).
The ``ind'' values are values of $p,q,r,s$, as well as the sign needed, and a qubit distinguishing between the one- and two-body terms.
The ``alt'' values are alternate values of these quantities, and ``keep'' governs the probability of swapping these registers for the state preparation via coherent alias sampling.
Since we need a bit to flag whether the amplitude being produced is for the real or imaginary part, that would indicate we need two extra bits output, one for the ``ind'' value and one for the ``alt'' value.
However, we can use one bit in the register indexing the nonzero entries to flag between real and imaginary parts.
It is most convenient to make this register the least significant bit.
Then we just need to produce ``alt'' values of this register, so the output size is only increased by 1 bit instead of 2.
A requirement for this approach is that the non-zero entries of $V$ that are retained are the same for the real and imaginary parts.

A further increase in the size of the output register is because we need to output values of $\K$, $\Kp$, and $\Q$.
The number of bits needed to store $\K$ is not simply $\lceil \log N_k \rceil$ because $\K$ is a vector.
The number of bits will be denoted $n_k$.
If we assume that the number of values is given by the product of numbers in the three dimensions $N_k=N_x N_y N_z$, then
\begin{equation}
n_k =  \lceil \log N_x \rceil + \lceil \log N_y \rceil + \lceil \log N_z \rceil.
\end{equation}
Therefore $\K$, $\Kp$, and $\Q$ increase the size of both the ind and alt registers by $3n_k$, for a total of $6n_k$.
The size of the output is given in Eq.~(A13) of \cite{Lee2020} as $m=\aleph+8\lceil \log(N/2) \rceil + 4$, and would here be increased to
\begin{equation}
\aleph+8\lceil \log(N/2) \rceil +6n_k + 5,
\end{equation}
where we have also increased the size of the output by 1 to account for selecting between real and imaginary parts, as discussed above.
The quantity $\aleph$ is the number of bits for the ``keep'' register.

The remaining consideration for the sparse state preparation is the symmetry.
In prior work there were three symmetries, with swap of $p,q$ with $r,s$ as well as swaps within the $p,q$ and $r,s$ pairs.
The method to take advantage of this was described from about Eq.~(49) on in \cite{Berry2019B}.
There you only perform the preparation for a restricted range of $p,q,r,s$, then use three qubits to control swaps to generate the symmetries.

Here we have the symmetry with swap of $p,q$ with $r,s$, but we can only swap the $p,q$ and $r,s$ pairs simultaneously.
We also need to take the complex conjugate when performing that swap.
In order to implement the symmetries here, we will have two control qubits.
One qubit will be in a $\ket{+}$ state and control swap of the $p,q$ with $r,s$ as before, except we now have the registers containing $\K,\Kp,\K\modmin\Q,\Kp\modmin\Q$ to swap.
That qubit is only set to $\ket{+}$ for the two-body term, since that symmetry does not make sense for the one-body term.
The second qubit is used to simultaneously swap the $p,q$ and $r,s$ pairs, as well as the registers containing $\K$, etc.
It will also be used as a control for a $Z$ phase gate on a qubit flagging imaginary components.
That is a Clifford gate and is not included in the Toffoli count.

The net result is that the cost of the swaps to produce these symmetries is unchanged from that in \cite{Berry2019B}, except in that we are counting the qubits needed to store $\K$, etc, as well as $p,q,r,s$.
Since a controlled swap of two qubits can be performed with a single Toffoli (and Clifford gates), the Toffoli cost of the two controlled swaps of registers is the total number of qubits used to store $p,q,r,s$ as well as $\K,\Kp,\K\modmin\Q,\Kp\modmin\Q$, which is $4\lceil \log(N/2) \rceil + 4n_k$.
Note that in the state preparation we will be producing $\K,\Kp,\Q$, and need to compute $\K\modmin\Q$ and $\Kp\modmin\Q$ before performing the swaps for these symmetries.

Assuming for the moment that $N_x,N_y,N_z$ are all powers of two, then the number of Toffolis needed for the modular subtractions of the three components will be $n_k-3$, unless one or more of $N_x,N_y,N_z$ is equal to 1.
It is simpler to give the cost as $n_k$ Toffolis, to avoid needing to address special cases.
A further complication is when one or more of $N_x,N_y,N_z$ are not powers of two.
In this case, the subtraction can be performed in the usual way for two's complement binary.
Then you can check if the result for any component is negative, and if it is then add the appropriate $N_x,N_y,N_z$ to make it non-negative.
The controlled addition of a classically given number has complexity $n_k$, so this at worst doubles the complexity to $2n_k$ for the modular subtraction.

The other major feature that we need to account for is the modified {\sc select} operation needed.
The basic circuit primitive was given in Figure~13 of \cite{Lee2020}, in order to apply $\vec Z Y_{p,\sigma}$ followed by $\vec Z X_{q,\sigma}$.
A more complicated circuit primitive was given in Figure~1 of \cite{Berry2019B}, which included testing $p=q$ which is not needed in the approach of \cite{Lee2020}.
Here the scheme is more complicated, because instead of having a fixed sequence where we need to apply $Y$ followed by $X$ we have every combination.
This can be achieved by simply performing each twice; once with a controlled $\vec Z Y$ and once with a controlled $\vec Z X$, with a doubling of the Toffoli complexity.
That can be seen easily from the diagram in Figure~9 of \cite{BabbushSpectra}.
There a control qubit is used, so that can be used to control application of this circuit with $Y$, then to control application of this circuit with $X$.

To understand how $X$ versus $Y$ is selected, note that there are effectively five bits controlling here.
Let us call the bit selecting between the one- and two-body terms $b_0$; this is created in the sparse state preparation.
Let us call the bit selecting real versus imaginary parts $b_1$; this is again created in the state preparation.
There also needs to be a bit $b_2$ for selecting between the two lines for real and the two lines for imaginary in the expression in \eq{twobody}.
Then we have $b_3$ to select between the two terms in the first set of square brackets in each line of \eq{twobody}, and a bit $b_4$ selecting between the two terms in the second set of square brackets.

Now, considering the operators indexed by $r,s$ first, these are applied for the two-body terms but not the one-body term.
This control of the operations adds only one Toffoli to the cost.
For the first operation, $\vec Z X_{s\Kp\tau}$ or $\vec Z Y_{s\Kp\tau}$, we can see that the selection between $X$ and $Y$ depends only on bit $b_4$.
For the second operation, the selection is independent of whether we have the real or imaginary part.
We select $X$ if we have $b_4=0$ (the first term) and $b_2=0$ (the first line), or if we have $b_4=b_2=1$.
To create a bit selecting between $X$ and $Y$ we can simply perform a CNOT between these bits, with no Toffoli cost.

Next, consider the operators indexed by $p,q$.
For simplicity we will first consider just the two-body terms.
Again the first operation can select between $X$ and $Y$ just by using the bit $b_3$ selecting between the terms.
Then for the second operation, we select $X$ if we have $b_1,b_2,b_3$ equal to $0,0,0$, or $0,1,1$, or $1,0,1$, or $1,1,0$.
It is easily seen that if we apply CNOTs with $b_1$ then $b_2$ as control and $b_3$ as target, then we should apply $X$ if we have $b_3=0$.
This selection can be performed without Toffolis again.

Now to take account of how the one-body terms are applied, it is convenient to rewrite the first line of \eq{simpleham} as
\begin{equation}
-\frac i4 \sum_{p,q=1}^{N/2} {\rm Re}(h_{p\K,q\K}) \left\{
\vec Z Y_{p\K\sigma} \vec Z X_{q\K\sigma} -
\vec Z X_{p\K\sigma} \vec Z Y_{q\K\sigma} \right\}.
\end{equation}
Then the selection between the operations is identical to that for $b_2=1$ (second lines) for the two-body part.
Therefore, for the above analysis of the two-body implementation, we can replace $b_2$ with a bit that is 1 if $b_2=1$ OR $b_0=0$.
This operation requires one more Toffoli.

Another modification we need to make is to compute $\Kp\modmin\Q$ and $\K\modmin\Q$ to use in the selection for the two-body operations.
As explained above, these modular subtractions have complexity at worst $2n_k$.
The calculation $\K\modmin\Q$ needs to be controlled on the bit $b_0$ selecting between the one- and two-body terms, which increases its complexity by $n_k$.
Therefore the complexity of this arithmetic is $3\lceil \log N_k\rceil$ Toffolis.
We can keep the working qubits in order to uncompute this arithmetic with Clifford gates.

Finally, we should account for the phase factors needed in the implementation.
The phase factors needed are as follows.
\begin{enumerate}
\item We should apply an $i$ phase factor on the one-body term. That can be implemented with an $S$ gate which is Clifford.
\item If we have the one-body term (flagged by $b_0=0$) we should flip the sign of the real part (flagged by $b_1=0$).
This can be done with a controlled phase, which is again Clifford.
\item For the two-body term ($b_0=1$), real ($b_1=0$), and second line ($b_2=1$) we should flip the sign.
This doubly controlled phase has a cost of one Toffoli.
\item We should flip the sign with $b_3=1$ if we have the two-body term ($b_0=1$) and the second line for real ($b_1=0,b_2=1$) or the first line for imaginary ($b_1=1,b_2=0$).
We should also flip the sign with $b_3=1$ if we have the one-body term ($b_0=0$) and real ($b_1=0$).
To achieve this we can first perform a CNOT with $b_1$ as control and $b_2$ as target.
Then, if $b_0=0,b_1=0$ OR $b_0=1,b_2=1$ we should apply a $Z$ gate to the qubit containing $b_3$.
This can be achieved with two double controlled phase gates, so has Toffoli cost 2.
\item We should flip the sign for $b_4=1$ if we have the second line $b_2=1$. That is just a controlled phase with no non-Clifford cost.
\end{enumerate}
As a result, the total complexity of implementing these phase factors is 3 Toffoli gates.

The total additional complexity is therefore $2$ Toffolis for the selection of $X$ versus $Y$ when we account for needing to perform the one-or two-body term, $3\lceil \log N_k\rceil$ Toffolis for subtractions, 3 Toffolis for phase factors, and doubling the selection cost to select between $X$ and $Y$.
The two Toffolis to account for the one-body term were one for selecting performing the operators indexed by $r,s$, and another Toffoli to perform an OR between $b_0$ and $b_2$ for the operators indexed by $p,q$.

A further complication arises where the $h$ and $V$ are dependent on the spins $\sigma$ and $\tau$.
This is easily accounted for by outputting the values of $\sigma,\tau$ as part of the state preparation.
This means that the size of both the ``ind'' and ``alt'' outputs are increased by 2, making the total size of the output increase by 4 to be
\begin{equation}
\aleph+8\lceil \log(N/2) \rceil +6n_k+ 9.
\end{equation}
Often there is the symmetry that for $V$ the value with $\sigma=\uparrow,\tau=\downarrow$ are the same as for $\sigma=\downarrow,\tau=\uparrow$.
This means that we can omit the case $\sigma=\downarrow,\tau=\uparrow$, and use a swap of these two qubits controlled by an ancilla qubit in the usual way for obtaining symmetries.
In the detailed costing below, we give results for the case where $h$ and $V$ are not dependent on spin for simplicity.

The QROM output size is
\begin{equation}
\wid = \zetabits +8n_N + 6n_k + 5,
\end{equation}
where $n_N=\lceil \log  (N/2) \rceil$. This output size is increased above that analysed in \cite{Lee2020}.
Then, using that output size, the formula for the cost of the preparation with $d$ unique nonzero entries is
\begin{equation}
\lceil\dm/\chunk_1\rceil+\wid(\chunk_1-1)
\end{equation}
and of the inverse preparation is
\begin{equation}
\lceil\dm/\chunk_2\rceil+\chunk_2 .
\end{equation}
Here $k_1$ and $k_2$ must be chosen as powers of 2.
This formula is the same as in \cite{Lee2020}, but with the modified value of $m$.

To begin the state preparation, we need to prepare an equal superposition state over $d$ basis states.
The analysis is described in \cite{Lee2020}, which gives the costing
$3\lceil \log d\rceil-3\factor+2\rotprec-9$ Toffoli gates.
Here $\eta$ is a number such that $2^\eta$ is a factor of $d$, and $b_r$ is a number of bits used for rotation of an ancilla qubit to improve the amplitude of success.
This is a cost needed both for the preparation and inverse preparation.

Other minor Toffoli costs are as follows.
We use extra ancillas to save cost, because a large number of ancillas were used for the QROM, and can be reused here without increasing the maximum number of ancillas needed.
In the following we use the notation $n_N=\lceil\log(N/2)\rceil$.
\begin{enumerate}
\item Perform \sel\ as shown in Figure~13 of \cite{Lee2020} twice,
but controlling between $X$ and $Y$. This complexity is $4NN_k-6$, since we have $8$ times a complexity of $NN_k/2-1$ for each of the selected operations, plus 2 Toffolis to generate the qubits we need for the control.
There were two Toffolis needed to account for selecting between one- and two- body terms, and otherwise the selection to account for the various terms can be performed using Clifford gates.
In addition to this, we need to perform swaps controlled by spin qubits twice for each of the two spin qubits, with a complexity $2NN_k$.
That then gives a total complexity of this step $6NN_k-6$.
\item The state preparation needs an inequality test on $\zetabits$ qubits, as well as controlled swaps.
The controlled swaps are on $n_N+3n_k + 2$ qubits.
Here $4\lceil\log(N/2)\rceil$ are for the values of $p$, $q$, $r$, and $s$, the $3n_k$ is for the $\K$, $\K'$, and $\Q$ values, a $+1$ is for the qubit which distinguishes between the one- and two-electron terms, and a further $+1$ comes from the qubit for selecting between the real and imaginary parts.
There are also ind and alt values of the sign, but the correct phase can be applied with Clifford gates, so this does not add to the Toffoli cost.
The cost of the inequality test on $\zetabits$ qubits is $\zetabits$.
As in \cite{Lee2020}, we can eliminate the non-Clifford cost of the inverse preparation using ancillas and measurements, so the Toffoli cost is $\zetabits+4n_N+3n_k +2$.
%For the spin-dependent case we also need to swap the two pairs of qubits for the spin, increasing the cost to $\zetabits+4n_N+3\lceil\log N_k\rceil+4$.
\item The controlled swaps used to generate the symmetries have a cost of $4n_N+4n_k$.
This is increased by $4n_k$ over that in \cite{Lee2020}, since we need to swap $\K$ registers as well.
Although there are only two controlled swaps rather than three in \cite{Lee2020}, two of the controlled swaps in \cite{Lee2020} together act on as many qubits as one controlled swap here, so the factor of 4 is the same as in \cite{Lee2020}.
A further $4n_k$ cost is for computing $\K\modmin\Q$ and $\Kp\modmin\Q$ (or $2n_k$ if $N_x,N_y,N_z$ are powers of 2), and an extra $n_k$ is needed to make the computation of $\K\modmin\Q$ controlled.
Again these controlled swaps can be inverted for the inverse preparation with measurements and Clifford gates.
Thus the total Toffoli cost here is $4n_N+9n_k$.
\item For the qubitization construction a reflection on the ancilla is needed as well.
The qubits that need to be reflected on are
\begin{enumerate}
    \item the $\lceil \log d \rceil$ qubits for preparing the state,
    \item  $\zetabits$ qubits for the equal superposition state in coherent alias sampling,
    \item two qubits that are used for controlled swaps to generate the symmetries of the state,
    \item the two spin qubits,
    \item the ancilla qubit that is rotated to produce the equal superposition state,
    \item and the qubits storing $b_2,b_3,b_4$, which are also used in the linear combination of unitaries.
\end{enumerate}
There is no non-Clifford Toffoli cost for the preparation on $b_2,b_3,b_4$, since an equal superposition may be prepared with a Hadamard.
They are control qubits that need to be reflected upon for the qubitization, so add a cost of 3 Toffolis to the reflection giving a total cost $\lceil \log d\rceil+\zetabits+6$.
\item As before, the control for the phase estimation uses unary iteration on the control registers, with one more Toffoli for each step.
The control by these registers is implemented simply by controlling the reflection, which needs just one Toffoli per step.
\item An extra three Toffolis are needed for the phase factors.
\end{enumerate}

Adding all these minor costs together gives, in the spin-independent case
\begin{align}
&2(3\lceil\log d\rceil - 3\factor+2\rotprec-9) + (6NN_k-6) + (\zetabits + 4n_N+3n_k +2)+4n_N+9n_k+\lceil\log d\rceil+\zetabits+6+2+3 \nn
&= 6NN_k + 8n_N + 10\lceil\log N_k\rceil + 2\zetabits + 7\lceil\log d\rceil - 6\factor +4\rotprec - 8.
\end{align}
The total cost for a single step is then
\begin{equation}\label{eq:sparseToffoli}
\left\lceil\frac{\dm}{\chunk_1}\right\rceil+\wid(\chunk_1-1) + \left\lceil\frac{\dm}{\chunk_2}\right\rceil+\chunk_2 + 6NN_k + 8n_N +12n_k + 2\zetabits + 7\lceil\log d\rceil - 6\factor +4\rotprec - 8 ,
\end{equation}
with $\wid = \zetabits +8n_N + 6\lceil\log N_k\rceil+5$, $n_N=\lceil\log(N/2)\rceil$, $\factor$ an integer such that $2^\factor$ is a factor of $d$, and $b_r$ the number of bits used for rotation of an ancilla qubit.

We may count the qubit costs by considering the maximum used during the QROM, as the advanced QROM has a high qubit usage that will not be exceeded in other parts of the algorithm.
The qubit costs are therefore as follows.
\begin{enumerate}
    \item The control register for the phase estimation uses $\lceil\log(\mathcal{I}+1)\rceil$ qubits, and there are $\lceil\log(\mathcal{I}+1)\rceil-1$ qubits for the unary iteration.
    \item The system uses $NN_k$ qubits.
    \item The $\lceil \log d\rceil+\zetabits+8$ qubits that need to be reflected upon listed above.
    \item A qubit is needed to flag success of the equal superposition state preparation.
    \item The phase gradient state uses $\rotprec$ qubits.
    \item The QROM uses qubits (including the output) $m\chunk_1+\lceil \log(d/\chunk_1)\rceil$.
\end{enumerate}
This gives a total number of logical qubits
\begin{equation}\label{eq:sparsequbits}
2\lceil\log(\mathcal{I}+1)\rceil + NN_k + \lceil \log d \rceil + \rotprec + \zetabits + m\chunk_1+\lceil \log(d/\chunk_1)\rceil+8,
\end{equation}
with $\wid = \zetabits +8n_N + 6\lceil \log N_k \rceil+ 5$.

\section{Single-factorization derivations}
\subsection{One-body correction for single factorization}
\label{app:singleonecor}

For the single factorized form of the Hamiltonian, we may use the same expressions for $\hat A$ and $\hat B$ for the case $\Q=0$ as for $\Q\ne 0$, with an additional correction proportional to the identity.
This yields a one-body correction in the case of $\hat{A}$ but not $\hat B$.
For $\hat{A}_{n}(\Q = 0)$ we obtain a term proportional to the identity, as follows
\begin{align}
\hat{A}_{n}(\Q = 0) &= \frac{1}{2}\sum_{\sigma \in \{\uparrow,\downarrow\}}\sum_{\K}\sum_{p \neq q}\left(L_{p\K, q\K, n}a_{p\K\sigma}^{\dagger}a_{q\K \sigma} + L_{p\K, q\K, n}^{*}a_{q\K\sigma}^{\dagger}a_{p\K\sigma} \right) + \sum_{\sigma \in \{\uparrow,\downarrow\}}\sum_{\K}\sum_{p}L_{p\K p\K, n}a_{p\K\sigma}^{\dagger}a_{p\K\sigma} \\
&= \sum_{\sigma \in \{\uparrow,\downarrow\}}\sum_{\K}^{N_{k}}\sum_{p\ne q}^{N/2}\left(\frac{i \Rea[L_{p\K q(\K\modmin\Q),n}]}{4}\left(\vec{Z}X_{p\K\sigma}\vec{Z}Y_{q(\K\modmin\Q)\sigma} - \vec{Z}Y_{p\K \sigma}\vec{Z}X_{q(\K\modmin\Q)\sigma}\right) \right. \nonumber \\
& \quad + \left. \frac{i\Ima[L_{p\K q(\K\modmin\Q),n}]}{4}\left(\vec{Z}X_{p\K\sigma}\vec{Z}X_{q(\K\modmin\Q)\sigma} + \vec{Z}Y_{p\K \sigma}\vec{Z}Y_{q(\K\modmin\Q)\sigma}\right)  \right) + \sum_{\sigma \in \{\uparrow,\downarrow\}}\sum_{\K}^{N_{k}}\sum_{p}^{N/2}\frac{L_{p\K,p\K,n}}{2}(\openone-Z) \nn
&= \sum_{\sigma \in \{\uparrow,\downarrow\}}\sum_{\K}^{N_{k}}\sum_{pq}^{N/2}\left(\frac{i \Rea[L_{p\K q(\K\modmin\Q),n}]}{4}\left(\vec{Z}X_{p\K\sigma}\vec{Z}Y_{q(\K\modmin\Q)\sigma} - \vec{Z}Y_{p\K \sigma}\vec{Z}X_{q(\K\modmin\Q)\sigma}\right) \right. \nonumber \\
& \quad + \left. \frac{i\Ima[L_{p\K q(\K\modmin\Q),n}]}{4}\left(\vec{Z}X_{p\K\sigma}\vec{Z}X_{q(\K\modmin\Q)\sigma} + \vec{Z}Y_{p\K \sigma}\vec{Z}Y_{q(\K\modmin\Q)\sigma}\right)  \right) + \sum_{\sigma \in \{\uparrow,\downarrow\}}\sum_{\K}^{N_{k}}\sum_{p}^{N/2}\frac{L_{p\K p\K,n}}{2}\openone .
\end{align}
Here we have used the symmetry of $L$, so $L_{p\K p\K, n}$ is real.
This derivation is similar to that for the one-body term in \app{onebodytermderv}.

Because $\hat{A}_{n}(\Q = 0)$ is squared, the identity term gives rise to a one-body correction
\begin{align}
&\frac i4 \sum_{\sigma \in \{\uparrow,\downarrow\}}\sum_{n}^{M}\sum_{\K}^{N_{k}}\sum_{p,q}^{N/2}\left(\Rea[L_{p\K q\K,n}]\left(\vec{Z}X_{p\K\sigma}\vec{Z}Y_{q\K \sigma} - \vec{Z}Y_{p\K \sigma}\vec{Z}X_{q\K \sigma}\right) \right. \nonumber \\
& \quad + \left. \Ima[L_{p\K q\K,n}]\left(\vec{Z}X_{p\K\sigma}\vec{Z}X_{q\K \sigma} + \vec{Z}Y_{p\K \sigma}\vec{Z}Y_{q\K\sigma}\right)  \right)\sum_{\Kp}^{N_{k}}\sum_{r=1}^{N/2}L_{r\Kp r\Kp,n} \nn
&= \frac i4 \sum_{\sigma \in \{\uparrow,\downarrow\}}\sum_{\K}^{N_{k}}\sum_{p,q}^{N/2}\sum_{\Kp}^{N_{k}}\sum_{r=1}^{N/2}\left(\Rea[V_{p\K,q\K,r\Kp,r\Kp}]\left(\vec{Z}X_{p\K\sigma}\vec{Z}Y_{q\K \sigma} - \vec{Z}Y_{p\K \sigma}\vec{Z}X_{q\K \sigma}\right) \right. \nonumber \\
& \quad + \left. \Ima[V_{p\K,q\K,r\Kp,r\Kp}]\left(\vec{Z}X_{p\K\sigma}\vec{Z}X_{q\K \sigma} + \vec{Z}Y_{p\K \sigma}\vec{Z}Y_{q\K\sigma}\right)  \right) .
\end{align}
Here there was a factor of $1/2$ on the square of $\hat{A}_{n}(\Q = 0)$, a factor of 2 from the cross term in the square, a factor of 2 from the sum over the spin on the identity, and so a factor of $1/2$ has been cancelled.
The form of this correction is identical to that for the one-body term, except $h_{p\K,q\K}$ is replaced with
\begin{equation}\label{eq:singfaconecor}
\sum_{r=1}^{N/2}\sum_{\Kp}^{N_{k}} V_{p\K,q\K,r\Kp,r\Kp}.
\end{equation}

For $\hat{B}_{n}(\Q = 0)$, it is easily seen that the symmetry $L_{p\K q\K,n}=L^*_{q\K p\K,n}$ implies that $\hat{B}_{n}(\Q = 0)=0$.
If we use the form for $\hat{B}_{n}(\Q = 0)$ in terms of Pauli operators given in \eq{Bform}, then it will be proportional to the identity due to the case $p=q$.
Squaring then just gives a correction proportional to the identity (which can be ignored in the implementation because it is just an energy shift), and it gives no one-body correction.
As a result we add the expression in \eq{singfaconecor} to $\hpq$ to obtain the complete one-body Hamiltonian given in \eq{H1Vp}.

\subsection{Complexity for single-factorized representation}
\label{app:single}

To see the changes we need to make to the algorithm for the single-factorized representation, recall that the two-body term was of the form \cite{Lee2020}
\begin{equation}
W' = \frac{1}{8} \sum_{\ell=1}^L \left(\sum_{\sigma \in \{\uparrow,\downarrow\}}\sum_{p,q=1}^{N/2} W^{(\ell)}_{pq} Q_{pq\sigma}\right)^2,
\end{equation}
where $Q_{pq\sigma}$ was an individual Pauli string.
So the changes in the representation are
\begin{itemize}
\item The sum over $\ell$ up to $L$ has been replaced with a sum over $\Q$ and $n$, as well as a sum over the squares of $A$ and $B$.
\item Inside the square, the sum over just $\sigma,p,q$ now also has a sum over $\K$.
\item Inside the sum, instead of just having a single Pauli string, we have a sum over $4$, with real and imaginary parts of $L_{p\K q(\K\modmin\Q),n}$.
\end{itemize}

The amendments we will make to the original algorithm (according to the description in \cite{Lee2020}) to implement the block encoding are as follows.
\begin{itemize}
\item For the sum over $\Q$ and $n$ we can combine them into $\ell$, and use the same state preparation method as before.
The value of $\Q$ will need to be used in the {\sc select} operation, so needs to be output as part of that state preparation.
\item In the preparation for the block encoding of $A$ and $B$, the index $\K$ will be needed as well as $p$ and $q$.
\item We no longer take advantage of $p,q$ symmetry.
\item We need to perform arithmetic to compute $\K\modmin\Q$ and $\Kp\modmin\Q$, with a cost of $4n_k$ (or $2n_k$ if $N_x,N_y,N_z$ are powers of 2).
\item A number of qubits can be used for selecting between the parts of the linear combination of unitaries, similar to the sparse case.
We have $b_0$ to select between the one- and two-body terms, $b_1$ for selecting between the real and imaginary parts, and $b_3$ selecting between the two terms in one application of $A$ or $B$.
The qubit $b_2$ can be used for selecting between $A$ and $B$, which is a change from the sparse case, where it was used for selecting between lines.
We do not need $b_4$ because we are implementing $A$ or $B$ twice (and creating the bit $b_3$ both times).
\item There needs to be a doubling of the selection cost to select between $X$ and $Y$ as in the sparse case.
\item The creation of the qubits for controlling between $X$ and $Y$ can be performed with one additional Toffoli.
Note first that the terms in $A$ are equivalent to the one-body part, and the terms in $B$ are the same except with the real and imaginary lines swapped around.
This means that we can use $b_0$ and $b_2$ as a control to flip $b_1$, which effectively swaps the real and imaginary parts for $B$ so it can be implemented in the same way.
Now, for the first selection of $X$ versus $Y$, we can apply a CNOT with $b_1$ as control and $b_3$ as target, and use that as control
For the second selection we can simply use $b_3$ as control.
\item For the phase factors, we just need a sign flip if $b_1=0$ and $b_3=1$, which is a Clifford controlled phase.
\end{itemize}

To explain the modifications needed for the costings, here we give the sequence of steps with the same numbering as in \cite{Lee2020}, explaining the differences.
\begin{enumerate}
\item We first prepare a state as
\begin{equation}
\frac 1{\sqrt{\lambda}} \left(\ket{0,0,0,0}\sqrt{\sum_{p, q}\left(|{\rm Re}(h'_{pq})|+|{\rm Im}(h'_{pq})|\right)} + \frac 1{\sqrt 2} \sum_{\Q,n}\ket{\ell,\Q,n,1}\sum_{\K,pq}(|\Rea[L_{p\K q(\K\modmin\Q),n}]|+|\Ima[L_{p\K q(\K\modmin\Q),n}]|) \right),
\end{equation}
where $\ket{\ell,\Q,n}$ indicates $\ell$ which starts from 1 indexing values of $\Q,n$, but $\Q$ and $n$ are also output in registers.
That is, we will be preparing $\ell$ while outputting values of $\Q,n$.
We are assuming the more difficult case where the number of values of $\Q$ or $n$ are not powers of 2, but if they are then further simplifications are possible.
This has complexity as follows.
\begin{enumerate}
\item Preparing an equal superposition on $MN_k+1$ basis states has complexity $3n_{MN}+2b_r-9$, where $b_r$ is the number of bits used for the rotation on the ancilla,
\begin{equation}
n_{MN} = \lceil\log(MN_k+1)\rceil .
\end{equation}
\item A QROM is applied with output size
\begin{equation}
b_{MN}= \zetabits_1+n_{MN}+2n_k+ 2 ,
\end{equation}
with $\zetabits_1$ being the number of bits used for the keep values (which govern the precision of the state preparation via the inequality test).
Here $n_{MN}$ and $2n_k$ are for $\ell$ and $\Q$, with the factor of 2 accounting for ind and alt values of $\Q$.
The extra 2 qubits are for outputting a qubit showing if $\ell=0$ (for selecting between the one- and two-body parts).
The complexity is
\begin{equation}
\left\lceil \frac{MN_k+1}{k_{MN}} \right\rceil + b_{MN}(k_{MN}-1).
\end{equation}
\item An inequality test is performed with complexity $\zetabits_1$.
\item A controlled swap is performed with complexity $n_k +\lceil \log M \rceil+1$.
\end{enumerate}
\item Next, we prepare a state on the second register as
\begin{align}
&\frac 1{\sqrt{\lambda}} \left(\ket{0,0,0,0}\sum_{p,q}\left[ \sqrt{2{|{\rm Re}(h'_{pq})|}}\ket{\theta_{pq0}^{(0)}}\ket{0,p,q,0}+\sqrt{2{|{\rm Im}(h'_{pq})|}}\ket{\theta_{pq1}^{(0)}}\ket{0,p,q,1}\right] \right. \nn
& 
+ \frac 1{\sqrt 2} \sum_{\Q,n}\ket{\ell,\Q,n,1}\sqrt{\sum_{\K,rs}(|\Rea[L_{r\K s(\K\modmin\Q),n}]|+|\Ima[L_{r\K s(\K\modmin\Q),n}]|)} \nn
& \times\left.
\sum_{\K,p,q}\left[\sqrt{|{\rm Re}(L_{p\K q(\K\modmin\Q),n})|}\ket{\theta_{\K pq0}^{(\ell)}}\ket{\K,p,q,0}+\sqrt{|{\rm Im}(L_{p\K q(\K\modmin\Q),n})|}\ket{\theta_{\K pq1}^{(\ell)}}\ket{\K,p,q,1}\right]\right)\ket{+}\ket{+},
\end{align}
where $\theta_{\K pq0}^{(\ell)}$, $\theta_{\K pq1}^{(\ell)}$ are used to obtain the correct signs on the terms, and the $\ket{+}$ states at the end are used to select the spin and control the swap between the $p$ and $q$ registers.

Now we have a distinction from \cite{Lee2020} in that we have separate real and imaginary parts, and a separate prepared qubit to flag between the real and imaginary parts.
Because of the large number of variables, we will again use a single variable for iteration, and use it to output $\K,p,q$.
The complexity of this state preparation is then as follows.
\begin{enumerate}
\item First, prepare an equal superposition over the variable for iteration.  There are $\L=N_k N^2/2$ values to take, which includes a factor of $2$ for the real and imaginary parts, $N_k$ for $\K$, and $N^2/4$ for the values of $p,q$.
Then the complexity of preparing the equal superposition is $3n_{\L}-3\factor+2b_r-9$, where $n_{\L}=\lceil \log \L \rceil$, with $\factor$ being the largest number such that $2^\factor$ is a factor of $\L$.
\item The size of the QROM output is
\begin{equation}
b_p = 2n_k + 4n_N + \zetabits_2+3,
\end{equation}
where the first term is for the three components of $\K$, the second is for $p$ and $q$.
The third is for ind and alt values of the qubit to store the correct sign, as well as an alt value of the extra qubit for selecting between the real and imaginary parts.
We do not include an ind value for that qubit, because it is part of the register we are iterating over.
The complexity of this QROM will be
\begin{equation}\label{eq:qromprep2a}
\left\lceil \frac{MN_k+1}{k_{p1}} \right\rceil 
\left\lceil \frac{\L}{k_{p2}} \right\rceil + b_p(k_{p1}k_{p2}-1),
\end{equation}
where we are accounting for the cost to select based on both the index from the factorization and the index for $\K,p,q$, and using the result for the complexity of QROM on two registers from Appendix G of \cite{Lee2020}.
\item Perform the inequality test with cost $\zetabits_2$, which is the bits of precision for this state preparation.
\item Perform the controlled swap with the alt values with cost $n_k +2n_N+1$.
Here we are swapping the ind and alt values of $\K,p,q$, as well as the qubit selecting between real and imaginary parts.
The sign required for the sign qubits can be implemented with Cliffords as in \cite{Lee2020}, so does not add to this Toffoli cost.
\end{enumerate}
\item We no longer perform swaps of $p$ and $q$ for symmetry, but we do need to perform arithmetic to compute $\K\modmin\Q$ and $\Kp\modmin\Q$, with a cost of $4n_k$.
\item Perform \sel\ by performing the sequence of four controlled $\vec Z X_{p,\sigma}$ or $\vec Z Y_{p,\sigma}$ operations.
The cost is $4(NN_k/2-1)$ Toffolis since it must be controlled, and there is a cost of one more Toffoli to create the qubits to control on.
In order to select the spin we also perform a swap controlled by the spin selection qubit before and after, with a cost of $NN_k$ Toffolis.
\item Reverse steps 2 and 3, where the complexities are the same except the QROM complexity which is changed to
\begin{equation}\label{eq:qromprep2b}
\left\lceil \frac{MN_k+1}{k'_{p1}} \right\rceil 
\left\lceil \frac{\L}{k'_{p2}} \right\rceil
+ k'_{p1}k'_{p2}.
\end{equation}
\item Reflect on the qubits that were prepared in step 2.
The qubits we need to reflect on are as follows.
\begin{enumerate}
\item The $n_{\L}$ qubits for the variable of iteration.
\item We need to reflect on the $\zetabits_2$ registers that are used for the equal superposition state for the state preparation.
\item One that is rotated for the preparation of the equal superposition state.
\item One for the spin.
\item One for controlling the swap between the $p$ and $q$ registers.
\item One for selecting between the real and imaginary part.
\item One for selecting between $A$ and $B$.
\end{enumerate}
That gives a total of $n_{\L}+\zetabits_2+5$ qubits.
The reflection needs to be controlled on the success of the preparation on the $\ell$ register, and $\ell\ne 0$, making the total cost
$n_{\L}+\zetabits_2+5$ Toffolis.
\item Perform steps 2 to 5 again, but this time $MN_k+1$ is replaced with $MN_k$ in \eq{qromprep2a} and \eq{qromprep2b}.
Also, the \sel\ operation needs to be controlled on $\ell\ne 0$, which flags the one-body term.
That requires another 4 Toffolis.
\item Invert the state preparation on the $\ell$ register, where the complexity of the QROM is reduced to
\begin{equation}
\left\lceil \frac{MN_k+1}{k'_{\L}} \right\rceil + k'_{\L}.
\end{equation}
\item To complete the step of the quantum walk, perform a reflection on the ancillas used for the state preparation.
There are $n_{MN}+n_{\L}+\zetabits_1+\zetabits_2+5$, where the qubits we need to reflect on are as follows.
\begin{enumerate}
\item The $n_{MN}$ qubits for the $\ell$ register.
\item The $n_{\L}$ qubits for the registers in the state preparation for $A$ and $B$.
\item The $\zetabits_1$ qubits for the equal superposition state used for preparing the state on the $\ell$ register using the coherent alias sampling.
\item The $\zetabits_2$ qubits for the equal superposition state for preparing the state for $A$ and $B$.
\item Two qubits rotated for the boosting the success probability for the equal superposition states.
\item One qubit for the spin.
\item One qubit for controlling the swap of the $p$ and $q$ registers.
\item One for selecting between the real and imaginary part.
\item One for selecting between $A$ and $B$.
\end{enumerate}
This reflection has cost $n_{MN}+n_{\L}+\zetabits_1+\zetabits_2+4$.
\item The steps of the walk are made controlled by using unary iteration on an ancilla used for the phase estimation.
Each step requires another two Toffolis for the unary iteration and making the reflection controlled.
\end{enumerate}

In this list of steps we have not explicitly included the part for applying the phase factors, but that has no non-Clifford cost.

Next we consider the total number of logical qubits needed for the simulation via this method.
\begin{enumerate}
\item The control register for the phase estimation, and the ancillas for the unary iteration, together need $2\lceil\log\mathcal{I}\rceil-1$ qubits.
\item There are $NN_k$ qubits for the target system.
\item There are $n_{MN}+2$ qubits for the $\ell$ register, the qubit rotated in preparing the equal superposition, and the qubit flagging success of preparing the equal superposition.
\item The state preparation on the $\ell$ register uses $b_{MN}=2n_k+2\lceil\log M\rceil + 2\zetabits_1+2$ qubits.
Here $2n_k+2\lceil\log M\rceil$ is for the ind and alt values of $\Q$ and $n$, $\zetabits_1$ are for keep values, $\zetabits_1$ are for the equal superposition state, 1 is for the output of the inequality test, and 2 are for the qubit flagging $\ell\ne 0$ and its alternate value.
\item There are $n_{\L}+2$ qubits needed for the register preparing $p,q,\K$ values, a qubit that is rotated for the equal superposition, and a qubit flagging success of preparing the equal superposition.
\item The equal superposition state used for the second preparation uses $\zetabits_2$ qubits.
\item The phase gradient register uses $b_r$ qubits.
\item The qubits for the spin, controlling the swap of $p$ and $q$, selection between the real and imaginary parts, and selection between $A$ and $B$ for a total of 4.
\item The QROM needs a number of qubits $b_pk_{p1}k_{p2}+\lceil\log[(MN_k+1)/k_{p1}]\rceil+\lceil\log[L/k_{p2}]\rceil$.
\end{enumerate}
The QROM for the state preparation on the second register uses a large number of temporary ancillas, which can be reused by later parts of the algorithm, so those later parts of the algorithm do not need the number of qubits counted.
The total number of qubits used is then
\begin{equation}\label{eq:lowranklog}
2\lceil\log\mathcal{I}\rceil+NN_k+n_{MN}+n_{\L}+2n_k+2\lceil\log M\rceil+2\zetabits_1+\zetabits_2+b_r+9+b_pk_{p1}k_{p2}+\lceil\log[(MN_k+1)/k_{p1}]\rceil+\lceil\log[L/k_{p2}]\rceil
\end{equation}
with $b_{p}=2n_k+2n_N+\zetabits_2+3$, $n_N=\lceil \log(N/2)\rceil$, $n_{\L}=\lceil\log \L\rceil$, $L=N_kN(N+2)/4$.
This completes the costing of the low rank factorization method.

\section{Double-factorization derivations}
\label{app:df_lambda_one_larger_derivation}
\subsection{One-body correction}
Here we derive the correction for the one-body Hamiltonian as given in Eq.~\eqref{eq:df_one_body_finalb}.
The lambda value for the Hamiltonian can be calculated by determining the total L1-norm using the second factorization
\begin{equation}
\hat{H}'_2 = \frac{1}{2}  \sum_{\Q}^{N_{k}}\sum_{n}^{M} \left(\hat{A}^2_{n}(\Q) + \hat{B}^2_{n}(\Q)\right)
\end{equation}
with
\begin{align}\label{eq:Afact}
2\hat{A}_{n}(\Q) &= \sum_{\K} \left[ U^A_{n}(\Q, \K) \left( \sum_{\sigma}\sum_{p}^{\ranktwo_{\Q,n,\K,A}}f^A_{p}(\Q, n, \K)(\mathbb{1} - Z_{p\K\sigma}) \right) U^A_{n}(\Q, \K)^{\dagger} \right]  \nn
&= \sum_{\K} U^A_{n}(\Q, \K)\hat{\mathbb{1}}^A_{\K}U^A_{n}(\Q,\K)^{\dagger} - \sum_{\K} U^A_{n}(\Q, \K)\hat{Z}^A_{\K}U^A_{n}(\Q,\K)^{\dagger} 
\end{align}
where $\hat{\mathbb{1}}^A_{\K} = \sum_{\sigma}\sum_{p}^{\ranktwo_{\Q,n,\K,A}}f^A_{p}(\Q, n, \K)\mathbb{1}$
and $\hat{Z}^A_{\K} = \sum_{\sigma}\sum_{p}^{\ranktwo_{\Q,n,\K,A}}f^A_{p}(\Q, n, \K)Z_{p\K\sigma}$, and
\begin{align}\label{eq:Bfact}
2\hat{B}_{n}(\Q) &= \sum_{\K} \left[ U^B_{n}(\Q, \K) \left( \sum_{\sigma}\sum_{p}^{\ranktwo_{\Q,n,\K,B}}f^B_{p}(\Q, n, \K)(\mathbb{1}- Z_{p\K\sigma}) \right) U^B_{n}(\Q, \K)^{\dagger} \right] \nn
&= \sum_{\K} U^B_{n}(\Q, \K)\hat{\mathbb{1}}^B_{\K}U^B_{n}(\Q,\K)^{\dagger} - \sum_{\K} U^B_{n}(\Q, \K)\hat{Z}^B_{\K}U^B_{n}(\Q,\K)^{\dagger} 
\end{align}
where $\hat{\mathbb{1}}^B_{\K} = \sum_{\sigma}\sum_{p}^{\ranktwo_{\Q,n,\K,B}}f^B_{p}(\Q, n, \K)\mathbb{1}$
and $\hat{Z}^B_{\K} = \sum_{\sigma}\sum_{p}^{\ranktwo_{\Q,n,\K,B}}f^B_{p}(\Q, n, \K)Z_{p\K\sigma}$.
The factor of 1/2 from the Jordan-Wigner transform is squared to 1/4, which is moved outside each term and combined with the prefactor 1/2 to produce a prefactor of 1/8.  
We note that $\hat{A}_{n}(\Q)^{2}$ can be written as
\begin{align}
4\hat{A}_{n}(\Q)^{2} &= \left( \sum_{\K} U^A_{n}(\Q, \K)\hat{\mathbb{1}}^A_{\K}U^A_{n}(\Q,\K)^{\dagger} - \sum_{\K} U^A_{n}(\Q, \K)\hat{Z}^A_{\K}U^A_{n}(\Q,\K)^{\dagger} \right) \nonumber \\
& \quad  \times \left( \sum_{\Kp} U^A_{n}(\Q, \Kp)\hat{\mathbb{1}}^A_{\Kp}U^A_{n}(\Q,\Kp)^{\dagger} - \sum_{\Kp} U^A_{n}(\Q, \Kp)\hat{Z}^A_{\Kp}U^A_{n}(\Q,\Kp)^{\dagger} \right) \nonumber \\
&= 2\sum_{\K} U^A_{n}(\Q, \K)\hat{\mathbb{1}}^A_{\K}U^A_{n}(\Q,\K)^{\dagger}\hat{A}_{n}(\Q) + 2\hat{A}_{n}(\Q)\sum_{\K} U^A_{n}(\Q, \K)\hat{\mathbb{1}}^A_{\K}U^A_{n}(\Q,\K)^{\dagger} \nonumber \\
& \quad + \sum_{\K,\Kp}U^A_{n}(\Q, \K)\hat{Z}^A_{\K}U^A_{n}(\Q,\K)^{\dagger} U^A_{n}(\Q, \Kp)\hat{Z}^A_{\Kp}U^A_{n}(\Q,\Kp)^{\dagger}  \nonumber \\
& \quad -  \sum_{\K} U^A_{n}(\Q, \K)\hat{\mathbb{1}}^A_{\K}U^A_{n}(\Q,\K)^{\dagger}\sum_{\Kp} U^A_{n}(\Q, \Kp)\hat{\mathbb{1}}^A_{\Kp}U^A_{n}(\Q,\Kp)^{\dagger} .
\label{eq:A2derv}
\end{align}
The last term in the above equation is proportional to the identity and is ignored. A similar expression can be derived for $\hat{B}_{n}(\Q)^{2}$ and thus the component of the two-body term involving two Pauli $Z$ operators is written as
\begin{align}
V &= \frac{1}{8}\sum_{\Q,n,\K,\Kp}U^A_{n}(\Q, \K)\hat{Z}^A_{\K}U^A_{n}(\Q,\K)^{\dagger} U^A_{n}(\Q, \Kp)\hat{Z}^A_{\Kp}U^A_{n}(\Q,\Kp)^{\dagger} \nn
& \quad + \frac{1}{8}\sum_{\Q,n,\K,\Kp}U^B_{n}(\Q, \K)\hat{Z}^B_{\K}U^B_{n}(\Q,\K)^{\dagger} U^B_{n}(\Q, \Kp)\hat{Z}^B_{\Kp}U^B_{n}(\Q,\Kp)^{\dagger} 
\end{align}
which implies the two-body L1-norm, $\lambda_{\mathrm{DF},2}$, is
\begin{align}
\lambda_{\mathrm{DF},2} = \frac{1}{4}\sum_{\Q,n}\left[ \left(\sum_{\K,p}^{N_{k}\ranktwo_{\Q,n,\K,A}}|f^A_{n}(p, \Q,\K)|\right)^{2} + \left(\sum_{\K,p}^{N_{k}\ranktwo_{\Q,n,\K,B}}|f^B_{n}(p, \Q,\K)|\right)^{2} \right]
\end{align}
where the factor of $1/8$ becomes a factor of $1/2$ accounting for spin. This factor of $1/2$ is further divided by two because we perform oblivious amplitude amplification--\textit{i.e.}\ the inner step of qubitization evolving by $2\hat{A}_{n}(\Q)^{2} - \mathbb{1}$ and $2\hat{B}_{n}(\Q)^{2} - \mathbb{1}$. 

Next, the one-body terms in the third line of Eq.~\eqref{eq:A2derv} can be rewritten as
\begin{equation}
    2\sum_{\K} \hat{\mathbb{1}}^A_{\K}\hat{A}_{n}(\Q) + 2\hat{A}_{n}(\Q)\sum_{\K} \hat{\mathbb{1}}^A_{\K} \, .
\end{equation}
This expression needs to be divided by 8 to give the contribution to the Hamiltonian, and there is a similar contribution from $\hat{B}_{n}(\Q)^{2}$ to give the overall contribution to the one-body Hamiltonian
\begin{align}
\frac{1}{2}\sum_{n,\Q}\left(\sum_{\K}\hat{\mathbb{1}}^A_{\K}\hat{A}_{n}(\Q) + \sum_{\K} \hat{\mathbb{1}}^B_{\K}\hat{B}_{n}(\Q) \right).
\end{align}
Taking the trace of Eq.~\eqref{eq:Afact} and \eqref{eq:Bfact} then implies
\begin{align}
    \sum_{\K} \hat{\mathbb{1}}^A_{\K} &= \mathbb{1} \tr(\hat{A}_{n}(\Q)) = \mathbb{1} \, \frac 12 \left[ \tr(\hat{\rho}_{n}(\Q)) + \tr(\hat{\rho}^\dagger_{n}(\Q))\right] , \\
    \sum_{\K} \hat{\mathbb{1}}^B_{\K} &= \mathbb{1} \tr(\hat{B}_{n}(\Q)) = \mathbb{1} \, \frac i2 \left[ \tr(\hat{\rho}_{n}(\Q)) - \tr(\hat{\rho}^\dagger_{n}(\Q))\right] .
\end{align}
The trace of $\hat{\rho}_{n}(\Q)$ is non-zero only for $\Q=0$.
In that case
\begin{equation}
    \tr(\hat{\rho}_{n}(0)) = 2\sum_{\K} \left(\sum_{r}^{N/2}L_{r \K r \K, n} \right)
\end{equation}
which is real.
Moreover, it is easily seen that $\hat{\rho}(0)$ is Hermitian using the symmetry $L_{p\K q\K,n} = L_{q\K p\K,n}^{*}$, so $\hat{A}_{n}(0)=\hat{\rho}_{n}(0)$ and $\hat{B}_{n}(0)=0$.
Therefore
\begin{align}
\sum_{\K}\hat{\mathbb{1}}^A_{\K}\hat{A}_{n}(0) + \sum_{\K}\hat{\mathbb{1}}^B_{\K}\hat{B}_{n}(0) = 2 \sum_{\K,p,q,\sigma} \left(\sum_{\Kp,r}
L_{p \K q \K, n}L_{r \Kp r \Kp, n} \right) a_{p\K\sigma}^\dagger a_{q\K\sigma}
.
\end{align}
Therefore the contribution to the one-body Hamiltonian becomes
\begin{align}
\frac{1}{2}\sum_{n,\Q}\left(\sum_{\K}\hat{\mathbb{1}}^A_{\K}\hat{A}_{n}(\Q) + \sum_{\K} \hat{\mathbb{1}}^B_{\K}\hat{B}_{n}(\Q) \right) = \sum_{\K,p,q,\sigma} \left(\sum_{\Kp,r}V_{ p \K , q \K, r \Kp, r \Kp} \right) a_{p\K\sigma}^\dagger a_{q\K\sigma}.
\end{align}
As a result, the complete one-body Hamiltonian is
\begin{align}\label{eq:df_one_body_final}
H_{1}' = \sum_{\K,p,q,\sigma}\left(h_{p\K, q\K} + \sum_{\Kp,r}V_{p\K,q\K,r\Kp,r\Kp}\right)a_{p\K\sigma}^{\dagger}a_{q\K\sigma} .
\end{align}
This is identical to the result that was obtained in the single-factorization case as in Eq.~\eqref{eq:singfaconecor}.
Thus the L1-norm of $H_{1}'$ is the sum
\begin{align}
\lambda_{\mathrm{DF},1} = \sum_{\K}\sum_{p}|\lambda_{\K,p}|
\end{align}
where $\lambda_{\K,p}$ is an eigenvalue of the matrix representing $H_{1}'(\K)$ which are the coefficients in the parenthesis of Eq.~\eqref{eq:df_one_body_final}.

\subsection{Complexity of the double-factorized representation}
Our form of the two-body part of the Hamiltonian is
\begin{equation}
\hat{H}'_2 = \frac{1}{2}  \sum_{\Q}^{N_{k}}\sum_{n}^{M} \left(\hat{A}^2_{n}(\Q) + \hat{B}^2_{n}(\Q)\right).
\end{equation}
with
\begin{align}
\hat{A}_{n}(\Q) = \sum_{\K} \left[ U^A_{n}(\Q, \K) \left( \sum_{\sigma}\sum_{r}^{\ranktwo_{\Q, n,\K,A}}f^A_{r}(\Q, n, \K)n_{r,\K,\sigma} \right) U^A_{n}(\Q, \K)^{\dagger} \right]
\end{align}
and similarly for $\hat{B}_{n}(\Q)$.
In comparison, the double-factorized Hamiltonian from \cite{vonBurg2020,Lee2020} is
\begin{equation}\label{eq:Fprime}
F' = \frac{1}{8}\sum_{\ell=1}^{\rankone} U_\ell \left( \sum_{\sigma \in \{\uparrow, \downarrow\}} \sum_{p=1}^{\ranktwo^{(\ell)}} f_{p}^{(\ell)} Z_{p,\sigma} \right)^2 U_\ell^\dagger.
\end{equation}
So, in contrast to the decomposition before, instead of a sum over $\ell$, we have a sum over $\Q,n$, and a qubit indexing over $\hat{A},\hat{B}$.
This difference can be accounted for easily in the method as presented in \cite{Lee2020}.
That method may be summarized as follows.
\begin{enumerate}
\item Perform a state preparation over $\ell$ for the first factorisation.
\item Use a QROM on $\ell$ to output some parameters needed for the state preparation for the second factorisation (the operator that is squared).
\item Perform the inner state preparation over $p$.
\item Apply a QROM to output the sequence of rotations dependent on $\ell$ and $p$.
\item Apply the Givens rotations.
\item Apply a controlled $Z$.
\item Invert the Givens rotations, QROM, and state preparation over $p$.
\item Perform a reflection on the ancilla qubits used for the state preparation over $p$.
\item Perform steps 3 to 7 again.
\item Invert the QROM from step 2.
\item Invert the state preparation from step 1.
\end{enumerate}
Note that this is distinct from the procedure in \cite{vonBurg2020} which combined the $\ell$ and $p$ preparations.

To account for the changes here, the index $\ell$ can be used to iterate through all possible values of $\Q,n$, and the qubit indexing over $\hat{A},\hat{B}$.
Most of the steps can be performed ignoring these values, but we will need to know $\Q$ before performing the Givens rotations.
It is convenient to output this value in the QROM used in step 2, which slightly increases the output size of this QROM.
We will also need to output $\K$ values, and these will be given in the second state preparation used in step 3.
But, that preparation will produce a joint index of $p$ and $\K$ without giving $\K$ explicitly (similar to our preparation over $\ell$ not giving $\Q$ explicitly.
This can be output by the QROM in step 4.

In order to apply the Givens rotations, we will need to perform controlled swaps of system registers $\K,\K\modmin\Q$ into working registers, then apply the Givens rotations on those working registers.
Since $\K\modmin\Q$ is not given directly by the state preparation, it needs to be computed with cost $2n_k$ (or $n_k$ if $N_x,N_y,N_z$ are powers of 2).
The controlled swaps have a Toffoli cost of $2n_k$ for the unary iteration, and $NN_k$ for the controlled swaps.
The cost of $NN_k$ is because we need to run through $NN_k/2$ system qubits twice.
These controlled swaps are performed 4 times, because they need to be performed before and after each application of the Givens rotations.
That gives a total cost from this part
\begin{equation}
4  NN_k + 12n_k .
\end{equation}

Then for the QROM outputting the Givens rotations, the number of items of data can be given as
\begin{equation}
    \sum_{\Q,n,\K} (\ranktwo_{\Q,n,\K,A} + \ranktwo_{\Q,n,\K,B}),
\end{equation}
where $\ranktwo_{\Q,n,\K,A}$ and $\ranktwo_{\Q,n,\K,B}$ are the cutoffs in the sums for $\hat A$ and $\hat B$.
As per Eq.~\eqref{eq:Xidef}, we define $\ranktwo$ as this quantity divided by $L=2N_k M$, so we can write the number of items of data as $L\ranktwo$.

The Givens rotations need to be on $2N$ orbitals, so there are $2N$ Givens rotations.
For each of these rotations two angles need to be specified, in contrast to one in \cite{vonBurg2020,Lee2020}.
The size of the data output for the QROM for the Givens rotations is increased to $2N\rotbits$, because there are 2 registers of size $N/2$, and there are 2 rotations of $\rotbits$ of precision for each Givens rotation.
The total complexity of applying the Givens rotations is increased to $16N(\rotbits-2)$.
This is an increase of a factor of 4 over that in \cite{Lee2020}, with a factor of 2 from using 2 working registers, and a factor of 2 because there are two rotations for each Givens rotation.

The other changes in the cost are relatively trivial.
There is a swap on the system registers controlled on the spin register.
Since this is now on $NN_k$ qubits instead of $N$, the cost is multiplied by $N_k$.

So, to summarize the complexity using the same numbering of steps as in \cite{Lee2020}, we have the following.
\begin{enumerate}
\item The cost of the state preparation over $\ell$ is
\begin{equation}
(3n_\rankone-3\factor+2\rotprec-9) % Step 1a
+\left\lceil\frac{\rankone+1}{k_{p1}}\right\rceil + b_{p1}(k_{p1}-1) % Step 1b
+\zetabits_1+n_\rankone,
\end{equation}
where $\rankone$ is now $2N_k M$ and as before $b_{p1} = n_\rankone+\zetabits_1$, $n_\rankone=\lceil \log \rankone\rceil$.
\item The complexity of the QROM on $\ell$ is now
\begin{equation}
\left\lceil\frac{\rankone+1}{k_o}\right\rceil + b_o(k_o-1),
\end{equation}
with
\begin{equation}
b_o = n_k + n_\ranktwo 
+n_{\rankone,\ranktwo}+\rotprec +1,
\end{equation}
with the extra $n_k$ being to output $\Q$.
Here $n_\ranktwo$ is the number of bits needed for $\ranktwo_{\K,p}$ values of $p$, and $n_{\rankone,\ranktwo}$
\begin{equation}
    n_{\rankone,\ranktwo} = \lceil \log(\rankone\ranktwo + N_k N/2) \rceil
\end{equation}
is the number of bits needed for the offset.
\item The cost of the second stage of state preparation is
\begin{align}
&4(7n_\ranktwo+2\rotprec-6) % Step 3(a)
+4(n_{\rankone,\ranktwo}-1) % Step 3(b)
+\left( \left\lceil \frac {\rankone \ranktwo+ NN_k/2}{k_{p2}}\right\rceil+\left\lceil \frac {\rankone \ranktwo}{k_{p2}}\right\rceil + 2b_{p2}(k_{p2}-1) 
\right) % Step 3(c)
+4(\zetabits_2+n_\ranktwo), % Step 3(d)
\end{align}
where the brackets are used to indicate the cost of parts (a) to (d) of step 3.
As well as using our modified definition of $\ranktwo$,
the only change over the costing in \cite{Lee2020} is replacing $N/2$ with $NN_k /2$ for the range of values for the one-body term.
In this cost we are including the second use of the preparation in part 7.
\item The cost of the number operators via QROM is
\begin{equation}
\left\lceil\frac {\rankone\ranktwo+NN_k/2}{k_r}\right\rceil+\left\lceil\frac {\rankone\ranktwo}{k_r}\right\rceil + (4N\rotbits+n_k)(k_r-1) 
+ \left\lceil\frac {\rankone \ranktwo+NN_k/2}{k'_r}\right\rceil+ \left\lceil\frac {\rankone \ranktwo}{k'_r}\right\rceil + 2k'_r % The cost of the QROM for the rotations in step 4b.
+4(n_{\rankone,\ranktwo}-1) % Steps 4(a) and 4(h).
+16N(\rotbits-2)+2NN_k+2. % Steps 4c to 4f.
\end{equation}
Here the term $4N\rotbits(k_r-1)$ has been increased by a factor of 4 over that in \cite{Lee2020}, because we have 2 times as many qubits that the Givens rotations need to act on, and there are twice as many rotations needed for each Givens rotation. (This term is corresponding to the output size for the QROM.)
We have also added $n_k$ for the output size so we can output the value of $\K$ needed to select the register.
Again $N/2$ is replaced with $NN_k /2$ for the one-body term.
The quantity $16N(\rotbits-2)$ is for the cost of the Givens rotations, and is also multiplied by a factor of 4 over that in \cite{Lee2020}.
The $2NN_k$ for the controlled swaps for spin, and is increased over $2N$ in \cite{Lee2020} because we now have $\K$.
\item The inversion of the state preparation has cost
\begin{align}
&2(7n_\ranktwo+2\rotprec-6)
+2(n_{\rankone,\ranktwo}-1)
+\left(\left\lceil\frac {\rankone \ranktwo+NN_k/2}{k'_{p2}}\right\rceil  + \left\lceil\frac {\rankone \ranktwo}{k'_{p2}}\right\rceil + 2k'_{p2}\right)
+2(\zetabits_2+n_\ranktwo).
\end{align}
This cost is the same as in part 3, except the cost of erasing the QROM is reduced.
We are again including both uses (with the second described in step 7).
\item The reflection for the oblivious amplitude amplification has an unchanged cost
\begin{equation}
n_\ranktwo+\zetabits_2+2.
\end{equation}
\item The cost of the second use of the block encoding to give the square are already accounted for above.
\item The cost of inverting step 1 is
\begin{equation}
(3n_\rankone-3\factor+2\rotprec-9) + \left\lceil\frac{\rankone+1}{k'_{p1}}\right\rceil + k'_{p1}+\zetabits_1+n_\rankone ,
\end{equation}
and for inverting step 2 is
\begin{equation}
\left\lceil\frac{\rankone+1}{k'_o}\right\rceil + k'_o, % Inverting the QROMs from step 2.
\end{equation}
where we are using the improved cost for erasing QROM.
\item The reflection cost is unchanged at
\begin{equation}
n_\rankone+n_\ranktwo+\zetabits_1+\zetabits_2+1.
\end{equation}
\item The extra cost of unary iteration on the control register and of controlling the reflection on that register is 2 Toffolis.
\item The new costs of performing controlled swaps into working registers and arithmetic to compute $\K\modmin\Q$ and $\Kp\modmin\Q$ are
\begin{equation}
4 NN_k + 12 n_k .
\end{equation}
\end{enumerate}
Adding all these costs together gives the total cost for block encoding the Hamiltonian.

The cost in terms of logical qubits is very similar to that for the original double-factorized approach.
The differences are as follows.
\begin{enumerate}
    \item There are registers needed to store $\K,\Q,\K\modmin\Q$.
    Because $\K\modmin\Q$ can be computed in place in the $\Q$ register, we only need storage for 2.
    Moreover, because $\K$ is given in the QROM output in part 4 above, it does not need to be added to that qubit costing.
    \item There are $N$ qubits used for the working registers (2 of size $N/2$).
    \item A number of parameters are changed, in particular the number of system qubits is now $NN_k$, and $\rankone$ is computed from the number of values of $\Q$ and $n$.
    \item The size of the output for the Givens rotations is multiplied by a factor of 4.
\end{enumerate}

\section{Tensor hypercontraction derivations}
\label{app:THCcomplex}
\subsection{THC symmetries}
In this section we derive the symmetry relationships for the central tensor based on the four-fold symmetry of the two-electron integral tensor as used in \eq{thc_ham_restricted_sum}.
Recall an element of the two-electron integral tensor can be represented in THC form as
\begin{align}
V_{p\K, q(\K\modmin\Q),r\Kp \modmin\Q, s\Kp} = \sum_{\mu,\nu}\chi_{p\K,\mu}^{*} \chi_{q(\K\modmin\Q),\mu}\zeta_{\mu\nu}^{\Q,\G_{1},\G_{2}} \chi_{r(\Kp\modmin\Q),\nu}^{*} \chi_{s\Kp,\nu}
\end{align}
where $\G_{1}$ is shorthand for $\G_{\K,\K-\Q}$ and $\G_{2}$ is shorthand for $G_{\Kp,\Kp-\Q}$.  The four fold symmetry of the complex valued two-electron integral tensor is reflected in the central tensor $\zeta$.  We recover the symmetry by first noting the four equivalent two-electron integrals
\begin{align}
V_{p\K, q(\K\modmin\Q),r\Kp -\Q, s\Kp} =& \sum_{\mu,\nu}\chi_{p\K,\mu}^{*} \chi_{q(\K\modmin\Q),\mu}\zeta_{\mu\nu}^{\Q,\G_{1},\G_{2}} \chi_{r(\Kp\modmin\Q),\nu}^{*} \chi_{s\Kp,\nu} \nonumber \\
V_{q(\K\modmin\Q), p\K,s\Kp, r\Kp -\Q}^{*} =& \left(\sum_{\mu,\nu}\chi_{q(\K\modmin\Q),\mu}^{*} \chi_{p\K,\mu}\zeta_{\mu\nu}^{(\modmin\Q),!\G_{1},!\G_{2}} \chi_{s\Kp,\nu}^{*} \chi_{r(\Kp\modmin\Q),\nu}\right)^{*} \nonumber \\
V_{r(\Kp\modmin\Q),s\Kp,p\K,q(\K\modmin\Q)} =& \sum_{\mu, \nu}\chi_{r(\Kp\modmin\Q),\mu}^{*}\chi_{s\Kp,\mu}\zeta_{\mu\nu}^{(\modmin\Q),!\G_{2},!\G_{1}}\chi_{p\K,\nu}^{*}\chi_{q(\K\modmin\Q),\nu} \nonumber \\\
V_{s\Kp,r(\Kp\modmin\Q),q(\K\modmin\Q),p\K} =& \left(\sum_{\mu, \nu}\chi_{s\Kp,\mu}^{*}\chi_{r(\Kp\modmin\Q),\mu}\zeta_{\mu\nu}^{\Q,\G_{2},\G_{1}}\chi_{q(\K\modmin\Q),\nu}^{*}\chi_{p\K,\nu}\right)^{*} \nonumber 
\end{align}
which, implies
\begin{align}
\zeta_{\mu\nu}^{\Q,\G_{1},\G_{2}} = \left(\zeta_{\mu\nu}^{(\modmin\Q),!\G_{1},!\G_{2}} \right)^{*} = \zeta_{\nu\mu}^{(\modmin\Q),!\G_{2},!\G_{1}} = \left(\zeta_{\nu\mu}^{\Q,\G_{2},\G_{1}}\right)^{*}.
\end{align}
Here $(\modmin\Q)$ is used to indicate a modular negative of $\Q$, similar to modular subtraction.
In the above expression the complement of $\G_{1}$, $!\G_{1}$, is defined through 
\begin{align}
\K_{p} - \K_{q} = \Q + \G_{1} \nonumber \\
\K_{q} - \K_{p} = (\modmin\Q) + !\G_{1} \nonumber \\
!\G_{1} = -\left(\Q + \G_{1} + (\modmin\Q)\right),\label{eq:thc_complement_G}
\end{align}
and it is important to note that $(\modmin\Q)$ is defined to be in the original set of $k$-points and it is useful as we only build $\zeta^{\Q}$. 
A similar expression can be derived for $!\G_{2}$.  
It is helpful to consider some concrete examples, which are given in \cref{tab:thc_gmapping_examples}. 
\begin{table}[H]
\centering
\begin{minipage}{0.85\textwidth}
\begin{ruledtabular}
\begin{tabular}{lcccccccc}
$k$-mesh & $\K_p$ & $\K_q$ & $\K_p-\K_q$ & $\Q$ & $\G$ & $\K_q-\K_p$& $(\modmin\Q)$ & $!\G$ \\ \hline
$[1, 1, 4]$ & (0, 0, 3) & (0, 0, 1) & (0, 0, 2) & (0, 0, 2)& (0,0,0) & $(0, 0, -2)$ & (0, 0, 2) & $(0, 0, -4)$ \\
\hline
$[1, 4, 4]$ & (0, 2, 1) & (0, 3, 1) & $(0, -1, 0)$ & (0, 3, 0)& $(0,-4,0)$ & $(0, 1, 0)$ & (0, 1, 0) & $(0, 0, 0)$ \\
$[1, 4, 4]$ & (0, 2, 1) & (0, 3, 3) & $(0, -1, -2)$ & (0, 3, 2)& $(0,-4,-4)$ & $(0, 1, 2)$ & (0, 1, 2) & $(0, 0, 0)$ \\
$[1, 4, 4]$ & (0, 1, 2) & (0, 1, 3) & $(0, 0, -1)$ & (0, 0, 3)& $(0,0,-4)$ & $(0, 0, 1)$ & (0, 0, 2) & $(0, 0, 0)$ \\
$[1, 4, 4]$ & (0, 1, 3) & (0, 1, 2) & (0, 0, 1) & (0, 0, 1)& (0,0,0) & $(0, 0, -1)$ & (0, 0, 3) & $(0, 0, -4)$ \\
\hline
$[4, 4, 4]$ & (2, 1, 3) & (3, 1, 2) & $(-1, 0, 1)$ & (3, 0, 1)& $(-4,0,0)$ & $(1, 0, -1)$ & (0, 0, 3) & $(0, 0, -4)$ \\
$[4, 4, 4]$ & (2, 1, 2) & (3, 3, 3) & $(-1, -2, -1)$ & (3, 2, 3)& $(-4,-4,-4)$ & $(1, 2, 1)$ & (1, 2, 1) & $(0, 0, 0)$ \\
\end{tabular}
\end{ruledtabular}
\caption{Some examples of the values that the different momentum labels can take in \cref{eq:thc_complement_G}. We restrict $\K, \Q, (\modmin\Q)$ to be in the original $k$-point set. \label{tab:thc_gmapping_examples}}
\end{minipage}
\end{table}
\subsection{Complexity of the tensor hypercontraction representation}
The following is a detailed costing for the qubitization oracles using the THC LCU.
In the initial state preparation, we need to prepare a superposition over $\Q,\G_1,\G_2,\mu,\nu$ with weights $\sqrt{|\zeta_{\mu\nu}^{\Q,\G_1,\G_2}|}$.
    The state can be prepared via the coherent alias sampling procedure, starting with QROM to output keep and alt values.
    One option here is to produce an equal superposition over $\Q,\G_1,\G_2,\mu,\nu$, then calculate a contiguous register from these values to use for the QROM.
    That procedure is fairly complicated, because it requires preparing equal superpositions over three components of $\Q$ as well as $\G_1,\G_2,\mu$ and $\nu$, then arithmetic for the contiguous register.
    To simplify the procedure we give the complexity for giving ind values like for sparse state preparation.
    That is, we prepare the contiguous register, and use the QROM to output both ind (index) and alt values of $\Q,\G_1,\G_2,\mu,\nu$.
    
    There is the symmetry $\zeta_{\mu\nu}^{\Q,\G_1,\G_2}  = (\zeta_{\nu,\mu}^{\Q,\G_2,\G_1})^{*}$, which indicates only half the range of $\mu,\nu,\G_1,\G_2$ values need be prepared.
    It is convenient to prepare the full range, but use part of the range for real and part for imaginary components.
    If we only were considering $\mu,\nu$, we could use $\mu\le \nu$ for real components and $\mu>\nu$ for imaginary components.
    To account for $\G_1,\G_2$ as well, we can combine them with $\mu,\nu$ as least-significant bits for combined integers to use in inequality tests.
    This inequality test between $\mu,\G_1$ and $\nu,\G_2$ is used to give a qubit flagging that the component should be imaginary.
    A further qubit in a $\ket{+}$ state is used to control a swap of $\mu,\G_1$ with $\nu,\G_2$ registers, and a controlled $Z$ gate on the qubit flagging the imaginary component gives the desired complex conjugate.
    As a result the range for $\mu,\nu$ is $M^2$ taking account of giving real and imaginary components.
    
    For $\Q$ there is also the symmetry where $\zeta_{\mu\nu}^{\Q,\G_1,\G_2}  = (\zeta_{\mu\nu}^{(\modmin\Q),!\G_1,!\G_2})^{*}$, so it is only necessary to produce approximately half as many values of $\Q$.
    This is complicated by the cases where $\Q=\modmin\Q$.
    If $N_x,N_y,N_z$ are odd, then the only case where this can be true is that $\Q=0$, so the number of values of $\Q$ that need be considered is $(N_xN_yN_z+1)/2$.
    If one of $N_x,N_y,N_z$ is even and the other two are odd, then for the one that is even there will be a second values of that component of $\Q$ that is equal to its negative.
    That means there are two value of $\Q$ overall satisfying $\Q=\modmin\Q$, and the number of unique values is $N_xN_yN_z/2+1$.
    Similarly, if there are two even values of $N_x,N_y,N_z$, then there are four values of $\Q$ satisfying $\Q=\modmin\Q$, and the number of unique values is $N_xN_yN_z/2+2$.
    For all three of $N_x,N_y,N_z$ even the number of unique values is $N_xN_yN_z/2+4$.
    We also need $NN_k/2$ values for the one-body term.
    The number of values is then
    \begin{equation}
        d = 32[N_xN_yN_z+2^v]M^2 + NN_k/2,
    \end{equation}
    where $v$ is the number of even values of $N_x,N_y,N_z$.

The size of the output is then
\begin{equation}
    m = 2(2\lceil \log M \rceil + n_k + 8) + \zetabits .
\end{equation}
where $\zetabits$ is the number of bits for the keep register.
There is a factor of 2 at the front to account for ind and alt values, then $\lceil \log M \rceil$ for each of $\mu$ and $\nu$, and $n_k$ for the components of $\Q$.
There is a further qubit distinguishing between the one and two-electron terms, a qubit giving the sign of the real or imaginary component of $\zeta$, and 6 qubits for $\G_1,\G_2$, for a total $+8$.

\begin{enumerate}
    \item There is a cost of $N_k N/2$ for controlled swaps for the spin. In principle this is performed four times, because it is performed before and after the two $c^\dagger c$ operators.
    The middle pair can be combined, with the single controlled swap being controlled by the parity of the two spin qubits, for a total cost of $3 N_k N/2$.
    
    \item Before the state preparation, we need to prepare an equal superposition over $d$ basis states, with costing
$3\lceil \log d\rceil-3\factor+2\rotprec-9$ Toffoli gates.
As before, $\eta$ is a number such that $2^\eta$ is a factor of $d$, and $b_r$ is a number of bits used for rotation of an ancilla qubit to improve the amplitude of success.
This cost is incurred twice, once for the preparation and once for the inverse preparation.
    
\item The complexity of the QROM being used for the state preparation is
\begin{equation}
    \left\lceil\frac{d}{k_p}\right\rceil + m (k_p-1),
\end{equation}
with $k_p$ being a power of 2.
The inverse preparation then has a cost
\begin{equation}
    \left\lceil\frac{d}{k'_p}\right\rceil + k'_p.
\end{equation}

\item We perform an inequality test with cost $\zetabits$.
Accounting for the inverse of the preparation gives a total cost $2\zetabits$.

\item The controlled swap based on the result of the inequality test is on
\begin{equation}
    2\lceil \log M \rceil + n_k + 7
\end{equation}
pairs of qubits, so has this Toffoli cost.
Note that we have $+7$ here rather than $+8$.
This is because we do not need to swap the sign qubits; the sign can be applied with $Z$ gates controlled on the result of the inequality test, not adding to the Toffoli cost (as usual).
This cost is incurred again in the inverse preparation for a total of $4\lceil \log M \rceil + 2n_k + 14$.

\item As described above, we perform an inequality test between $\mu,\G_1$ and $\nu,\G_2$ to give the qubit flagging whether we have a real or imaginary component.
Then we perform a controlled swap of $\mu,\G_1$ with $\nu,\G_2$ to generate one symmetry for the state preparation, with the complex conjugate applied using a Clifford gate.
This part therefore has Toffoli cost $2\lceil\log M\rceil+6$.
This cost is incurred again in the inverse preparation giving a total cost $4\lceil\log M\rceil+12$.
In addition to this controlled swap, we perform a controlled swap in the middle, but it is not controlled so does not add to the Toffoli complexity.

\item For the symmetry where $\zeta_{\mu\nu}^{\Q,\G_1,\G_2}  = (\zeta_{\mu\nu}^{(\modmin\Q),!\G_1,!\G_2})^{*}$, we can use a second control qubit to flip the sign on $\Q$, negate $\G_1$ and $\G_2$, and apply the complex conjugate.
The complex conjugate again can be applied with a Clifford gate, and so can the controlled-NOT gates on $\G_1,\G_2$.
A controlled sign flip of $\Q$ can be performed with $2n_k$ Toffolis, simply by flipping the sign in usual two's complement binary, then controlling addition of  $N_x,N_y,N_z$ in each component.

\item Next we need to prepare a superposition over allowed values of $\K$, because $\K-\Q-\G$ needs to be in the allowed range of $\K$ values (using $\G$ to indicate $\G_1$ or $\G_2$ depending on which part we are performing).
In particular, for the $x$-component we have an allowed range for $k_x$ from $Q_x$ to $N_x-1$ when $G_x=0$, or
$0$ to $Q_x-1$ when $G_x\neq 0$.
It is similar for the other two components.
We can therefore prepare a superposition over the appropriate range then add $Q_x$ if $G_x=0$.

Creating an equal superposition requires Hadamards on the appropriate subset of qubits, as well as a $Q_x,G_x$-dependent rotation to give a high success probability for the amplitude amplification.
This information can be output with Toffoli cost $2N_x-2$ on the qubits representing $Q_x,G_x$.
The complexity of the controlled Hadamards is then $\lceil \log N_x \rceil$ Toffolis, assuming we use a catalytic T state as in \cite{Lee2020}.
The complexity of preparing the equal superposition is then
$6\lceil \log N_x \rceil+2\rotprec-6$,
including $3\lceil \log N_x \rceil$ for three rounds of $\lceil \log N_x \rceil$ controlled Hadamards.
The reason why there is $-6$ rather than $-9$ is the inequality test is with a value in a quantum register (in each of three tests), which requires one more Toffoli than an inequality test with a classically given value.

The controlled addition of $Q_x$ has complexity $2\lceil \log N_x\rceil$.
The total complexity of the preparation of the superposition for the three components of $\K$ is therefore
\begin{equation}
    N_x+N_y+N_z + 8n_k + 6\rotprec - 24.
\end{equation}
This cost is incurred four times for the preparation and inverse preparation of $\K$ and $\Kp$.

\item In order to account for the one-body term, we note than the one-body term has a single $\mu$ and $\K$ rather than $\Q$.
We also do not want the operations we perform in the two-body part for the symmetry to affect the one-body part.
We can therefore output $\mu=\nu$ for the one-body part in the QROM, so the swap of $\mu$ and $\nu$ has no effect.
The value of $\K$ for the one-body part can be stored in the same register as used for $\Q$ for the two-body part.
To prevent the operations used to generate the symmetry $\zeta_{\mu\nu}^{\Q,\G_1,\G_2}  = (\zeta_{\mu\nu}^{(\modmin\Q),!\G_1,!\G_2})^{*}$ being applied for the one-body part, we can simply apply a Toffoli to produce a new control qubit.
The remaining part above is the preparation of the superposition over the $\K$ values controlled on $\Q$; this does not need to be amended to account for the one-body part because there we will not be using this value in the extra register.

\item Now that we have prepared the register that is in an equal superposition over the appropriate range of $\K$, we need to use that in combination with $\Q$ and $\mu$ to prepare a superposition with the correct weights.
To do this, we will use coherent alias sampling in the usual way, but will need to construct an appropriate register to iterate over from registers $\K,\Q,\G,\mu$.
First we compute $\K-\Q-\G$ in an ancilla register.
These two subtractions have cost $2n_k$.
Since it needs to be computed and uncomputed for each of the two factors in the Hamiltonian, the total cost is $8n_k$.

Now, because $\K$ and $\K\modmin\Q$ uniquely specify $\Q,\G$, these two registers can be used for the iteration instead of $\Q$, with the additional advantage that they are both over the full range of the Brillouin zone.
Now we need to compute a contiguous register
\begin{equation}
    (((((\K_xN_y + \K_y)N_z + \K_z)N_x + \Kp_x)N_y + \Kp_y)N_z + \Kp_z)M + \mu ,
\end{equation}
where we are using $\Kp$ for $\K\modmin\Q$.
This contiguous register includes many multiplications by classically chosen constants, which has complexity depending on how many ones are in these constants.
The worst case is where these numbers are all ones, so we will give the cost for that case even though it is rare.

As discussed in \cite{Sanders2020} the cost of multiplying two integers when one is given classically is no more than the product of the numbers of bits.
For the additions, the cost is no more than the number of bits on the larger number.
So, we have a cost as follows.
\begin{enumerate}
    \item For multiplying $\K_xN_y$ a cost of $\lceil\log N_x\rceil \lceil \log N_y \rceil$.
    Here $N_y$ would have more bits if it were a power of 2, but then the multiplication cost would be zero.
    \item For adding $+\K_y$ a cost of $\lceil\log N_x N_y \rceil$.
    \item For multiplying $\times N_z$ the cost is $\lceil\log N_x N_y \rceil \lceil \log N_z \rceil$.
    \item For adding $+\K_z$ the cost is $\lceil\log N_k \rceil$.
    \item For multiplying $\times N_x$ the cost is $\lceil\log N_k \rceil \lceil \log N_x \rceil$.
    \item For adding $+\Kp_x$ the cost is $\lceil\log N_x N_k \rceil$.
    \item For multiplying $\times N_y$ the cost is $\lceil\log N_x N_k \rceil \lceil \log N_y \rceil$.
    \item For adding $+\Kp_y$ the cost is $\lceil\log N_x N_y N_k \rceil$.
    \item For multiplying $\times N_z$ the cost is $\lceil\log N_x N_y N_k \rceil \lceil \log N_z \rceil$.
    \item For adding $+\Kp_z$ the cost is $\lceil\log N_k^2\rceil$.
    \item For multiplying $\times M$ the cost is $\lceil\log N_k^2 \rceil \lceil \log M\rceil$.
    \item For finally adding $+\mu$ the cost is $\lceil\log N_k^2 M\rceil$.
\end{enumerate}
We need to add all these items together to give the total cost,
and it needs to be multiplied by 4 because we compute and uncompute for each of the two factors in the Hamiltonian.

Next we have a QROM on this contiguous register with cost
\begin{equation}
    \left\lceil \frac{N_k^2M}{k_{\rm nrm}} \right\rceil + (k_{\rm nrm}-1) (n_k + \zetabits).
\end{equation}
with $k_{\rm nrm}$ a power of 2.
This is because there are $N_k^2M$ items to iterate over, and we need to output $n_k$ bits for the alternate value of $k$ and $\zetabits$ for the keep value.
We have twice this cost because of the two factors in the Hamiltonian, but the erasure cost for each factor is
\begin{equation}
    \left\lceil \frac{N_k^2M}{k_{\rm era}} \right\rceil + k_{\rm era}.
\end{equation}

The last two steps of the coherent alias sampling are an inequality test with cost $\zetabits$ and a controlled swap with cost $n_k$.
These costs are incurred 4 times, once for preparation and once for inverse preparation for each of the two factors for the Hamiltonian.

\item We will need to prepare a register that is $\K-\Q-\G$ again.
We previously computed this, but we need to compute it again because we have performed a state preparation on $\K$.
This has a cost of $2n_k$ again, and needs to be done 4 times for a total cost of $8n_k$.

A further subtlety is that we are storing the value of $\K$ to use in the $\Q$ register in the one-body case.
We can perform a controlled swap into the working register for $\K$ or $\K\modmin\Q$, which has a total cost of $4n_k$.
Combined with the arithmetic cost this is $12n_k$.

\item To use the register with $\K$ or $\K\modmin\Q$ to control the swap of system registers into working registers, we can use each qubit to control swaps of the system registers in a similar way as is used for advanced QROM.
The cost for selecting each qubit out of $N_k$ is $N_k-1$, similar to the use in advanced QROM, despite the use of multiple components.
In particular, we can perform swaps of system registers based on the $x$-component of $\K$ with cost $(N_x-1)N_y N_z$.
Then swapping the registers based on the $y$-component out of the subset of $N_y N_z$ has cost $(N_y-1)N_z$.
Then the cost of swapping based on the $z$-component has cost $N_z-1$.
Adding these three costs together gives $N_xN_yN_z-1=N_k-1$.
This is performed for each of the $N/2$ qubits we need, for cost $N(N_k-1)/2$.
We need to swap and inverse swap 8 times on $NN_k/2$ system registers, for a total cost of $4N(N_k-1)$.

\item Next, we consider the output of the rotations for the $c$ modes.
These will be controlled by the registers with $\mu$ and $\K$ (or $\K\modmin\Q$), as well as the qubit selecting between the one- and two-body terms.
A difficulty is that we would need a contiguous register in order to be able to effectively apply the advanced QROM.
A method around this is to use a QROM on the selection qubit and $\K$ to output an offset, then add $\mu$.
The complexity of that QROM is $2N_k$, then the complexity of the addition is $\lceil \log N_k(M+N/2)\rceil$.

That is in the case where we need to apply the one-body term as part of the implementation.
Recall that we only need to do that once when we are applying $c^\dagger c$ twice for the two-body term.
In the part where we are not applying the one-body term we instead have complexity $N_k+\lceil \log N_kM \rceil$.

The size of the QROM output for the rotations is then $N\rotbits$.
That is again because we need Givens rotations on $N/2$ qubits, and need two angles for each Givens rotation with $\rotbits$ each.
The complexity is then, in the case with the one-body term,
\begin{equation}
    \left\lceil \frac{N_k(M+N/2)}{k_r} \right\rceil + N\rotbits(k_r-1),
\end{equation}
and for the case where we do not need the one-body term
\begin{equation}
    \left\lceil \frac{N_kM}{k_r} \right\rceil + N\rotbits(k_r-1),
\end{equation}
with $k_r$ being a power of 2.
The cost of the Givens rotations is $2N(\rotbits-2)$, because we have $N/2$ qubits and 2 angles for each Givens rotation.

The cost of erasing the QROM is for the two cases is
\begin{align}
    &\left\lceil \frac{N_k(M+N/2)}{k'_r} \right\rceil + k'_r, \\
    &\left\lceil \frac{N_kM}{k'_r} \right\rceil + k'_r.
\end{align}
Lastly we note that we need to apply the sequence of Givens rotations 8 times to account for the four $c^\dagger$ and $c$ operators, and similarly we need to apply the QROM and invert it four times.
That gives a total complexity for the QROM-based basis rotations
\begin{align}
    &2\left\lceil \frac{N_k(M+N/2)}{k_r} \right\rceil + 2N\rotbits(k_r-1)
+2\left\lceil \frac{N_kM}{k_r} \right\rceil + 2N\rotbits(k_r-1)
+2\left\lceil \frac{N_k(M+N/2)}{k'_r} \right\rceil + 2k'_r \nn
&+2\left\lceil \frac{N_kM)}{k'_r} \right\rceil + 2k'_r
+16N(\rotbits-2) + 12N_k + 4\lceil \log N_k(M+N/2)\rceil+4\lceil \log N_kM \rceil ,
\end{align}
where $16N(\rotbits-2)$ is for the Givens rotations themselves, and the terms from $12N_k$ on are for creating and erasing contiguous registers.

\item The next part of the complexity that needs to be accounted for is the selection of $\vec Z X$ and $\vec Z Y$ for the implementation of the $c^\dagger$ and $c$ operators.
We only need to select between $\vec Z X$ and $\vec Z Y$, but not select the location the $X$ or $Y$ is performed since these are applied in the working registers.
This selection can therefore be performed entirely with Clifford gates.
However, we do need additional control to avoid performing these operators for one of the $c^\dagger c$ for the one-body component.
That is a cost of just one Toffoli for each of $c^\dagger$ and $c$.

A complication is that we need to perform $Z$ gates on remaining system registers (that have not been swapped into the working registers).
We perform unary iteration on the register containing $\K$ (or $\K\modmin\Q$), and use that to apply the appropriate $Z$ operators with a Toffoli cost $N_k-1$.
The Toffoli complexity is independent of $N$, because we only have a Toffoli cost for the unary iteration, not for the controlled-$Z$ gates.

\item The last part to the complexity when constructing the qubitised operator is the reflection on the control ancillas.
The qubits we need to reflect on are as follows.
\begin{enumerate}
    \item The $\lceil \log d\rceil$ qubits used for the equal superposition state for the initial preparation, and another qubit rotated for this preparation.
    \item The $\zetabits$ used for the equal superposition state for the inequality test in state preparation.
    \item The two qubits for the two spins in the sum.
    \item There are $n_k$ qubits that an equal superposition of $\K$ values is prepared in, and this is done twice for $2n_k$ qubits.
    To save on qubit use we can also reuse these qubits and flag that they are equal to zero in between preparing $\K$ and $\Kp$, but that does not affect the Toffoli count.
    \item There are $\zetabits$ used for the equal superposition state for two rounds of state preparation for $\K$.
    Again these can be reused but that does not affect the qubit count.
    \item There are 4 qubits to select between $X$ and $Y$ for each of the $c^\dagger,c$.
    \item There are two qubits used to generate the symmetries.
\end{enumerate}
This is a total of
\begin{equation}
    \lceil \log d\rceil + 3\zetabits + 2n_k + 9
\end{equation}
qubits for the control, and the reflection cost is 2 less than this.
We do require an additional Toffoli for the control by the phase estimation registers, and another for unary iteration on the phase estimation registers.
Therefore this expression can be used for the additional Toffoli cost.
\end{enumerate}

Next we account for the qubit costs.

\begin{enumerate}
    \item There are $N N_k$ system qubits.
\item The control register for the phase estimation, and the ancillas for the unary iteration, together need $2\lceil\log\mathcal{I}\rceil-1$ qubits.
    \item There are all the qubits needed for controls as described in the last item in the Toffoli costing above.
    Taking the inversion of the equal superposition over $\K$ to be flagged by a single qubit, the $2n_k$ can be replaced with $n_k+1$.
    Similarly flag qubits can be used on the qubits used for the equal superposition state for the inequality test to replace $3\zetabits$ with $\zetabits+2$ qubits.
    That gives a number of qubits
\begin{equation}
    \lceil \log d\rceil + \zetabits + n_k + 12.
\end{equation}
\item A phase gradient state of size $\rotbits$ is used for rotations.
\item A single T state is used for the controlled Hadamards.
    \item The QROM used for the first state preparation outputs $m$ qubits.
    \item This first state preparation also uses $m(k_p-1)+\lceil \log ( d/k_p )\rceil-1$ temporary qubits.
    \item A single qubit is used for the result of the inequality test for the coherent alias sampling.
    \item A single qubit is used for the result of the inequality test between $\mu,\G_1$ and $\nu,\G_2$.
    \item There are $n_k+3b_r$ qubits used as the output of the QROM used to give the information needed for the preparation of the equal superposition over $\K$, as well as $b_r$ for $\K$.
    \item In the preparation of the state for $\K$ we compute $\K\modmin\Q$ in an ancilla needing $n_k$ qubits, and compute a contiguous register that needs $\lceil \log(N_k^2 M)\rceil$ qubits.
    \item The state preparation for $\K$ uses $n_k +\rotbits$ output qubits.
    \item The state preparation uses
        \begin{equation}
(k_{\rm nrm}-1)(n_k +\rotbits)
        +\left\lceil \log \left(\frac{N_k^2M}{k_{\rm nrm}}  \right) \right\rceil -1
    \end{equation}
    temporary qubits.
    \item We also use $\zetabits$ in this state preparation for a superposition state, and another qubit for the result of the inequality test.
    \item There are $\lceil \log N_k(M+N/2)\rceil$ qubits used for the contiguous register for the QROM for the qubit rotations.
    \item There are $N\rotbits k_r$ used for the QROM for the rotations, with another
    \begin{equation}
        \left\lceil \log \left(\frac{N_k(M+N/2)}{k_r}  \right) \right\rceil -1
    \end{equation}
    temporary qubits.
\end{enumerate}
Accounting for the maximum total involves determining the maximum number used which depends on the use of temporary ancillas.
We first need to determine whether the temporary qubits described in part 13 or the total qubits in parts 14 to 16 are larger.
We take the maximum of these, and add it to the qubits used in parts 8 to 12.
Then we compare that to the number of temporary qubits in part 7.
The maximum of that is added to the qubits in parts 1 to 6.

\begin{figure}[tbh]
    \centering
\includegraphics[scale=0.925]{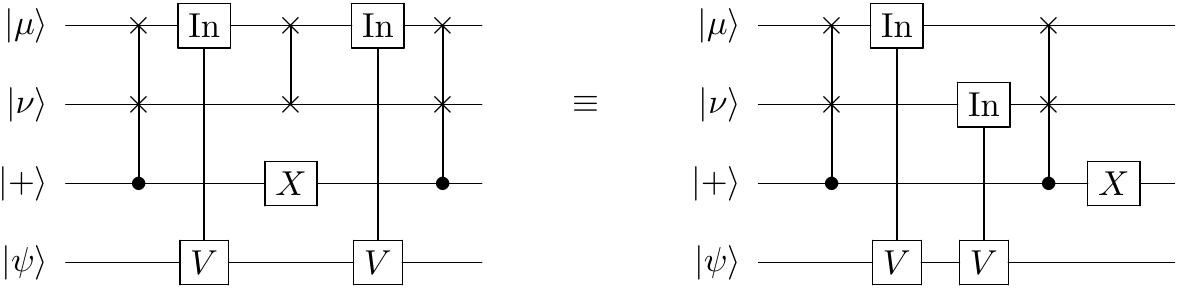}
    \caption{This shows how to construct a self-inverse procedure for block encoding two-electron terms, as in Fig. 6 of \cite{Lee2020}.
    The left side is the manifestly self-inverse form, and the right side is a more intuitive form where the $\ket{+}$ state is used to generate the symmetry between $\mu$ and $\nu$, and the two $V$ operations are controlled by $\mu$ and $\nu$ in succession.}
    \label{fig:selfinverse}
\end{figure}

Next we give a little more detail on how the construction is made self-inverse.
As explained in Fig. 6 of \cite{Lee2020} (shown above as Figure~\ref{fig:selfinverse}) the THC construction may be made self-inverse by using the qubit in the $\ket{+}$ state which controls the swap of the $\ket{\mu}$ and $\ket{\nu}$ registers.
As can be seen in Figure~\ref{fig:df_kmesh_vs_toff}, we are using a similar procedure, where the $\ket{+}$ state controls the swap of $\mu$ and $\nu$ at the top left and at the lower right.
The $X$ gate and swaps on the lower left corresponds to the $X$ gate and swap in the middle of Figure~\ref{fig:selfinverse}.

We have currently just drawn the quantum circuit as having $c$ and $c^\dagger$, but these would be implemented using ancilla qubits to control the election between $X$ and $Y$ in $X\pm iY$.
For the implementation to be self-inverse, we would want the qubit used to control the first $c$ to be the same as that for the final $c^\dagger$.
This can be achieved by taking the four qubits for control of each of the $c$ and $c^\dagger$ so the top one can be used as the control each time.
In particular, after the first $c$, swap the first two qubits, then after the $c^\dagger$ swap the first pair with the second pair, then after the next $c$ swap the first two qubits again.
As a result, the first qubit can be used as the control each time.
Moreover, this arrangement of swaps is obviously self-inverse.

The application of the controlled $X$ and $iY$ gates is also automatically self-inverse.
The reason is that $iY$ squared is $-\openone$.
In the block encoding we perform $Z$ gates for the control qubits for $c$ but not $c^\dagger$.
Then performing the unitaries for the block encoding twice, we have first that the final controlled $c^\dagger$ in the first block encoding is matched with the first $c$ for the second block encoding.
The same control qubit is used, so the two controlled $X$ operations cancel, and the two controlled $iY$ operations give $-\openone$.
This cancels with the $Z$ gate on that control qubit.
In this way all the operations can be cancelled, and because each time we have controlled $c$ and $c^\dagger$ matched, which cancel the $Z$ gate on the control qubit.
There is no additional Toffoli cost for these swaps and phase gates, because they are all Clifford gates.

\section{Correlation diagnostics for LNO}\label{app:LNOCorrelation}

\begin{table}[H]
\centering
\begin{minipage}{0.85\textwidth}
\label{tab:LNODiagnostics}
\begin{ruledtabular}
\begin{tabular}{cccccc}
structure& max($|t_1|$)& max($|t_2|$)& T1& D1& UHF $S^2$\\ \hline
C2/m& 0.2538& 0.0330& 0.0482& 0.1912& 0.7783\\
P2$_1$/c& 0.2313& 0.0322&  0.0472& 0.2178& 0.8027\\
P2/c& 0.1688& 0.0571& 0.0371& 0.2089& 1.0447\\
\end{tabular}
\end{ruledtabular}
\caption{\label{tab:LNOcorr} Some common diagnostics of strong correlation from the ROHF-CCSD calculations for each of the distorted LiNiO$_2$ structures. The UHF $S^2$ values in the final column are given per formula unit and should be compared with the exact doublet ($\langle S^2\rangle = 0.75$). The basis set (GTH-DZVP) and other details of the calculations are described in Section~\ref{sec:LNOCor}}
\end{minipage}
\end{table}
For insulators like the distorted structures of LiNiO$_2$, there are several common diagnostics that are used in molecular calculations to identify cases of ``strong correlation." Here we examine the maximum elements of the CCSD T$_1$ and T$_2$ tensors (max($|t_1|$) and max($|t_2|$)), which are commonly used as a measure of correlation\cite{DeYonker2007}, as well as the T1\cite{Lee1989} and D1\cite{Nielsen1999,Lee2003} diagnostics computed from the ROHF $k$-point CCSD calculations. We also show the expectation value of the $S^2$ operator for the UHF solution because spin contamination in the UHF calculation can be a signature of strong correlation. The results are shown in Table~\ref{tab:LNOcorr}. As expected, these results do not suggest particularly strong correlation. Only the max($|t_1|$) values for the C2/m and P2$_1$/c structures and the UHF $S^2$ for the P2/c structure are larger than might be expected. The larger max($|t_1|$) is likely an indication that the ROHF orbitals are far from optimal, and the symmetry breaking in the UHF calculations does not mean that CCSD cannot provide reliable energies.

\section{Classical timing benchmarks \label{app:ccsd_timing}}
\begin{figure}
    \centering
    \includegraphics[width=8.5cm]{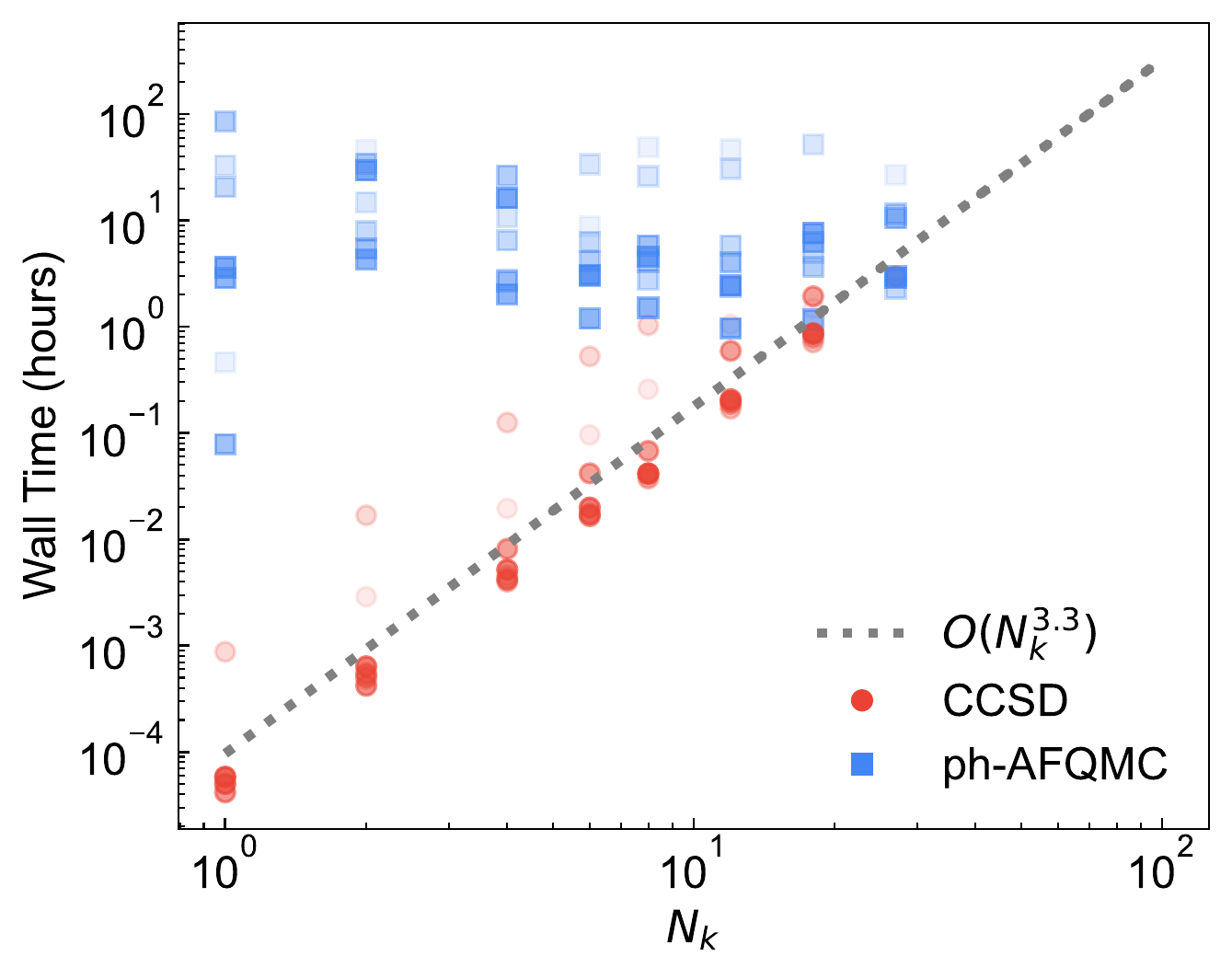}
    \caption{CCSD and ph-AFQMC timings for the benchmark systems in \cref{tab:benchmark_systems} (hydrogen ommitted) in the cc-pVDZ basis set. The CCSD data follow a $\mathcal{O}(N_k^{3.4})$ scaling which is consistent with the expected $\mathcal{O}(N_k^4)$ scaling\cite{Mcclain2017}.}
    \label{fig:ccsd_afqmc_timing}
\end{figure}

In order to compare the quantum algorithm run time to state of the art classical algorithms we measured the cost to compute the CCSD and ph-AFQMC total energy for the benchmark systems listed in \cref{tab:benchmark_systems} in double- and triple-zeta basis sets.
The results of these timings are presented in \cref{fig:ccsd_afqmc_timing}.

For CCSD we used pyscf \cite{Sun2018,Sun2020} and timed the data on a node with 30 3.1 GHz Intel Xeon CPUs (30 OpenMP threads).

For ph-AFQMC, which is considerably more expensive than CCSD for small system sizes, we estimated the run time by performing a short ph-AFQMC calculation for each system using 8 Nvidia V100 GPUs.
From this data we can then estimate the run time to achieve an statistical error bar (per atom) of $1\times10^{-4}$ Ha through the assumption that the statistical error of ph-AFQMC decays like $N_s^{-1/2}$, where $N_s$ is the number of Monte Carlo samples.
Formally, ph-AFQMC should asymptotically scale like $\mathcal{O}(N_k^3)$ \cite{PhysRevB.100.045127} assuming the number of samples required to reach the desired precision does not scale with the system size.
Interestingly, we found that the statistical error bar per atom for a fixed number of samples actually decreased with $N_k$, which implies the variance of ph-AFQMC is increasing sub-linearly with the system size.
For smaller system sizes it is important to note that practically one can saturate the GPU with walkers with nearly no loss speed \cite{malone2020accelerating}, thus reducing the error bar given a fixed wall time. As a result the small $N_k$ AFQMC numbers represent a large overestimation in runtime one would practically need to obtain the desired statistical error.
Another confounding factor which may affect the scaling of ph-AFQMC is the time step error (we fixed the time step at 0.005 Ha$^{-1}$ for all systems).
Recent results suggest that ph-AFQMC suffers from a size extensivity error \cite{lee2022twenty}, which is practically remedied through time step extrapolation.
This may break the assumption of a constant number of Monte Carlo steps required to reach the desired precision (with a fixed time step) \cite{lee2022twenty}. Due to these factors we decided not to fit a trend line to the  ph-AFQMC data.

We should stress that these comparisons are for illustrative purposes only, the absolute timings will differ with different classical implementations, codes and architectures and the run times should be considered with these factors in mind.

\section{Generating the THC factors\label{app:kthc}}
To generate the $k$-point dependent THC factors we used the interpolative separable density fitting approach (ISDF) \cite{Lu2016,Hu2017,Dong2018} as a starting point, which has recently been adapted to incorporate translational symmetry \cite{Lu2016,wu2021low}. 
Recall from \cref{eq:thc_cell_periodic} that the goal is to approximate the cell periodic orbital product evaluated on a dense real space grid using a much reduced set of grid points.
In particular we aim to solve
\begin{align}
Z = u_{p\K_p}^*(\mathbf{r}) u_{q \K_q}(\mathbf{r}) &\approx \sum_\mu  \xi_\mu(\mathbf{r})u_{p\K_p}^*(\mathbf{r}_\mu) u_{q \K_q}(\mathbf{r}_\mu) = \Theta C,
\end{align}
for $\xi_\mu(\mathbf{r})$, where $N_\mu = c_\mathrm{THC} N/2$,  $\{\mathbf{r}_\mu\}$ is a subset of the real space points $\{\mathbf{r}\}$, and $c_\mathrm{THC}$ is the THC rank parameter.
We used the K-means clustering centroidal Veronoi tesselation (CVT) algorithm to find the initial grid points \cite{Dong2018}.
The interpolating vectors $\Theta_{\mu \mathbf{r}} = \xi_\mu(\mathbf{r})$ can be found via
\begin{equation}
\Theta = ZC^{\dagger} (C C^{\dagger})^{-1}
\end{equation}
where both $ZC^\dagger$ and $C C^\dagger$ can be efficiently formed due to their product separable form \cite{Hu2017}.

Once the interpolating points ($\{\mathbf{r}_\mu\}$) and vectors  $(\xi_\mu(\mathbf{r}))$ have been found we can form any ERI block via
\begin{equation}
V_{p\K_p,q\K_q,r\K_r, s\K_s} 
= \sum_{\mu\nu} \chi_{p\K_p}^{(\mu)*} \chi_{q \K_q}^{(\mu)}
\zeta_{\mu\nu}^{\Q,\G_{pq}, \G_{sr}}
\chi_{r\K_r}^{(\nu)*} \chi_{s \K_s}^{(\nu)},
\end{equation}
where
\begin{align}
\zeta_{\mu\nu}^{\Q, \G_{pq}, \G_{sr}} &= \int d \R \int d\R' e^{-i(\Q + \G_{pq})\cdot\R} \xi_\mu(\R) V(\R,\R') \xi_\nu(\R')e^{i (\Q + \G_{sr})\cdot\R'}. \label{eq:thc_central_eval}
\end{align}
To evaluate \cref{eq:thc_central_eval} we follow the Gaussian plane wave density fitting recipe \cite{lippert1997hybrid,Vandevondele2005} and perform the following steps:

\begin{enumerate}
\item $\xi_\mu(\G) = \texttt{FFT}\left[ \xi_\mu (\mathbf{r})\right]$
\item $\tilde{V}_{\mu}^{\Q, \G_{pq}} (\mathbf{r}') = e^{-i(\Q + \G_{pq})\cdot \mathbf{r}'} \texttt{IFFT}\left[\xi_\mu(\G) V(|\G-(\Q+\G_{pq})|)\right]$ 
\item $\zeta_{\mu\nu}^{\Q,\G_{pq}\G_{sr}} = \sum_{\mathbf{r}'}  \Delta \mathbf{r}' \xi_\nu(\mathbf{r}')
e^{i(\G_{sr}-\G_{pq})\cdot\mathbf{r}'}
\tilde{V}_\mu^{\Q\G_{pq}}(\mathbf{r}')$ 
\end{enumerate}
where $\Delta \mathbf{r}' = \frac{\Omega}{N_g}$, $\Omega$ is the unit cell volume, $N_g$ is the number of real space grid points and $V(\G)$ is the Coulomb kernel in reciprocal space.

An important consideration for the quantum implementation of THC is the minimization of the L1-norm of the central tensor $\zeta^\mathbf{Q}_{\mu\nu}$ as this value directly affects the scaling of the algorithm.
Following previous work for molecular systems \cite{Lee2020,Goints_pnas.2203533119}, we attempt to reduce $\lambda$ by further compressing the THC factors through a regularized optimization scheme.
This is an important step as $\lambda$ can grow considerably with the THC rank.
We use the ISDF solution for $\chi_{p\K_p}(\R_\mu)$, and $\zeta_{\mu\nu}^{\Q\G\G'}$ as an initial guess and perform a subsequent optimization of the following cost function:
\begin{align}
L(\chi, \zeta) =& \sum_{\Q}\sum_{\K\K'}\sum_{pqrs}\left( \left|(\K p \K \modmin \Q q| r \Kp\modmin\Q s \K') 
- \sum_{\mu\nu} \chi_{p\K}^{(\mu)*} \chi_{q \K\modmin\Q}^{(\mu)}
\zeta_{\mu\nu}^{\Q,\G, \G'}
\chi_{r(\Kp\modmin\Q)}^{(\nu)*} \chi_{s \K'}^{(\nu)}\right|^2
+ p \left|\sum_{\mu\nu}\mathcal{N}^\mu_{p\K q(\K\modmin\Q)} \zeta_{\mu\nu}^{\Q\G\G'}\mathcal{N}^\nu_{r(\Kp\modmin\Q) s\K'} \right| \right),
\end{align}
where the penalty parameter $p$ was set such that the $L_1$ part of the objective function was of similar magnitude to the $L_2$ part.
To optimize the cost function we used the L-BFGS-B implemented in Scipy with gradients evaluated using JAX.
Like Ref.~\cite{Lee2020} we found it necessary to perform an additional optimization of the $L_2$ part of the loss function after the BFGS optimization in order to obtain good MP2 correlation energies.
As these subsequent optimizations are quite costly we limited the number of optimization steps to 3000 for both the L-BFGS-B and AdaGrad stages.
With these additional steps we found a rank paramter of $c_\mathrm{THC} = 6-8$ was sufficient to obtain an MP2 error of roughly 0.1 mHa error per cell across a range of systems, which is consistent with that found in molecular systems \cite{Goints_pnas.2203533119}.

\end{document}